\documentclass[10pt]{article}

\usepackage{authblk}
\usepackage{natbib}
\usepackage{mathtools}
\usepackage{amsmath}
\usepackage{amsthm}
\usepackage{amssymb}
\usepackage{amsbsy}
\usepackage{amsfonts}
\usepackage{amscd}
\usepackage{mathrsfs}
\usepackage{bm}
\usepackage{breqn}
\usepackage{color, soul}
\usepackage{enumerate}
\usepackage{empheq}
\usepackage{rotating}
\usepackage{ftnxtra}
\usepackage{fnpos}
\usepackage[titletoc,toc,title]{appendix}
\usepackage{euscript}
\usepackage{graphicx}
\usepackage{epsfig}
\usepackage{epstopdf}
\DeclareGraphicsExtensions{.pdf,.png,.jpg,.eps}
\usepackage{pstool}
\usepackage{upgreek}
\usepackage{mathrsfs}

\usepackage{xcolor}

\usepackage{psfrag}

\numberwithin{equation}{section}

\theoremstyle{plain}	
\newtheorem{thm}{Theorem}[section]

\newtheorem{prop}[thm]{Proposition}
\newtheorem*{prop*}{Proposition} 
\theoremstyle{definition}

\usepackage{graphicx}
\usepackage{lipsum}

\usepackage{caption}
\usepackage{subcaption}

\usepackage{hyperref}
\hypersetup{colorlinks=true, linkcolor=blue}
\hypersetup{colorlinks=true,citecolor=blue}

\DeclareMathAlphabet{\mathpzc}{OT1}{pzc}{m}{it}

\usepackage{amsmath, amsthm, amssymb}

\usepackage{cleveref}

\DeclarePairedDelimiter\abs{\lvert}{\rvert}

\makeatletter
\newsavebox{\@brx}
\newcommand{\llangle}[1][]{\savebox{\@brx}{\(\m@th{#1\langle}\)}%
  \mathopen{\copy\@brx\mkern2mu\kern-0.9\wd\@brx\usebox{\@brx}}}
\newcommand{\rrangle}[1][]{\savebox{\@brx}{\(\m@th{#1\rangle}\)}%
  \mathclose{\copy\@brx\mkern2mu\kern-0.9\wd\@brx\usebox{\@brx}}}%
\let\oldabs\abs
\def\abs{\@ifstar{\oldabs}{\oldabs*}}
\makeatother

\usepackage{accents}

    {\end{bmatrix}}%

\usepackage{enumitem}

\usepackage[utf8]{inputenc}
\usepackage{dutchcal}
\usepackage{multirow}

\renewcommand{\arraystretch}{2.0}

\usepackage{longtable}
\usepackage{booktabs}

\begin{document}


\title{\textbf{Universal Displacements in Linear \\Strain-Gradient Elasticity}}

\author[1]{Dimitris Sfyris}
\author[2,3]{Arash Yavari\thanks{Corresponding author, e-mail: arash.yavari@ce.gatech.edu}}
\affil[1]{\small \textit{Institute of Applied and Computational Mathematics (IACM), Foundation for Research and
Technology (FORTH), Heraklion, Greece}}
\affil[2]{\small \textit{School of Civil and Environmental Engineering, Georgia Institute of Technology, Atlanta, GA 30332, USA}}
\affil[3]{\small \textit{The George W. Woodruff School of Mechanical Engineering, Georgia Institute of Technology, Atlanta, GA 30332, USA}}

\maketitle

\begin{abstract}
We study universal displacement fields in three-dimensional linear strain-gradient elasticity. Following the approach of \citet{Yavari2020}, we extend the method to the Toupin--Mindlin first strain-gradient elasticity. For each material symmetry class, by requiring the equilibrium equations to hold for any material in that class in the absence of body forces, we derive the corresponding universality constraints on the displacement field and obtain the complete set of universal displacements. Using the full symmetry classification and compact matrix representations of the elasticity tensors \citep{Auffrayetal2013,Auffrayetal2019}, we give explicit characterizations for all $48$ strain-gradient symmetry classes, including centrosymmetric and chiral classes. For several high-symmetry classes, the strain-gradient universality constraints do not impose restrictions beyond the classical ones, so the universal displacement families coincide with those of classical linear elasticity (e.g., the isotropic classes $\mathbb{SO}(3)$ and $\mathbb{O}(3)$). For lower symmetry classes, the strain-gradient universality constraints can be stricter than their classical counterpart: the universal displacement fields form a proper subset of the classical universal displacements, obtained by imposing additional higher-order differential constraints that eliminate some of the classical families.
\end{abstract}

\begin{description}
\item[Keywords:] Universal deformations, universal displacements, linear elasticity, strain-gradient elasticity, anisotropic elasticity.
\end{description}

\tableofcontents

\section{Introduction} \label{Sec:Introduction}

A universal motion, whether a deformation in nonlinear elasticity or a displacement in linear elasticity, is one that can be maintained in the absence of body forces for all materials within a prescribed class. Equivalently, a universal motion can be realized by applying only boundary tractions, independently of the specific constitutive choice within that class, for example homogeneous compressible isotropic solids or homogeneous incompressible isotropic solids. In nonlinear elasticity, universal deformations have served as a source of exact solutions that are valuable both for experiments and for theory, see, for example, \citet{Rivlin1951,Tadmor2012,Goriely2017}.

The systematic study of universal deformations was initiated in the seminal works of Jerry Ericksen \citep{Ericksen1954,Ericksen1955}, motivated in part by the earlier exact solutions of Ronald Rivlin \citep{Rivlin1948,Rivlin1949a,Rivlin1949b}. For homogeneous compressible isotropic hyperelastic solids, \citet{Ericksen1955} showed that universal deformations must be homogeneous. The incompressible isotropic case is substantially more subtle, as internal constraints introduce additional structure and the resulting classification problem is significantly more complex \citep{Saccomandi2001}. In his celebrated analysis, \citet{Ericksen1954} identified four families of universal deformations for incompressible isotropic solids beyond isochoric homogeneous deformations. A fifth family was later discovered independently by \citet{SinghPipkin1965} and \citet{KlingbeilShield1966}, and it provided a counterexample to Ericksen's conjecture that constant-principal-invariant universal deformations must be homogeneous \citep{Fosdick1966}. The question of whether further inhomogeneous universal deformations with constant principal invariants exist remains open.

Ericksen's universality framework has been extended in multiple directions, including inhomogeneous isotropic elasticity, anisotropic elasticity, isotropic and anisotropic Cauchy elasticity, Cauchy elasticity with inextensible fibers, anelasticity, and implicit elasticity \citep{Yavari2021,YavariGoriely2021,Yavari2022Universal,YavariGoriely2016,Goodbrake2020,Yavari2024UniversalCauchy,MotaghianYavari2026Universal,Yavari2025UniversalDeformations,YavariGoriely2024Implicit,YavariMerodioShariff2025Universal}. In linear elasticity, the analogue of universal deformations is the concept of universal displacements \citep{Truesdell1966,Gurtin1972,Yavari2020}. A systematic classification of universal displacements in compressible anisotropic linear elasticity for all eight symmetry classes was carried out in \citep{Yavari2020}. One conclusion of that analysis is that the admissible space of universal displacements depends explicitly on the symmetry class, with larger symmetry groups admitting larger spaces of universal displacements. Related extensions include inhomogeneous linear elasticity, linear anelasticity, linear Cauchy elasticity, and linear elasticity with inextensible fibers \citep{YavariGoriely2022,Yavari2022Anelastic-Universality,Yavari2024UniversalDisplacements,YavariSfyris2025UniversalDisplacements}.

The objective of the present work is to extend the universality program to linear strain-gradient elasticity. We focus on the Toupin--Mindlin first strain-gradient theory and seek to identify all universal displacement fields in three dimensions for each material symmetry class. Our analysis follows that of \citet{Yavari2020}: within each symmetry class we require the strain-gradient equilibrium equations, in the absence of body forces, to hold for arbitrary choices of the independent elastic constants. 
This requirement imposes differential universality constraints on the displacement field.
By carrying out this procedure across the full symmetry classification, including centrosymmetric and chiral classes, and by using compact matrix representations of the elasticity tensors \citep{Auffrayetal2013,Auffrayetal2019}, we obtain explicit class-by-class characterizations and, in particular, determine when the strain-gradient universality constraints introduce restrictions beyond those of classical linear elasticity.

While the presence of strain-gradient terms suggests additional constraints, it is not evident a priori how these constraints affect the structure of universal displacements. The associated universality PDEs are higher-order and strongly coupled, and there is no guarantee that classical universal families persist or that a complete classification is possible. The present work shows that a complete classification can be carried out across all symmetry classes and determines when the strain-gradient universality constraints introduce additional restrictions on the admissible displacement families.

This paper is organized as follows. In \S\ref{Sec:Classical-linear-elasticity} we outline the required results in classical linear elasticity following the approach of~\citet{Yavari2020}, and we summarize the corresponding universality constraints and universal displacements for each symmetry class. These results serve as the point of departure for the subsequent strain-gradient analysis. In \S\ref{Sec:Strain-gradient-linear-elasticity} we review the strain-gradient equilibrium equations and derive, for each symmetry class, the corresponding universality constraints, leading to a complete classification of universal displacement fields within the Toupin--Mindlin first strain-gradient theory. Conclusions are given in \S\ref{Sec:Conclusions}.

\section{Classical linear elasticity} \label{Sec:Classical-linear-elasticity}

In this section we review the main findings of \citet{Yavari2020} on universal displacements in classical linear elasticity. A universal displacement field is one that satisfies the equilibrium equations, in the absence of body forces, for every material within a prescribed class. 
This requirement yields a system of partial differential equations (PDEs) for the displacement field (the universality constraints), which we summarize below for each of the eight symmetry classes.

\subsection{Triclinic class}

For the triclinic class, there are $21$ independent elastic constants. The corresponding universality constraints are
\begin{equation}\label{triclinic-linear-constraints1}
\begin{dcases}
	\frac{\partial^2 u_1}{\partial x_1^2} = \frac{\partial^2 u_1}{\partial x_1 \partial x_2} 
	= \frac{\partial^2 u_1}{\partial x_1 \partial x_3} = 0\,,\\
	\frac{\partial^2 u_2}{\partial x_2^2} = \frac{\partial^2 u_2}{\partial x_1 \partial x_2} 
	= \frac{\partial^2 u_2}{\partial x_2 \partial x_3} = 0\,,\\
	\frac{\partial^2 u_3}{\partial x_3^2} = \frac{\partial^2 u_3}{\partial x_1 \partial x_3} 
	= \frac{\partial^2 u_3}{\partial x_2 \partial x_3} = 0\,,
\end{dcases}
\end{equation}
together with
\begin{equation} \label{triclinic-linear-constraints2}
\begin{dcases} 
	\frac{\partial^2 u_{\alpha}}{\partial x_{\beta}^2} 
	= 0\,,\quad \alpha \neq \beta\,,\quad \alpha, \beta \in \{1,2,3\}\,,   \\
	\frac{\partial^2 u_1}{\partial x_2 \partial x_3} 
	= \frac{\partial^2 u_2}{\partial x_1 \partial x_3} 
	= \frac{\partial^2 u_3}{\partial x_1 \partial x_2} = 0\,. 
\end{dcases}
\end{equation}
Solving \eqref{triclinic-linear-constraints1} and \eqref{triclinic-linear-constraints2} one concludes that the only universal displacements in the triclinic class are homogeneous displacement fields.

\subsection{Monoclinic class} \label{Sec:Monoclinic-class-classical}

For the monoclinic class, there are $13$ independent elastic constants. A subset of the universality constraints coincides with \eqref{triclinic-linear-constraints1}, and its general solution can be written as
\begin{equation} \label{monoclinic-linear-constraints}
\begin{aligned} 
	& u_1 (x_1, x_2, x_3)=cx_1+\hat{u}_1 (x_2, x_3)  \,, \\
	& u_2 (x_1, x_2, x_3)=cx_2+\hat{u}_2 (x_1, x_3) \,, \\
	& u_3 (x_1, x_2, x_3)=cx_3+\hat{u}_3(x_1, x_2) \,. 
\end{aligned}
\end{equation}
The remaining universality constraints are
\begin{equation} \label{monoclinic-linear-constraints-1}
\begin{aligned} 
	&\frac{\partial^2 \hat{u}_1}{\partial x_2^2}=\frac{\partial^2 \hat{u}_1}{\partial x_3^2}=0\,,\\
	&\frac{\partial^2 \hat{u}_2}{\partial x_1^2}=\frac{\partial^2 \hat{u}_2}{\partial x_3^2}=0\,,\\
	&\frac{\partial^2 \hat{u}_3}{\partial x_1^2}=\frac{\partial^2 \hat{u}_3}{\partial x_2^2}=0\,, 
\end{aligned}
\end{equation}
together with
\begin{equation} \label{monoclinic-linear-constraints-2}
	\frac{\partial^2 \hat{u}_1}{\partial x_2 \partial x_3}
	+\frac{\partial^2 \hat{u}_2}{\partial x_1 \partial x_3}=0\,, \qquad 
	\frac{\partial^2 \hat{u}_3}{\partial x_1 \partial x_2}=0\,.
\end{equation}
These constraints imply that the universal displacements in the monoclinic class are a superposition of a homogeneous displacement field and the one-parameter inhomogeneous displacement field $(c x_2 x_3\,,-c x_1 x_3\,,0)$.

\subsection{Orthotropic class}

In this case there are $9$ independent elastic constants and the universality constraints read
\begin{equation}
\begin{dcases}
\label{orthotropic-linear-constraints}
	\frac{\partial^2 u_1}{\partial x_1^2} = \frac{\partial^2 u_1}{\partial x_1 \partial x_2} 
	= \frac{\partial^2 u_1}{\partial x_1 \partial x_3} = 0\,, \\
	\frac{\partial^2 u_2}{\partial x_2^2} = \frac{\partial^2 u_2}{\partial x_2 \partial x_1} 
	= \frac{\partial^2 u_2}{\partial x_2 \partial x_3} = 0\,, \\
	\frac{\partial^2 u_3}{\partial x_3^2} = \frac{\partial^2 u_3}{\partial x_3 \partial x_1} 
	= \frac{\partial^2 u_3}{\partial x_2 \partial x_3} = 0\,, \\
	\frac{\partial^2 u_1}{\partial x_2^2} = \frac{\partial^2 u_1}{\partial x_3^2} = 0\,, \\
	\frac{\partial^2 u_2}{\partial x_1^2} = \frac{\partial^2 u_2}{\partial x_3^2} = 0\,, \\
	\frac{\partial^2 u_3}{\partial x_1^2} = \frac{\partial^2 u_3}{\partial x_2^2} = 0\,. 
\end{dcases}
\end{equation}
Solving these PDEs, one concludes that universal displacements are a superposition of a homogeneous displacement field and the following three-parameter inhomogeneous displacement field: $(\alpha_1 x_2 x_3, \alpha_2 x_1 x_3, \alpha_3 x_1 x_2)$.

\subsection{Tetragonal class}

In this case there are $6$ independent elastic constants and the universality constraints read
\begin{equation}\label{tetragonal-linear-constraints}
\begin{dcases}
	\frac{\partial^2 u_1}{\partial x_1^2} = \frac{\partial^2 u_1}{\partial x_1 \partial x_2} 
	= \frac{\partial^2 u_1}{\partial x_2^2} = \frac{\partial^2 u_1}{\partial x_3^2} = 0\,,\\
	\frac{\partial^2 u_2}{\partial x_1 \partial x_2} = \frac{\partial^2 u_2}{\partial x_2^2} 
	= \frac{\partial^2 u_2}{\partial x_1^2} = \frac{\partial^2 u_2}{\partial x_3^2} = 0\,,\\
	\frac{\partial^2 u_1}{\partial x_1 \partial x_3} + \frac{\partial^2 u_2}{\partial x_2 \partial x_3} = 0\,,\\
	\frac{\partial^2 u_3}{\partial x_1 \partial x_3} = \frac{\partial^2 u_3}{\partial x_2 \partial x_3} 
	= \frac{\partial^2 u_3}{\partial x_3^2} = 0\,,\\
	\frac{\partial^2 u_3}{\partial x_1^2} + \frac{\partial^2 u_3}{\partial x_2^2} = 0\,.
\end{dcases}
\end{equation}
Solving these PDEs, one concludes that universal displacements are a superposition of homogeneous displacement fields and the following inhomogeneous displacement field:
\begin{equation}\label{tetragonal-linear-inhomogeneous}
\begin{aligned}
	&u_1^{\text{inh}}(x_1, x_2, x_3) = c_1 x_2 x_3 + c_2 x_1 x_3\,,\\
	&u_2^{\text{inh}}(x_1, x_2, x_3) = -c_2 x_2 x_3 + c_3 x_1 x_3\,,\\
	&u_3^{\text{inh}}(x_1, x_2, x_3) = g(x_1, x_2)\,,
\end{aligned}
\end{equation}
where $c_1, c_2, c_3$ are constants and $g(x_1, x_2)$ is an arbitrary harmonic function.

\subsection{Trigonal class}

For the trigonal class, there are $6$ independent elastic constants and the universality constraints are 
\begin{equation}\label{trigonal-linear-constraints}
\begin{dcases}
	\frac{\partial^2 u_3}{\partial x_1 \partial x_3} = \frac{\partial^2 u_3}{\partial x_2 \partial x_3} 
	= \frac{\partial^2 u_3}{\partial x_3^2}= 0\,,\\
	\frac{\partial^2 u_1}{\partial x_1^2} + \frac{\partial^2 u_1}{\partial x_2^2} 
	= \frac{\partial^2 u_2}{\partial x_1^2} + \frac{\partial^2 u_2}{\partial x_2^2} 
	= \frac{\partial^2 u_3}{\partial x_1^2} + \frac{\partial^2 u_3}{\partial x_2^2} = 0\,,\\
	\frac{\partial^2 u_1}{\partial x_3^2} = \frac{\partial^2 u_2}{\partial x_3^2} = 0\,,\\
	 2\frac{\partial^2 u_1}{\partial x_1 \partial x_3} -2 \frac{\partial^2 u_2}{\partial x_2 \partial x_3} 
	+ \frac{\partial^2 u_3}{\partial x_1^2} - \frac{\partial^2 u_3}{\partial x_2^2} = 0\,,\\
	\frac{\partial^2 u_1}{\partial x_1 \partial x_3} + \frac{\partial^2 u_2}{\partial x_1 \partial x_3} 
	+ \frac{\partial^2 u_3}{\partial x_1 \partial x_2} = 0\,,\\
	\frac{\partial^2 u_1}{\partial x_1 \partial x_3} + \frac{\partial^2 u_2}{\partial x_2 \partial x_3} = 0\,,\\
	\frac{\partial^2 u_1}{\partial x_1^2}=3\frac{\partial^2 u_1}{\partial x_2^2} \,.
\end{dcases}
\end{equation}
Solving these PDEs, one concludes that universal displacements are a superposition of homogeneous displacement fields and the following inhomogeneous displacement fields:
\begin{equation}\label{trigonal-linear-inhomogeneous}
\begin{aligned}
	& u_1^{\text{inh}}(x_1, x_2, x_3) = a_{123} x_1 x_2 x_3 + a_{12} x_1 x_2  +a_{13} x_1 x_3 +a_{23} x_2 x_3   \,,\\
	& u_2^{\text{inh}}(x_1, x_2, x_3) = \frac{1}{2} (a_{12}+a_{123} x_3)(x_1^2-x_2^2) + b_{13} x_1 x_3 -a_{13} x_2 x_3            
	\,,\\
	& u_3^{\text{inh}}(x_1, x_2, x_3) = -a_{123} x_1^2 x_2 -(a_{23}+b_{13}) x_1 x_2 
	+\frac{1}{3} a_{123} x_2^3-a_{13} (x_1^2-x_2^2)         \,.
\end{aligned}
\end{equation}

\subsection{Transversely isotropic class}

In this case there are $5$ independent elastic constants and the universality constraints read
\begin{equation}\label{transverse-isotropy-linear-constraints}
\begin{dcases}
	\frac{\partial^2 u_3}{\partial x_1 \partial x_3} = \frac{\partial^2 u_3}{\partial x_2 \partial x_3} 
	= \frac{\partial^2 u_3}{\partial x_3^2}= 0\,,\\
	\frac{\partial^2 u_3}{\partial x_1^2} + \frac{\partial^2 u_3}{\partial x_2^2} 
	= \frac{\partial^2 u_1}{\partial x_3^2} + \frac{\partial^2 u_2}{\partial x_3^2} = 0\,,\\
	\frac{\partial^2 u_1}{\partial x_1^2} + \frac{\partial^2 u_1}{\partial x_2^2} 
	= \frac{\partial^2 u_1}{\partial x_2^2} + \frac{\partial^2 u_2}{\partial x_2^2} = 0\,,\\
	\frac{\partial^2 u_1}{\partial x_2^2}= \frac{\partial^2 u_2}{\partial x_1 \partial x_2} \,,\\
	\frac{\partial^2 u_2}{\partial x_1^2}= \frac{\partial^2 u_1}{\partial x_1 \partial x_2}  \,,\\
	\frac{\partial^2 u_1}{\partial x_1 \partial x_3} + \frac{\partial^2 u_2}{\partial x_2 \partial x_3} = 0\,.
\end{dcases}
\end{equation}
Thus, one concludes that the universal displacements are written as
\begin{equation}\label{transverse-isotropy-linear-solution}
\begin{aligned}
	& u_1(x_1, x_2, x_3) = c_1 x_1 + c_2 x_2 + c x_2 x_3 
	+ x_3\, h_1(x_1, x_2) + k_1(x_1, x_2)\,,\\
	& u_2(x_1, x_2, x_3) = -c_2 x_1 + c_1 x_2 - c x_1 x_3 
	+ x_3\, h_2(x_1, x_2) + k_2(x_1, x_2)\,,\\
	& u_3(x_1, x_2, x_3) = c_3 x_3 + \hat{u}_{33}(x_1, x_2)\,,
\end{aligned}
\end{equation}
where
\begin{equation}\label{transverse-isotropy-linear-solution-2}
\begin{aligned}
	& \xi(x_2 + i x_1) = h_1(x_1, x_2) + i\, h_2(x_1, x_2)\,,\\
	& \eta(x_2 + i x_1) = k_1(x_1, x_2) + i\, k_2(x_1, x_2)\,,
\end{aligned}
\end{equation}
are holomorphic functions, and $\hat{u}_{33}$ is an arbitrary harmonic function.

\subsection{Cubic class}

For the cubic class there are $3$ independent elastic constants. The universality PDEs are
\begin{equation}\label{cubic-linear-constraints}
\begin{dcases}
	\frac{\partial^2 u_1}{\partial x_1^2}=\frac{\partial^2 u_2}{\partial x_2^2}=\frac{\partial^2 u_3}{\partial x_3^2}=0\,,\\
	\frac{\partial^2 u_1}{\partial x_2^2}+\frac{\partial^2 u_1}{\partial x_3^2}=0\,,\\
	\frac{\partial^2 u_2}{\partial x_1^2}+\frac{\partial^2 u_2}{\partial x_3^2}=0\,,\\
	\frac{\partial^2 u_3}{\partial x_1^2}+\frac{\partial^2 u_3}{\partial x_2^2}=0\,,\\
	\frac{\partial^2 u_1}{\partial x_1 \partial x_3}+\frac{\partial^2 u_1}{\partial x_2 \partial x_3}=0\,,\\
	\frac{\partial^2 u_2}{\partial x_1 \partial x_2}+\frac{\partial^2 u_3}{\partial x_1 \partial x_3}=0\,,\\
	\frac{\partial^2 u_1}{\partial x_1 \partial x_2}+\frac{\partial^2 u_3}{\partial x_2 \partial x_3}=0\,.
\end{dcases}
\end{equation}
The universal displacements are of the following form:
\begin{equation}\label{cubic-linear-solution}
\begin{aligned}
	& u_1(x_1,x_2,x_3)=\tfrac{a}{2} x_1 (x_3^2-x_2^2)+c_1 x_1 x_3 + b_1 x_1 x_2 + d_1 x_1 + g_1(x_2,x_3)\,,\\
	& u_2(x_1,x_2,x_3)=\tfrac{a}{2} x_2 (x_1^2-x_3^2)+a_1 x_1 x_2 - c_1 x_2 x_3 + d_2 x_2 + g_2(x_1,x_3)\,,\\
	& u_3(x_1,x_2,x_3)=\tfrac{a}{2} x_3 (x_2^2-x_1^2) - a_1 x_1 x_3 - b_1 x_2 x_3 + d_3 x_3 + g_3(x_1,x_2)\,,
\end{aligned}
\end{equation}
where $g_i$, $i=1,2,3$, are arbitrary harmonic functions.

\subsection{Isotropy class}

For isotropic solids the universality constraints read
\begin{equation}\label{isotropy-linear-constraints}
\begin{dcases}
	\tfrac{\partial^2 u_1}{\partial x_3^2}+\tfrac{\partial^2 u_1}{\partial x_2^2}
	+\tfrac{\partial^2 u_3}{\partial x_1 \partial x_3}+\tfrac{\partial^2 u_2}{\partial x_1 \partial x_2}
	+2\tfrac{\partial^2 u_1}{\partial x_1^2}=0\,,\\
	-\tfrac{\partial^2 u_1}{\partial x_3^2}-\tfrac{\partial^2 u_1}{\partial x_2^2}
	+\tfrac{\partial^2 u_3}{\partial x_1 \partial x_3}+\tfrac{\partial^2 u_2}{\partial x_1 \partial x_2}=0\,,\\
	\tfrac{\partial^2 u_2}{\partial x_3^2}+\tfrac{\partial^2 u_3}{\partial x_2 \partial x_3}
	+2\tfrac{\partial^2 u_2}{\partial x_2^2}+\tfrac{\partial^2 u_1}{\partial x_1 \partial x_2}
	+\tfrac{\partial^2 u_2}{\partial x_1^2}=0\,,\\
	-\tfrac{\partial^2 u_2}{\partial x_3^2}+\tfrac{\partial^2 u_3}{\partial x_2 \partial x_3}
	+\tfrac{\partial^2 u_1}{\partial x_1 \partial x_2}-\tfrac{\partial^2 u_2}{\partial x_1^2}=0\,,\\
	2\tfrac{\partial^2 u_3}{\partial x_3^2}+\tfrac{\partial^2 u_2}{\partial x_2 \partial x_3}
	+\tfrac{\partial^2 u_3}{\partial x_2^2}+\tfrac{\partial^2 u_1}{\partial x_1 \partial x_3}
	+\tfrac{\partial^2 u_3}{\partial x_1^2}=0\,,\\
	\tfrac{\partial^2 u_2}{\partial x_2 \partial x_3}-\tfrac{\partial^2 u_3}{\partial x_2^2}
	+\tfrac{\partial^2 u_1}{\partial x_1 \partial x_3}-\tfrac{\partial^2 u_3}{\partial x_1^2}=0\,.
\end{dcases}
\end{equation}
By subtracting the first two equations we obtain
\begin{equation}\label{isotropy-1}
	\frac{\partial^2 u_1}{\partial x_3^2}
	+\frac{\partial^2 u_1}{\partial x_2^2}+\frac{\partial^2 u_1}{\partial x_1^2}=0\,, 	
	\qquad \text{or} \qquad \Delta u_1=0\,.
\end{equation}
Subtracting the third and fourth equations we obtain $\Delta u_2=0$, while subtracting the fifth and sixth equations one finds $\Delta u_3=0$. By adding the first two equations we find
\begin{equation}\label{isotropy-4}
	\frac{\partial^2 u_3}{\partial x_1 \partial x_3}+\frac{\partial^2 u_2}{\partial x_1 \partial x_2}
	+\frac{\partial^2 u_1}{\partial x_1^2}=0\,.
\end{equation}
Adding the third and fourth equations we obtain
\begin{equation} \label{isotropy-5}
	\frac{\partial^2 u_3}{\partial x_2 \partial x_3}+\frac{\partial^2 u_1}{\partial x_1 \partial x_2}
	+\frac{\partial^2 u_2}{\partial x_2^2}=0\,.
\end{equation}
Adding the fifth and sixth equations we find
\begin{equation}\label{isotropy-6}
	\frac{\partial^2 u_3}{\partial x_3^2}+\frac{\partial^2 u_2}{\partial x_2 \partial x_3}
	+\frac{\partial^2 u_1}{\partial x_1 \partial x_3}=0\,.
\end{equation}
Each of the above equations \eqref{isotropy-4}--\eqref{isotropy-6} is one of the components of the vector $\operatorname{grad}\circ\operatorname{div}{\bf u}$ (with components $u_{i,ij}$); the first corresponds to $j=1$, the second to $j=2$, and the third to $j=3$. This provides an alternative path to the argument of \citet{Yavari2020} for establishing that
\begin{equation}\label{isotropy-7}
	\Delta {\bf u}=\mathbf{0}\,,\qquad \operatorname{grad}\circ\operatorname{div}{\bf u}=\mathbf{0}\,,
\end{equation}
for the classical linear isotropic elastic case.

The validity of these equations implies (in a similar vein with \citet{Yavari2020}) that the universal displacement fields can be expressed as a superposition of homogeneous displacement fields and a non-homogeneous one, which is the divergence of an antisymmetric matrix with components that are solutions of a Poisson's equation. In Cartesian coordinates one has
\begin{equation}\label{displacementisotropy}
\begin{aligned}
& u_1=\frac{\partial \alpha}{\partial x_2}+ \frac{\partial \beta}{\partial x_3} +\frac{c}{3}x_1+k_1\,,\\
& u_2=-\frac{\partial \alpha}{\partial x_1}+\frac{\partial \gamma}{\partial x_3}+\frac{c}{3}x_2+k_2\,,\\
& u_3=-\frac{\partial \beta}{\partial x_1}-\frac{\partial \gamma}{\partial x_2}+\frac{c}{3}x_3+k_3\,,
\end{aligned}
\end{equation}
where $\alpha(x_1, x_2, x_3)$, $\beta(x_1, x_2, x_3)$, and $\gamma(x_1, x_2, x_3)$ are arbitrary functions.

\section{Linear strain gradient elasticity}  \label{Sec:Strain-gradient-linear-elasticity}

Linear strain-gradient elasticity extends classical linear elasticity by allowing the strain energy density to depend not only on the infinitesimal strain, but also on its spatial gradient, thereby introducing intrinsic material length scales and higher-order stresses. The nonlinear strain-gradient theory was formulated by \citet{Toupin1962}. Linear versions of this theory, in three equivalent forms, are given in \citep{Mindlin1964}, and their equivalence is proved in \citep{MindlinEshel1968}. Closely related linear couple-stress formulations, emphasizing curvature measures and additional stress quantities, are given in \citep{MindlinTiersten1962}. Formulations in which the strain energy depends on the gradient of the rotation are also given in \citep{AeroKuvshinskii1961,Toupin1964}. More recent work includes analytical solutions in isotropic strain-gradient elasticity \citep{Enakoutsa2015,Iesan2013,IesanQuintanilla2016}, Green tensors and regularized fundamental solutions for isotropic first strain-gradient elasticity \citep{LazarPo2018}, and comparisons with experiments for the isotropic linear strain-gradient case \citep{Lametal2003}. Fundamental questions related to positive definiteness and strong ellipticity are studied in \citep{LazarAgiasofitouBoehlke2022,EremeyevLazar2022}. In particular, \citet{LazarAgiasofitouBoehlke2022} specialize the Toupin--Mindlin first strain-gradient theory to cubic and isotropic materials, derive the corresponding gradient-elastic constants and characteristic lengths, and compute them for selected materials. A brief summary of these results is also given in \citep{LazarAgiasofitou2023}. A broad overview of gradient elasticity formulations, including length-scale identification and numerical implementations, is provided in \citep{AskesAifantis2011}.

In gradient elasticity, the balance of linear momentum in the absence of inertial and body forces reads \citep{MindlinEshel1968}
\begin{equation}
          \label{momentum}
	\sigma_{ik, i}-\tau_{ijk,ij}=0\,.
\end{equation}
The constitutive equations for the Cauchy stress $\boldsymbol{\sigma}$ and hyper-stress $\boldsymbol{\tau}$ are written as
\begin{equation}
          \label{constitutive}
	\sigma_{ik}=\mathsf{C}_{iklm} e_{lm}+\mathsf{M}_{iklmn} e_{lm,n}\,,\qquad
	\tau_{ijk}=\mathsf{M}_{lmijk} e_{lm}+\mathsf{A}_{ijklmn} e_{lm,n}\, ,
\end{equation}
where $e_{lm}=\frac{1}{2}(u_{l,m}+u_{m,l})$ is the strain tensor of linear elasticity. These correspond to the following quadratic elastic energy density
\begin{equation}
           \label{energy}
	W=\frac{1}{2} {\bf e} : \boldsymbol{\mathsf{C}}: {\bf e}
	+ {\bf e}:\boldsymbol{\mathsf{M}} : \nabla {\bf e} + \frac{1}{2} \nabla {\bf e} :  \boldsymbol{\mathsf{A}} :  \nabla {\bf e}\,.
\end{equation} 
The elasticity tensors $\boldsymbol{\mathsf{C}}$, $\boldsymbol{\mathsf{M}}$, and $\boldsymbol{\mathsf{A}}$ have the following symmetries
\begin{align}
 \label{symmetries}
	& \mathsf{C}_{ijlm} =\mathsf{C}_{jilm} =\mathsf{C}_{lmij}\,, \\
	& \mathsf{M}_{ijklm} =\mathsf{M}_{jiklm} =\mathsf{M}_{ijlkm}\,, \\
	 & \mathsf{A}_{ijklmn} =\mathsf{A}_{jiklmn} =\mathsf{A}_{lmnijk}\,. 
\end{align}
The tensor $\boldsymbol{\mathsf{A}}$ has $17$ symmetry classes~\citep[Table~1, p.~1205]{Auffrayetal2013}, the tensor $\boldsymbol{\mathsf{M}}$ has $29$ symmetry classes~\citep[Theorem~3.5, p.~203]{Auffrayetal2019}, and the classical elasticity tensor $\boldsymbol{\mathsf{C}}$ has $8$ symmetry classes~\citep[see, e.g.,][]{ForteVianello1996}.

All symmetry classes of strain gradient elasticity are defined as \citep{Auffrayetal2019}
\begin{equation}
           \label{symclass1}
	G_{\mathsf{L}}=G_{\boldsymbol{\mathsf{A}}} \cap G_{\boldsymbol{\mathsf{M}}} \cap G_{\boldsymbol{\mathsf{C}}}\,,
\end{equation}
where
\begin{align}
\label{symclass2}
	&G_{\boldsymbol{\mathsf{C}}}=\left\{ {\bf Q} \in O(3) : Q_{io} Q_{jp} Q_{kq} Q_{lr} {\mathsf C}_{opqr}	
	={\mathsf C}_{ijkl}  \right\}, \\
	&G_{\boldsymbol{\mathsf{M}}}=\left\{ {\bf Q} \in O(3) : Q_{io} Q_{jp} Q_{kq} Q_{lr} Q_{ms} 
	{\mathsf M}_{opqrs}={\mathsf M}_{ijklm}  \right\}, \\
	&G_{\boldsymbol{\mathsf{A}}}=\left\{ {\bf Q} \in O(3) : Q_{io} Q_{jp} Q_{kq} Q_{lr} Q_{ms}  
	Q_{mt} {\mathsf A}_{opqrst}={\mathsf A}_{ijklmn}  \right\} \,. 
\end{align}
In total there are $48$ symmetry classes defined in \citep{Auffrayetal2019}.

In the two-dimensional setting, the explicit matrix representations were derived by \citet{Auffrayetal2009} for the sixth-order tensor and by \citet{Auffrayetal2015} for the fifth-order tensor. The role of centrosymmetry in the fifth-order coupling tensor was emphasized by \citet{LakesBenedict1982} and \citet{Lakes2001}, who noted that this tensor vanishes for centrosymmetric microstructures and may be nonzero otherwise. In three dimensions, an explicit isotropic form of the sixth-order tensor was given by \citet{dell'Isolaetal2009}, and the fifth-order tensor was examined for specific cases by \citet{Papanicolopoulos2011}. A general framework for counting and classifying the symmetry classes of the fifth- and sixth-order tensors was developed by \citet{OliveAuffray2013,OliveAuffray2014}. Finally, complete expressions for the sixth- and fifth-order tensors, including their full symmetry classification, are given in \citep{Auffrayetal2013,Auffrayetal2019}.

The subdivision of three-dimensional strain-gradient elasticity into $48$ classes by Auffray and co-workers is based on requiring the least symmetric triplet $(\boldsymbol{\mathsf{C}}, \boldsymbol{\mathsf{M}}, \boldsymbol{\mathsf{A}})$ of physical symmetry classes to be compatible with the material symmetries. It is worth noting that a classification based on plane symmetries has only recently been proposed \citep{Quangetal2021}.

Universal displacements satisfy the equilibrium equations, in the absence of body forces, for any material within a prescribed class. 
Considering arbitrariness of the classical elastic constants, one recovers the universal displacements of classical linear elasticity \citep{Yavari2020}. When additional elastic constants are present, one expects further universality constraints beyond those of the classical theory. Consequently, for a given symmetry class, the set of universal displacements may be a proper subset of the classical one. This possibility is examined carefully, class by class, in the following sections. 
The matrix representations of $\boldsymbol{\mathsf M}$ and $\boldsymbol{\mathsf A}$ in \S\ref{Sec:Strain-gradient-linear-elasticity} are taken from \citet{Auffrayetal2013,Auffrayetal2019}, who provided the complete matrix representation and full symmetry classification for strain-gradient linear elastic solids.

\paragraph{General strategy:} To obtain the universal displacements, we first use the independence of the material constants in the classical linear elastic part, which yields potential universal displacements from the known results of linear elasticity \citep{Yavari2020}. For these potential universal displacements, we then examine whether the independence of the material constants associated with the fifth- and sixth-order tensors of strain-gradient elasticity \citep{Auffrayetal2013,Auffrayetal2019} imposes additional constraints. When this occurs, we report the resulting simplifications; otherwise, we simply report the additional constraints.

\subsection{Triclinic classes}

Triclinic classes are the richest material classes in terms of the number of independent elastic constants. There are $300$ independent elastic constants for the triclinic $\mathsf{1}$ class and $192$ for the $\mathbb{Z}_2^c$ class. For both classes, the elasticity tensor $\boldsymbol{\mathsf{C}}$ has the same form as in the triclinic class of classical linear elasticity. Therefore, we examine whether the classical universal displacements also satisfy the additional universality constraints imposed by the independent elastic constants of the generalized elasticity tensors $\boldsymbol{\mathsf{A}}$ and $\boldsymbol{\mathsf{M}}$.

\subsubsection{Triclinic class $\mathsf{1}$} 

The tensor $\boldsymbol{\mathsf{A}}_{\mathbb{Z}_2^c}$ has $171$ independent components; it is an $18\times 18$ symmetric matrix of the form
\begin{equation}
\label{A_Z_2^c}
    \boldsymbol{\mathsf{A}}_{\mathbb{Z}_2^c}(\mathbf{x})=\begin{bmatrix}
    A^{(15)} & B^{(25)} & C^{(25)} & D^{(15)}  \\
    B^{(25)} & E^{(15)} & F^{(25)} & G^{(15)}   \\
    C^{(25)} & F^{(25)} & H^{(15)} & I^{(15)}   \\
    D^{(15)} & G^{(15)} & I^{(15)} & J^{(6)}
    \end{bmatrix}\,,
\end{equation}
where $A^{(15)}, E^{(15)}, H^{(15)} \in M^s(5)$, $B^{(25)}, C^{(25)}, F^{(25)} \in M(5)$, $D^{(15)}, G^{(15)}, I^{(15)} \in M(5,3)$, and $J^{(6)} \in M^s(3)$. Here, $M^s(n)$, $M(n)$, and $M(n,m)$ denote the sets of symmetric $n \times n$, all $n \times n$, and all $n \times m$ matrices, respectively.
The tensor $\boldsymbol{\mathsf{M}}_{\mathsf{1}}$ has $108$ independent components; it is a $6 \times 18$ matrix of the form
\begin{equation}
\label{M_1}
    \boldsymbol{\mathsf{M}}_{\mathsf{1}}(\mathbf{x})=\begin{bmatrix}
    \bar{A}^{(15)} & \bar{B}^{(15)} & \bar{C}^{(15)} & \bar{D}^{(9)}  \\
    \bar{E}^{(5)} & \bar{F}^{(5)} & \bar{G}^{(5)} & \bar{H}^{(3)}   \\
    \bar{I}^{(5)} & \bar{J}^{(5)} & \bar{K}^{(5)} & \bar{L}^{(3)}   \\
    \bar{M}^{(5)} & \bar{N}^{(5)} & \bar{O}^{(5)} & \bar{P}^{(3)}
    \end{bmatrix}\,,
\end{equation}
where $\bar{A}^{(15)}, \bar{B}^{(15)}, \bar{C}^{(15)} \in M(3,5)$, $\bar{D}^{(9)} \in M(3)$, $\bar{E}^{(5)}, \bar{F}^{(5)}, \bar{G}^{(5)}, \bar{I}^{(5)}, \bar{J}^{(5)}, \bar{K}^{(5)}, \bar{M}^{(5)}, \bar{N}^{(5)}, \bar{O}^{(5)} \in M(1,5)$, and $\bar{H}^{(3)}, \bar{L}^{(3)}, \bar{P}^{(3)} \in M(1,3)$.
The remaining $21$ components stem from the classical form of the elasticity tensor $\boldsymbol{\mathsf{C}}$ for the triclinic class (Hermann--Maugin symbol $1$ \citep{Haussuhl2007,Newnkam2005}), giving a total of $300$ independent components in this case. From the classical linear elastic part we obtain the universality constraints \eqref{triclinic-linear-constraints1} and \eqref{triclinic-linear-constraints2}. Thus, the only potential universal displacements are homogeneous displacement fields.  

The universality constraints corresponding to the fifth-order tensor are as follows (third-order PDEs):
\begin{equation}
\begin{dcases}
\label{triclinicfifthorder}
\frac{\partial^3 u_1}{\partial x_1^3} &= \frac{\partial^3 u_1}{\partial x_1 \partial x_2^2} = \frac{\partial^3 u_1}{\partial x_1^2 \partial x_2} = \frac{\partial^3 u_1}{\partial x_1^2 \partial x_3} = 0\,, \\
\frac{\partial^3 u_1}{\partial x_2^2 \partial x_3} &= \frac{\partial^3 u_1}{\partial x_3^3} = \frac{\partial^3 u_1}{\partial x_1 \partial x_2 \partial x_3} = \frac{\partial^3 u_1}{\partial x_1 \partial x_3^2} = \frac{\partial^3 u_1}{\partial x_2^3} = 0\,, \\
\frac{\partial^3 u_2}{\partial x_1 \partial x_2^2} &= \frac{\partial^3 u_2}{\partial x_1^3} = \frac{\partial^3 u_2}{\partial x_1 \partial x_3^2} = \frac{\partial^3 u_2}{\partial x_1 \partial x_2 \partial x_3} = 0\,,\\
\frac{\partial^3 u_2}{\partial x_1^2 \partial x_2} &= \frac{\partial^3 u_2}{\partial x_2^3} = \frac{\partial^3 u_2}{\partial x_2^2 \partial x_3} = \frac{\partial^3 u_2}{\partial x_2 \partial x_3^2} = \frac{\partial^3 u_2}{\partial x_3^3} = \frac{\partial^3 u_2}{\partial x_1^2 \partial x_3} = 0\,,\\
\frac{\partial^3 u_3}{\partial x_1 \partial x_3^2} &= \frac{\partial^3 u_3}{\partial x_1^3} = \frac{\partial^3 u_3}{\partial x_1 \partial x_2^2} = \frac{\partial^3 u_3}{\partial x_1^2 \partial x_2} = 0\,,\\
\frac{\partial^3 u_3}{\partial x_1^2 \partial x_3} &= \frac{\partial^3 u_3}{\partial x_2^2 \partial x_3} = \frac{\partial^3 u_3}{\partial x_3^3} = \frac{\partial^3 u_3}{\partial x_1 \partial x_2 \partial x_3} = \frac{\partial^3 u_3}{\partial x_2 \partial x_3^2} = 0\,. 
\end{dcases}
\end{equation}
The universality constraints corresponding to the sixth-order tensor are (fourth-order PDEs):
\begin{equation}
\begin{dcases}
\label{triclinicsixthorder}
\frac{\partial^4 u_1}{\partial x_1^4} = \frac{\partial^4 u_1}{\partial x_1^3 \partial x_2} = \frac{\partial^4 u_1}{\partial x_1^3 \partial x_3} = \frac{\partial^4 u_1}{\partial x_2^4} = \frac{\partial^4 u_1}{\partial x_2^2 \partial x_3^2} = 0\,, \\
\frac{\partial^4 u_1}{\partial x_1 \partial x_2^3} = \frac{\partial^4 u_1}{\partial x_1 \partial x_2^2 \partial x_3} = \frac{\partial^4 u_1}{\partial x_2^3 \partial x_3} = \frac{\partial^4 u_1}{\partial x_3^4} = \frac{\partial^4 u_1}{\partial x_1 \partial x_2 \partial x_3^2} = 0\,,\\
\frac{\partial^4 u_1}{\partial x_2 \partial x_3^3} = \frac{\partial^4 u_1}{\partial x_1^2 \partial x_2^2} = \frac{\partial^4 u_1}{\partial x_1^2 \partial x_2 \partial x_3}  = \frac{\partial^4 u_1 }{\partial x_2^3 \partial x_3} = 0\,,\\
\frac{\partial^4 u_2}{\partial x_1^3 \partial x_2} = \frac{\partial^4 u_2}{\partial x_1^2 \partial x_2^2} = \frac{\partial^4 u_2}{\partial x_1^4} = \frac{\partial^4 u_2}{\partial x_1^2 \partial x_3^2} = \frac{\partial^4 u_2}{\partial x_1^2 \partial x_2 \partial x_3} = 0\,, \\
\frac{\partial^4 u_2}{\partial x_1^3 \partial x_3} = \frac{\partial^4 u_2}{\partial x_1 \partial x_2^3} = \frac{\partial^4 u_2}{\partial x_2^4} = \frac{\partial^4 u_2}{\partial x_2^3 \partial x_3} = \frac{\partial^4 u_2}{\partial x_2^2 \partial x_3^2} = 0\,, \\
\frac{\partial^4 u_2}{\partial x_1 \partial x_3^3} = \frac{\partial^4 u_2}{\partial x_1 \partial x_2^2 \partial x_3} = \frac{\partial^4 u_2}{\partial x_1 \partial x_2 \partial x_3^2} = \frac{\partial^4 u_2}{\partial x_3^4} = 0\,,\\
\frac{\partial^4 u_3}{\partial x_1^3 \partial x_2} = \frac{\partial^4 u_3}{\partial x_1^2 \partial x_3^2} = \frac{\partial^4 u_3}{\partial x_1^4} = \frac{\partial^4 u_3}{\partial x_1^2 \partial x_2^2} = \frac{\partial^4 u_3}{\partial x_1^3 \partial x_2} = 0\,, \\
\frac{\partial^4 u_3}{\partial x_1 \partial x_2^2 \partial x_3} = \frac{\partial^4 u_3}{\partial x_2^2 \partial x_3^2} = \frac{\partial^4 u_3}{\partial x_2^3 \partial x_3} = \frac{\partial^4 u_3}{\partial x_2^4} = \frac{\partial^4 u_3}{\partial x_1 \partial x_2^3} = 0\,, \\
\frac{\partial^4 u_3}{\partial x_1 \partial x_3^3} = \frac{\partial^4 u_3}{\partial x_3^4} = \frac{\partial^4 u_3}{\partial x_1 \partial x_2 \partial x_3^2} = \frac{\partial^4 u_3}{\partial x_1^2 \partial x_2 \partial x_3} = \frac{\partial^4 u_3}{\partial x_2 \partial x_3^3} = 0\,. 
\end{dcases}
\end{equation}
Third-order universality PDEs are associated with the matrix $\boldsymbol{\mathsf{M}}_{\mathsf{1}}$, while fourth-order universality PDEs are associated with the matrix $\boldsymbol{\mathsf{A}}_{\mathbb{Z}_2^c}$. The additional PDEs induced by $\boldsymbol{\mathsf{M}}_{\mathsf{1}}$ and $\boldsymbol{\mathsf{A}}_{\mathbb{Z}_2^c}$ are therefore of third and fourth order, respectively, and are satisfied trivially by any homogeneous displacement field. 
Thus, one has the following result.

\begin{prop}
The only universal displacements in triclinic $\mathsf{1}$ class linear strain-gradient elastic solids are homogeneous displacement fields.
\end{prop}

\subsubsection{Triclinic class $\mathbb{Z}_2^c$} 

Tensor $\boldsymbol{\mathsf{A}}_{\mathbb{Z}_2^c}$ has $171$ independent components and has the same form as that of class $\mathsf{1}$. Tensor $\boldsymbol{\mathsf{M}}_{\mathbb{Z}_2^c}$ is null. Tensor $\boldsymbol{\mathsf{C}}_{\mathbb{Z}_2^c}$ has the same form as that of class $\mathsf{1}$, with the only difference that it corresponds to the Hermann--Maugin symbol $\bar{1}$. Since the triclinic $\mathbb{Z}_2^c$ class is obtained from the triclinic $\mathsf{1}$ class by setting $\boldsymbol{\mathsf{M}}_{\mathbb{Z}_2^c}=\mathbf{0}$, the following result follows immediately.

\begin{prop}
The only universal displacements in triclinic $\mathbb{Z}_2^c$ class linear strain-gradient elastic solids are homogeneous displacement fields.
\end{prop}

\renewcommand{\arraystretch}{1.15}
\begin{longtable}{p{0.30\textwidth} p{0.66\textwidth}}
\caption{Summary of universal displacements: Triclinic classes.}\label{Table-UD-summary-triclinic}\\
\hline
\textbf{Symmetry class} & \textbf{Universal displacement family} \\
\hline
\endfirsthead
\hline
Symmetry class & Universal displacement family (continued) \\
\hline
\endhead
\hline
\endfoot
\hline
\endlastfoot
Triclinic class $\mathsf{1}$ & Homogeneous displacement fields \\ \hline
Triclinic class $\mathbb{Z}_2^c$ & Homogeneous displacement fields \\ \hline
\end{longtable}
\renewcommand{\arraystretch}{2.0}

\subsection{Monoclinic classes}

For the monoclinic classes $\mathbb{Z}_2$, $\mathbb{Z}_2 \oplus \mathbb{Z}_2^c$, and $\mathbb{Z}_2^-$, the matrix $\boldsymbol{\mathsf{C}}$ has $13$ independent elastic constants and coincides with that of the classical monoclinic linear elastic case. By contrast, for the monoclinic class $\mathbb{D}_2^{v}$, the matrix $\boldsymbol{\mathsf{C}}$ has $9$ independent components and coincides with that of the classical orthotropic case. Accordingly, for the classes $\mathbb{Z}_2$, $\mathbb{Z}_2 \oplus \mathbb{Z}_2^c$, and $\mathbb{Z}_2^-$ we examine whether the additional universality constraints stemming from the fifth- and sixth-order tensors are satisfied by the universal displacements of the classical monoclinic case. For the class $\mathbb{D}_2^{v}$, the same procedure is followed, but the candidate universal displacements are those of the classical orthotropic case.

\subsubsection{Monoclinic class $\mathbb{Z}_2$} 

Tensor $\boldsymbol{\mathsf{A}}_{\mathbb{Z}_2 \oplus \mathbb{Z}_2^c}$ has $91$ independent components\,; it is of the form
\begin{equation}
\label{A_Z_2-1}
\boldsymbol{\mathsf{A}}_{\mathbb{Z}_2^{e_3} \oplus \mathbb{Z}_2^c}(\mathbf{x})=
\begin{bmatrix}
A^{(15)} & B^{(25)} & 0 & 0  \\
B^{(25)} & E^{(15)} & 0 & 0   \\
0 & 0 & H^{(15)} & I^{(15)}   \\
0 & 0 & I^{(15)} & J^{(6)}
\end{bmatrix}\,,
\end{equation}
or
\begin{equation}
\label{A_Z_2-2}
\boldsymbol{\mathsf{A}}_{\mathbb{Z}_2^{e_1} \oplus \mathbb{Z}_2^c}(\mathbf{x})=
\begin{bmatrix}
A^{(15)} &  & 0 & D^{(15)}  \\
0 & E^{(15)} & F^{(25)} & 0   \\
0 & F^{(25)} & H^{(15)} & 0   \\
D^{(15)} & 0 & 0 & J^{(6)}
\end{bmatrix}\,,
\end{equation}
depending on whether the $\pi$ rotation is taken to be around ${\bf e}_3$ or ${\bf e}_1$, respectively. As previously, $A^{(15)}$, $E^{(15)}$, $H^{(15)} \in M^s(5)$, $B^{(25)}$, $F^{(25)} \in M(5)$, $D^{(15)}$, $I^{(15)} \in M(5, 3)$, and $J^{(6)} \in M^s(3)$. 
The expressions $\boldsymbol{\mathsf{A}}_{\mathbb{Z}_2^{e_3} \oplus \mathbb{Z}_2^c}(\mathbf{x})$ and $\boldsymbol{\mathsf{A}}_{\mathbb{Z}_2^{e_1} \oplus \mathbb{Z}_2^c}(\mathbf{x})$ provide two conjugate representations of the same symmetry class $\mathbb{Z}_2$: in the first representation, $\boldsymbol{\mathsf{A}}_{\mathbb{Z}_2^{e_3} \oplus \mathbb{Z}_2^c}(\mathbf{x})$, the $\pi$ rotation is taken about ${\bf e}_3$, whereas in the second representation, $\boldsymbol{\mathsf{A}}_{\mathbb{Z}_2^{e_1} \oplus \mathbb{Z}_2^c}(\mathbf{x})$, it is taken about ${\bf e}_1$. In our calculations, we take ${\bf e}_3$ to be normal to the plane of material symmetry folloiwng \citet{Yavari2020}.

The tensor $\boldsymbol{\mathsf{M}}_{\mathbb{Z}_2}$ has $52$ components; it is written as
\begin{equation}
\label{M_Z_2}
	\boldsymbol{\mathsf{M}}_{\mathbb{Z}_2}=\boldsymbol{\mathsf{M}}_{\mathbb{D}_2}
	+\boldsymbol{\mathsf{M}_{\mathbb{D}_2^v}}\,,
\end{equation}
where
\begin{equation}
\label{M_D_2}
\boldsymbol{\mathsf{M}_{\mathbb{D}_2}} (\mathbf{x})=
\begin{bmatrix}
0 & 0 & 0 & \bar{D}^{(9)}  \\
\bar{E}^{(5)} & 0 & 0 & 0   \\
0 & \bar{J}^{(5)} & 0 & 0   \\
0 & 0 & \bar{O}^{(5)} & 0
\end{bmatrix}\,,\qquad
\end{equation}
and
\begin{equation}
\label{M_D_2^v}
\boldsymbol{\mathsf{M}_{\mathbb{D}_2^v}}(\mathbf{x})=
\begin{bmatrix}
0 & 0 & \bar{C}^{(15)} & 0  \\
0 & \bar{F}^{(5)} & 0 & 0   \\
\bar{I}^{(5)} & 0 & 0 & 0   \\
0 & 0 & 0 & \bar{P}^{(3)}
\end{bmatrix}\,.
\end{equation}
As before, $\bar{C}^{(15)} \in M(3, 5)$, $\bar{D}^{(9)} \in M(3)$, $\bar{E}^{(5)}$, $\bar{F}^{(5)}$, $\bar{J}^{(5)}$, $\bar{O}^{(5)} \in M(1, 5)$, and $\bar{P}^{(3)} \in M(1, 3)$. The remaining $13$ components stem from the classical form of the elasticity tensor for the monoclinic case (Hermann--Maugin symbol $2$). Thus, in total, this case has $156$ independent components. From the classical linear elastic part, we obtain constraints of the monoclinic type (see \S\ref{Sec:Monoclinic-class-classical}), which admit as candidate universal displacements the superposition of a homogeneous displacement field and the one-parameter inhomogeneous field $(c x_2 x_3, -c x_1 x_3, 0)$.

From the components of the fifth-order tensor we obtain the following third-order universality PDEs:
\begin{equation}
\begin{dcases}
\label{monoclinicZ-2fifthorder}
\frac{\partial^3 u_1}{\partial x_1^2 \partial x_3} = \frac{\partial^3 u_1}{\partial x_2^2 \partial x_3} = \frac{\partial^3 u_1}{\partial x_3^3} = \frac{\partial^3 u_1}{\partial x_1 \partial x_2 \partial x_3} = 0\,, \\
\frac{\partial^3 u_1}{\partial x_1 \partial x_3^2} = \frac{\partial^3 u_1}{\partial x_1 \partial x_2^2} = \frac{\partial^3 u_1}{\partial x_1^3} = \frac{\partial^3 u_1}{\partial x_2 \partial x_3^2} = \frac{\partial^3 u_1}{\partial x_1^2 \partial x_2} = \frac{\partial^3 u_1}{\partial x_2^3} = 0\,, \\
\frac{\partial^3 u_2}{\partial x_1 \partial x_2 \partial x_3} = \frac{\partial^3 u_2}{\partial x_1^2 \partial x_3} = \frac{\partial^3 u_2}{\partial x_2^2 \partial x_3} = \frac{\partial^3 u_2}{\partial x_3^3} = \frac{\partial^3 u_2}{\partial x_1 \partial x_3^2} = 0\,, \\
\frac{\partial^3 u_2}{\partial x_2 \partial x_3^2} = \frac{\partial^3 u_2}{\partial x_1^2 \partial x_2} = \frac{\partial^3 u_2}{\partial x_2^3} = \frac{\partial^3 u_2}{\partial x_1 \partial x_2^2} = \frac{\partial^3 u_2}{\partial x_1^3} = 0\,, \\
\frac{\partial^3 u_3}{\partial x_1 \partial x_3^2} = \frac{\partial^3 u_3}{\partial x_1^3} = \frac{\partial^3 u_3}{\partial x_1 \partial x_2^2} = \frac{\partial^3 u_3}{\partial x_1^2 \partial x_2} = 0\,, \\
\frac{\partial^3 u_3}{\partial x_2 \partial x_3^2} = \frac{\partial^3 u_3}{\partial x_2^3} = \frac{\partial^3 u_3}{\partial x_3^3} = \frac{\partial^3 u_3}{\partial x_1^2 \partial x_3}  = 0\,. 
\end{dcases}
\end{equation}
From the components of the sixth-order tensor we obtain the following fourth-order PDEs:
\begin{equation}
\begin{dcases}
\label{monoclinicZ-2sixthorder}
\frac{\partial^4 u_1}{\partial x_1^4} = \frac{\partial^4 u_1}{\partial x_1^3 \partial x_2} = \frac{\partial^4 u_1}{\partial x_2^4} = \frac{\partial^4 u_1}{\partial x_2^2 \partial x_3^2} = \frac{\partial^4 u_1}{\partial x_1 \partial x_2^3} = 0\,, \\
\frac{\partial^4 u_1}{\partial x_1 \partial x_3^3} = \frac{\partial^4 u_1}{\partial x_3^4} = \frac{\partial^4 u_1}{\partial x_1 \partial x_2 \partial x_3^2} = \frac{\partial^4 u_1}{\partial x_1^2 \partial x_2^2} = \frac{\partial^4 u_1}{\partial x_1^2 \partial x_3^2} = 0\,, \\
\frac{\partial^4 u_1}{\partial x_1^3 \partial x_3} = \frac{\partial^4 u_1}{\partial x_1^2 \partial x_2 \partial x_3} = \frac{\partial^4 u_1}{\partial x_1 \partial x_2^2 \partial x_3} = \frac{\partial^4 u_1}{\partial x_2^3 \partial x_3} = \frac{\partial^4 u_1}{\partial x_2 \partial x_3^3} = 0\,, \\
\frac{\partial^4 u_2}{\partial x_1^3 \partial x_2} = \frac{\partial^4 u_2}{\partial x_1^2 \partial x_2^2} = \frac{\partial^4 u_2}{\partial x_1^2 \partial x_3^2} = \frac{\partial^4 u_2}{\partial x_1 \partial x_2^3} = \frac{\partial^4 u_2}{\partial x_2^4} = 0\,, \\
\frac{\partial^4 u_2}{\partial x_2^2 \partial x_3^2} = \frac{\partial^4 u_2}{\partial x_3^4} = \frac{\partial^4 u_2}{\partial x_1 \partial x_2 \partial x_3^2} = \frac{\partial^4 u_2}{\partial x_1^4} = \frac{\partial^4 u_2}{\partial x_1^2 \partial x_2 \partial x_3} = 0\,, \\
\frac{\partial^4 u_2}{\partial x_1 \partial x_2^2 \partial x_3} = \frac{\partial^4 u_2}{\partial x_1^3 \partial x_3} = \frac{\partial^4 u_2}{\partial x_2 \partial x_3^3} = \frac{\partial^4 u_2}{\partial x_2^3 \partial x_3} = 0\,, \\
\frac{\partial^4 u_3}{\partial x_1^3 \partial x_3} = \frac{\partial^4 u_3}{\partial x_1^2 \partial x_3^2} = \frac{\partial^4 u_3}{\partial x_1 \partial x_2^2 \partial x_3} = \frac{\partial^4 u_3}{\partial x_2^4} = \frac{\partial^4 u_3}{\partial x_3^4} = 0\,, \\
\frac{\partial^4 u_3}{\partial x_1 \partial x_2 \partial x_3^2} = \frac{\partial^4 u_3}{\partial x_1 \partial x_3^3} = \frac{\partial^4 u_3}{\partial x_2 \partial x_3^3} = \frac{\partial^4 u_3}{\partial x_2^3 \partial x_3} = \frac{\partial^4 u_3}{\partial x_1^4} = 0\,, \\
\frac{\partial^4 u_3}{\partial x_1^2 \partial x_2^2} = \frac{\partial^4 u_3}{\partial x_1^3 \partial x_2}  = \frac{\partial^4 u_3}{\partial x_1 \partial x_2^3} = 0\,. 
\end{dcases}
\end{equation}
Substituting the superposition of homogeneous displacement fields and the one-parameter inhomogeneous field $(c x_2 x_3\,,\,-c x_1 x_3\,,\,0)$ into \eqref{monoclinicZ-2fifthorder} and \eqref{monoclinicZ-2sixthorder}, we find that these constraints are satisfied trivially. 
Thus, we have the following result.
\begin{prop}
The only universal displacements in monoclinic $\mathbb{Z}_2$ class linear strain-gradient elastic solids are the superposition of homogeneous displacement fields and the one-parameter inhomogeneous field $(c x_2 x_3\,,\,-c x_1 x_3\,,\,0)$.
\end{prop}

\subsubsection{Monoclinic class $\mathbb{Z}_2 \oplus \mathbb{Z}_2^c$} 

The tensor $\boldsymbol{\mathsf{A}}_{\mathbb{Z}_2 \oplus \mathbb{Z}_2^c}$ has $91$ independent components and has the same form as that of class $\mathbb{Z}_2$. Tensor $\boldsymbol{\mathsf{M}}_{\mathbb{Z}_2 \oplus \mathbb{Z}_2^c}$ is null. Tensor $\boldsymbol{\mathsf{C}}_{\mathbb{Z}_2 \oplus \mathbb{Z}_2^c}$ has the same form as that of class $\mathbb{Z}_2$, with $13$ independent components, with the only difference that it corresponds to the Hermann--Maugin symbol $2/m$.
We thus arrive at the following result. 
\begin{prop}
The only universal displacements in monoclinic $\mathbb{Z}_2 \oplus \mathbb{Z}_2^c$ class linear strain-gradient elastic solids are the superposition of homogeneous displacement fields and the one-parameter inhomogeneous field $(c x_2 x_3\,,\,-c x_1 x_3\,,\,0)$.
\end{prop}

\subsubsection{Monoclinic class $\mathbb{Z}_2^-$} 

Tensor $\boldsymbol{\mathsf{A}}_{\mathbb{Z}_2 \oplus \mathbb{Z}_2^c}$ has $91$ independent components and has the same form as that of class $\mathbb{Z}_2$. Tensor $\boldsymbol{\mathsf{M}}_{\mathbb{Z}_2^-}$ has $56$ independent components and has the form
\begin{equation}
	\label{M_Z_2^-}
	\boldsymbol{\mathsf{M}}_{\mathbb{Z}_2^-}=\begin{bmatrix}
		\bar{A}^{(15)} & \bar{B}^{(15)} & 0 & 0  \\
		0 & 0 & \bar{G}^{(5)} & \bar{H}^{(13)}   \\
		0 & 0  & \bar{K}^{(5)} & \bar{L}^{(3)}     \\
		\bar{M}^{(5)} & \bar{N}^{(5)} & 0 & 0   
	\end{bmatrix}\,.
\end{equation}
Tensor $\boldsymbol{\mathsf{C}}_{\mathbb{Z}_2^-}$ has $13$ independent components and corresponds to the Hermann--Maugin symbol $m$.

The fourth-order universality constraints coincide with those of class $\mathbb{Z}_2$; the difference lies in the third-order constraints on the displacement field, which in this case are
\begin{equation}
\begin{dcases}
\frac{\partial^3 u_1}{\partial x_3^3} = \frac{\partial^3 u_1}{\partial x_1 \partial x_2^2} = \frac{\partial^3 u_1}{\partial x_1 \partial x_3^2} = \frac{\partial^3 u_1}{\partial x_1^2 \partial x_2} = 0\,, \\
\frac{\partial^3 u_1}{\partial x_2 \partial x_3^2} = \frac{\partial^3 u_1}{\partial x_2^3} = \frac{\partial^3 u_1}{\partial x_1^2 \partial x_3} = \frac{\partial^3 u_1}{\partial x_1 \partial x_2 \partial x_3} = \frac{\partial^3 u_1}{\partial x_2^2 \partial x_3}  = 0\,, \\
\frac{\partial^3 u_2}{\partial x_1 \partial x_2^2} = \frac{\partial^3 u_2}{\partial x_1^3} = \frac{\partial^3 u_2}{\partial x_1 \partial x_3^2} = \frac{\partial^3 u_2}{\partial x_1^2 \partial x_2} = \frac{\partial^3 u_2}{\partial x_2^3} = 0\,, \\
\frac{\partial^3 u_2}{\partial x_2 \partial x_3^2} = \frac{\partial^3 u_2}{\partial x_1 \partial x_2 \partial x_3} = \frac{\partial^3 u_2}{\partial x_2^2 \partial x_3} = \frac{\partial^3 u_2}{\partial x_1^2 \partial x_3} = \frac{\partial^3 u_2}{\partial x_3^3} = 0\,, \\
\frac{\partial^3 u_3}{\partial x_1 \partial x_3^2} = \frac{\partial^3 u_3}{\partial x_1^2 \partial x_3} = \frac{\partial^3 u_3}{\partial x_2^2 \partial x_3} = \frac{\partial^3 u_3}{\partial x_3^3} = 0\,, \\
\frac{\partial^3 u_3}{\partial x_1 \partial x_2 \partial x_3} = \frac{\partial^3 u_3}{\partial x_2 \partial x_3^2} = \frac{\partial^3 u_3}{\partial x_1^2 \partial x_2} = \frac{\partial^3 u_3}{\partial x_2^3} = \frac{\partial^3 u_3}{\partial x_1 \partial x_2^2} = \frac{\partial^3 u_3}{\partial x_1^3} = 0\,. 
\end{dcases}
\end{equation}
Substituting the superposition of a homogeneous displacement field and the one-parameter inhomogeneous displacement field $(c x_2 x_3\,,-c x_1 x_3\,,\,0)$, we obtain the following result.
\begin{prop}
The only universal displacements in monoclinic $\mathbb{Z}_2^-$ class linear strain-gradient elastic solids are the superposition of homogeneous displacement fields and the one-parameter inhomogeneous field $(c x_2 x_3\,,-c x_1 x_3\,,\,0)$.
\end{prop}

\subsubsection{Monoclinic class $\mathbb{D}_2^v$} 

Tensor $\boldsymbol{\mathsf{A}}_{\mathbb{D}_2 \oplus \mathbb{Z}_2^c}$ has $51$ independent components and has the same form as that of class $\mathbb{D}_2$. Tensor $\boldsymbol{\mathsf{M}}_{\mathbb{D}_2^{v}}$ has $28$ independent components and has the same form as \eqref{M_D_2^v}. Tensor $\boldsymbol{\mathsf{C}}_{\mathbb{D}_2 \oplus \mathbb{Z}_2^c}$ has $9$ independent components and corresponds to the Hermann--Maugin symbol $2mm$. Thus, although it belongs to the monoclinic strain-gradient class, $\mathbb{D}_2^{v}$ has the same classical elasticity tensor as the orthotropic class of linear elasticity. Consequently, from the classical linear elastic part we obtain constraints of the form \eqref{orthotropic-linear-constraints}, and the candidate universal displacements are the superposition of a homogeneous displacement field and the three-parameter inhomogeneous displacement field $(\alpha_1 x_2 x_3\,,\,\alpha_2 x_1 x_3\,,\,\alpha_3 x_1 x_2)$.

The fourth-order universality constraints coincide with those of class $\mathbb{D}_2$ and are given in the next subsection in \eqref{orthotropic-D2-sixthorder}. The third-order universality constraints for this case are
\begin{equation}
\begin{dcases}
\label{D^v_2-thirdorder}
\frac{\partial^3 u_1}{\partial x_3^3} = \frac{\partial^3 u_1}{\partial x_1^2 \partial x_3} = \frac{\partial^3 u_1}{\partial x_2^2 \partial x_3} = \frac{\partial^3 u_1}{\partial x_3^3} = 0\,, \\
\frac{\partial^3 u_1}{\partial x_1 \partial x_2 \partial x_3} = \frac{\partial^3 u_1}{\partial x_1 \partial x_3^2} = \frac{\partial^3 u_1}{\partial x_1^3} = \frac{\partial^3 u_1}{\partial x_1 \partial x_2^2} = 0\,, \\
\frac{\partial^3 u_2}{\partial x_1 \partial x_2 \partial x_3} = \frac{\partial^3 u_2}{\partial x_2^2 \partial x_3} = \frac{\partial^3 u_2}{\partial x_1^2 \partial x_3} = \frac{\partial^3 u_2}{\partial x_3^3} = \frac{\partial^3 u_2}{\partial x_2 \partial x_3^2} = 0\,, \\
\frac{\partial^3 u_2}{\partial x_1^2 \partial x_2} = \frac{\partial^3 u_2}{\partial x_2^3} = 0\,, \\
\frac{\partial^3 u_3}{\partial x_1 \partial x_3^2} = \frac{\partial^3 u_3}{\partial x_1^3} = \frac{\partial^3 u_3}{\partial x_1 \partial x_2^2} = \frac{\partial^3 u_3}{\partial x_2 \partial x_3^2} = 0\,, \\
\frac{\partial^3 u_3}{\partial x_1^2 \partial x_2} = \frac{\partial^3 u_3}{\partial x_2^3} = \frac{\partial^3 u_3}{\partial x_3^3} = \frac{\partial^3 u_3}{\partial x_1^2 \partial x_3} = \frac{\partial^3 u_3}{\partial x_2^2 \partial x_3} = 0\,. 
\end{dcases}
\end{equation}
Substituting the superposition of a homogeneous displacement field and the three-parameter inhomogeneous displacement field $(\alpha_1 x_2 x_3\,,\,\alpha_2 x_1 x_3\,,\,\alpha_3 x_1 x_2)$ into \eqref{D^v_2-thirdorder} and into the fourth-order constraints of the orthotropic class $\mathbb{D}_2$ given in \eqref{orthotropic-D2-sixthorder}, we obtain the following result.
\begin{prop}
The only universal displacements in monoclinic $\mathbb{D}_2^v$ class linear strain-gradient elastic solids are the superposition of homogeneous displacement fields and the following three-parameter inhomogeneous displacement field: $(\alpha_1 x_2 x_3\,,\,\alpha_2 x_1 x_3\,,\,\alpha_3 x_1 x_2)$.
\end{prop}

\renewcommand{\arraystretch}{1.15}
\begin{longtable}{p{0.30\textwidth} p{0.66\textwidth}}
\caption{Summary of universal displacements: Monoclinic classes.}\label{Table-UD-summary-monoclinic}\\
\hline
\textbf{Symmetry class} & \textbf{Universal displacement family} \\
\hline
\endfirsthead
\hline
Symmetry class & Universal displacement family (continued) \\
\hline
\endhead
\hline
\endfoot
\hline
\endlastfoot
Monoclinic class $\mathbb{Z}_2$ & Superposition of homogeneous fields and $(c x_2 x_3\,,\,-c x_1 x_3\,,\,0)$ \\ \hline
Monoclinic class $\mathbb{Z}_2 \oplus \mathbb{Z}_2^c$ & Superposition of homogeneous fields and $(c x_2 x_3\,,\,-c x_1 x_3\,,\,0)$ \\ \hline
Monoclinic class $\mathbb{Z}_2^{-}$ & Superposition of homogeneous fields and $(c x_2 x_3\,,\,-c x_1 x_3\,,\,0)$ \\ \hline
Monoclinic class $\mathbb{D}_2^{v}$ & Superposition of homogeneous fields and $(\alpha_1 x_2 x_3\,,\,\alpha_2 x_1 x_3\,,\,\alpha_3 x_1 x_2)$ \\ \hline
\end{longtable}
\renewcommand{\arraystretch}{2.0}

\subsection{Orthotropic classes}

There are five orthotropic classes in total. For the classes $\mathbb{D}_2$ and $\mathbb{D}_2 \oplus \mathbb{Z}_2^c$, the elasticity matrix $\boldsymbol{\mathsf{C}}$ coincides with that of the orthotropic class of classical linear elasticity. For the classes $\mathbb{Z}_4^-$ and $\mathbb{D}_4^h$, the elasticity tensor $\boldsymbol{\mathsf{C}}$ coincides with that of the tetragonal class of classical linear elasticity. Finally, for the class $\mathbb{D}_3^v$, the elasticity matrix $\boldsymbol{\mathsf{C}}$ coincides with that of the trigonal class of classical linear elasticity.

\subsubsection{Orthotropic class $\mathbb{D}_2$} 

Tensor $\boldsymbol{\mathsf{A}}_{\mathbb{D}_2 \oplus \mathbb{Z}_2^c}$ has $51$ independent components and has the form
\begin{equation}
    \label{A_D_2}
    \boldsymbol{\mathsf{A}}_{\mathbb{D}_2 \oplus \mathbb{Z}_2^c}(\mathbf{x})=
    \begin{bmatrix}
        A^{(15)} & 0 & 0 & 0 \\
        0 & E^{(15)} & 0 & 0 \\
        0 & 0 & H^{(15)} & 0 \\
        0 & 0 & 0 & J^{(6)}
    \end{bmatrix}_S\,,
\end{equation}
where, as previously, $A^{(15)}$, $E^{(15)}$, $H^{(15)} \in M^s(5)$ and $J^{(6)} \in M^s(3)$, and the superscript $S$ indicates symmetrization. Tensor $\boldsymbol{\mathsf{M}}_{\mathbb{D}_2}$ has the form \eqref{M_D_2} and has $24$ independent components. Thus, in total, this case has $84$ independent components, since $9$ components stem from the classical linear elasticity tensor for the orthorhombic case (Hermann--Maugin symbol $222$). The sixth-order tensor for this class coincides with that of the monoclinic class $\mathbb{D}_2^{v}$. Hence, from the classical linear elastic part we infer that the candidate universal displacements are the superposition of  homogeneous displacement fields and the following three-parameter inhomogeneous displacement field: $(\alpha_1 x_2 x_3\,,\, \alpha_2 x_1 x_3\,,\, \alpha_3 x_1 x_2)$. From the components of the fifth-order tensor we obtain the following third-order universality PDEs:
\begin{equation}
\begin{dcases}
\label{orthotropic-D2-fifthorder}
\frac{\partial^3 u_1}{\partial x_1 \partial x_2 \partial x_3}=\frac{\partial^3 u_1}{\partial x_1^2 \partial x_3}=\frac{\partial^3 u_1}{\partial x_2^2 \partial x_3}=\frac{\partial^3 u_1}{\partial x_1 \partial x_3^2}=0\,, \\
\frac{\partial^3 u_1}{\partial x_1^2 \partial x_2}=\frac{\partial^3 u_1}{\partial x_2 \partial x_3^2}=\frac{\partial^3 u_1}{\partial x_2^3 }=0\,, \\
\frac{\partial^3 u_2}{\partial x_1^2 \partial x_3}=\frac{\partial^3 u_2}{\partial x_2^2 \partial x_3}=\frac{\partial^3 u_2}{\partial x_3^3}=\frac{\partial^3 u_2}{\partial x_1 \partial x_2 \partial x_3}=\frac{\partial^3 u_2}{\partial x_1 \partial x_2^2}=0\,, \\
\frac{\partial^3 u_2}{\partial x_1 \partial x_3^2}=\frac{\partial^3 u_2}{\partial x_1^3}=0\,, \\
\frac{\partial^3 u_3}{\partial x_1^2 \partial x_2}=\frac{\partial^3 u_3}{\partial x_2 \partial x_3^2}=\frac{\partial^3 u_3}{\partial x_2^3}=\frac{\partial^3 u_3}{\partial x_3^3}=0\,, \\
\frac{\partial^3 u_3}{\partial x_1 \partial x_2^2}=\frac{\partial^3 u_3}{\partial x_1^3}=\frac{\partial^3 u_3}{\partial x_1 \partial x_3^2}=\frac{\partial^3 u_3}{\partial x_1 \partial x_2 \partial x_3}=0\,. 
\end{dcases}
\end{equation}
From the components of the sixth-order tensor we obtain the following fourth-order universality PDEs:
\begin{equation}
\begin{dcases}
\label{orthotropic-D2-sixthorder}
\frac{\partial^4 u_1}{\partial x_1^4} = \frac{\partial^4 u_1}{\partial x_2^4} = \frac{\partial^4 u_1}{\partial x_2^2 \partial x_3^2} = \frac{\partial^4 u_1}{\partial x_3^4} = \frac{\partial^4 u_1}{\partial x_1^2 \partial x_2^2} = 0\,, \\
\frac{\partial^4 u_1}{\partial x_1^2 \partial x_3^2} = \frac{\partial^4 u_1}{\partial x_1^3 \partial x_3} = \frac{\partial^4 u_1}{\partial x_1 \partial x_3^3} = \frac{\partial^4 u_1}{\partial x_1 \partial x_2 \partial x_3^2} 
= \frac{\partial^4 u_1}{\partial x_1 \partial x_2^2 \partial x_3} = 0\,, \\
\frac{\partial^4 u_2}{\partial x_1^3 \partial x_3} = \frac{\partial^4 u_2}{\partial x_1 \partial x_2^3} = \frac{\partial^4 u_2}{\partial x_1 \partial x_2 \partial x_3^2} = \frac{\partial^4 u_2}{\partial x_1^2 \partial x_2^2} 
= \frac{\partial^4 u_2}{\partial x_2^4} = 0\,, \\
\frac{\partial^4 u_2}{\partial x_1^4} = \frac{\partial^4 u_2}{\partial x_2^2 \partial x_3^2} = \frac{\partial^4 u_2}{\partial x_3^4} = \frac{\partial^4 u_2}{\partial x_1^2 \partial x_2 \partial x_3} = 0\,, \\
\frac{\partial^4 u_2}{\partial x_2^3 \partial x_3} = \frac{\partial^4 u_2}{\partial x_2 \partial x_3^3} = 0\,, \\
\frac{\partial^4 u_3}{\partial x_1^3 \partial x_3} = \frac{\partial^4 u_3}{\partial x_1 \partial x_3^3} = \frac{\partial^4 u_3}{\partial x_1 \partial x_2 \partial x_3^2} = \frac{\partial^4 u_3}{\partial x_1 \partial x_2^2 \partial x_3} 
= \frac{\partial^4 u_3}{\partial x_1^2 \partial x_2 \partial x_3} = 0\,, \\
\frac{\partial^4 u_3}{\partial x_2^3 \partial x_3} = \frac{\partial^4 u_3}{\partial x_1^2 \partial x_3^2} = \frac{\partial^4 u_3}{\partial x_3^4} = \frac{\partial^4 u_3}{\partial x_2 \partial x_3^3} = \frac{\partial^4 u_3}{\partial x_1^4} = 0\,, \\
\frac{\partial^4 u_3}{\partial x_1^2 \partial x_2^2} = \frac{\partial^4 u_3}{\partial x_2^4} = 0\,. 
\end{dcases}
\end{equation}
Substituting the superposition of homogeneous displacement fields and the three-parameter inhomogeneous displacement field $(\alpha_1 x_2 x_3\,,\, \alpha_2 x_1 x_3\,,\, \alpha_3 x_1 x_2)$ into the above third- and fourth-order PDEs, we find that they are satisfied trivially. This yields the following result.
\begin{prop}
The only universal displacements in orthotropic $\mathbb{D}_2$ class linear strain-gradient elastic solids are the superposition of homogeneous displacement fields and the following three-parameter inhomogeneous displacement field: $(\alpha_1 x_2 x_3\,,\, \alpha_2 x_1 x_3\,,\, \alpha_3 x_1 x_2)$.
\end{prop}

\subsubsection{Orthotropic class $\mathbb{D}_2 \oplus \mathbb{Z}_2^c$} 

Tensor $\boldsymbol{\mathsf{A}}_{\mathbb{D}_2 \oplus \mathbb{Z}_2^c}$ has $51$ independent components and has the same form as that of class $\mathbb{D}_2$. Tensor $\boldsymbol{\mathsf{M}}_{\mathbb{D}_2 \oplus \mathbb{Z}_2^c}$ is null. Tensor $\boldsymbol{\mathsf{C}}_{\mathbb{D}_2 \oplus \mathbb{Z}_2^c}$ has the same form as that of class $\mathbb{D}_2$, with $9$ independent components, with the only difference that it corresponds to the Hermann--Maugin symbol $mmm$. This yields the following result:
\begin{prop}
The only universal displacements in orthotropic $\mathbb{D}_2 \oplus \mathbb{Z}_2^c$ class linear strain-gradient elastic solids are the superposition of homogeneous displacement fields and the following three-parameter inhomogeneous displacement field: $(\alpha_1 x_2 x_3\,,\, \alpha_2 x_1 x_3\,,\, \alpha_3 x_1 x_2)$.
\end{prop}

\subsubsection{Orthotropic class $\mathbb{Z}_4^-$} 

Tensor $\boldsymbol{\mathsf{A}}_{\mathbb{Z}_4 \oplus \mathbb{Z}_2^c}$ has $45$ independent components and has the same form as that of class $\mathbb{Z}_4$. Tensor $\boldsymbol{\mathsf{M}}_{\mathbb{Z}_4^-}$ has $26$ independent components and has the form
\begin{equation}
	\label{M_Z_a^-}
	\boldsymbol{\mathsf{M}}_{\mathbb{Z}_4^-} = 
	\begin{bmatrix}
		0 & 0 & \bar{C}^{(7)} & \bar{D}^{(5)} \\
		\bar{E}^{(5)} & \bar{F}^{(5)} & 0 & 0 \\
		-\bar{F}^{(5)} & \bar{E}^{(5)} & 0 & 0 \\
		0 & 0 & \bar{O}^{(3)} & \bar{P}^{(1)}
	\end{bmatrix}\,,
\end{equation}
where
\begin{equation}
	\label{barC^7-barO^3}
	\bar{C}^{(7)} \in 
	\begin{bmatrix}
		\bar{c}_{11} & \bar{c}_{12} & \bar{c}_{13} & \bar{c}_{14} & \bar{c}_{15} \\
		-\bar{c}_{11} & -\bar{c}_{14} & -\bar{c}_{15} & -\bar{c}_{12} & -\bar{c}_{13} \\
		\bar{c}_{31} & \bar{c}_{32} & \bar{c}_{33} & -\bar{c}_{32} & -\bar{c}_{33}
	\end{bmatrix}\,,\qquad
	\bar{O}^{(3)} \in 
	\begin{bmatrix}
		\bar{o}_{11} & \bar{o}_{12} & \bar{o}_{13} & \bar{o}_{12} & \bar{o}_{13}
	\end{bmatrix}\,,
\end{equation}
and
\begin{equation}
	\label{barD^5}
	\bar{D}^{(5)} \in 
	\begin{bmatrix}
		\bar{d}_{11} & \bar{d}_{12} & \bar{d}_{13} \\
		\bar{d}_{12} & \bar{d}_{11} & \bar{d}_{13} \\
		\bar{d}_{31} & \bar{d}_{31} & \bar{d}_{33}
	\end{bmatrix}\, ,\qquad
	\bar{P}^{(1)} \in 
	\begin{bmatrix}
		\bar{p}_{11} & -\bar{p}_{11} & 0
	\end{bmatrix}\,.
\end{equation}
Tensor $\boldsymbol{\mathsf{C}}_{\mathbb{D}_4 \oplus \mathbb{Z}_2^c}$ has $6$ independent components and corresponds to the Hermann--Maugin symbol $4mm$. The classical linear elastic part coincides with the tetragonal case, so we obtain constraints of the form \eqref{tetragonal-linear-constraints}, and the candidate universal displacements are the superposition of a homogeneous displacement field and the inhomogeneous field \eqref{tetragonal-linear-inhomogeneous}.

The third-order universality PDEs induced by the fifth-order tensor read
\begin{equation}
\begin{dcases}
\label{Z^-_4-third-order-constraints}
\frac{\partial^3 u_2}{\partial x_1^2 \partial x_3} = \frac{\partial^3 u_3}{\partial x_1^2 \partial x_2} = \frac{\partial^3 u_3}{\partial x_2 \partial x_3^2} = \frac{\partial^3 u_3}{\partial x_1 \partial x_3^2} = \frac{\partial^3 u_1}{\partial x_1^2 \partial x_3} = 0\,, \\
\frac{\partial^3 u_3}{\partial x_1^3} = \frac{\partial^3 u_2}{\partial x_1 \partial x_2 \partial x_3} = \frac{\partial^3 u_3}{\partial x_1 \partial x_2^2} = \frac{\partial^3 u_3}{\partial x_2^3} = \frac{\partial^3 u_3}{\partial x_3^3} = 0\,, \\
\frac{\partial^3 u_2}{\partial x_3^3} = \frac{\partial^3 u_1}{\partial x_2^2 \partial x_3} = \frac{\partial^3 u_1}{\partial x_3^3} = \frac{\partial^3 u_1}{\partial x_1 \partial x_2 \partial x_3} = \frac{\partial^3 u_3}{\partial x_1 \partial x_2 \partial x_3} = 0\,, \\
\frac{\partial^3 u_3}{\partial x_1^2 \partial x_3} = \frac{\partial^3 u_3}{\partial x_2 \partial x_3^2} = 0\,, \\
\frac{\partial^3 u_2}{\partial x_1 \partial x_2^2} + \frac{\partial^3 u_1}{\partial x_1^2 \partial x_2} = 0\,, \\
-\frac{\partial^3 u_2}{\partial x_2 \partial x_3^2} + \frac{\partial^3 u_1}{\partial x_1 \partial x_3^2} = 0\,, \\
-\frac{\partial^3 u_2}{\partial x_2^3} + \frac{\partial^3 u_1}{\partial x_1^3} = 0\,, \\ 
\frac{\partial^3 u_1}{\partial x_1 \partial x_2^2} - \frac{\partial^3 u_2}{\partial x_1^2 \partial x_2} = 0\,, \\
\frac{\partial^3 u_2}{\partial x_1 \partial x_2^2} + \frac{\partial^3 u_1}{\partial x_1^2 \partial x_2} = 0\,, \\
\frac{\partial^3 u_1}{\partial x_2^3} + \frac{\partial^3 u_2}{\partial x_1^3} = 0\,, \\
\frac{\partial^3 u_1}{\partial x_2 \partial x_3^2} + \frac{\partial^3 u_2}{\partial x_1 \partial x_3^2} = 0\,. 
\end{dcases}
\end{equation}
The fourth-order universality PDEs induced by the sixth-order tensor are
\begin{equation}
\begin{dcases}
\label{Z^-_4-fourth-order-constraints}
\frac{\partial^4 u_1}{\partial x_1^4} = \frac{\partial^4 u_3}{\partial x_1^3 \partial x_3} = \frac{\partial^4 u_2}{\partial x_1 \partial x_2^3} = \frac{\partial^4 u_3}{\partial x_1 \partial x_2^2 \partial x_3} = \frac{\partial^4 u_1}{\partial x_2^2 \partial x_3^2} = 0\,, \\
\frac{\partial^4 u_3}{\partial x_1 \partial x_2 \partial x_3^2} = \frac{\partial^4 u_3}{\partial x_1 \partial x_3^3} = \frac{\partial^4 u_1}{\partial x_3^4} = \frac{\partial^4 u_2}{\partial x_1^4} = \frac{\partial^4 u_2}{\partial x_1^2 \partial x_3^2} = 0\,, \\
\frac{\partial^4 u_3}{\partial x_1^2 \partial x_2 \partial x_3} = \frac{\partial^4 u_1}{\partial x_1 \partial x_2 \partial x_3^2} = \frac{\partial^4 u_2}{\partial x_3^4} = \frac{\partial^4 u_1}{\partial x_1^2 \partial x_3^2} = \frac{\partial^4 u_2}{\partial x_1 \partial x_2 \partial x_3^2} = 0\,, \\
\frac{\partial^4 u_2}{\partial x_2^2 \partial x_3^2} = \frac{\partial^4 u_2}{\partial x_1^4} = \frac{\partial^4 u_1}{\partial x_1^3 \partial x_2} = \frac{\partial^4 u_3}{\partial x_2^2 \partial x_3^2} = \frac{\partial^4 u_3}{\partial x_2^3 \partial x_3} = 0\,, \\
\frac{\partial^4 u_3}{\partial x_2^2 \partial x_3^2} = \frac{\partial^4 u_2}{\partial x_1^2 \partial x_3^2} = \frac{\partial^4 u_3}{\partial x_3^4} = \frac{\partial^4 u_3}{\partial x_2 \partial x_3^3} = \frac{\partial^4 u_3}{\partial x_1^3 \partial x_3} = 0\,, \\
\frac{\partial^4 u_3}{\partial x_1^2 \partial x_2^2} = \frac{\partial^4 u_2}{\partial x_1^2 \partial x_2 \partial x_3} = \frac{\partial^4 u_1}{\partial x_1 \partial x_2^2 \partial x_3}=0\,, \\
\frac{\partial^4 u_1}{\partial x_1^2 \partial x_2^2} + \frac{\partial^4 u_2}{\partial x_1^3 \partial x_2} = 0\,,  \frac{\partial^4 u_1}{\partial x_2^4} + 2 \frac{\partial^4 u_1}{\partial x_1^2 \partial x_2^2} = 0\,, \\
\frac{\partial^4 u_1}{\partial x_1 \partial x_2^3} + 3 \frac{\partial^4 u_2}{\partial x_1^2 \partial x_2}= 0\,,  \frac{\partial^4 u_1}{\partial x_1 \partial x_2^3} + \frac{\partial^4 u_2}{\partial x_1^2 \partial x_2^2} = 0\,, \\
2 \frac{\partial^4 u_2}{\partial x_1^2 \partial x_2^2} + \frac{\partial^4 u_2}{\partial x_1^4} = 0\,,  \frac{\partial^4 u_2}{\partial x_1 \partial x_2^3} + \frac{\partial^4 u_1}{\partial x_1^4 } = 0\,, \\
- \frac{\partial^4 u_1}{\partial x_2^4} - 3  \frac{\partial^4 u_2}{\partial x_1 \partial x_2^3} = 0\,,  - \frac{\partial^4 u_3}{\partial x_1 \partial x_2^2 \partial x_3} +  \frac{\partial^4 u_1}{\partial x_1^2 \partial x_2^3} = 0\,, \\
- 3 \frac{\partial^4 u_1}{\partial x_1^2 \partial x_2^2} -  \frac{\partial^4 u_2}{\partial x_1^3 \partial x_2} = 0\,,  \frac{\partial^4 u_2}{\partial x_2^3 \partial x_3} + \frac{\partial^4 u_1}{\partial x_1^3 \partial x_3} = 0\,, \\
\frac{\partial^4 u_2}{\partial x_2 \partial x_3^3} + \frac{\partial^4 u_1}{\partial x_1 \partial x_3^3} = 0\,,  -\frac{\partial^4 u_2}{\partial x_1 \partial x_2^2 \partial x_3} + \frac{\partial^4 u_1}{\partial x_1^2 \partial x_2 \partial x_3} = 0\,, \\
\frac{\partial^4 u_1}{\partial x_2^3 \partial x_3} - \frac{\partial^4 u_2}{\partial x_1^3 \partial x_3} = 0\,,  \frac{\partial^4 u_3}{\partial x_2^4} + \frac{\partial^4 u_3}{\partial x_1^4} = 0\,, \\
\frac{\partial^4 u_2}{\partial x_1 \partial x_2^3} + \frac{\partial^4 u_3}{\partial x_1 \partial x_2 \partial x_3^2} = 0\,,  \frac{\partial^4 u_2}{\partial x_1^3 \partial x_3} + \frac{\partial^4 u_3}{\partial x_1^3 \partial x_2} = 0\,, \\
\frac{\partial^4 u_1}{\partial x_2^3 \partial x_3} -3 \frac{\partial^4 u_2}{\partial x_1 \partial x_2^2 \partial x_3} = 0\,, \\
-\frac{\partial^4 u_1}{\partial x_2^2 \partial x_3} - \frac{\partial^4 u_3}{\partial x_1 \partial x_2^3} + 3 \frac{\partial^4 u_1}{\partial x_1^2 \partial x_2 \partial x_3} + \frac{\partial^4 u_2}{\partial x_1^3 \partial x_3} = 0\,, \\
\frac{\partial^4 u_2}{\partial x_1 \partial x_2^2 \partial x_3} + \frac{\partial^4 u_3}{\partial x_1 \partial x_2^3} = 0\,,  \frac{\partial^4 u_1}{\partial x_1^2 \partial x_2 \partial x_3} + \frac{\partial^4 u_3}{\partial x_1^3 \partial x_2} = 0\,. 
\end{dcases}
\end{equation}

Substituting the tetragonal linear elastic candidate displacement field \eqref{tetragonal-linear-inhomogeneous} into the above third- and fourth-order constraints, we obtain the following additional constraints on the function $g(x_1, x_2)$:
\begin{equation}\label{Z^-_4-constraints-for-g}
\begin{aligned}
	&\frac{\partial^3 g}{\partial x_1^3}
	=\frac{\partial^3 g}{\partial x_1 \partial x_2^2}
	=\frac{\partial^3 g}{\partial x_2^3}
	=\frac{\partial^3 g}{\partial x_1^2 \partial x_2}
	=\frac{\partial^4 g}{\partial x_1^2 \partial x_2^2}
	=\frac{\partial^4 g}{\partial x_1 \partial x_2^3}
	=\frac{\partial^4 g}{\partial x_1^3 \partial x_2}
	=0\,,\\
	&\frac{\partial^4 g}{\partial x_1^4}+\frac{\partial^4 g}{\partial x_2^4}=0\,,\\
	&\frac{\partial^2 g}{\partial x_1^2}+\frac{\partial^2 g}{\partial x_2^2}=0\,.
\end{aligned}
\end{equation}
After straightforward manipulations, these constraints yield $g(x_1, x_2)$ in the form
\begin{equation} \label{Z^-_4-g-final-form}
	g(x_1, x_2)=c_1 +c_2 x_1 +c_3 x_2 +c_4 x_1 x_2 +c_5(x^2_2-x_1^2)\,.
\end{equation}
This gives the following characterization:
\begin{prop}
The universal displacements in trigonal $\mathbb{Z}_4^-$ class linear strain-gradient elastic solids are the superposition of homogeneous displacement fields and the inhomogeneous displacement field \eqref{tetragonal-linear-inhomogeneous}, with $g(x_1, x_2)$ given by \eqref{Z^-_4-g-final-form}.
\end{prop}

\subsubsection{Orthotropic class $\mathbb{D}_3 ^v$} 

Tensor $\boldsymbol{\mathsf{A}}_{\mathbb{D}_3 \oplus \mathbb{Z}_2^c}$ has $34$ independent components and has the same form as that of class $\mathbb{D}_3$. Tensor $\boldsymbol{\mathsf{M}}_{\mathbb{D}_3^v}$ has $20$ independent components and has the same form as \eqref{M_D_3^v}. Tensor $\boldsymbol{\mathsf{C}}_{\mathbb{D}_3 \oplus \mathbb{Z}_2^c}$ has $6$ independent components and corresponds to the Hermann--Maugin symbol $3m$. The classical linear elastic part coincides with the trigonal case, so we have the universality constraints \eqref{trigonal-linear-constraints}, and the candidate universal displacements are the superposition of homogeneous displacement fields and the inhomogeneous field \eqref{trigonal-linear-inhomogeneous}.

The third-order universality constraints for this case read
\begin{equation}
\begin{dcases}
\label{D^v_3-third-order-constraints-1}
-\frac{\partial^3 u_2}{\partial x_2^3}+\frac{1}{\sqrt{2}}
\left(2\frac{\partial^3 u_2}{\partial x_1 \partial x_2^2}
+\frac{\partial^3 u_1}{\partial x_1^2 \partial x_2}+\frac{\partial^3 u_2}{\partial x_1^3} \right) =0 \,,\\
\frac{1}{\sqrt{2}} \frac{\partial^3 u_1}{\partial x_2^3}
+\sqrt{2} \frac{\partial^3 u_2}{\partial x_1 \partial x_2^2}
+\left(1-\frac{1}{\sqrt{2}}\right) \frac{\partial^3 u_1}{\partial x_1^2 \partial x_2}=0 \,,\\
-\frac{1}{2}\left(2+\sqrt{2}\right) \frac{\partial^3 u_1}{\partial x_2^3}
-\left(1+\sqrt{2}\right) \frac{\partial^3 u_2}{\partial x_1 \partial x_2^2}
+\frac{1}{\sqrt{2}}
\left(2 \frac{\partial^3 u_1}{\partial x_1^2 \partial x_2}+3\frac{\partial^3 u_2}{\partial x_1^3}\right)=0
\,,\\
\frac{\partial^3 u_1}{\partial x_2 \partial x_3^2} = \frac{\partial^3 u_2}{\partial x_1 \partial x_3^2}=0\,,\qquad
\frac{\partial^3 u_3}{\partial x_1 \partial x_2 \partial x_3}
= \frac{\partial^3 u_3}{\partial x_1 \partial x_3^2}=0 \,,\\
\frac{\partial^3 u_1}{\partial x_1^2 \partial x_3}=0\,,\qquad
\sqrt{2} \frac{\partial^3 u_3}{\partial x_1 \partial x_2^2}+\frac{\partial^3 u_3}{\partial x_1^3}=0\,,\qquad
\frac{\partial^3 u_1}{\partial x_2^2 \partial x_3}=0 \,,\\
\frac{\partial^3 u_2}{\partial x_1 \partial x_2 \partial x_3}=0\,,\qquad
\frac{\partial^3 u_3}{\partial x_1 \partial x_2^2}=0\,,\qquad
\frac{\partial^3 u_3}{\partial x_3^3}=0 \,, \qquad \frac{\partial^3 u_1}{\partial x_3^3}=0 \,.
\end{dcases}
\end{equation}
and
\begin{equation}
\begin{dcases}
\label{D^v_3-third-order-constraints-2}
\frac{\partial^3 u_3}{\partial x_2 \partial x_3^2}=\frac{\partial^3 u_2}{\partial x_1^2 \partial x_3}=\frac{\partial^3 u_3}{\partial x_1^2 \partial x_2}=0 \,, \\
\frac{\partial^3 u_3}{\partial x_2^3}=\frac{\partial^3 u_3}{\partial x_3^3}=\frac{\partial^3 u_2}{\partial x_3^3}=0 \,, \\
\frac{1}{\sqrt{2}} \frac{\partial^3 u_1}{\partial x_1 \partial x_2^2}-\frac{1}{2}\left(-2+\sqrt{2}\right)\frac{\partial^3 u_2}{\partial x_1^2 \partial x_2}-\frac{2}{\sqrt{2}} \frac{\partial^3 u_1}{\partial x_1^3}=0 \,, \\
-\frac{1}{\sqrt{2}} \frac{\partial^3 u_2}{\partial x_1^2 \partial x_2} -\frac{1}{2}\left(-2+\sqrt{2}\right)\frac{\partial^3 u_1}{\partial x_1^3}=0 \,, \\
\left(2+\frac{1}{\sqrt{2}}\right)\frac{\partial^3 u_1}{\partial x_1 \partial x_2^2}+\frac{2}{\sqrt{2}} \frac{\partial^3 u_2}{\partial x_1^2 \partial x_2}- \frac{1}{\sqrt{2}}\frac{\partial^3 u_1}{\partial x_1^3}=0 \,, \\
-\frac{\partial^3 u_2}{\partial x_2 \partial x_3^2}+\frac{\partial^3 u_1}{\partial x_1 \partial x_3^2}-\sqrt{2} \frac{\partial^3 u_3}{\partial x_1^2 \partial x_3}=0 \,, \\
-\frac{\partial^3 u_2}{\partial x_2 \partial x_3^2}-\frac{\partial^3 u_3}{\partial x_2^2 \partial x_3}-\sqrt{2}\left(\frac{\partial^3 u_1}{\partial x_1 \partial x_3^2}+\frac{\partial^3 u_3}{\partial x_1^2 \partial x_3}\right)=0 \,, \\
\left(1-\frac{1}{\sqrt{2}}\right)\frac{\partial^3 u_2}{\partial x_1^2 \partial x_3}-\left(-1+\sqrt{2}\right)\frac{\partial^3 u_3}{\partial x_1^2 \partial x_2}=0 \,, \\
3 \left(\frac{\partial^3 u_2}{\partial x_2 \partial x_3^2}+\frac{\partial^3 u_3}{\partial x_2^2 \partial x_3}\right)+\frac{\partial^3 u_1}{\partial x_1 \partial x_3^2}+\frac{\partial^3 u_3}{\partial x_1^2 \partial x_3}=0 \,, \\
\frac{\partial^3 u_2}{\partial x_2 \partial x_3^2}-\frac{\partial^3 u_1}{\partial x_1 \partial x_2^3}=0\,,
\end{dcases}
\end{equation}
as well as
\begin{equation}
\begin{dcases}
\label{D^v_3-third-order-constraints-2}
\frac{\partial^3 u_1}{\partial x_1 \partial x_2 \partial x_3}=\frac{\partial^3 u_2}{\partial x_2^2 \partial x_3}=0 \,, \\
\frac{\partial^3 u_2}{\partial x_2 \partial x_3^2}+\frac{\partial^3 u_1}{\partial x_1 \partial x_3^2}=0 \,, \\
\frac{\partial^3 u_2}{\partial x_2^3}+\frac{\partial^3 u_1}{\partial x_1^3}=0 \,, \\
\frac{\partial^3 u_1}{\partial x_1 \partial x_2^2}+\frac{\partial^3 u_2}{\partial x_1^2 \partial x_2}=0 \,, \\
\frac{\partial^3 u_3}{\partial x_2^2 \partial x_3}+\frac{\partial^3 u_3}{\partial x_1^2 \partial x_3}=0 \,, \\
\frac{\partial^3 u_2}{\partial x_2^3}+\frac{\partial^3 u_1}{\partial x_1^3}=0 \,, \\
\frac{\partial^3 u_3}{\partial x_2 \partial x_3^2}+\frac{\partial^3 u_3}{\partial x_1^2 \partial x_3}=0 \,. 
\end{dcases}
\end{equation}

The fourth-order universality constraints for this case are written as
\begin{equation}
\begin{dcases}
\label{D^v_3-fourth-order-constraints-1}
\frac{\partial^4 u_3}{\partial x_1 \partial x_3^3}=\frac{\partial^4 u_1}{\partial x_3^4}=\frac{\partial^4 u_2}{\partial x_1 \partial x_3^3}=\frac{\partial^4 u_1}{\partial x_2 \partial x_3^3}=0 \,, \\
\frac{1}{\sqrt{2}} \frac{\partial^4 u_1}{\partial x_2^4}+\frac{1}{\sqrt{2}}\frac{\partial^4 u_2}{\partial x_1 \partial x_2^3}+\left(1+\sqrt{2}\right)\frac{\partial^4 u_1}{\partial x_1^2 \partial x_2^2}+\frac{\partial^4 u_2}{\partial x_1^3 \partial x_2}+\frac{\partial^4 u_1}{\partial x_1^4}=0 \,, \\
\frac{\partial^4 u_1}{\partial x_1^2 \partial x_2^2}+\frac{\partial^4 u_2}{\partial x_1^3 \partial x_2}=0 \,, \\
\frac{\partial^4 u_2}{\partial x_1 \partial x_2 \partial x_3^2}+\frac{\partial^4 u_3}{\partial x_1 \partial x_2^2 \partial x_3}+\frac{\partial^4 u_1}{\partial x_1^2 \partial x_3^2}+\frac{\partial^4 u_3}{\partial x_1^3 \partial x_3}=0 \,, \\
\frac{\partial^4 u_1}{\partial x_2^2 \partial x_3^2}+\frac{\partial^4 u_3}{\partial x_1 \partial x_2^2 \partial x_3}+\frac{\partial^4 u_1}{\partial x_1^2 \partial x_3^2}+\frac{\partial^4 u_3}{\partial x_1^3 \partial x_3}=0 \,, \\
-\frac{1}{\sqrt{2}}\frac{\partial^4 u_1}{\partial x_2^2}+\left(1-\frac{1}{\sqrt{2}}\right)\frac{\partial^4 u_2}{\partial x_1 \partial x_2^3}-\left(-1+\sqrt{2}\right) \frac{\partial^4 u_1}{\partial x_1^2 \partial x_2^2}+\frac{\partial^4 u_2}{\partial x_1^3 \partial x_2}=0  \,, \\
\frac{\partial^4 u_1}{\partial x_2^2 \partial x_3^2}-\left(-1+\sqrt{2}\right)\frac{\partial^4 u_2}{\partial x_1 \partial x_2 \partial x_3^2}+2 \frac{\partial^4 u_3}{\partial x_1 \partial x_2 \partial x_3^2}-\sqrt{2}\frac{\partial^4 u_3}{\partial x_1 \partial x_2^2 \partial x_3}=0 \,, \\
-\sqrt{2}\left(\frac{\partial^4 u_1}{\partial x_2^2 \partial x_3^2}+\frac{\partial^4 u_3}{\partial x_1 \partial x_2^2 \partial x_3}\right)+2\left(\frac{\partial^4 u_2}{\partial x_1 \partial x_2 \partial x_3^2}+\frac{\partial^4 u_3}{\partial x_1 \partial x_2^2 \partial x_3}\right)=0 \,, \\
-\sqrt{2}\frac{\partial^4 u_1}{\partial x_2^2 \partial x_3^2}-\left(-4+\sqrt{2}\right)\frac{\partial^4 u_2}{\partial x_1\partial x_2^2 \partial x_3}+\frac{\partial^4 u_1}{\partial x_1 \partial x_2^3}-\frac{\partial^4 u_1}{\partial x_1^2 \partial x_2 \partial x_3}=0 \,, \\
\frac{\partial^4 u_2}{\partial x_1 \partial x_2^2 \partial x_3}-\frac{\partial^4 u_3}{\partial x_1 \partial x_2^3}-\frac{\partial^4 u_3}{\partial x_1^3 \partial x_3}+\frac{\partial^4 u_3}{\partial x_1^3 \partial x_2}=0 \,, \\
\frac{\partial^4 u_1}{\partial x_2^3 \partial x_3}+\left(1-\frac{3}{\sqrt{2}}\right) \frac{\partial^4 u_2}{\partial x_1 \partial x_2^2 \partial x_3}+\frac{1}{\sqrt{2}}\left(\frac{\partial^4 u_1}{\partial x_1^2 \partial x_2 \partial x_3}+\frac{\partial^4 u_2}{\partial x_1^3 \partial x_3}+\frac{\partial^4 u_3}{\partial x_1^3 \partial x_2}\right)=0 \,, \\
2\frac{\partial^4 u_1}{\partial x_2^3 \partial x_3}+\left(5-2\sqrt{2}\right)\frac{\partial^4 u_2}{\partial x_1 \partial x_2^2 \partial x_3} \\
-\left(-1+2\sqrt{2}\right)\left(\frac{\partial^4 u_1}{\partial x_1 \partial x_2^3}+2\frac{\partial^4 u_1}{\partial x_1 ^2 \partial x_2 \partial x_3}+\frac{\partial^4 u_2}{\partial x_1^3 \partial x_3}+\frac{\partial^4 u_3}{\partial x_1^3 \partial x_2}\right)=0 \,, \\
-\frac{\partial^4 u_1}{\partial x_2^3 \partial x_3}-\frac{1}{2}\left(2+\sqrt{2}\right)\frac{\partial^4 u_2}{\partial x_1 \partial x_2^2 \partial x_3}-\frac{5}{\sqrt{2}}\left(5\frac{\partial^4 u_1}{\partial x_1 ^2 \partial x_2 \partial x_3}+\frac{\partial^4 u_2}{\partial x_1^3 \partial x_3}+\frac{\partial^4 u_3}{\partial x_1^3 \partial x_2}\right)=0 \,, \\
\frac{\partial^4 u_1}{\partial x_2^3 \partial x_3}+\frac{\partial^4 u_2}{\partial x_1 \partial x_2^2 \partial x_3}-\left(1+4\sqrt{2}\right)\frac{\partial^4 u_1}{\partial x_1 ^2 \partial x_2 \partial x_3}+\left(1+2\sqrt{2}\right)\frac{\partial^4 u_2}{\partial x_1^3 \partial x_3}-2\sqrt{2}\frac{\partial^4 u_3}{\partial x_1^3 \partial x_2}=0 \,, \\
4 \frac{\partial^4 u_1}{\partial x_2^3 \partial x_3}+\left(3+2\sqrt{2}\right)\frac{\partial^4 u_2}{\partial x_1 \partial x_2^2 \partial x_3}+\left(1+2\sqrt{2}\right)\frac{\partial^4 u_3}{\partial x_1 ^2 \partial x_2^3} \\
-4\left(-1+\sqrt{2}\right)\frac{\partial^4 u_1}{\partial x_1^2 \partial x_2 \partial x_3}-\left(1+2\sqrt{2}\right)\frac{\partial^4 u_2}{\partial x_1^3 \partial x_3}+\left(1-2\sqrt{2}\right)\frac{\partial^4 u_3}{\partial x_1^3 \partial x_2}=0 \,, \\
\frac{\partial^4 u_2}{\partial x_1 \partial x_2 \partial x_3^2}+\frac{\partial^4 u_1}{\partial x_1^2 \partial x_3^2}=0 \,, \\
\frac{\partial^4 u_1}{\partial x_1^2 \partial x_3^2}+\frac{\partial^4 u_3}{\partial x_1^3 \partial x_3}=0 \,, \\
2\frac{\partial^4 u_1}{\partial x_1^2 \partial x_3^2}-\left(-1+2\sqrt{2}\right)\frac{\partial^4 u_2}{\partial x_1 \partial x_2 \partial x_3^2}+\frac{\partial^4 u_3}{\partial x_1 \partial x_2^2 \partial x_3}-\sqrt{2}\left(\frac{\partial^4 u_1}{\partial x_1^2 \partial x_3^2}+\frac{\partial^4 u_3}{\partial x_1^3 \partial x_3}\right)=0 \,, \\
\frac{\partial^4 u_1}{\partial x_1 \partial x_2 \partial x_3^2}+\frac{\partial^4 u_3}{\partial x_1 \partial x_2^2 \partial x_3}=0 \,, \\
\sqrt{2}\frac{\partial^4 u_1}{\partial x_2^4}+\left(1+\sqrt{2}\right)\frac{\partial^4 u_2}{\partial x_1 \partial x_2^3}-\left(1+\sqrt{2}\right)\frac{\partial^4 u_1}{\partial x_1^2 \partial x_2^2}-\left(1+2\sqrt{2}\right)\frac{\partial^4 u_2}{\partial x_1^3 \partial x_2}+\frac{\partial^4 u_1}{\partial x_1^4}=0 \,,
\end{dcases}
\end{equation}
and
\begin{equation}
\begin{dcases}
\label{D^v_3-fourth-order-constraints-2}
\frac{\partial^4 u_1}{\partial x_1 \partial x_2 \partial x_3^2}=\frac{\partial^4 u_3}{\partial x_3^4}=\frac{\partial^4 u_2}{\partial x_3^4}=\frac{\partial^4 u_3}{\partial x_2 \partial x_3^3}=\frac{\partial^4 u_3}{\partial x_2^3 \partial x_3}=\frac{\partial^4 u_1}{\partial x_1 \partial x_2^3}=0 \,, \\
\frac{\partial^4 u_2}{\partial x_2^4}+\frac{\partial^4 u_1}{\partial x_1 \partial x_2^3}+\left(1+\sqrt{2}\right) \frac{\partial^4 u_2}{\partial x_1^2 \partial x_2^2}+\frac{1}{\sqrt{2}}\frac{\partial^4 u_2}{\partial x_1^4}=0 \,, \\
\frac{\partial^4 u_1}{\partial x_1 \partial x_2^3}+\frac{\partial^4 u_2}{\partial x_1^2 \partial x_2^2}=0 \,, \\
\frac{\partial^4 u_1}{\partial x_1 \partial x_2^3}-\frac{\partial^4 u_2}{\partial x_1^2 \partial x_2^2}-\frac{\partial^4 u_2}{\partial x_1^4}=0 \,, \\
\frac{\partial^4 u_2}{\partial x_2^2 \partial x_3^2}+\frac{\partial^4 u_2}{\partial x_1^2 \partial x_3^2}+\frac{\partial^4 u_3}{\partial x_1^2 \partial x_2 \partial x_3}=0 \,, \\
\frac{\partial^4 u_1}{\partial x_1 \partial x_2^3}-\left(-1+\sqrt{2}\right)\frac{\partial^4 u_2}{\partial x_1^2 \partial x_2^2}+\left(1-\frac{1}{\sqrt{2}}\right)\frac{\partial^4 u_1}{\partial x_1^3 \partial x_2}-\frac{1}{\sqrt{2}}\frac{\partial^4 u_2}{\partial x_1^4} =0 \,, \\
\sqrt{2}\frac{\partial^4 u_3}{\partial x_1^2 \partial x_3^2}+2\frac{\partial^4 u_1}{\partial x_1 \partial x_2 \partial x_3^2}=0 \,, \\
-3\frac{\partial^4 u_2}{\partial x_1^2 \partial x_2 \partial x_3}-2\frac{\partial^4 u_3}{\partial x_1^2 \partial x_2^2}+\frac{\partial^4 u_1}{\partial x_1^3 \partial x_3}=0 \,, \\
\sqrt{2}\left(\frac{\partial^4 u_1}{\partial x_1^2 \partial x_2 \partial x_3}+\frac{\partial^4 u_3}{\partial x_1^2 \partial x_2^2}\right)+\frac{\partial^4 u_1}{\partial x_1^3 \partial x_3}=0 \,, \\
-\frac{\partial^4 u_2}{\partial x_2 \partial x_3^3}+\frac{\partial^4 u_1}{\partial x_1 \partial x_3^3}=0 \,, \\
\frac{\partial^4 u_3}{\partial x_2^2 \partial x_3^2}+\frac{\partial^4 u_3}{\partial x_1^2 \partial x_3^2}=0 \,, \\
\frac{\partial^4 u_1}{\partial x_1 \partial x_2^2 \partial x_3}+\frac{1}{2}\frac{\partial^4 u_2}{\partial x_1^2 \partial x_2 \partial x_3}+\frac{\partial^4 u_3}{\partial x_1^2 \partial x_2^2}-\frac{1}{2}\frac{\partial^4 u_1}{\partial x_1^3 \partial x_3}=0 \,, \\
-\left(1+\sqrt{2}\right)\frac{\partial^4 u_1}{\partial x_1 \partial x_2^2 \partial x_3}-\sqrt{2}\frac{\partial^4 u_2}{\partial x_1^2 \partial x_2 \partial x_3}+\frac{\partial^4 u_1}{\partial x_1^3 \partial x_3}=0 \,, \\
\frac{\partial^4 u_2}{\partial x_2^3 \partial x_3}-\frac{3}{2}\frac{\partial^4 u_2}{\partial x_1^2 \partial x_2 \partial x_3}-\frac{\partial^4 u_3}{\partial x_1^2 \partial x_2^2}-\frac{1}{2}\frac{\partial^4 u_1}{\partial x_1^3 \partial x_3}=0 \,, \\
3\frac{\partial^4 u_2}{\partial x_2^3 \partial x_3}+\frac{\partial^4 u_3}{\partial x_2^4}-\left(1+2\sqrt{2}\right)\left(2\frac{\partial^4 u_1}{\partial x_1 \partial x_2^2 \partial x_3}+\frac{\partial^4 u_2}{\partial x_1^2 \partial x_2 \partial x_3}+\frac{\partial^4 u_3}{\partial x_1^2 \partial x_2^2}\right)=0 \,, \\
-\left(2+2\sqrt{2}\right)\frac{\partial^4 u_1}{\partial x_1 \partial x_2^2 \partial x_3}-\sqrt{2}\frac{\partial^4 u_2}{\partial x_1^2 \partial x_2 \partial x_3}-\left(1+2\sqrt{2}\right)\frac{\partial^4 u_3}{\partial x_1^2 \partial x_2^2}+2\frac{\partial^4 u_1}{\partial x_1^3 \partial x_3}+\frac{\partial^4 u_3}{\partial x_1^4}=0 \,, \\
-\frac{\partial^4 u_2}{\partial x_2 \partial x_3^3}-2\frac{\partial^4 u_3}{\partial x_2 \partial x_3^3}-\frac{\partial^4 u_3}{\partial x_2^2 \partial x_3^2}+\frac{\partial^4 u_1}{\partial x_1 \partial x_3^3}+3 \frac{\partial^4 u_3}{\partial x_1^2 \partial x_3^2}=0 \,, \\
-\frac{\partial^4 u_2}{\partial x_2 \partial x_3^3}-\frac{\partial^4 u_3}{\partial x_2^2 \partial x_3^2}+\frac{\partial^4 u_1}{\partial x_1 \partial x_3^3}+\frac{\partial^4 u_3}{\partial x_1^2 \partial x_3^2}=0 \,, \\
\frac{\partial^4 u_2}{\partial x_2^2 \partial x_3^2}+\frac{\partial^4 u_1}{\partial x_1 \partial x_2 \partial x_3^2}=0 \,,
\end{dcases}
\end{equation}
as well as
\begin{equation}
\begin{dcases}
\label{D^v_3-fourth-order-constraints-3}
\frac{\partial^4 u_2}{\partial x_1^2 \partial x_3^2}=\frac{\partial^4 u_2}{\partial x_1^2 \partial x_2^2}=\frac{\partial^4 u_1}{\partial x_1^3 \partial x_2}=\frac{\partial^4 u_3}{\partial x_1^2 \partial x_2 \partial x_3}=\frac{\partial^4 u_3}{\partial x_3^4}=0 \,, \\
\frac{\partial^4 u_2}{\partial x_2^3 \partial x_3}+\frac{\partial^4 u_1}{\partial x_1 \partial x_2^2 \partial x_3}+\frac{\partial^4 u_2}{\partial x_1^2 \partial x_2 \partial x_3}+\frac{\partial^4 u_1}{\partial x_1^3 \partial x_3}=0 \,, \\
\frac{\partial^4 u_1}{\partial x_1 \partial x_2^2 \partial x_3}+\frac{\partial^4 u_2}{\partial x_1^2 \partial x_2 \partial x_3}=0 \,, \\
\frac{\partial^4 u_3}{\partial x_2 \partial x_3^3}+\frac{\partial^4 u_3}{\partial x_1^2 \partial x_3^2}=0 \,, \\
\frac{\partial^4 u_2}{\partial x_2 \partial x_3^3}+\frac{\partial^4 u_3}{\partial x_2^2 \partial x_3^2}+\frac{\partial^4 u_1}{\partial x_1 \partial x_3^3}+\frac{\partial^4 u_3}{\partial x_1^2 \partial x_3^2}=0 \,, \\
-\left(-1+\sqrt{2}\right)\frac{\partial^4 u_1}{\partial x_1 \partial x_2^3}+\left(-1+\sqrt{2}\right)\frac{\partial^4 u_1}{\partial x_1^3 \partial x_3}-\frac{\partial^4 u_2}{\partial x_1^4}=0 \,, \\
\sqrt{2}\frac{\partial^4 u_1}{\partial x_1 \partial x_2^3}-\frac{\partial^4 u_1}{\partial x_1^3 \partial x_2}=0 \,, \\
-\frac{\partial^4 u_3}{\partial x_2^2 \partial x_3^2}+\frac{\partial^4 u_3}{\partial x_1^2 \partial x_3^2}=0 \,, \\
-\frac{3}{\sqrt{2}} \frac{\partial^4 u_2}{\partial x_2^2 \partial x_3^2}-\frac{1}{\sqrt{2}}\frac{\partial^4 u_3}{\partial x_2^3 \partial x_3}+\sqrt{2}\frac{\partial^4 u_1}{\partial x_1 \partial x_2 \partial x_3^2}=0 \,, \\
\frac{\partial^4 u_2}{\partial x_2^2 \partial x_3^2}-\left(1+\sqrt{2}\right)\frac{\partial^4 u_1}{\partial x_1 \partial x_2 \partial x_3^2}=0 \,, \\
-\left(2+3\sqrt{2}\right) \frac{\partial^4 u_1}{\partial x_1 \partial x_2^3}-\sqrt{2}\frac{\partial^4 u_2}{\partial x_1^4}=0 \,, \\
4 \frac{\partial^4 u_2}{\partial x_2^4}-\left(2+\sqrt{2}\right)\frac{\partial^4 u_1}{\partial x_1 \partial x_2^3}-\sqrt{2}\frac{\partial^4 u_2}{\partial x_1^2 \partial x_2^2}-\left(2+3\sqrt{2}\right)\frac{\partial^4 u_1}{\partial x_1^3 \partial x_2}-3\sqrt{2}\frac{\partial^4 u_2}{\partial x_1^4}=0 \,, \\
\frac{\partial^4 u_2}{\partial x_2^2 \partial x_2^2}+\frac{\partial^4 u_3}{\partial x_2^3 \partial x_3}=0 \,, \\
\frac{\partial^4 u_2}{\partial x_2^3 \partial x_3}+\frac{\partial^4 u_3}{\partial x_2^2 \partial x_3^2}+\frac{\partial^4 u_1}{\partial x_1 \partial x_3^3}=0 \,, \\
\frac{\partial^4 u_2}{\partial x_2^3 \partial x_3}+\frac{\partial^4 u_3}{\partial x_3^4}+\frac{\partial^4 u_1}{\partial x_1 \partial x_2^2 \partial x_3}+\frac{\partial^4 u_2}{\partial x_1^2 \partial x_2 \partial x_3}=0 \,, \\
2\frac{\partial^4 u_3}{\partial x_1^2 \partial x_2^2}+\frac{\partial^4 u_1}{\partial x_1^3 \partial x_3}+\frac{\partial^4 u_3}{\partial x_1^4}=0 \,, \\
\frac{\partial^4 u_2}{\partial x_2^3 \partial x_3}+\frac{\partial^4 u_1}{\partial x_1 \partial x_2^2 \partial x_3}+\frac{\partial^4 u_2}{\partial x_1^2 \partial x_2 \partial x_3}+\frac{\partial^4 u_1}{\partial x_1^3 \partial x_3}=0 \,, \\
\frac{\partial^4 u_1}{\partial x_1 \partial x_2^2 \partial x_3}+\frac{\partial^4 u_2}{\partial x_1^2 \partial x_2 \partial x_3^2}=0 \,, \\
2\frac{\partial^4 u_1}{\partial x_1 \partial x_2^2 \partial x_3}+\frac{\partial^4 u_2}{\partial x_1^2 \partial x_2 \partial x_3}+\frac{\partial^4 u_3}{\partial x_1^2 \partial x_3^2}=0 \,, \\
\sqrt{2}\frac{\partial^4 u_1}{\partial x_1 \partial x_2^2 \partial x_3}+\left(-2+\sqrt{2}\right)\frac{\partial^4 u_2}{\partial x_1^2 \partial x_2 \partial x_3}+2\left(-1+\sqrt{2}\right)\frac{\partial^4 u_3}{\partial x_1^2 \partial x_3^2}=0 \,. 
\end{dcases}
\end{equation}

Substituting the inhomogeneous displacement field \eqref{trigonal-linear-inhomogeneous} into the above universality PDEs, we find that they are satisfied provided that $a_{123}=0$. Thus, we have proved the following result.
\begin{prop}
The universal displacements in orthotropic $\mathbb{D}_3^v$ class linear strain-gradient elastic solids are the superposition of homogeneous displacement fields and the inhomogeneous displacement field \eqref{trigonal-linear-inhomogeneous} with $a_{123}=0$.
\end{prop}

\subsubsection{Orthotropic class $\mathbb{D}_4^h$} 

Tensor $\boldsymbol{\mathsf{A}}_{\mathbb{D}_4 \oplus \mathbb{Z}_2^c}$ has $28$ independent components and has the same form as that of class $\mathbb{D}_4$. Tensor $\boldsymbol{\mathsf{M}}_{\mathbb{D}_4^h}$ has $13$ independent components and has the form
\begin{equation}
	\label{M_D_4^h}
	\boldsymbol{\mathsf{M}}_{\mathbb{D}_4^h} = \begin{bmatrix}
		0 & 0 & \bar{C}^{(7)} & 0 \\
		0 & \bar{F}^{(5)} & 0 & 0 \\
		-\bar{F}^{(5)} & 0 & 0 & 0 \\
		0 & 0 & 0 & \bar{P}^{(1)}
	\end{bmatrix} \,.
\end{equation}
Tensor $\boldsymbol{\mathsf{C}}_{\mathbb{D}_4 \oplus \mathbb{Z}_2^c}$ has $6$ independent components and corresponds to the Hermann--Maugin symbol $4mm$.
The classical linear elastic part coincides with the tetragonal case. Thus, from the classical linear elastic part we have the universality constraints \eqref{tetragonal-linear-constraints}, and the candidate universal displacements are the superposition of a homogeneous displacement field and the inhomogeneous displacement field \eqref{tetragonal-linear-inhomogeneous}.

The third-order universality PDEs induced by the fifth-order tensor read
\begin{equation}
\begin{dcases}
\label{D^h_4-third-order-constraints}
\frac{\partial^3 u_3}{\partial x_1 \partial x_3^2} = \frac{\partial^3 u_1}{\partial x_1^2 \partial x_3} = \frac{\partial^3 u_3}{\partial x_1^3} = \frac{\partial^3 u_2}{\partial x_1 \partial x_2 \partial x_3} = \frac{\partial^3 u_3}{\partial x_1 \partial x_2^2} = 0\,, \\
\frac{\partial^3 u_1}{\partial x_2 \partial x_3^2} = \frac{\partial^3 u_1}{\partial x_3^3} = \frac{\partial^3 u_3}{\partial x_2 \partial x_3^2} = \frac{\partial^3 u_2}{\partial x_2^2 \partial x_3} = \frac{\partial^3 u_3}{\partial x_2^3} = 0\,, \\
\frac{\partial^3 u_1}{\partial x_1 \partial x_2 \partial x_3} = \frac{\partial^3 u_3}{\partial x_1^2 \partial x_2} = \frac{\partial^3 u_2}{\partial x_1^2 \partial x_3} = \frac{\partial^3 u_3}{\partial x_3^3} = \frac{\partial^3 u_2}{\partial x_3^3} = 0\,, \\
\frac{\partial^3 u_2}{\partial x_2 \partial x_3^2} = \frac{\partial^3 u_2}{\partial x_2^3}=\frac{\partial^3 u_1}{\partial x_1 \partial x_2} = \frac{\partial^3 u_1}{\partial x_1 \partial x_2^2}=\frac{\partial^3 u_1}{\partial x_1 \partial x_3^2} = \frac{\partial^3 u_1}{\partial x_1^3} = 0\,, \\
\frac{\partial^3 u_2}{\partial x_1^2 \partial x_2} = \frac{\partial^3 u_3}{\partial x_1^2 \partial x_3}=\frac{\partial^3 u_3}{\partial x_2^2 \partial x_3} = 0\,. 
\end{dcases}
\end{equation}
The fourth-order universality PDEs induced by the sixth-order tensor are
\begin{equation}
\begin{dcases}
\label{D^h_4-fourth-order-constraints}
\frac{\partial^4 u_1}{\partial x_1^4} = \frac{\partial^4 u_3}{\partial x_1^3 \partial x_3} = \frac{\partial^4 u_2}{\partial x_1 \partial x_2^3} = \frac{\partial^4 u_3}{\partial x_1 \partial x_2^2 \partial x_3} = \frac{\partial^4 u_1}{\partial x_2^2 \partial x_3^2} = 0\,, \\
\frac{\partial^4 u_3}{\partial x_1 \partial x_2 \partial x_3^2} = \frac{\partial^4 u_3}{\partial x_1 \partial x_3^3} = \frac{\partial^4 u_1}{\partial x_1^2 \partial x_3^2} = \frac{\partial^4 u_2}{\partial x_1 \partial x_2 \partial x_3^2} = 0\,, \\
\frac{\partial^4 u_2}{\partial x_2^2} = \frac{\partial^4 u_1}{\partial x_1^3 \partial x_2} = \frac{\partial^4 u_3}{\partial x_2^2 \partial x_3^2} = \frac{\partial^4 u_3}{\partial x_1^2 \partial x_3^2} = \frac{\partial^4 u_3}{\partial x_1^2 \partial x_2 \partial x_3} = 0\,, \\
\frac{\partial^4 u_3}{\partial x_3^4} = \frac{\partial^4 u_2}{\partial x_3^4} = \frac{\partial^4 u_3}{\partial x_2 \partial x_3^3} = \frac{\partial^4 u_2}{\partial x_2^2 \partial x_3^2} = \frac{\partial^4 u_1}{\partial x_1 \partial x_2 \partial x_3^2} = 0\,, \\
\frac{\partial^4 u_3}{\partial x_2^3 \partial x_3} = \frac{\partial^4 u_2}{\partial x_1^2 \partial x_3^2} = \frac{\partial^4 u_3}{\partial x_1^2 \partial x_3^2} = \frac{\partial^4 u_2}{\partial x_1^2 \partial x_2 \partial x_3} = \frac{\partial^4 u_1}{\partial x_1 \partial x_2^2 \partial x_3} = 0\,, \\
\frac{\partial^4 u_1}{\partial x_1^2 \partial x_2^2} + \frac{\partial^4 u_2}{\partial x_1^3 \partial x_2} = 0\,,  \frac{\partial^4 u_1}{\partial x_2^4} + 2 \frac{\partial^4 u_1}{\partial x_1^2 \partial x_2^2} = 0\,, \\
\frac{\partial^4 u_1}{\partial x_3^4} + \frac{\partial^4 u_3}{\partial x_1 \partial x_3^3}= 0\,,  \frac{\partial^4 u_1}{\partial x_1 \partial x_2^3} + \frac{\partial^4 u_2}{\partial x_1^2 \partial x_2^2} = 0\,, \\
2 \frac{\partial^4 u_2}{\partial x_1^2 \partial x_2^2} + \frac{\partial^4 u_2}{\partial x_1^4} = 0\,,  \frac{\partial^4 u_2}{\partial x_2^3 \partial x_3} + \frac{\partial^4 u_1}{\partial x_1^3 \partial x_3 } = 0\,, \\
\frac{\partial^4 u_3}{\partial x_2^3 \partial x_3} +  \frac{\partial^4 u_3}{\partial x_1^2 \partial x_3^2} = 0\,,  \frac{\partial^4 u_2}{\partial x_2 \partial x_3^3} +  \frac{\partial^4 u_1}{\partial x_1 \partial x_3^3} = 0\,, \\
\frac{\partial^4 u_3}{\partial x_2^4} +  \frac{\partial^4 u_3}{\partial x_1^4} = 0\,. 
\end{dcases}
\end{equation}

Substituting the tetragonal linear elastic candidate displacement field \eqref{tetragonal-linear-inhomogeneous} into the above third- and fourth-order universality PDEs, we obtain the following additional universality PDEs for the function $g(x_1, x_2)$:
\begin{equation}\label{D^h_4-constraints-for-g}
\begin{aligned}
	&\frac{\partial^3 g}{\partial x_1^3}
	=\frac{\partial^3 g}{\partial x_1 \partial x_2^2}
	=\frac{\partial^3 g}{\partial x_2^3}
	=\frac{\partial^3 g}{\partial x_1^2 \partial x_2}
	=\frac{\partial^4 g}{\partial x_1^2 \partial x_2^2}
	=0\,,\\
	&\frac{\partial^4 g}{\partial x_1^4}+\frac{\partial^4 g}{\partial x_2^4}=0\,,\\
	&\frac{\partial^2 g}{\partial x_1^2}+\frac{\partial^2 g}{\partial x_2^2}=0\,.
\end{aligned}
\end{equation}
After straightforward manipulations, one finds that $g(x_1, x_2)$ has the same form as in \eqref{Z^-_4-g-final-form}. This gives the following characterization:
\begin{prop}
The universal displacements in orthotropic $\mathbb{D}_4^h$ class linear strain-gradient elastic solids are the superposition of homogeneous displacement fields and the inhomogeneous displacement field \eqref{tetragonal-linear-inhomogeneous}, with $g(x_1, x_2)$ given by \eqref{Z^-_4-g-final-form}.
\end{prop}

\renewcommand{\arraystretch}{1.15}
\begin{longtable}{p{0.30\textwidth} p{0.66\textwidth}}
\caption{Summary of universal displacements: Orthotropic classes.}\label{Table-UD-summary-orthotropic}\\
\hline
\textbf{Symmetry class} & \textbf{Universal displacement family} \\
\hline
\endfirsthead
\hline
\textbf{Symmetry class} & \textbf{Universal displacement family} (continued) \\
\hline
\endhead
\hline
\endfoot
\hline
\endlastfoot
Orthotropic class $\mathbb{D}_2$ & Superposition of homogeneous fields and $(\alpha_1 x_2 x_3\,,\, \alpha_2 x_1 x_3\,,\, \alpha_3 x_1 x_2)$ \\ \hline
Orthotropic class $\mathbb{D}_2 \oplus \mathbb{Z}_2^c$ & Superposition of homogeneous fields and $(\alpha_1 x_2 x_3\,,\, \alpha_2 x_1 x_3\,,\, \alpha_3 x_1 x_2)$ \\ \hline
Orthotropic class $\mathbb{Z}_4^-$ & Superposition of a homogeneous field and \eqref{tetragonal-linear-inhomogeneous} with $g(x_1, x_2)$ given by \eqref{Z^-_4-g-final-form} \\ \hline
Orthotropic class $\mathbb{D}_3^{v}$ & Superposition of a homogeneous field and \eqref{trigonal-linear-inhomogeneous} with $a_{123}=0$ \\ \hline
Orthotropic class $\mathbb{D}_4^{h}$ & Superposition of a homogeneous field and \eqref{tetragonal-linear-inhomogeneous} with $g(x_1, x_2)$ given by \eqref{Z^-_4-g-final-form} \\ \hline
\end{longtable}
\renewcommand{\arraystretch}{2.0}

\subsection{Trigonal classes}

Out of the seven trigonal classes, $\mathbb{Z}_3$ and $\mathbb{Z}_3 \oplus \mathbb{Z}_2^c$, as well as $\mathbb{D}_3$ and $\mathbb{D}_3 \oplus \mathbb{Z}_2^c$, have an elasticity matrix $\boldsymbol{\mathsf{C}}$ that coincides with that of the trigonal classical linear elasticity. The classes $\mathbb{Z}_6^-$ and $\mathbb{D}_6^h$ have an elasticity tensor $\boldsymbol{\mathsf{C}}$ that coincides with that of the transversely isotropic classical linear elasticity. Finally, the class $\mathbb{D}_4^v$ has an elasticity matrix $\boldsymbol{\mathsf{C}}$ that coincides with that of the tetragonal classical linear elasticity.

\subsubsection{Trigonal class $\mathbb{Z}_3$} 

Tensor $\boldsymbol{\mathsf{A}}_{\mathbb{Z}_3 \oplus \mathbb{Z}_2^c}$ has $57$ independent components\,; it has the form
\begin{equation}
\label{A_Z_3}
    \boldsymbol{\mathsf{A}}_{\mathbb{Z}_3 \oplus \mathbb{Z}_2^c}(\mathbf{x}) =
    \begin{bmatrix}
    A^{(11)}+\eta A_c & B^{(6)}+\theta B_C & C^{(3)} & D^{(4)}  \\
    & A^{(11)} & F^{(8)} & G^{(9)} \\
    & & H^{(6)} & I^{(4)} \\
    & & & J^{(4)}
    \end{bmatrix}_S
    +
    \begin{bmatrix}
    0 & 0 & f(G^{(9)}) & f(F^{(8)}) \\
    & 0 & f(D^{(4)}) & 0 \\
    & & f(J^{(4)}) & 0 \\
    & & & 0
    \end{bmatrix}_S\,.
\end{equation}
Matrix $A^{(11)}$ has $11$ independent components and has the following form
\begin{equation}
\label{A^11}
    A^{(11)} = \begin{bmatrix}
    a_{11} & a_{12} & a_{13} & a_{14} & a_{15}  \\
     & a_{22} & -a_{13} + \sqrt{2}\, a_{III} & a_I & a_{II}   \\
     &  & -a_{12} + a_{III}^* & a_{34} & a_{35} \\
     &  &  & a_{44} & a_{45} \\
     &  &  &  & a_{55}      
    \end{bmatrix}_S\,,
\end{equation}
where
\begin{equation}
\label{A^11rest}
	a_I = a_{14} - \sqrt{2}\, a_{34} \,,\qquad 
	a_{II} = a_{15} - \sqrt{2}\, a_{35} \,,\qquad
	a_{III} = \frac{a_{11} - a_{22}}{2} \,,\qquad 
	a_{III}^* = \frac{a_{11} + a_{22}}{2} \,.
\end{equation}
The expressions for $B^{(6)}$, $C^{(3)}$, $D^{(4)}$, $G^{(9)}$, $H^{(6)}$, and $I^{(4)}$ are
\begin{equation}
\label{B^6}
    B^{(6)} = \begin{bmatrix}
    0 & b_{12} & -\frac{\sqrt{2}}{2} b_{12} & b_{24} + \sqrt{2}\, b_{34} & b_{25} + \sqrt{2}\, b_{35} \\
     & 0 & -\frac{\sqrt{2}}{2} b_{12} & b_{24} & b_{25} \\
     &  & 0 & b_{34} & b_{35} \\
     &  &  & 0 & b_{45} \\
     &  &  &  & 0
    \end{bmatrix}_A \,,
\end{equation}
\begin{equation}
\label{C^3}
    C^{(3)} = \begin{bmatrix}
    c_{11} & c_{12} & c_{13} & c_{12} & c_{13} \\
    -c_{11} & -c_{12} & -c_{13} & -c_{12} & -c_{13} \\
    -\sqrt{2}\, c_{11} & -\sqrt{2}\, c_{12} & -\sqrt{2}\, c_{13} & -\sqrt{2}\, c_{12} & -\sqrt{2}\, c_{13} \\
    0 & 0 & 0 & 0 & 0 \\
    0 & 0 & 0 & 0 & 0
    \end{bmatrix} \,,
\end{equation}
\begin{equation}
\label{D^4}
   D^{(4)} = \begin{bmatrix}
    0 & d_{12} & -d_{12} \\
    d_{21} & -d_{12} & d_{12} \\
    -\frac{d_{21}}{\sqrt{2}} & -\sqrt{2}\, d_{12} & \sqrt{2}\, d_{12} \\
    d_{41} & 0 & 0 \\
    d_{51} & 0 & 0
    \end{bmatrix} \,,
\end{equation}
\begin{equation}\label{F^8}
\begin{aligned}
& F^{(8)} =
\begin{bmatrix}
f_{11} & f_{12} & f_{13} & f_{14} & f_{15} \\
-f_{11} & -f_{12} + \beta_I & f_{23} & -f_{12} + \beta_I & -f_{15} - 2\, \beta_{II} \\
-\sqrt{2}\, f_{11} & -\sqrt{2}(f_{12} - \tfrac{3}{2} \beta_I) & -\sqrt{2}(f_{15} + \beta_{II}) & -\sqrt{2}(f_{12} - \tfrac{1}{2} \beta_I) & -\sqrt{2}(f_{13} - \beta_{II}) \\
0 & 0 & f_{43} & 0 & -f_{43} \\
0 & 0 & f_{53} & 0 & -f_{53}
\end{bmatrix} \,,\\
\\
&\beta_I = \frac{f_{12} - f_{14}}{2} \,, \qquad
\beta_{II} = \frac{f_{13} + f_{23}}{2} \,,
\end{aligned}
\end{equation}
\begin{equation}\label{G^9}
\begin{aligned}
& G^{(9)} =
\begin{bmatrix}
g_{11} & g_{12} & g_{13} \\
g_{21} & g_{23} - 2\, \gamma_{III} & g_{23} \\
\sqrt{2}\, \gamma_I & \sqrt{2}\, \gamma_{II} & \sqrt{2}(2\, \gamma_{III} + \gamma_{II}) \\
g_{41} & g_{42} & g_{42} \\
g_{51} & g_{52} & g_{52}
\end{bmatrix} \,,\\
&\gamma_I = \frac{g_{11} - g_{21}}{2} \,, \qquad 
\gamma_{II} = \frac{g_{13} - g_{23}}{2} \,, \qquad
\gamma_{III} = \frac{g_{12} - g_{13}}{2} \,,
\end{aligned}
\end{equation}
\begin{equation}\label{H^6-I^4}
H^{(6)}=
\begin{bmatrix}
h_{11} & h_{12} & h_{13} & h_{12} & h_{13} \\
 & h_{22} & h_{23} & h_{22} & h_{23} \\
 &  & h_{33} & h_{23} & h_{33} \\
 &  &  & h_{22} & h_{23} \\
 &  &  &  & h_{33}
\end{bmatrix}_S \,,\qquad
I^{(4)}=
\begin{bmatrix}
0 & i_{12} & -i_{12} \\
0 & i_{22} & -i_{22} - \sqrt{2}\, i_{31} \\
i_{31} & i_{32} & -i_{32} \\
0 & i_{22} + \sqrt{2}\, i_{31} & -i_{22} \\
-i_{31} & i_{32} & -i_{32}
\end{bmatrix} \,,
\end{equation}
and matrix $J^{(4)}$ is given in \eqref{H^9-I^7-J^4}.
Matrices $A_c$ and $B_c$ are independent of material parameters and are given as
\begin{equation}\label{A_c-B_c}
A_c=
\begin{bmatrix}
1 & -1 & -\sqrt{2} & 0 & 0 \\
 & 1 & \sqrt{2} & 0 & 0 \\
 &  & 2 & 0 & 0 \\
 &  &  & 0 & 0 \\
 &  &  &  & 0
\end{bmatrix}_S \,,\qquad
B_c=
\begin{bmatrix}
1 & 0 & -\frac{3\sqrt{2}}{2} & 0 & 0 \\
-2 & 1 & \frac{\sqrt{2}}{2} & 0 & 0 \\
-\frac{\sqrt{2}}{2} & \frac{3\sqrt{2}}{2} & 2 & 0 & 0 \\
0 & 0 & 0 & 0 & 0 \\
0 & 0 & 0 & 0 & 0
\end{bmatrix} \,.
\end{equation}
The matrix-valued functions $f(G^{(9)})$, $f(F^{(8)})$, $f(D^{(4)})$ and $f(J^{(4)})$ are defined as
\begin{equation}
\label{f(G^9)}
   f(G^{(9)})=\begin{bmatrix}
    0 & -\frac{\sqrt{2}}{2} \gamma_I^* & -g_{12}-\gamma_{II} & \frac{\sqrt{2}}{2} (g_{11} +\gamma_I) & \gamma_{II}^*  \\
    0 & -\frac{\sqrt{2}}{2} \gamma_I^* & -g_{12}+3\gamma_{II}+4\gamma_{III} & \frac{\sqrt{2}}{2} (g_{11} +\gamma_I) & \gamma_{II}^*   \\
    0 & -2\gamma_I  & 0 & 0 & 2\sqrt{2}(\gamma_{III}+\gamma_{II})    \\
    0 & -\frac{\sqrt{2}}{2} g_{41} & -g_{42} & \frac{\sqrt{2}}{2} g_{41} & g_{42} \\
    0 & -\frac{\sqrt{2}}{2} g_{51} & -g_{52} & \frac{\sqrt{2}}{2} g_{51} & g_{52}      
    \end{bmatrix}\,,
\end{equation}
\begin{equation}
\label{f(F^8)}
   f(F^{(8)})=\begin{bmatrix}
    \sqrt{2} \beta_I & \beta_{II} & 2\beta_{III}-\beta_{II}   \\
    0 & \beta_{II} & 3\beta_{II}-2\beta_{III}   \\
    \beta_I & -2\sqrt{2}(\beta_{II}-\beta_{III}) & 0  \\
    0 & f_{43} &f_{43} \\
    0 & f_{53} &f_{53}       
    \end{bmatrix}\,,
\end{equation} 
where 
\begin{equation}
\label{f(F^8)rest}
	\gamma_I^*=\frac{g_{11}+g_{21}}{2}\,,\qquad 
	\gamma_{II}^*=\frac{g_{13}+g_{23}}{2}\,, \qquad
	\beta_{III}=\frac{f_{13}-f_{15}}{2}\,,
\end{equation}
and
\begin{equation}
\label{f(D^4)-f(J^4)}
   f(D^{(4)})=\begin{bmatrix}
    0 & 0 & 0 & 0 & 0  \\
    0 & \frac{d_{21}}{\sqrt{2}} & 0 & -\frac{d_{21}}{\sqrt{2}} & 0   \\
    0 & -\frac{d_{21}}{2}  & 0 & \frac{d_{21}}{2} & 0    \\
    0 & \frac{\sqrt{2}}{2} d_{41} & 0 & -\frac{\sqrt{2}}{2} d_{41} & 0 \\
    0 & \frac{\sqrt{2}}{2} d_{51} & 0 & -\frac{\sqrt{2}}{2} d_{51} & 0      
    \end{bmatrix}\,,\qquad
    f(J^{(4)})=\begin{bmatrix}
    0 & 0 & 0 & 0 & 0  \\
     & 0 & 0 & -j_{11} & -\sqrt{2}j_{12}   \\
     &   & 0 & -\sqrt{2}j_{12} & -(j_{22}+j_{23})    \\
     &  &  & 0 & 0 \\
     &  &  &  & 0      
    \end{bmatrix}_S\,.
\end{equation}

Matrix $\boldsymbol{\mathsf{M}}_{\mathbb{Z}_3}$ has 36 independent components and it is of the form 
\begin{equation}
\label{M_Z_3}
\boldsymbol{\mathsf{M}}_{\mathbb{Z}_3}=\boldsymbol{\mathsf{M}}_{\mathbb{D}_3}+\boldsymbol{\mathsf{M}}_{\mathbb{D}_3^v}\,,
\end{equation}
where
\begin{equation}
\label{M_D_3}
   \boldsymbol{\mathsf{M}}_{\mathbb{D}_3}=\begin{bmatrix}
    \bar{A}^{(6)} & 0 & 0 & \bar{D}^{(4)}  \\
    \bar{E}^{(4)} & 0 & 0 & \bar{H}^{(2)}   \\
    0 & -\bar{E}^{(4)}  & 0 & 0     \\
    0 & 0 & 0 & 0   
    \end{bmatrix}+
\begin{bmatrix}
    0 & 0 & 0 &   \\
    0 & 0 & 0 & 0   \\
    0 & 0  & g(\bar{H}^{(2)}) & 0     \\
    0 & g(\bar{A}^{(6)}) & g(\bar{D}^{(4)}) & 0   
    \end{bmatrix}\,,
\end{equation}
\begin{equation}
\label{M_D_3^v}
   \boldsymbol{\mathsf{M}}_{\mathbb{D}_3^v}=\begin{bmatrix}
    0 & \bar{B}^{(6)} & \bar{C}^{(8)} & 0  \\
    0 & \bar{F}^{(4)} & 0 & 0   \\
    \bar{F}^{(4)} & 0  & 0 & \bar{L}^{(2)}     \\
    0 & 0 & 0 & 0   
    \end{bmatrix}+
\begin{bmatrix}
    0 & 0 & 0 &  0 \\
    0 & 0 & -g(\bar{L}^{(2)}) & 0   \\
    0 & 0  & 0 & 0     \\
    -g(\bar{B}^{(6)}) & 0 & 0 & g(\bar{C}^{(8)})   
    \end{bmatrix}\,,
\end{equation}
\begin{equation}
\label{barA^6-barB^6}
   \bar{A}^{(6)}, \bar{B}^{(6)} \in \begin{bmatrix}
    \bar{a}_{11} & \bar{a}_{12} & \frac{\sqrt{2}}{2}(\bar{a}_{11}+\bar{a}_{12}+2\bar{a}_{22}) & \bar{a}_{14} & \bar{a}_{15} \\
    -(\bar{a}_{11}+\bar{a}_{12}+\bar{a}_{22}) & \bar{a}_{22} & -\frac{\sqrt{2}}{2}(\bar{a}_{11}-2\bar{a}_{22}) & -\bar{a}_{14} & -\bar{a}_{15}   \\
    -\frac{\sqrt{2}}{2}\bar{a}_{33} & \frac{\sqrt{2}}{2}\bar{a}_{33}  & \bar{a}_{33} & 0 & 0 
    \end{bmatrix}\,,
\end{equation}
\begin{equation}
\label{barC^8-barD^4}
   \bar{C}^{(8)} \in \begin{bmatrix}
    \bar{c}_{11} & \bar{c}_{12} & \bar{c}_{13} & \bar{c}_{22} & \bar{c}_{23} \\
    \bar{c}_{11} & \bar{c}_{22} & \bar{c}_{23} & \bar{c}_{12} & \bar{c}_{13}   \\
    \bar{c}_{31} & \bar{c}_{32} & \bar{c}_{33} & \bar{c}_{32} & \bar{c}_{33}
    \end{bmatrix}\,,\qquad
   \bar{D}^{(4)} \in \begin{bmatrix}
    \bar{d}_{11} & \bar{d}_{12} & \bar{d}_{13}  \\
    -\bar{d}_{12} & -\bar{d}_{11} & -\bar{d}_{13}   \\
    \bar{d}_{31} & -\bar{d}_{31} & 0
    \end{bmatrix}\,,
\end{equation}
\begin{equation}
\label{barE^4-barF^4-barH^2-barL^2}
   \bar{E}^{(4)}, \bar{F}^{(4)} \in \begin{bmatrix}
    \bar{e}_{11} & \bar{e}_{12} & \frac{\sqrt{2}}{2}(\bar{e}_{11}-\bar{e}_{12}) & \bar{e}_{14} & \bar{e}_{15} 
    \end{bmatrix}\,,\qquad
   \bar{H}^{(2)}, \bar{L}^{(2)} \in \begin{bmatrix}
    \bar{h}_{11} & \bar{h}_{11} & \bar{h}_{13} 
    \end{bmatrix}\,,
\end{equation}
\begin{equation}
\label{g(barA^6)-g(barB^6)}
   g(\bar{A}^{(6)}), g(\bar{B}^{(6)}) = \begin{bmatrix}
    -\frac{\sqrt{2}}{2}(2\bar{a}_{11}+\bar{a}_{12}+\bar{a}_{22}) & -\frac{\sqrt{2}}{2}(\bar{a}_{12}-\bar{a}_{22}) & -(\bar{a}_{11}+\bar{a}_{22}) & -\sqrt{2} \bar{a}_{14} & -\sqrt{2} \bar{a}_{15} 
    \end{bmatrix}\,,
\end{equation}
\begin{equation}
\label{g(barC^8)}
   g(\bar{C}^{(8)}) = \begin{bmatrix}
    \frac{\sqrt{2}}{2}(\bar{c}_{13}-\bar{c}_{23}) & \frac{\sqrt{2}}{2}(\bar{c}_{13}-\bar{c}_{23}) & (\bar{c}_{12}-\bar{c}_{22}) 
    \end{bmatrix}\,,
\end{equation}
\begin{equation}
\label{g(barD^4)}
   g(\bar{D}^{(4)}) = \begin{bmatrix}
    0 & \bar{d}_{13} &  -\frac{\sqrt{2}}{2}(\bar{d}_{11}+\bar{d}_{12}) & \bar{d}_{13} & \frac{\sqrt{2}}{2}(\bar{d}_{11}+\bar{d}_{12}) 
    \end{bmatrix}\,,
\end{equation}
\begin{equation}
\label{g(barH^2)}
   g(\bar{H}^{(2)}), g(\bar{L}^{(2)}) = \begin{bmatrix}
    0 & -\frac{\sqrt{2}}{2}\bar{h}_{13} & -\bar{h}_{11} & \frac{\sqrt{2}}{2}\bar{h}_{13} & \bar{h}_{11} 
    \end{bmatrix}\,.
\end{equation}
Tensor $\boldsymbol{\mathsf{C}}_{\mathbb{D}_3 \oplus \mathbb{Z}_2^c}$ corresponds to the trigonal class with Hermann--Maugin symbol $32$ and has $6$ independent components. Thus, from the classical linear elastic part we obtain universality PDEs of the form \eqref{trigonal-linear-constraints}, and the candidate universal displacements are the superposition of homogeneous displacement fields and the inhomogeneous field \eqref{trigonal-linear-inhomogeneous}.

The third-order universality PDEs induced by the fifth-order tensor read
\begin{equation}
\begin{dcases}
\label{Z_3-third-order-constraints-1}
\frac{\partial^3 u_3}{\partial x_2 \partial x_3^2} = \frac{\partial^3 u_3}{\partial x_1^2 \partial x_3}=\frac{\partial^3 u_3}{\partial x_3^3}=\frac{\partial^3 u_2}{\partial x_3^3}=\frac{\partial^3 u_3}{\partial x_1 \partial x_3^2}=\frac{\partial^3 u_3}{\partial x_1 \partial x_2 \partial x_3}=0 \,, \\
\frac{\partial^3 u_2}{\partial x_1 \partial x_2 \partial x_3}=\frac{\partial^3 u_3}{\partial x_1^3}=\frac{\partial^3 u_1}{\partial x_2^2 \partial x_3}=\frac{\partial^3 u_3}{\partial x_1 \partial x_2^2}=\frac{\partial^3 u_1}{\partial x_3^3}=\frac{\partial^3 u_1}{\partial x_2 \partial x_3^2}=0 \,, \\
\frac{\partial^3 u_2}{\partial x_1 \partial x_3^2}=\frac{\partial^3 u_3}{\partial x_1 \partial x_2 \partial x_3}=\frac{\partial^3 u_1}{\partial x_1^2 \partial x_3}=\frac{\partial^3 u_2}{\partial x_2^2 \partial x_3}=\frac{\partial^3 u_3}{\partial x_2^3}=\frac{\partial^3 u_2}{\partial x_1^2 \partial x_3}=0 \,, \\
\frac{\partial^3 u_1}{\partial x_1 \partial x_2 \partial x_3}=\frac{\partial^3 u_1}{\partial x_1^2 \partial x_3}=0 \,,
\end{dcases}
\end{equation}
as well as
\begin{equation}
\begin{dcases}
\label{Z_3-third-order-constraints-2}
- \sqrt{2} \frac{\partial^3 u_2}{\partial x_2^3}+\left(1+\frac{1}{\sqrt{2}}\right) \frac{\partial^3 u_1}{\partial x_1 \partial x_2^2}+\frac{1}{2}\left(4+\sqrt{2}\right)\frac{\partial^3 u_2}{\partial x_1^2 \partial x_2}=0 \,, \\
-\frac{ 1}{\sqrt{2}}  \frac{\partial^3 u_2}{\partial x_2^3}-\frac{\partial^3 u_1}{\partial x_1 \partial x_2^2}+\frac{1}{2} \left(4+\sqrt{2}\right)   \frac{\partial^3 u_2}{\partial x_1^2 \partial x_2}=0 \,, \\
-\frac{1}{2} \left(2+\frac{1}{\sqrt{2}}\right)\frac{\partial^3 u_2}{\partial x_2^3}+\left(1+\frac{3}{\sqrt{2}}\right)\frac{\partial^3 u_1}{\partial x_1 \partial x_2^2}+\left(2+\sqrt{2}\right)\frac{\partial^3 u_2}{\partial x_1^2 \partial x_2}=0 \,,\\
-\frac{\partial^3 u_2}{\partial x_2 \partial x_3^2}-\frac{\partial^3 u_1}{\partial x_1 \partial x_3^2}+\frac{\partial^3 u_3}{\partial x_1^2 \partial x_3}=0 \,, \\
-\sqrt{2}\left(\frac{\partial^3 u_2}{\partial x_2 \partial x_3^2}+\frac{\partial^3 u_3}{\partial x_2^2 \partial x_3}\right)+\frac{\partial^3 u_1}{\partial x_1 \partial x_3^2}+\frac{\partial^3 u_3}{\partial x_1^2 \partial x_3}=0 \,, \\
\frac{\partial^3 u_3}{\partial x_2^2 \partial x_3}-\frac{\partial^3 u_3}{\partial x_1^2 \partial x_3}=0 \,, \\
-\frac{\partial^3 u_2}{\partial x_2 \partial x_3^2}-\frac{\partial^3 u_3}{\partial x_2^2 \partial x_3}+\frac{\partial^3 u_1}{\partial x_1 \partial x_3^2}+\frac{\partial^3 u_3}{\partial x_1^2 \partial x_3}=0 \,, \\
\frac{\partial^3 u_2}{\partial x_2 \partial x_3^2}-\frac{\partial^3 u_1}{\partial x_1 \partial x_3^2}=0 \,,
\end{dcases}
\end{equation}
and 
\begin{equation}
\begin{dcases}
\label{Z_3-third-order-constraints-3}
-\frac{\partial^3 u_1}{\partial x_2^3}+\frac{1}{\sqrt{2}}\left(2\frac{\partial^3 u_2}{\partial x_1 \partial x_2^2}-\frac{\partial^3 u_1}{\partial x_1^2 \partial x_2}+\frac{\partial^3 u_2}{\partial x_1^3}\right)=0 \,, \\
\frac{1}{\sqrt{2}} \frac{\partial^3 u_1}{\partial x_2^3} +\left(1-\frac{3}{\sqrt{2}}\right) \frac{\partial^3 u_2}{\partial x_1 \partial x_2^2}+\sqrt{2} \frac{\partial^3 u_1}{\partial x_1^2 \partial x_2}=0 \,, \\
-\frac{1}{2}\left(2+\frac{1}{\sqrt{2}}\right) \frac{\partial^3 u_1}{\partial x_2^3}-\left(1+\sqrt{2}\right) \frac{\partial^3 u_2}{\partial x_1 \partial x_2^2}-\frac{2}{\sqrt{2}} \frac{\partial^3 u_1}{\partial x_1^2 \partial x_2}+\frac{3}{\sqrt{2}} \frac{\partial^3 u_2}{\partial x_1^3}=0 \,. 
\end{dcases}
\end{equation}
We also have
\begin{equation}
\begin{dcases}
\label{Z_3-third-order-constraints-4}
\frac{1}{\sqrt{2}} \frac{\partial^3 u_1}{\partial x_2^3} + \frac{1}{\sqrt{2}} \frac{\partial^3 u_2}{\partial x_1 \partial x_2^2}-\left(2+\sqrt{2}\right) \frac{\partial^3 u_1}{\partial x_1^2 \partial x_2}-\frac{\partial^3 u_2}{\partial x_1^3}=0 \,, \\
\frac{1}{\sqrt{2}} \frac{\partial^3 u_1}{\partial x_2^3} -\left(1+\sqrt{2}\right)\frac{\partial^3 u_1}{\partial x_1^2 \partial x_2}+\frac{1}{\sqrt{2}} \frac{\partial^3 u_2}{\partial x_1^3} =0 \,, \\
\sqrt{2} \frac{\partial^3 u_1}{\partial x_2^3} +\left(1-\frac{1}{\sqrt{2}}\right) \frac{\partial^3 u_2}{\partial x_1 \partial x_2^2}-2\left(1+\sqrt{2}\right) \frac{\partial^3 u_1}{\partial x_1^2 \partial x_2}-\frac{1}{2}\left(2+\sqrt{2}\right)  \frac{\partial^3 u_2}{\partial x_1^3}=0 \,, 
\end{dcases}
\end{equation}
and
\begin{equation}
\begin{dcases}
\label{Z_3-third-order-constraints-5}
- \frac{1}{\sqrt{2}} \frac{\partial^3 u_1}{\partial x_1 \partial x_2^2}-\frac{1}{2}\left(-2+\sqrt{2}\right)\frac{\partial^3 u_2}{\partial x_1^2 \partial x_2}+\frac{2}{\sqrt{2}}\frac{\partial^3 u_1}{\partial x_1^3}=0 \,, \\
-\frac{1}{\sqrt{2}} \frac{\partial^3 u_2}{\partial x_1^2 \partial x_2}-\frac{1}{2} \left(2+\sqrt{2}\right) \frac{\partial^3 u_1}{\partial x_1^3}=0 \,, \\
\left(2+\frac{1}{\sqrt{2}}\right)\frac{\partial^3 u_1}{\partial x_1 \partial x_2^2}+ \frac{2}{\sqrt{2}} \frac{\partial^3 u_2}{\partial x_1^2 \partial x_2}-\frac{1}{\sqrt{2}}\frac{\partial^3 u_1}{\partial x_1^3}=0 \,, \\
\frac{\partial^3 u_2}{\partial x_2 \partial x_3^2}-\frac{\partial^3 u_3}{\partial x_2 \partial x_3^2}-\frac{\partial^3 u_1}{\partial x_1 \partial x_3^2}-\sqrt{2} \frac{\partial^3 u_3}{\partial x_1^2 \partial x_3}=0 \,, \\
-\frac{\partial^3 u_2}{\partial x_2 \partial x_3^2}-\frac{\partial^3 u_3}{\partial x_2 \partial x_3^2}-\sqrt{2} \left( \frac{\partial^3 u_1}{\partial x_1 \partial x_3^2}-\sqrt{2} \frac{\partial^3 u_3}{\partial x_1^2 \partial x_3} \right) =0 \,, \\
-\frac{\partial^3 u_3}{\partial x_2^2 \partial x_3}+\frac{\partial^3 u_3}{\partial x_1^2 \partial x_3}=0 \,, \\
-\frac{\partial^3 u_2}{\partial x_2 \partial x_3^2}-\frac{\partial^3 u_3}{\partial x_2^2 \partial x_3}+3\left(\frac{\partial^3 u_1}{\partial x_1 \partial x_3^2}+\frac{\partial^3 u_3}{\partial x_1^2 \partial x_3}\right)=0 \,, \\
\frac{\partial^3 u_2}{\partial x_2 \partial x_3^2}-\frac{\partial^3 u_1}{\partial x_1 \partial x_3^2}=0 \,. 
\end{dcases}
\end{equation}
We furthermore find
\begin{equation}
\begin{dcases}
\label{Z_3-third-order-constraints-6}
-\frac{\partial^3 u_1}{\partial x_2^3}+\frac{\partial^3 u_2}{\partial x_1^2 \partial x_2}-\frac{\partial^3 u_1}{\partial x_1^2 \partial x_2}+\frac{\partial^3 u_2}{\partial x_1^3}=0 \,, \\
\frac{\partial^3 u_1}{\partial x_1 \partial x_2^2}+\frac{1}{\sqrt{2}}\left(\frac{\partial^3 u_1}{\partial x_2^3}+\frac{\partial^3 u_2}{\partial x_1 \partial x_2^2}\right)+\frac{\partial^3 u_1}{\partial x_1^2 \partial x_2}+\frac{1}{\sqrt{2}}\left(\frac{\partial^3 u_1}{\partial x_1^2 \partial x_2}+\frac{\partial^3 u_2}{\partial x_1^3}\right)=0 \,, \\
\frac{\partial^3 u_1}{\partial x_1 \partial x_2^2}-\frac{1}{\sqrt{2}}\left(\frac{\partial^3 u_1}{\partial x_2^3}+\frac{\partial^3 u_2}{\partial x_1 \partial x_2^2}\right)+\frac{\partial^3 u_1}{\partial x_1^2 \partial x_2}-\frac{1}{\sqrt{2}}\left(\frac{\partial^3 u_1}{\partial x_1^2 \partial x_2}+\frac{\partial^3 u_2}{\partial x_1^3}\right)=0 \,,
\end{dcases}
\end{equation}
and 
\begin{equation}
\begin{dcases}
\label{Z_3-third-order-constraints-7}
\frac{\partial^3 u_2}{\partial x_1 \partial x_2^2}+\frac{\partial^3 u_1}{\partial x_1 \partial x_3^2}=0 \,, \\
\frac{\partial^3 u_2}{\partial x_2^3}+\frac{\partial^3 u_1}{\partial x_1^3}=0 \,, \\
\frac{\partial^3 u_1}{\partial x_1 \partial x_2^2}+\frac{\partial^3 u_2}{\partial x_1^2 \partial x_2}=0 \,, \\
\frac{\partial^3 u_3}{\partial x_2^2 \partial x_3}+\frac{\partial^3 u_1}{\partial x_1 \partial x_3^2}+\frac{\partial^3 u_3}{\partial x_1^2 \partial x_3}=0 \,, \\
\frac{\partial^3 u_3}{\partial x_2 \partial x_3^2}+\frac{\partial^3 u_3}{\partial x_1^2 \partial x_3}=0 \,, \\
\frac{\partial^3 u_2}{\partial x_2 \partial x_3^2}+\frac{\partial^3 u_3}{\partial x_2^2 \partial x_2}+\frac{\partial^3 u_1}{\partial x_1 \partial x_3^2}+\frac{\partial^3 u_3}{\partial x_1^2 \partial x_3}=0 \,, \\
\frac{\partial^3 u_2}{\partial x_2^2 \partial x_3}+\frac{\partial^3 u_3}{\partial x_2^3}+\frac{\partial^3 u_2}{\partial x_1^2 \partial x_3}+\frac{\partial^3 u_3}{\partial x_1^2 \partial x_3}=0 \,, \\
\sqrt{2}\frac{\partial^3 u_3}{\partial x_2^3}+\left(2+\sqrt{2}\right)\frac{\partial^3 u_2}{\partial x_1^2 \partial x_3}+\left(-4+\sqrt{2}\right)\frac{\partial^3 u_3}{\partial x_1^2 \partial x_2}=0 \,. 
\end{dcases}
\end{equation}

The fourth-order universality PDEs read
\begin{equation}
\begin{dcases}
\label{Z_3-fourth-order-constraints-1}
\frac{\partial^4 u_3}{\partial x_1 \partial x_2^3}=\frac{\partial^4 u_1}{\partial x_3^4}=\frac{\partial^4 u_2}{\partial x_1 \partial x_2^3}=\frac{\partial^4 u_1}{\partial x_2 \partial x_2^3}=0 \,, \\
\frac{\partial^4 u_1}{\partial x_1 \partial x_3^3}=\frac{\partial^4 u_2}{\partial x_2^4}=\frac{\partial^4 u_1}{\partial x_1 \partial x_2^3}=\frac{\partial^4 u_2}{\partial x_1^4}=\frac{\partial^4 u_3}{\partial x_3^4}=\frac{\partial^4 u_3}{\partial x_2 \partial x_3^3}=0 \,, \\
\frac{\partial^4 u_3}{\partial x_1 \partial x_3^3}=\frac{\partial^4 u_3}{\partial x_1^2 \partial x_3^2}=\frac{\partial^4 u_2}{\partial x_2 \partial x_3^3}=\frac{\partial^4 u_3}{\partial x_2^2 \partial x_2^2}=0 \,, \\
\frac{\partial^4 u_1}{\partial x_1^3 \partial x_2}=\frac{\partial^4 u_2}{\partial x_1^2 \partial x_2^2}=\frac{\partial^4 u_3}{\partial x_1 \partial x_2 \partial x_3^2}=\frac{\partial^4 u_3}{\partial x_1^2 \partial x_2 \partial x_3}=\frac{\partial^4 u_3}{\partial x_3^4}=0 \,, 
\end{dcases}
\end{equation}
as well as
\begin{equation}
\begin{dcases}
\label{Z_3-fourth-order-constraints-2}
\frac{1}{\sqrt{2}} \frac{\partial^4 u_1}{\partial x_2^4}+\frac{1}{\sqrt{2}} \frac{\partial^4 u_2}{\partial x_1 \partial x_2^3}+\left(1+\sqrt{2}\right) \frac{\partial^4 u_1}{\partial x_1^2 \partial x_2^2}+\frac{\partial^4 u_2}{\partial x_1^3 \partial x_2}+\frac{\partial^4 u_3}{\partial x_1^4} = 0 \,, \\
\frac{\partial^4 u_1}{\partial x_1^2 \partial x_2^2}+\frac{\partial^4 u_2}{\partial x_1^3 \partial x_2}=0 \,, \\
-\frac{\partial^4 u_1}{\partial x_2^4}+\frac{\partial^4 u_2}{\partial x_1 \partial x_2^3}-\frac{\partial^4 u_1}{\partial x_1^2 \partial x_2^2}+\frac{\partial^4 u_2}{\partial x_1^3 \partial x_2}=0 \,, \\
\frac{\partial^4 u_2}{\partial x_1 \partial x_2 \partial x_3^2}+\frac{\partial^4 u_3}{\partial x_1 \partial x_2^2 \partial x_3}+\frac{\partial^4 u_1}{\partial x_1^2 \partial x_3^2}+\frac{\partial^4 u_3}{\partial x_1^3 \partial x_3}=0 \,, \\
\frac{\partial^4 u_1}{\partial x_1^2 \partial x_3^2}+\frac{\partial^4 u_3}{\partial x_1 \partial x_2^2 \partial x_3}+\frac{\partial^4 u_1}{\partial x_1^2 \partial x_3^2}+\frac{\partial^4 u_3}{\partial x_1^3 \partial x_3}=0 \,, \\
-\frac{1}{\sqrt{2}} \frac{\partial^4 u_1}{\partial x_2^4}+\left(1-\frac{1}{\sqrt{2}}\right) \frac{\partial^4 u_2}{\partial x_1 \partial x_2^3}-\left(-1+\sqrt{2}\right) \frac{\partial^4 u_1}{\partial x_1^2 \partial x_2^2}+\frac{\partial^4 u_2}{\partial x_1^3 \partial x_2}=0 \,, \\
\frac{\partial^4 u_1}{\partial x_2^2 \partial x_3^2}-\left(-1+\sqrt{2}\right) \frac{\partial^4 u_2}{\partial x_1 \partial x_2 \partial x_3^2}-2\frac{\partial^4 u_3}{\partial x_1 \partial x_2 \partial x_3^2}-\sqrt{2}\frac{\partial^4 u_3}{\partial x_1 \partial x_2^2 \partial x_3} =0 \,, \\
-\sqrt{2}\left(\frac{\partial^4 u_1}{\partial x_2^2 \partial x_3^2}+\frac{\partial^4 u_3}{\partial x_1 \partial x_2^2 \partial x_3}\right) +2\left(\frac{\partial^4 u_2}{\partial x_1 \partial x_2 \partial x_3^2}+\frac{\partial^4 u_3}{\partial x_1 \partial x_2^2 \partial x_3}\right)=0 \,, \\
-\sqrt{2} \frac{\partial^4 u_1}{\partial x_2^3 \partial x_3}-\left(-4+\sqrt{2}\right) \frac{\partial^4 u_2}{\partial x_1 \partial x_2^2 \partial x_3}+\frac{\partial^4 u_3}{\partial x_1 \partial x_2^3}-\frac{\partial^4 u_1}{\partial x_1^2 \partial x_2 \partial x_3}=0 \,, \\
\frac{\partial^4 u_2}{\partial x_1 \partial x_2^2 \partial x_3}-\frac{\partial^4 u_3}{\partial x_1 \partial x_2^3} -\frac{\partial^4 u_2}{\partial x_1^3 \partial x_2}+\frac{\partial^4 u_3}{\partial x_1^3 \partial x_3}=0 \,, \\
\frac{\partial^4 u_1}{\partial x_1^2 \partial x_2 \partial x_3}+\left(1-\frac{3}{\sqrt{2}}\right) \frac{\partial^4 u_2}{\partial x_1 \partial x_2^2 \partial x_3}+\frac{1}{\sqrt{2}} \left(\frac{\partial^4 u_1}{\partial x_1^2 \partial x_2 \partial x_3}+\frac{\partial^4 u_2}{\partial x_1^3 \partial x_3}+\frac{\partial^4 u_3}{\partial x_1^3 \partial x_2}\right)=0 \,, \\
2 \frac{\partial^4 u_1}{\partial x_2^3 \partial x_3}+\left(5-2\sqrt{2}\right)\frac{\partial^4 u_2}{\partial x_1 \partial x_2^2 \partial x_3}-\left(-1+2\sqrt{2}\right)\left(\frac{\partial^4 u_3}{\partial x_1 \partial x_2^3}+2\frac{\partial^4 u_2}{\partial x_1^2 \partial x_2 \partial x_3}+\frac{\partial^4 u_2}{\partial x_1^3 \partial x_3}+\frac{\partial^4 u_3}{\partial x_1^3 \partial x_2}\right)=0 \,, \\
-\frac{\partial^4 u_1}{\partial x_2^3 \partial x_3}-\frac{1}{2}\left(2+\sqrt{2}\right)\frac{\partial^4 u_2}{\partial x_1 \partial x_2^2 \partial x_3}-\frac{1}{\sqrt{2}}\left(\frac{\partial^4 u_1}{\partial x_1^2 \partial x_2 \partial x_3}+\frac{\partial^4 u_2}{\partial x_1^3 \partial x_3}+\frac{\partial^4 u_3}{\partial x_1^3 \partial x_2}\right)=0 \,, \\
\frac{\partial^4 u_1}{\partial x_2^3 \partial x_3}+\frac{\partial^4 u_2}{\partial x_1 \partial x_2^2 \partial x_3}-\left(1+2\sqrt{2}\right)\frac{\partial^4 u_2}{\partial x_1^3 \partial x_3}-2\sqrt{2}\frac{\partial^4 u_3}{\partial x_1^3 \partial x_2}=0 \,, \\
4 \frac{\partial^4 u_1}{\partial x_2^3 \partial x_3}+\left(3+2\sqrt{2}\right)\frac{\partial^4 u_2}{\partial x_1 \partial x_2^2 \partial x_3}-4\left(-1+\sqrt{2}\right)\frac{\partial^4 u_1}{\partial x_1^2 \partial x_2 \partial x_3}-\left(1+2\sqrt{2}\right)\frac{\partial^4 u_2}{\partial x_1^3 \partial x_3} \\
\qquad +\left(1-2\sqrt{2}\right) \frac{\partial^4 u_3}{\partial x_1^3 \partial x_2}=0 \,, \\
\frac{\partial^4 u_2}{\partial x_1 \partial x_2 \partial x_3^2}+\frac{\partial^4 u_3}{\partial x_1 \partial x_2^2 \partial x_3}=0 \,, \\
2 \frac{\partial^4 u_1}{\partial x_2^2 \partial x_3^2}-\left(-1+\sqrt{2}\right)\frac{\partial^4 u_2}{\partial x_1 \partial x_2 \partial x_3^2}+\frac{\partial^4 u_3}{\partial x_1 \partial x_2^2 \partial x_3} -\sqrt{2}\left(\frac{\partial^4 u_1}{\partial x_1^2 \partial x_3^2}+\frac{\partial^4 u_3}{\partial x_1^3 \partial x_3}\right)=0 \,, \\
\sqrt{2}\frac{\partial^4 u_1}{\partial x_2^4}+\left(1+\sqrt{2}\right)\frac{\partial^4 u_2}{\partial x_1 \partial x_2^3}-\left(1+\sqrt{2}\right)\frac{\partial^4 u_1}{\partial x_1^2 \partial x_2^2}-\left(1+\sqrt{2}\right)\frac{\partial^4 u_2}{\partial x_1^3 \partial x_2}+\frac{\partial^4 u_1}{\partial x_1^4}=0 \,, 
\end{dcases}
\end{equation}
and
\begin{equation}
\begin{dcases}
\label{Z_3-fourth-order-constraints-3}
\frac{\partial^4 u_2}{\partial x_2^2 \partial x_3^2}+\frac{\partial^4 u_1}{\partial x_1 \partial x_2 \partial x_3^2}=0 \,, \\
-\sqrt{2} \frac{\partial^4 u_2}{\partial x_2^2 \partial x_3^2}-\frac{\partial^4 u_1}{\partial x_1 \partial x_2 \partial x_3^2}-\frac{\partial^4 u_2}{\partial x_1^2 \partial x_3^2}-2\frac{\partial^4 u_3}{\partial x_1^2 \partial x_2 \partial x_3}=0 \,, \\
\frac{\partial^4 u_2}{\partial x_2^2 \partial x_3^2}+\frac{\partial^4 u_3}{\partial x_2^3 \partial x_3}+\frac{\partial^4 u_2}{\partial x_1^2 \partial x_3^2}+\frac{\partial^4 u_3}{\partial x_1^2 \partial x_2 \partial x_3}=0 \,, \\
-2 \frac{\partial^4 u_2}{\partial x_1 \partial x_2 \partial x_3^2}+\sqrt{2} \frac{\partial^4 u_2}{\partial x_1^2 \partial x_3^2}+\left(-2+\sqrt{2}\right) \frac{\partial^4 u_3}{\partial x_1^2 \partial x_2 \partial x_3}=0 \,, \\
\frac{\partial^4 u_2}{\partial x_3^4}+\frac{\partial^4 u_3}{\partial x_1 \partial x_3^3}=0 \,, \\
\left(10+\sqrt{2}\right) \frac{\partial^4 u_1}{\partial x_1 \partial x_2^3}-3\sqrt{2}\frac{\partial^4 u_2}{\partial x_1^4}=0 \,, \\
-\frac{\partial^4 u_2}{\partial x_2^3 \partial x_3}-\frac{\partial^4 u_1}{\partial x_1 \partial x_2^2 \partial x_3}+\frac{\partial^4 u_2}{\partial x_1^2 \partial x_2 \partial x_3}+\frac{\partial^4 u_1}{\partial x_1^3 \partial x_3}=0 \,, \\
-\frac{\partial^4 u_2}{\partial x_2^3 \partial x_3}-\frac{\partial^4 u_3}{\partial x_2^4}-\left(1+2\sqrt{2}\right)\frac{\partial^4 u_1}{\partial x_1 \partial x_2^2 \partial x_3}-\left(1+2\sqrt{2}\right)\frac{\partial^4 u_2}{\partial x_1^2 \partial x_2 \partial x_3}+3\frac{\partial^4 u_3}{\partial x_1^3 \partial x_3}+\frac{\partial^4 u_3}{\partial x_1^4}=0 \,, \\
\left(8+3\sqrt{2}\right)\frac{\partial^4 u_2}{\partial x_2^3 \partial x_3}+\sqrt{2}\left(3\frac{\partial^4 u_1}{\partial x_1 \partial x_2^2 \partial x_3}+11\frac{\partial^4 u_2}{\partial x_1^2 \partial x_2 \partial x_3}+4\frac{\partial^4 u_3}{\partial x_1^2 \partial x_2^2}-\frac{\partial^4 u_1}{\partial x_1^3 \partial x_3}\right)=0 \,, \\
\left(1+2\sqrt{2}\right)\frac{\partial^4 u_1}{\partial x_1 \partial x_2^2 \partial x_3}+2\left(1+\sqrt{2}\right) \frac{\partial^4 u_2}{\partial x_1^2 \partial x_2 \partial x_3}+\frac{\partial^4 u_3}{\partial x_1^2 \partial x_2^2}-3\frac{\partial^4 u_1}{\partial x_1^3 \partial x_3}-\frac{\partial^4 u_3}{\partial x_1^4}=0 \,, \\
\frac{\partial^4 u_2}{\partial x_2^3 \partial x_3}+\frac{\partial^4 u_3}{\partial x_2^4}-\frac{\partial^4 u_1}{\partial x_1 \partial x_2^2 \partial x_3}-\left(1+2\sqrt{2}\right)\frac{\partial^4 u_2}{\partial x_1^2 \partial x_2 \partial x_3}+2\sqrt{2}\frac{\partial^4 u_3}{\partial x_1^2 \partial x_2^2}-3\frac{\partial^4 u_1}{\partial x_1^3 \partial x_3}-\frac{\partial^4 u_3}{\partial x_1^4}=0 \,, \\
-\sqrt{2} \frac{\partial^4 u_2}{\partial x_2^3 \partial x_3}+\left(8-5\sqrt{2}\right)\frac{\partial^4 u_1}{\partial x_1 \partial x_2^2 \partial x_3}-\sqrt{2}\left(\frac{\partial^4 u_2}{\partial x_1^2 \partial x_2 \partial x_3}+4 \frac{\partial^4 u_3}{\partial x_1^2 \partial x_2^2}+\frac{\partial^4 u_1}{\partial x_1^3 \partial x_3}\right)=0 \,, \\
\frac{\partial^4 u_2}{\partial x_2^3 \partial x_3}+\frac{\partial^4 u_3}{\partial x_2^4}-\left(3+4\sqrt{2}\right)\frac{\partial^4 u_1}{\partial x_1 \partial x_2^2 \partial x_3}-\left(5+2\sqrt{2}\right)\frac{\partial^4 u_2}{\partial x_1^2 \partial x_2 \partial x_3}-2\left(1+\sqrt{2}\right)\frac{\partial^4 u_3}{\partial x_1^2 \partial x_2^2}+3\frac{\partial^4 u_1}{\partial x_1^3 \partial x_3} \\
+\frac{\partial^4 u_3}{\partial x_1^4}=0\,, \\
\frac{1}{\sqrt{2}}\frac{\partial^4 u_2}{\partial x_2 \partial x_3^3}+2\frac{\partial^4 u_3}{\partial x_2 \partial x_3^3}-\frac{1}{\sqrt{2}}\frac{\partial^4 u_2}{\partial x_1 \partial x_3^3}=0 \,, \\
\frac{\partial^4 u_2}{\partial x_2 \partial x_3^3}-\frac{\partial^4 u_1}{\partial x_1 \partial x_3^3}=0 \,, \\
\frac{\partial^4 u_2}{\partial x_2 \partial x_3^3}+\frac{\partial^4 u_3}{\partial x_2^2 \partial x_3^2}=0 \,, \\
-\frac{\partial^4 u_2}{\partial x_2^2 \partial x_3^2}-\frac{\partial^4 u_3}{\partial x_2^3 \partial x_3}+\frac{\partial^4 u_1}{\partial x_1 \partial x_2 \partial x_3^2}+\frac{\partial^4 u_2}{\partial x_1^2 \partial x_3^2}+2\frac{\partial^4 u_3}{\partial x_1^2 \partial x_2 \partial x_3}=0 \,, \\
-\frac{\partial^4 u_2}{\partial x_1^2 \partial x_3^2}+\frac{\partial^4 u_2}{\partial x_1^2 \partial x_2 \partial x_3}=0 \,, 
\end{dcases}
\end{equation}
We also have
\begin{equation}
\begin{dcases}
\label{Z_3-fourth-order-constraints-4}
\frac{\partial^4 u_2}{\partial x_2^2 \partial x_3^2}+\frac{\partial^4 u_1}{\partial x_1 \partial x_2 \partial x_3^2}=0 \,, \\
\frac{\partial^4 u_2}{\partial x_2^2 \partial x_3^2}+\frac{\partial^4 u_3}{\partial x_2^3 \partial x_3}+\frac{\partial^4 u_2}{\partial x_1^2 \partial x_3^2}+\frac{\partial^4 u_3}{\partial x_1^2 \partial x_2 \partial x_3^2}=0 \,, \\
-\left(-1+\sqrt{2}\right)\frac{\partial^4 u_1}{\partial x_1 \partial x_2 \partial x_3^2}+\frac{\partial^4 u_2}{\partial x_1^2 \partial x_3^2}+2\frac{\partial^4 u_3}{\partial x_1^2 \partial x_2 \partial x_3^2}=0 \,, \\
2\left(\frac{\partial^4 u_1}{\partial x_1 \partial x_2 \partial x_3^2}+\frac{\partial^4 u_3}{\partial x_1^2 \partial x_2 \partial x_3}\right)-\sqrt{2}\left(\frac{\partial^4 u_2}{\partial x_1^2 \partial x_3^2}+\frac{\partial^4 u_3}{\partial x_1^2 \partial x_2 \partial x_3}\right)=0 \,, \\
\frac{\partial^4 u_2}{\partial x_3^4}+\frac{\partial^4 u_3}{\partial x_2 \partial x_3^3}=0 \,, \\
-3 \frac{\partial^4 u_2}{\partial x_1^2 \partial x_2 \partial x_3}-2\frac{\partial^4 u_3}{\partial x_1^2 \partial x_2^2}+\frac{\partial^4 u_1}{\partial x_1^3 \partial x_3}=0 \,, \\
-\frac{\partial^4 u_2}{\partial x_1^2 \partial x_2 \partial x_3^2}-\sqrt{2}\left(\frac{\partial^4 u_1}{\partial x_1 \partial x_2^2 \partial x_3^2}+\frac{\partial^4 u_3}{\partial x_1^2 \partial x_2^2}\right)+ \frac{\partial^4 u_1}{\partial x_1^3 \partial x_3}=0 \,, \\
-\frac{\partial^4 u_2}{\partial x_2 \partial x_3^3}+\frac{\partial^4 u_1}{\partial x_1 \partial x_3^3}=0 \,, \\
\frac{\partial^4 u_1}{\partial x_1 \partial x_2^2 \partial x_3}+\frac{1}{2}\frac{\partial^4 u_2}{\partial x_1^2 \partial x_2 \partial x_3^2}+\frac{\partial^4 u_3}{\partial x_1^2 \partial x_2^2}-\frac{1}{2}\frac{\partial^4 u_1}{\partial x_1^3 \partial x_2}=0 \,, \\
-\left(1+\sqrt{2}\right)\frac{\partial^4 u_1}{\partial x_1 \partial x_2^2 \partial x_3}-\sqrt{2} \frac{\partial^4 u_2}{\partial x_1^2 \partial x_2 \partial x_3}+\frac{\partial^4 u_1}{\partial x_1^3 \partial x_3}=0 \,, \\
\frac{\partial^4 u_2}{\partial x_2^3 \partial x_3}-\frac{3}{2}\frac{\partial^4 u_2}{\partial x_1^2 \partial x_2 \partial x_3}-\frac{\partial^4 u_3}{\partial x_1^2 \partial x_2^2}-\frac{1}{2}\frac{\partial^4 u_1}{\partial x_1^3 \partial x_3}=0 \,, \\
3 \frac{\partial^4 u_2}{\partial x_2^3 \partial x_3}+\frac{\partial^4 u_3}{\partial x_2^4}-\left(1+2\sqrt{2}\right)\left(2 \frac{\partial^4 u_1}{\partial x_1 \partial x_2^2 \partial x_3}+\frac{\partial^4 u_2}{\partial x_1^2 \partial x_2 \partial x_3}+\frac{\partial^4 u_3}{\partial x_1^2 \partial x_2^2}\right)=0 \,, \\
-\left(2+3\sqrt{2}\right)\frac{\partial^4 u_1}{\partial x_1 \partial x_2^2 \partial x_3}-\sqrt{2}\frac{\partial^4 u_2}{\partial x_1^2 \partial x_2 \partial x_3}-\left(1+2\sqrt{2}\right)\frac{\partial^4 u_3}{\partial x_1^2 \partial x_2^2}+2\frac{\partial^4 u_1}{\partial x_1^3 \partial x_3}+\frac{\partial^4 u_3}{\partial x_1^4}=0 \,, \\
-\frac{\partial^4 u_2}{\partial x_2 \partial x_3^3}-\frac{\partial^4 u_3}{\partial x_2^2 \partial x_3^2}+\frac{\partial^4 u_1}{\partial x_1 \partial x_3^3}=0 \,, \\
\frac{\partial^4 u_2}{\partial x_2^2 \partial x_3^2}+\frac{\partial^4 u_1}{\partial x_1 \partial x_2 \partial x_3^2}=0 \,, \\
\frac{\partial^4 u_3}{\partial x_2^3 \partial x_3}+\frac{\partial^4 u_2}{\partial x_1^2 \partial x_2 \partial x_3}=0 \,, \\
-\sqrt{2}\frac{\partial^4 u_1}{\partial x_1 \partial x_2 \partial x_3^2}+\frac{\partial^4 u_3}{\partial x_1^2 \partial x_2 \partial x_3}=0 \,, \\
\frac{\partial^4 u_2}{\partial x_1^2 \partial x_3^2}-\frac{\partial^4 u_3}{\partial x_1^2 \partial x_2 \partial x_3}=0 \,, \\
2 \left(2+\sqrt{2}\right)\frac{\partial^4 u_2}{\partial x_1^2 \partial x_2^2}-2\sqrt{2}\frac{\partial^4 u_1}{\partial x_1^3 \partial x_2}=0 \,, 
\end{dcases}
\end{equation}
and
\begin{equation}
\begin{dcases}
\label{Z_3-fourth-order-constraints-5}
\frac{\partial^4 u_1}{\partial x_1^4}+\left(-\sqrt{2}+3\right)\frac{\partial^4 u_2}{\partial x_1 \partial x_2^3}+3\frac{\partial^4 u_1}{\partial x_1^2 \partial x_2^2}+\frac{\partial^4 u_2}{\partial x_1^3 \partial x_2}+\sqrt{2}\frac{\partial^4 u_1}{\partial x_1^4}=0 \,, \\
\frac{\partial^4 u_2}{\partial x_1 \partial x_2 \partial x_3^2}-\frac{\partial^4 u_3}{\partial x_1 \partial x_2^2 \partial x_3}+\frac{\partial^4 u_1}{\partial x_1^2 \partial x_3^2}-\frac{\partial^4 u_3}{\partial x_1^3 \partial x_3}=0 \,, \\
\frac{\partial^4 u_1}{\partial x_2^2 \partial x_3^2}+\frac{\partial^4 u_2}{\partial x_1 \partial x_2 \partial x_3^2}+\sqrt{2}\left(-\frac{\partial^4 u_3}{\partial x_1 \partial x_2^2 \partial x_3}+\frac{\partial^4 u_1}{\partial x_1^2 \partial x_3^2}\right)=0 \,, \\
-\frac{\partial^4 u_1}{\partial x_2^2 \partial x_3^2}-\frac{\partial^4 u_3}{\partial x_1 \partial x_2^2 \partial x_3}-\frac{\partial^4 u_1}{\partial x_1^2 \partial x_3^2}-\frac{\partial^4 u_3}{\partial x_1^3 \partial x_3}=0 \,, \\
-\sqrt{2}\left(\frac{\partial^4 u_1}{\partial x_2^2 \partial x_3^2}+\frac{\partial^4 u_3}{\partial x_1 \partial x_2^2 \partial x_3}\right)+2\left(\frac{\partial^4 u_2}{\partial x_1 \partial x_2 \partial x_3^2}+\frac{\partial^4 u_3}{\partial x_1 \partial x_2^2 \partial x_3}\right)=0 \,, \\
-\sqrt{2}\frac{\partial^4 u_1}{\partial x_2^4}-\left(4+3\sqrt{2}\right)\frac{\partial^4 u_2}{\partial x_1 \partial x_2^3}+\left(10+9\sqrt{2}\right)\frac{\partial^4 u_1}{\partial x_1^2 \partial x_2^2}+\left(10+3\sqrt{2}\right)\frac{\partial^4 u_2}{\partial x_1^3 \partial x_2}=0 \,, \\
\frac{\partial^4 u_2}{\partial x_1 \partial x_2 \partial x_3^2}+\frac{\partial^4 u_1}{\partial x_1^2 \partial x_2 \partial x_3}=0 \,, \\
\sqrt{2}\frac{\partial^4 u_1}{\partial x_2^3 \partial x_3}+\left(1+2\sqrt{2}\right)\frac{\partial^4 u_2}{\partial x_1 \partial x_2^2 \partial x_3}+\sqrt{2}\frac{\partial^4 u_3}{\partial x_1 \partial x_2^3}+\left(-1+\sqrt{2}\right)\frac{\partial^4 u_1}{\partial x_1^2 \partial x_2 \partial x_3}+\sqrt{2}\frac{\partial^4 u_3}{\partial x_1^3 \partial x_2}=0 \,, \\
\frac{\partial^4 u_2}{\partial x_1 \partial x_2^2 \partial x_3}+\frac{\partial^4 u_3}{\partial x_1 \partial x_2^3}-2\frac{\partial^4 u_1}{\partial x_1^2 \partial x_2 \partial x_3}=0 \,, \\
\left(1+2\sqrt{2}\right)\frac{\partial^4 u_1}{\partial x_2^3 \partial x_3}+\left(2+4\sqrt{2}\right)\frac{\partial^4 u_2}{\partial x_1 \partial x_2^2 \partial x_3}+\left(1+2\sqrt{2}\right)\frac{\partial^4 u_3}{\partial x_1 \partial x_2^3}-3\frac{\partial^4 u_1}{\partial x_1^2 \partial x_2 \partial x_3}=0 \,, \\
\frac{\partial^4 u_1}{\partial x_2^3 \partial x_3}+2\frac{\partial^4 u_2}{\partial x_1 \partial x_2^2 \partial x_3}+\left(4+\sqrt{2}\right)\frac{\partial^4 u_1}{\partial x_1^2 \partial x_2 \partial x_3}+\sqrt{2}\frac{\partial^4 u_2}{\partial x_1^3 \partial x_3}+\frac{\partial^4 u_3}{\partial x_1^3 \partial x_2}=0 \,, \\
2 \frac{\partial^4 u_1}{\partial x_1^2 \partial x_2 \partial x_3}+\frac{\partial^4 u_2}{\partial x_1^3 \partial x_3}+\frac{\partial^4 u_3}{\partial x_1^3 \partial x_2}=0 \,, \\
-2\sqrt{2}\frac{\partial^4 u_1}{\partial x_2^3 \partial x_3}+\left(1-4\sqrt{2}\right)\frac{\partial^4 u_2}{\partial x_1 \partial x_2^2 \partial x_3}-2\sqrt{2}\frac{\partial^4 u_3}{\partial x_1 \partial x_2^3}-\left(-5+\sqrt{2}\right)\frac{\partial^4 u_1}{\partial x_1^2 \partial x_2 \partial x_3} \\
-\left(-1+\sqrt{2}\right)\frac{\partial^4 u_2}{\partial x_1^3 \partial x_2}+\frac{\partial^4 u_3}{\partial x_1^3 \partial x_2}=0 \,, \\
\frac{\partial^4 u_2}{\partial x_1 \partial x_2 \partial x_3^2}+\frac{\partial^4 u_1}{\partial x_1^2 \partial x_3^2}=0 \,, \\
\left(-1+\sqrt{2}\right)\frac{\partial^4 u_2}{\partial x_1 \partial x_2 \partial x_3^2}-\frac{\partial^4 u_3}{\partial x_1 \partial x_2 \partial x_3^2}=0 \,, \\
2 \frac{\partial^4 u_3}{\partial x_1 \partial x_2^2 \partial x_3}+\frac{\partial^4 u_1}{\partial x_1^2 \partial x_3^2}+\frac{\partial^4 u_3}{\partial x_1^3 \partial x_3}=0 \,, 
\end{dcases}
\end{equation}
as well as
\begin{equation}
\begin{dcases}
\label{Z_3-fourth-order-constraints-6}
\frac{\partial^4 u_2}{\partial x_2^3 \partial x_3}+\frac{\partial^4 u_1}{\partial x_1 \partial x_2 \partial x_3^2}+\frac{\partial^4 u_2}{\partial x_1^2 \partial x_2 \partial x_3}+\frac{\partial^4 u_1}{\partial x_1^3 \partial x_3}=0 \,, \\
\frac{\partial^4 u_1}{\partial x_1 \partial x_2^2 \partial x_3}+\frac{\partial^4 u_2}{\partial x_1^2 \partial x_2 \partial x_3}=0 \,, \\
-\frac{3}{\sqrt{2}}\frac{\partial^4 u_2}{\partial x_1^2 \partial x_3^2}-\frac{1}{\sqrt{2}}\frac{\partial^4 u_3}{\partial x_2^3 \partial x_3}+\sqrt{2}\frac{\partial^4 u_1}{\partial x_1 \partial x_2 \partial x_3^2}+\frac{1}{2}\left(4+\sqrt{2}\right)\left(\frac{\partial^4 u_2}{\partial x_1^2 \partial x_3^2}+\frac{\partial^4 u_3}{\partial x_1^2 \partial x_2 \partial x_3}\right)=0 \,, \\
\frac{\partial^4 u_2}{\partial x_2^2 \partial x_3^2}-\left(1+\sqrt{2}\right)\frac{\partial^4 u_1}{\partial x_1 \partial x_2 \partial x_3^2}-\sqrt{2}\frac{\partial^4 u_2}{\partial x_1^2 \partial x_3^2}=0 \,, \\
4 \frac{\partial^4 u_2}{\partial x_2^4}-\left(2+\sqrt{2}\right)\frac{\partial^4 u_1}{\partial x_1 \partial x_2^3}-3\sqrt{2}\frac{\partial^4 u_2}{\partial x_1^4}=0 \,, \\
\frac{\partial^4 u_2}{\partial x_2^2 \partial x_3^2}+\frac{\partial^4 u_3}{\partial x_2^3 \partial x_3}-4\frac{\partial^4 u_1}{\partial x_1 \partial x_2 \partial x_3^2}-\frac{\partial^4 u_2}{\partial x_1^2 \partial x_3^2}=0 \,, \\
2\frac{\partial^4 u_3}{\partial x_1^2 \partial x_2^2}+\frac{\partial^4 u_1}{\partial x_1^3 \partial x_3}+\frac{\partial^4 u_3}{\partial x_1^4}+\frac{\partial^4 u_2}{\partial x_2^3 \partial x_3}+\frac{\partial^4 u_3}{\partial x_24}+\frac{\partial^4 u_1}{\partial x_1 \partial x_2^2 \partial x_3}+\frac{\partial^4 u_2}{\partial x_1^2 \partial x_2 \partial x_3}=0  \,, \\
\frac{\partial^4 u_2}{\partial x_2^3 \partial x_3}+\frac{\partial^4 u_1}{\partial x_1 \partial x_2^2 \partial x_3}+\frac{\partial^4 u_2}{\partial x_1^2 \partial x_2 \partial x_3}+\frac{\partial^4 u_1}{\partial x_1^3 \partial x_3}=0 \,, \\
\frac{\partial^4 u_1}{\partial x_1 \partial x_2^2 \partial x_3}+\frac{\partial^4 u_2}{\partial x_1^2 \partial x_2 \partial x_3}=0 \,, \\
-\sqrt{2}\frac{\partial^4 u_1}{\partial x_1^2 \partial x_2 \partial x_3}-\left(-2+\sqrt{2}\right)\frac{\partial^4 u_2}{\partial x_1^2 \partial x_2 \partial x_3}-2\left(-1+\sqrt{2}\right)\frac{\partial^4 u_3}{\partial x_1^2 \partial x_2^2}=0 \,, \\
2\frac{\partial^4 u_1}{\partial x_1 \partial x_2^ \partial x_3} +\frac{\partial^4 u_2}{\partial x_1^2 \partial x_2 \partial x_3}+\frac{\partial^4 u_3}{\partial x_1^2 \partial x_2^2}=0 \,, 
\end{dcases}
\end{equation}
and finally
\begin{equation}
\begin{dcases}
\label{Z_3-fourth-order-constraints-7}
\frac{\partial^4 u_1}{\partial x_2^3 \partial x_3}-\left(-1+\sqrt{2}\right)\frac{\partial^4 u_2}{\partial x_1 \partial x_2^2 \partial x_3}+\left(-1+\sqrt{2}\right)\frac{\partial^4 u_1}{\partial x_1^2 \partial x_2 \partial x_3}-\frac{\partial^4 u_2}{\partial x_1^3 \partial x_3}=0 \,, \\
-\sqrt{2}\frac{\partial^4 u_1}{\partial x_2^2 \partial x_3^2}-\left(1+\sqrt{2}\right)\frac{\partial^4 u_2}{\partial x_1 \partial x_2 \partial x_3^2}+\frac{\partial^4 u_1}{\partial x_1^2 \partial x_3^2}=0 \,, \\
-\sqrt{2}\frac{\partial^4 u_1}{\partial x_2^4}-\left(1+\sqrt{2}\right)\frac{\partial^4 u_2}{\partial x_1 \partial x_2^3}-\left(-1+\sqrt{2}\right)\frac{\partial^4 u_1}{\partial x_1^2 \partial x_2^2}-\left(1+\sqrt{2}\right)\frac{\partial^4 u_2}{\partial x_1^3 \partial x_2}+\frac{\partial^4 u_1}{\partial x_1^4}=0 \,, \\
3 \sqrt{2}\frac{\partial^4 u_2}{\partial x_1 \partial x_2^3}+\left(-4+2\sqrt{2}\right)\frac{\partial^4 u_1}{\partial x_1^2 \partial x_2^2}-\left(4+\sqrt{2}\right)\frac{\partial^4 u_2}{\partial x_1^3 \partial x_2}-\sqrt{2} \frac{\partial^4 u_1}{\partial x_1^4}=0 \,, \\
\frac{\partial^4 u_1}{\partial x_1^2 \partial x_2^2}+\frac{\partial^4 u_2}{\partial x_1^3 \partial x_2}=0 \,, \\
2 \frac{\partial^4 u_2}{\partial x_1 \partial x_2^3}-\sqrt{2}\left(\frac{\partial^4 u_1}{\partial x_1^2 \partial x_2^2}+\frac{\partial^4 u_2}{\partial x_1^3 \partial x_2}\right)=0 \,, \\
-\sqrt{2} \frac{\partial^4 u_2}{\partial x_1 \partial x_2^3}-\left(-4+\sqrt{2}\right)\frac{\partial^4 u_1}{\partial x_1^2 \partial x_2^2}-\left(-4+\sqrt{2}\right)\frac{\partial^4 u_2}{\partial x_1^3 \partial x_2}-\sqrt{2}\frac{\partial^4 u_1}{\partial x_1^4}=0 \,, \\
-\left(-2+\sqrt{2}\right)\frac{\partial^4 u_1}{\partial x_1^2 \partial x_2^2}-\sqrt{2}\frac{\partial^4 u_2}{\partial x_1^3 \partial x_2}=0 \,, \\
\frac{\partial^4 u_3}{\partial x_1 \partial x_2^2 \partial x_3}-\frac{\partial^4 u_3}{\partial x_1^3 \partial x_3}=0 \,, \\
\frac{1}{\sqrt{2}}\frac{\partial^4 u_1}{\partial x_2^2 \partial x_3^2}+\left(2+\sqrt{2}\right)\frac{\partial^4 u_2}{\partial x_1 \partial x_2 \partial x_3^2}+\left(2+\frac{1}{\sqrt{2}}\right)\frac{\partial^4 u_3}{\partial x_1 \partial x_2^2 \partial x_3}-\frac{1}{\sqrt{2}}\left(\frac{\partial^4 u_1}{\partial x_1^2 \partial x_3^2}+\frac{\partial^4 u_3}{\partial x_1^3 \partial x_3}\right)=0 \,, \\
4\frac{\partial^4 u_1}{\partial x_2^2 \partial x_3^2}+6\frac{\partial^4 u_2}{\partial x_1 \partial x_2 \partial x_3^2}+6\frac{\partial^4 u_3}{\partial x_1 \partial x_2^2 \partial x_3}-2\left(\frac{\partial^4 u_1}{\partial x_1^2 \partial x_3^2}+\frac{\partial^4 u_3}{\partial x_1^3 \partial x_3}\right)=0 \,, \\
-3\frac{\partial^4 u_2}{\partial x_1 \partial x_2^2 \partial x_3}+\frac{\partial^4 u_3}{\partial x_1 \partial x_2^3}-2\frac{\partial^4 u_1}{\partial x_1^2 \partial x_2 \partial x_3}-\frac{\partial^4 u_2}{\partial x_1^3 \partial x_3}+\frac{\partial^4 u_3}{\partial x_1^3 \partial x_3}=0 \,, \\
\frac{\partial^4 u_1}{\partial x_2^3 \partial x_3}+\frac{\partial^4 u_3}{\partial x_1 \partial x_2^3}-3\frac{\partial^4 u_1}{\partial x_1^2 \partial x_2 \partial x_3}-\frac{\partial^4 u_2}{\partial x_1^3 \partial x_3}=0 \,, \\
\frac{\partial^4 u_2}{\partial x_1 \partial x_2^2 \partial x_3}+\frac{\partial^4 u_3}{\partial x_ \partial x_2^3}+\frac{\partial^4 u_1}{\partial x_1^2 \partial x_2 \partial x_3}+\frac{\partial^4 u_3}{\partial x_1^3 \partial x_2}=0 \,.  
\end{dcases}
\end{equation}

Substituting the inhomogeneous displacement field \eqref{trigonal-linear-inhomogeneous} into the above universality PDEs, we find that they are satisfied provided that $a_{123}=0$\,. We therefore arrive at the following result:
\begin{prop}
The universal displacements in trigonal $\mathbb{Z}_3$ class linear strain-gradient elastic solids are the superposition of homogeneous displacement fields and the inhomogeneous displacement field \eqref{trigonal-linear-inhomogeneous} with $a_{123}=0$. 
\end{prop}

\subsubsection{Trigonal class $\mathbb{D}_3$}

Tensor $\boldsymbol{\mathsf{A}}_{\mathbb{D}_3 \oplus \mathbb{Z}_2^c}$ has $34$ independent components; it is of the form 
\begin{equation}
\label{A_D_3}
    \boldsymbol{\mathsf{A}}_{\mathbb{D}_3 \oplus \mathbb{Z}_2^c}(\mathbf{x})=\begin{bmatrix}
    A^{(11)}+\eta A_c & 0 & 0 & D^{(4)}  \\
     & A^{(11)} & F^{(8)} & 0   \\
     &  & H^{(6)} & 0   \\
     &  &  & J^{(4)}      
    \end{bmatrix}_S+
    \begin{bmatrix}
    0 & 0 & 0 & f(F^{(8)})  \\
     & 0 & f(D^{(4)}) & 0   \\
     &  & f(J^{(4)}) & 0   \\
     &  &  & 0      
    \end{bmatrix}_S\,,
\end{equation}
where the coefficients of these two matrices are the same as those of $\boldsymbol{\mathsf{A}}_{\mathbb{Z}_3 \oplus \mathbb{Z}_2^c}$ in the trigonal $\mathbb{Z}_3$ class. 
Tensor $\boldsymbol{\mathsf{M}}_{\mathbb{D}_3}$ has $16$ independent components and has the form \eqref{M_D_3}. Thus, this case has $56$ independent components in total, since $6$ components stem from the classical linear elasticity tensor for the trigonal case (Hermann--Maugin symbol $32$).

The third-order universality PDEs induced by the fifth-order tensor read
\begin{equation}\label{D_3-third-order-constraints-1}
\begin{aligned}
	\frac{\partial^3 u_3}{\partial x_2 \partial x_3^2}
	=\frac{\partial^3 u_2}{\partial x_3^3}
	=\frac{\partial^3 u_3}{\partial x_1 \partial x_2 \partial x_3}
	=\frac{\partial^3 u_1}{\partial x_3^3}
	=\frac{\partial^3 u_3}{\partial x_1 \partial x_3^2}
	=\frac{\partial^3 u_1}{\partial x_2 \partial x_3^2}
	=\frac{\partial^3 u_2}{\partial x_1 \partial x_3^2}
	=0\,.
\end{aligned}
\end{equation}
as well as
\begin{equation}
\begin{dcases}
\label{D_3-third-order-constraints-2}
- \sqrt{2} \frac{\partial^3 u_2}{\partial x_2^3}+\left(1+\frac{1}{\sqrt{2}}\right) \frac{\partial^3 u_1}{\partial x_1 \partial x_2^2}+\frac{1}{2}\left(4+\sqrt{2}\right)\frac{\partial^3 u_2}{\partial x_1^2 \partial x_2}=0 \,, \\
-\frac{ 1}{\sqrt{2}}  \frac{\partial^3 u_2}{\partial x_2^3}-\frac{\partial^3 u_1}{\partial x_1 \partial x_2^2}+\frac{1}{2} \left(4+\sqrt{2}\right)   \frac{\partial^3 u_2}{\partial x_1^2 \partial x_2}=0 \,, \\
-\frac{1}{2} \left(2+\frac{1}{\sqrt{2}}\right)\frac{\partial^3 u_2}{\partial x_2^3}+\left(1+\frac{3}{\sqrt{2}}\right)\frac{\partial^3 u_1}{\partial x_1 \partial x_2^2}+\left(2+\sqrt{2}\right)\frac{\partial^3 u_2}{\partial x_1^2 \partial x_2}=0 \,,\\
\frac{\partial^3 u_2}{\partial x_2 \partial x_3^2}-\sqrt{2}\frac{\partial^3 u_3}{\partial x_2 \partial x_3^2}=0 \,, \\
-\sqrt{2}\left(\frac{\partial^3 u_2}{\partial x_2 \partial x_3^2}+\frac{\partial^3 u_3}{\partial x_2^2 \partial x_3}\right)+\frac{\partial^3 u_1}{\partial x_1 \partial x_3^2}+\frac{\partial^3 u_3}{\partial x_1^2 \partial x_3}=0 \,, \\
\frac{\partial^3 u_3}{\partial x_2^2 \partial x_3}-\frac{\partial^3 u_3}{\partial x_1^2 \partial x_3}=0 \,, \\
\sqrt{2}\frac{\partial^3 u_2}{\partial x_2^2 \partial x_3}+\sqrt{2}\frac{\partial^3 u_3}{\partial x_2^3}+\left(-4+\sqrt{2}\right)\frac{\partial^3 u_1}{\partial x_1 \partial x_2 \partial x_3}-2\left(1+\sqrt{2}\right) \frac{\partial^3 u_2}{\partial x_1^2 \partial x_3}-\left(-2+\sqrt{2}\right)\frac{\partial^3 u_3}{\partial x_1^2 \partial x_2}=0 \,, \\
2\left(2+\frac{1}{\sqrt{2}}\right)\frac{\partial^3 u_2}{\partial x_2^2 \partial x_3}+\sqrt{2}\frac{\partial^3 u_3}{\partial x_2^3}+\left(2+\sqrt{2}\right)\frac{\partial^3 u_1}{\partial x_1 \partial x_2 \partial x_3}+2\sqrt{2}\frac{\partial^3 u_2}{\partial x_1^2 \partial x_3}-\left(-2+\sqrt{2}\right)\frac{\partial^3 u_3}{\partial x_1^2 \partial x_3}=0 \,, \\
\frac{\partial^3 u_2}{\partial x_2^2 \partial x_3}+\frac{\partial^3 u_2}{\partial x_1^2 \partial x_3}=0 \,, \\
\frac{\partial^3 u_2}{\partial x_2^2 \partial x_3}+\frac{1}{\sqrt{2}}\frac{\partial^3 u_1}{\partial x_1 \partial x_2 \partial x_3}+\left(-1+\frac{1}{\sqrt{2}}\right)\frac{\partial^3 u_2}{\partial x_1^2 \partial x_3}-\left(1+\sqrt{2}\right)\frac{\partial^3 u_3}{\partial x_1^2 \partial x_2}=0 \,, \\
-\frac{\partial^3 u_2}{\partial x_2^2 \partial x_3}-\frac{\partial^3 u_3}{\partial x_2^3}+\left(1+\frac{1}{\sqrt{2}}\right) \frac{\partial^3 u_1}{\partial x_1 \partial x_2 \partial x_3}-\frac{1}{\sqrt{2}}\left(\frac{\partial^3 u_2}{\partial x_1^2 \partial x_3}2\frac{\partial^3 u_3}{\partial x_1^2 \partial x_2}\right)=0 \,, \\
\frac{\partial^3 u_3}{\partial x_3^3}-\frac{\partial^3 u_3}{\partial x_2 \partial x_3^2}=0 \,, \\
-\frac{\partial^3 u_2}{\partial x_2 \partial x_3^2}-\frac{\partial^3 u_3}{\partial x_2^2 \partial x_3}+\frac{\partial^3 u_1}{\partial x_1 \partial x_3^2}+\frac{\partial^3 u_3}{\partial x_1^2 \partial x_3}=0 \,, \\
\frac{\partial^3 u_2}{\partial x_2 \partial x_3^2}-\frac{\partial^3 u_1}{\partial x_1 \partial x_3^2}=0 \,,
\end{dcases}
\end{equation}
we also have
\begin{equation}
\begin{dcases}
\label{D_3-third-order-constraints-3}
\frac{1}{\sqrt{2}} \frac{\partial^3 u_1}{\partial x_2^3} + \frac{1}{\sqrt{2}} \frac{\partial^3 u_2}{\partial x_1 \partial x_2^2}-\left(2+\sqrt{2}\right) \frac{\partial^3 u_1}{\partial x_1^2 \partial x_2}-\frac{\partial^3 u_2}{\partial x_1^3}=0 \,, \\
\frac{1}{\sqrt{2}} \frac{\partial^3 u_1}{\partial x_2^3} -\left(1+\sqrt{2}\right)\frac{\partial^3 u_1}{\partial x_1^2 \partial x_2}-\frac{1}{\sqrt{2}} \frac{\partial^3 u_2}{\partial x_1^3} =0 \,, \\
\sqrt{2}  \frac{\partial^3 u_1}{\partial x_2^3} +\left(1\frac{1}{\sqrt{2}}\right) \frac{\partial^3 u_2}{\partial x_1 \partial x_2^2}-\left(1+\sqrt{2}\right) \frac{\partial^3 u_1}{\partial x_1^2 \partial x_2}-\frac{1}{2}\left(2+\sqrt{2}\right)  \frac{\partial^3 u_2}{\partial x_1^3}=0 \,, \\
-2\left(-1+\sqrt{2}\right)\frac{\partial^3 u_1}{\partial x_2^2 \partial x_3}-\left(-4+3\sqrt{2}\right)\frac{\partial^3 u_2}{\partial x_1 \partial x_2 \partial x_3}+\left(-2+\sqrt{2}\right)\frac{\partial^3 u_3}{\partial x_1 \partial x_2^2}-\sqrt{2}\left(\frac{\partial^3 u_1}{\partial x_1^2 \partial x_3}+\frac{\partial^3 u_3}{\partial x_1^3}\right)=0 \,, \\
-2 \sqrt{2}\frac{\partial^3 u_1}{\partial x_2^2 \partial x_3}-\left(-2+\sqrt{2}\right)\frac{\partial^3 u_2}{\partial x_1 \partial x_2 \partial x_3}+\left(-2+\sqrt{2}\right)\frac{\partial^3 u_3}{\partial x_1 \partial x_2^2}-\left(4+\sqrt{2}\right)\frac{\partial^3 u_1}{\partial x_1^2 \partial x_3}-\sqrt{2}\frac{\partial^3 u_3}{\partial x_1^3}=0 \,, \\
\frac{\partial^3 u_1}{\partial x_2^2 \partial x_3}+\frac{\partial^3 u_1}{\partial x_1^2 \partial x_3}=0 \,, \\
\left(-1+\frac{1}{\sqrt{2}}\right)\frac{\partial^3 u_1}{\partial x_2^2 \partial x_3}-\frac{1}{\sqrt{2}}\frac{\partial^3 u_2}{\partial x_1 \partial x_2 \partial x_3}-\left(1+\sqrt{2}\right)\frac{\partial^3 u_3}{\partial x_1 \partial x_2^2}+\frac{\partial^3 u_1}{\partial x_1^2 \partial x_3}=0 \,, \\
-\frac{1}{\sqrt{2}}\frac{\partial^3 u_1}{\partial x_2^2 \partial x_3}+\left(1+\frac{1}{\sqrt{2}}\right)\frac{\partial^3 u_2}{\partial x_1 \partial x_2 \partial x_3}+\sqrt{2}\frac{\partial^3 u_3}{\partial x_1 \partial x_2^2}-\frac{\partial^3 u_1}{\partial x_1^2 \partial x_3}-\frac{\partial^3 u_3}{\partial x_1^3}=0 \,, 
\end{dcases}
\end{equation}
and finally
\begin{equation}
\begin{dcases}
\label{D_3-third-order-constraints-4}
\sqrt{2} \frac{\partial^3 u_1}{\partial x_2^2 \partial x_3}+\left(1+\sqrt{2}\right)\frac{\partial^3 u_2}{\partial x_1 \partial x_2 \partial x_3}-\frac{\partial^3 u_1}{\partial x_1^2 \partial x_3}=0 \,, \\
\frac{\partial^3 u_1}{\partial x_2^2 \partial x_3}+\left(1+\frac{1}{\sqrt{2}}\right)\frac{\partial^3 u_2}{\partial x_1 \partial x_2 \partial x_3}-\frac{1}{\sqrt{2}}\frac{\partial^3 u_1}{\partial x_1^2 \partial x_3}=0 \,, \\
-\frac{\partial^3 u_1}{\partial x_2^3}+\frac{\partial^3 u_2}{\partial x_1 \partial x_2^2}-\frac{\partial^3 u_1}{\partial x_1^2 \partial x_2}+\frac{\partial^3 u_2}{\partial x_1^3}=0 \,, \\
\frac{\partial^3 u_2}{\partial x_1 \partial x_2^2}+\frac{1}{\sqrt{2}}\left(\frac{\partial^3 u_1}{\partial x_2^3}+\frac{\partial^3 u_2}{\partial x_1 \partial x_2^2}\right)+\frac{\partial^3 u_1}{\partial x_1^2 \partial x_2}+\frac{1}{\sqrt{2}}\left(\frac{\partial^3 u_1}{\partial x_1^2 \partial x_2}+\frac{\partial^3 u_2}{\partial x_1^3}\right)=0 \,, \\
\frac{\partial^3 u_2}{\partial x_1 \partial x_2^2}-\frac{1}{\sqrt{2}}\left(\frac{\partial^3 u_1}{\partial x_2^3}+\frac{\partial^3 u_2}{\partial x_1 \partial x_2^2}\right)+\frac{\partial^3 u_1}{\partial x_1^2 \partial x_2}-\frac{1}{\sqrt{2}}\left(\frac{\partial^3 u_1}{\partial x_1^2 \partial x_2}+\frac{\partial^3 u_2}{\partial x_1^3}\right)=0 \,, \\ 
\frac{\partial^3 u_1}{\partial x_2^2 \partial x_3}+2\frac{\partial^3 u_2}{\partial x_1 \partial x_2 \partial x_3}+3\frac{\partial^3 u_3}{\partial x_1 \partial x_2^2}-\frac{\partial^3 u_1}{\partial x_1^2 \partial x_3}-\frac{\partial^3 u_3}{\partial x_1^3}=0 \,, \\
2\left(1-\frac{1}{\sqrt{2}}\right)\frac{\partial^3 u_1}{\partial x_2^2 \partial x_3}+\left(-2+\sqrt{2}\right)\frac{\partial^3 u_2}{\partial x_1 \partial x_2 \partial x_3}-\left(4+\sqrt{2}\right)\frac{\partial^3 u_3}{\partial x_1 \partial x_2^2}+\sqrt{2}\left(2\frac{\partial^3 u_1}{\partial x_1^2 \partial x_3}+\frac{\partial^3 u_3}{\partial x_1^3}\right)=0 \,. 
\end{dcases}
\end{equation}
The fourth-order universality PDEs are written as
\begin{equation}
\begin{dcases}
\label{D_3-fourth-order-constraints-1}
\frac{\partial^4 u_3}{\partial x_1 \partial x_2^3}=\frac{\partial^4 u_3}{\partial x_1 \partial x_2 \partial x_3^2}=\frac{\partial^4 u_1}{\partial x_1 \partial x_2^3}=\frac{\partial^4 u_2}{\partial x_1^4}=\frac{\partial^4 u_2}{\partial x_2^2 \partial x_3^2}=\frac{\partial^4 u_2}{\partial x_1 \partial x_3^3}=0 \,, \\
\frac{\partial^4 u_3}{\partial x_1^2 \partial x_3^2}=\frac{\partial^4 u_2}{\partial x_1^2 \partial x_3^2}=\frac{\partial^4 u_3}{\partial x_3^4}=\frac{\partial^4 u_2}{\partial x_3^4}=\frac{\partial^4 u_3}{\partial x_2^2 \partial x_3^2}=\frac{\partial^4 u_3}{\partial x_2 \partial x_3^3}=\frac{\partial^4 u_1}{\partial x_1 \partial x_3^3}=0 \,, \\
\frac{\partial^4 u_3}{\partial x_2^3 \partial x_3}=\frac{\partial^4 u_1}{\partial x_1 \partial x_2 \partial x_3^2}=\frac{\partial^4 u_2}{\partial x_1^2 \partial x_3^2}=\frac{\partial^4 u_3}{\partial x_2 \partial x_3^3}=\frac{\partial^4 u_2}{\partial x_2^4}=0 \,, \\
\frac{\partial^4 u_2}{\partial x_1^2 \partial x_3^2}=\frac{\partial^4 u_1}{\partial x_1^3 \partial x_2}=\frac{\partial^4 u_3}{\partial x_1^2 \partial x_2 \partial x_3}=\frac{\partial^4 u_1}{\partial x_3^4}=0 \,, 
\end{dcases}
\end{equation}
as well as
\begin{equation}
\begin{dcases}
\label{D_3-fourth-order-constraints-2}
\frac{1}{\sqrt{2}} \frac{\partial^4 u_1}{\partial x_2^4}+\frac{1}{\sqrt{2}} \frac{\partial^4 u_2}{\partial x_1 \partial x_2^3}+\left(1+\sqrt{2}\right) \frac{\partial^4 u_1}{\partial x_1^2 \partial x_2^2}+\frac{\partial^4 u_2}{\partial x_1^3 \partial x_2}+\frac{\partial^4 u_3}{\partial x_1^4} = 0 \,, \\
\frac{\partial^4 u_1}{\partial x_1^2 \partial x_2^2}+\frac{\partial^4 u_2}{\partial x_1^3 \partial x_2}=0 \,, \\
-\frac{\partial^4 u_1}{\partial x_2^4}+\frac{\partial^4 u_2}{\partial x_1 \partial x_2^3}-\frac{\partial^4 u_1}{\partial x_1^2 \partial x_2^2}+\frac{\partial^4 u_2}{\partial x_1^3 \partial x_2}=0 \,, \\
\frac{\partial^4 u_2}{\partial x_1 \partial x_2 \partial x_3^2}+\frac{\partial^4 u_3}{\partial x_1 \partial x_2^2 \partial x_3}+\frac{\partial^4 u_1}{\partial x_1^2 \partial x_3^2}+\frac{\partial^4 u_3}{\partial x_1^3 \partial x_3}=0 \,, \\
\frac{\partial^4 u_1}{\partial x_1^2 \partial x_3^2}+\frac{\partial^4 u_3}{\partial x_1 \partial x_2^2 \partial x_3}+\frac{\partial^4 u_1}{\partial x_1^2 \partial x_3^2}+\frac{\partial^4 u_3}{\partial x_1^3 \partial x_3}=0 \,, \\
-\frac{1}{\sqrt{2}} \frac{\partial^4 u_1}{\partial x_2^4}+\left(1-\frac{1}{\sqrt{2}}\right) \frac{\partial^4 u_2}{\partial x_1 \partial x_2^3}-\left(-1+\sqrt{2}\right) \frac{\partial^4 u_1}{\partial x_1^2 \partial x_2^2}+\frac{\partial^4 u_2}{\partial x_1^3 \partial x_2}=0 \,, \\
\frac{\partial^4 u_1}{\partial x_2^2 \partial x_3^2}-\left(-1+\sqrt{2}\right) \frac{\partial^4 u_2}{\partial x_1 \partial x_2 \partial x_3^2}-2\frac{\partial^4 u_3}{\partial x_1 \partial x_2 \partial x_3^2}-\sqrt{2}\frac{\partial^4 u_3}{\partial x_1 \partial x_2^2 \partial x_3} =0 \,, \\
-\sqrt{2}\left(\frac{\partial^4 u_1}{\partial x_2^2 \partial x_3^2}+\frac{\partial^4 u_3}{\partial x_1 \partial x_2^2 \partial x_3}\right) +2\left(\frac{\partial^4 u_2}{\partial x_1 \partial x_2 \partial x_3^2}+\frac{\partial^4 u_3}{\partial x_1 \partial x_2^2 \partial x_3}\right)=0 \,, \\
-\sqrt{2} \frac{\partial^4 u_1}{\partial x_2^3 \partial x_3}-\left(-4+\sqrt{2}\right) \frac{\partial^4 u_2}{\partial x_1 \partial x_2^2 \partial x_3}+\frac{\partial^4 u_3}{\partial x_1 \partial x_2^3}-\frac{\partial^4 u_1}{\partial x_1^2 \partial x_2 \partial x_3}=0 \,, \\
\frac{\partial^4 u_2}{\partial x_1 \partial x_2^2 \partial x_3}-\frac{\partial^4 u_3}{\partial x_1 \partial x_2^3} -\frac{\partial^4 u_2}{\partial x_1^3 \partial x_2}+\frac{\partial^4 u_3}{\partial x_1^3 \partial x_3}=0 \,, \\
\frac{\partial^4 u_1}{\partial x_1^2 \partial x_2 \partial x_3}+\left(1-\frac{3}{\sqrt{2}}\right) \frac{\partial^4 u_2}{\partial x_1 \partial x_2^2 \partial x_3}+\frac{1}{\sqrt{2}} \left(\frac{\partial^4 u_1}{\partial x_1^2 \partial x_2 \partial x_3}+\frac{\partial^4 u_2}{\partial x_1^3 \partial x_3}+\frac{\partial^4 u_3}{\partial x_1^3 \partial x_2}\right)=0 \,, \\
2 \frac{\partial^4 u_1}{\partial x_2^3 \partial x_3}+\left(5-2\sqrt{2}\right)\frac{\partial^4 u_2}{\partial x_1 \partial x_2^2 \partial x_3}-\left(-1+2\sqrt{2}\right)\left(\frac{\partial^4 u_3}{\partial x_1 \partial x_2^3}+2\frac{\partial^4 u_2}{\partial x_1^2 \partial x_2 \partial x_3}+\frac{\partial^4 u_2}{\partial x_1^3 \partial x_3}+\frac{\partial^4 u_3}{\partial x_1^3 \partial x_2}\right)=0 \,, \\
-\frac{\partial^4 u_1}{\partial x_2^3 \partial x_3}-\frac{1}{2}\left(2+\sqrt{2}\right)\frac{\partial^4 u_2}{\partial x_1 \partial x_2^2 \partial x_3}-\frac{1}{\sqrt{2}}\left(\frac{\partial^4 u_1}{\partial x_1^2 \partial x_2 \partial x_3}+\frac{\partial^4 u_2}{\partial x_1^3 \partial x_3}+\frac{\partial^4 u_3}{\partial x_1^3 \partial x_2}\right)=0 \,, \\
\frac{\partial^4 u_1}{\partial x_2^3 \partial x_3}+\frac{\partial^4 u_2}{\partial x_1 \partial x_2^2 \partial x_3}-\left(1+2\sqrt{2}\right)\frac{\partial^4 u_2}{\partial x_1^3 \partial x_3}-2\sqrt{2}\frac{\partial^4 u_3}{\partial x_1^3 \partial x_2}=0 \,, \\
4 \frac{\partial^4 u_1}{\partial x_2^3 \partial x_3}+\left(3+2\sqrt{2}\right)\frac{\partial^4 u_2}{\partial x_1 \partial x_2^2 \partial x_3}-4\left(-1+\sqrt{2}\right)\frac{\partial^4 u_1}{\partial x_1^2 \partial x_2 \partial x_3}-\left(1+2\sqrt{2}\right)\frac{\partial^4 u_2}{\partial x_1^3 \partial x_3} \\
\qquad +\left(1-2\sqrt{2}\right) \frac{\partial^4 u_3}{\partial x_1^3 \partial x_2}=0 \,, \\
\frac{\partial^4 u_2}{\partial x_1 \partial x_2 \partial x_3^2}+\frac{\partial^4 u_3}{\partial x_1 \partial x_2^2 \partial x_3}=0 \,, \\
2 \frac{\partial^4 u_1}{\partial x_2^2 \partial x_3^2}-\left(-1+\sqrt{2}\right)\frac{\partial^4 u_2}{\partial x_1 \partial x_2 \partial x_3^2}+\frac{\partial^4 u_3}{\partial x_1 \partial x_2^2 \partial x_3} -\sqrt{2}\left(\frac{\partial^4 u_1}{\partial x_1^2 \partial x_3^2}+\frac{\partial^4 u_3}{\partial x_1^3 \partial x_3}\right)=0 \,, \\
\sqrt{2}\frac{\partial^4 u_1}{\partial x_2^4}+\left(1+\sqrt{2}\right)\frac{\partial^4 u_2}{\partial x_1 \partial x_2^3}-\left(1+\sqrt{2}\right)\frac{\partial^4 u_1}{\partial x_1^2 \partial x_2^2}-\left(1+\sqrt{2}\right)\frac{\partial^4 u_2}{\partial x_1^3 \partial x_2}+\frac{\partial^4 u_1}{\partial x_1^4}=0 \,, 
\end{dcases}
\end{equation}
also
\begin{equation}
\begin{dcases}
\label{D_3-fourth-order-constraints-3}
\frac{\partial^4 u_2}{\partial x_2^2 \partial x_3^2}+\frac{\partial^4 u_3}{\partial x_2^3 \partial x_3}+\frac{\partial^4 u_2}{\partial x_1^2 \partial x_3^2}+\frac{\partial^4 u_3}{\partial x_1^2 \partial x_2 \partial x_3^2}=0 \,, \\
-\left(-1+\sqrt{2}\right)\frac{\partial^4 u_1}{\partial x_1 \partial x_2 \partial x_3^2}+\frac{\partial^4 u_2}{\partial x_1^2 \partial x_3^2}+2\frac{\partial^4 u_3}{\partial x_1^2 \partial x_2 \partial x_3^2}=0 \,, \\
2\left(\frac{\partial^4 u_1}{\partial x_1 \partial x_2 \partial x_3^2}+\frac{\partial^4 u_3}{\partial x_1^2 \partial x_2 \partial x_3}\right)-\sqrt{2}\left(\frac{\partial^4 u_2}{\partial x_1^2 \partial x_3^2}+\frac{\partial^4 u_3}{\partial x_1^2 \partial x_2 \partial x_3}\right)=0 \,, \\
\frac{\partial^4 u_2}{\partial x_3^4}+\frac{\partial^4 u_3}{\partial x_2 \partial x_3^3}=0 \,, \\
-3 \frac{\partial^4 u_2}{\partial x_1^2 \partial x_2 \partial x_3}-2\frac{\partial^4 u_3}{\partial x_1^2 \partial x_2^2}+\frac{\partial^4 u_1}{\partial x_1^3 \partial x_3}=0 \,, \\
-\frac{\partial^4 u_2}{\partial x_1^2 \partial x_2 \partial x_3^2}-\sqrt{2}\left(\frac{\partial^4 u_1}{\partial x_1 \partial x_2^2 \partial x_3^2}+\frac{\partial^4 u_3}{\partial x_1^2 \partial x_2^2}\right)+ \frac{\partial^4 u_1}{\partial x_1^3 \partial x_3}=0 \,, \\
-\frac{\partial^4 u_2}{\partial x_2 \partial x_3^3}+\frac{\partial^4 u_1}{\partial x_1 \partial x_3^3}=0 \,, \\
\frac{\partial^4 u_1}{\partial x_1 \partial x_2^2 \partial x_3}+\frac{1}{2}\frac{\partial^4 u_2}{\partial x_1^2 \partial x_2 \partial x_3^2}+\frac{\partial^4 u_3}{\partial x_1^2 \partial x_2^2}-\frac{1}{2}\frac{\partial^4 u_1}{\partial x_1^3 \partial x_2}=0 \,, \\
-\left(1+\sqrt{2}\right)\frac{\partial^4 u_1}{\partial x_1 \partial x_2^2 \partial x_3}-\sqrt{2} \frac{\partial^4 u_2}{\partial x_1^2 \partial x_2 \partial x_3}+\frac{\partial^4 u_1}{\partial x_1^3 \partial x_3}=0 \,, \\
\frac{\partial^4 u_2}{\partial x_2^3 \partial x_3}-\frac{3}{2}\frac{\partial^4 u_2}{\partial x_1^2 \partial x_2 \partial x_3}-\frac{\partial^4 u_3}{\partial x_1^2 \partial x_2^2}-\frac{1}{2}\frac{\partial^4 u_1}{\partial x_1^3 \partial x_3}=0 \,, \\
3 \frac{\partial^4 u_2}{\partial x_2^3 \partial x_3}+\frac{\partial^4 u_3}{\partial x_2^4}-\left(1+2\sqrt{2}\right)\left(2 \frac{\partial^4 u_1}{\partial x_1 \partial x_2^2 \partial x_3}+\frac{\partial^4 u_2}{\partial x_1^2 \partial x_2 \partial x_3}+\frac{\partial^4 u_3}{\partial x_1^2 \partial x_2^2}\right)=0 \,, \\
-\left(2+3\sqrt{2}\right)\frac{\partial^4 u_1}{\partial x_1 \partial x_2^2 \partial x_3}-\sqrt{2}\frac{\partial^4 u_2}{\partial x_1^2 \partial x_2 \partial x_3}-\left(1+2\sqrt{2}\right)\frac{\partial^4 u_3}{\partial x_1^2 \partial x_2^2}+2\frac{\partial^4 u_1}{\partial x_1^3 \partial x_3}+\frac{\partial^4 u_3}{\partial x_1^4}=0 \,, \\
-\frac{\partial^4 u_2}{\partial x_2 \partial x_3^3}-\frac{\partial^4 u_3}{\partial x_2^2 \partial x_3^2}+\frac{\partial^4 u_1}{\partial x_1 \partial x_3^3}=0 \,, \\
\frac{\partial^4 u_2}{\partial x_2^2 \partial x_3^2}+\frac{\partial^4 u_1}{\partial x_1 \partial x_2 \partial x_3^2}=0 \,, \\
\frac{\partial^4 u_3}{\partial x_2^3 \partial x_3}+\frac{\partial^4 u_2}{\partial x_1^2 \partial x_2 \partial x_3}=0 \,, \\
-\sqrt{2}\frac{\partial^4 u_1}{\partial x_1 \partial x_2 \partial x_3^2}+\frac{\partial^4 u_3}{\partial x_1^2 \partial x_2 \partial x_3}=0 \,, \\
\frac{\partial^4 u_2}{\partial x_1^2 \partial x_3^2}-\frac{\partial^4 u_3}{\partial x_1^2 \partial x_2 \partial x_3}=0 \,, \\
2 \left(2+\sqrt{2}\right)\frac{\partial^4 u_2}{\partial x_1^2 \partial x_2^2}-2\sqrt{2}\frac{\partial^4 u_1}{\partial x_1^3 \partial x_2}=0 \,, 
\end{dcases}
\end{equation}
and finally
\begin{equation}
\begin{dcases}
\label{Z_3-fourth-order-constraints-4}
\frac{\partial^4 u_2}{\partial x_2^3 \partial x_3}+\frac{\partial^4 u_1}{\partial x_1 \partial x_2 \partial x_3^2}+\frac{\partial^4 u_2}{\partial x_1^2 \partial x_2 \partial x_3}+\frac{\partial^4 u_1}{\partial x_1^3 \partial x_3}=0 \,, \\
\frac{\partial^4 u_1}{\partial x_1 \partial x_2^2 \partial x_3}+\frac{\partial^4 u_2}{\partial x_1^2 \partial x_2 \partial x_3}=0 \,, \\
-\frac{3}{\sqrt{2}}\frac{\partial^4 u_2}{\partial x_1^2 \partial x_3^2}-\frac{1}{\sqrt{2}}\frac{\partial^4 u_3}{\partial x_2^3 \partial x_3}+\sqrt{2}\frac{\partial^4 u_1}{\partial x_1 \partial x_2 \partial x_3^2}+\frac{1}{2}\left(4+\sqrt{2}\right)\left(\frac{\partial^4 u_2}{\partial x_1^2 \partial x_3^2}+\frac{\partial^4 u_3}{\partial x_1^2 \partial x_2 \partial x_3}\right)=0 \,, \\
\frac{\partial^4 u_2}{\partial x_2^2 \partial x_3^2}-\left(1+\sqrt{2}\right)\frac{\partial^4 u_1}{\partial x_1 \partial x_2 \partial x_3^2}-\sqrt{2}\frac{\partial^4 u_2}{\partial x_1^2 \partial x_3^2}=0 \,, \\
4 \frac{\partial^4 u_2}{\partial x_2^4}-\left(2+\sqrt{2}\right)\frac{\partial^4 u_1}{\partial x_1 \partial x_2^3}-3\sqrt{2}\frac{\partial^4 u_2}{\partial x_1^4}=0 \,, \\
\frac{\partial^4 u_2}{\partial x_2^2 \partial x_3^2}+\frac{\partial^4 u_3}{\partial x_2^3 \partial x_3}-4\frac{\partial^4 u_1}{\partial x_1 \partial x_2 \partial x_3^2}-\frac{\partial^4 u_2}{\partial x_1^2 \partial x_3^2}=0 \,, \\
2\frac{\partial^4 u_3}{\partial x_1^2 \partial x_2^2}+\frac{\partial^4 u_1}{\partial x_1^3 \partial x_3}+\frac{\partial^4 u_3}{\partial x_1^4}+\frac{\partial^4 u_2}{\partial x_2^3 \partial x_3}+\frac{\partial^4 u_3}{\partial x_24}+\frac{\partial^4 u_1}{\partial x_1 \partial x_2^2 \partial x_3}+\frac{\partial^4 u_2}{\partial x_1^2 \partial x_2 \partial x_3}=0  \,, \\
\frac{\partial^4 u_2}{\partial x_2^3 \partial x_3}+\frac{\partial^4 u_1}{\partial x_1 \partial x_2^2 \partial x_3}+\frac{\partial^4 u_2}{\partial x_1^2 \partial x_2 \partial x_3}+\frac{\partial^4 u_1}{\partial x_1^3 \partial x_3}=0 \,, \\
\frac{\partial^4 u_1}{\partial x_1 \partial x_2^2 \partial x_3}+\frac{\partial^4 u_2}{\partial x_1^2 \partial x_2 \partial x_3}=0 \,, \\
-\sqrt{2}\frac{\partial^4 u_1}{\partial x_1^2 \partial x_2 \partial x_3}-\left(-2+\sqrt{2}\right)\frac{\partial^4 u_2}{\partial x_1^2 \partial x_2 \partial x_3}-2\left(-1+\sqrt{2}\right)\frac{\partial^4 u_3}{\partial x_1^2 \partial x_2^2}=0 \,, \\
2\frac{\partial^4 u_1}{\partial x_1 \partial x_2^ \partial x_3} +\frac{\partial^4 u_2}{\partial x_1^2 \partial x_2 \partial x_3}+\frac{\partial^4 u_3}{\partial x_1^2 \partial x_2^2}=0 \,. 
\end{dcases}
\end{equation}

Substituting the inhomogeneous displacement field \eqref{trigonal-linear-inhomogeneous} into the above universality PDEs, we find that they are satisfied provided that $a_{123}=0$. Accordingly, we have the following result.
\begin{prop}
The universal displacements in trigonal $\mathbb{D}_3$ class linear strain-gradient elastic solids are the superposition of homogeneous displacement fields and the inhomogeneous displacement field \eqref{trigonal-linear-inhomogeneous} with $a_{123}=0$. 
\end{prop}

\subsubsection{Trigonal class $\mathbb{Z}_3 \oplus \mathbb{Z}_2^c$} 

Tensor $\boldsymbol{\mathsf{A}}_{\mathbb{Z}_3 \oplus \mathbb{Z}_2^c}$ has $57$ independent components and has the same form as that of class $\mathbb{Z}_3$. Tensor $\boldsymbol{\mathsf{M}}_{\mathbb{Z}_3 \oplus \mathbb{Z}_2^c}$ is a null tensor. Tensor $\boldsymbol{\mathsf{C}}_{\mathbb{D}_3 \oplus \mathbb{Z}_2^c}$ has $6$ independent components and corresponds to the Hermann--Maugin symbol $\bar{3}m$.

The universality PDEs for this class are only of fourth order and coincide with those of the trigonal class $\mathbb{Z}_3$. Substituting the inhomogeneous displacement field \eqref{trigonal-linear-inhomogeneous} into these universality PDEs, we find that they are satisfied. Accordingly, we have the following result.
\begin{prop}
The universal displacements in trigonal $\mathbb{Z}_3 \oplus \mathbb{Z}_2^c$ class linear strain-gradient elastic solids are the superposition of homogeneous displacement fields and the inhomogeneous displacement field \eqref{trigonal-linear-inhomogeneous}.
\end{prop}

\subsubsection{Trigonal class $\mathbb{D}_3 \oplus \mathbb{Z}_2^c$} 

Tensor $\boldsymbol{\mathsf{A}}_{\mathbb{D}_3 \oplus \mathbb{Z}_2^c}$ has $34$ independent components and has the same form as that of class $\mathbb{D}_3$. Tensor $\boldsymbol{\mathsf{M}}_{\mathbb{D}_3 \oplus \mathbb{Z}_2^c}$ is a null tensor. Tensor $\boldsymbol{\mathsf{C}}_{\mathbb{D}_3 \oplus \mathbb{Z}_2^c}$ has the same form as that of class $\mathbb{D}_3$, with $6$ independent components, the only difference being that it corresponds to the Hermann--Maugin symbol $\bar{3}m$.

The universality PDEs for this class are only of fourth order and coincide with those of the trigonal class $\mathbb{D}_3$. Substituting the inhomogeneous displacement field \eqref{trigonal-linear-inhomogeneous} into these universality PDEs, we find that they are satisfied. Accordingly, we have the following result.
\begin{prop}
The universal displacements in trigonal $\mathbb{D}_3 \oplus \mathbb{Z}_2^c$ class linear strain-gradient elastic solids are the superposition of homogeneous displacement fields and the inhomogeneous displacement field \eqref{trigonal-linear-inhomogeneous}.
\end{prop}

\subsubsection{Trigonal class $\mathbb{Z}_6^-$} 

Tensor $\boldsymbol{\mathsf{A}}_{\mathbb{Z}_6 \oplus \mathbb{Z}_2^c}$ has $33$ independent components and has the same form as that of class $\mathbb{Z}_6$.
Tensor $\boldsymbol{\mathsf{M}}_{\mathbb{Z}_6^- }$ has $16$ independent components and is of the form
\begin{equation}
	\label{M_Z_6^-}
	\boldsymbol{\mathsf{M}}_{\mathbb{Z}_6^-}=\boldsymbol{\mathsf{M}}_{\mathbb{D}_6^h}+\begin{bmatrix}
		0 & \bar{B}^{(6)} & 0 & 0  \\
		0 & 0 & 0 & 0   \\
		0 & 0  & 0 & \bar{L}^{(2)}     \\
		0 & 0 & 0 & 0   
	\end{bmatrix} 
	+\begin{bmatrix}
		0 & 0 & 0 & 0  \\
		0 & 0 & -g(\bar{L}^{(2)}) & 0   \\
		0 & 0  & 0 & 0     \\
		-g(\bar{B}^{(6)}) & 0 & 0 & 0   
	\end{bmatrix}\,,
\end{equation}
where
\begin{equation}
	\label{M_D_6^h}
	\boldsymbol{\mathsf{M}}_{\mathbb{D}_6^h}=\begin{bmatrix}
		\bar{A}^{(6)} & 0 & 0 & 0  \\
		0 & 0 & 0 & \bar{H}^{(2)}   \\
		0 & 0  & 0 & 0     \\
		0 & 0 & 0 & 0   
	\end{bmatrix} 
	+\begin{bmatrix}
		0 & 0 & 0 & 0  \\
		0 & 0 & 0 & 0   \\
		0 & 0  & g(\bar{H}^{(2)}) & 0     \\
		0 & g(\bar{A}^{(6)}) & 0 & 0   
	\end{bmatrix}\,.
\end{equation}
Tensor $\boldsymbol{\mathsf{C}}_{\mathbb{O}(2) \oplus \mathbb{Z}_2^c}$ has $5$ independent components and corresponds to the Curie group with Hermann--Maugin symbol $\infty m$. Consequently, from the classical linear elastic part we obtain the universality constraints \eqref{transverse-isotropy-linear-constraints}, and the candidate universal displacements are those given in \eqref{transverse-isotropy-linear-solution}.

We obtain for this class the following third-order universality PDEs:
\begin{equation}
\begin{dcases}
\label{Z_6^-third-order-constraints-1}
- \sqrt{2} \left( \frac{\partial^3 k_2}{\partial x_2^3} + x_3 \frac{\partial^3 h_2}{\partial x_2^3}\right)
+\left(1+\frac{1}{\sqrt{2}}\right) \left( \frac{\partial^3 k_1}{\partial x_1 \partial x_2^2} + x_3 \frac{\partial^3 h_1}{\partial x_1 \partial x_2^2}\right)
+\left(-8+\sqrt{2}\right) \left( \frac{\partial^3 k_2}{\partial x_1^2 \partial x_2} + x_3 \frac{\partial^3 h_2}{\partial x_1^2 \partial x_2}\right)=0  \\
-\frac{1}{\sqrt{2}} \left(\frac{\partial^3 k_2}{\partial x_2^3}+x_3  \frac{\partial^3 h_2}{\partial x_2^3}  \right)
-  \frac{\partial^3 k_1}{\partial x_1 \partial x_2^2} -x_3  \frac{\partial^3 h_1}{\partial x_1 \partial x_2^2}
+ \frac{1}{2}\left(4+\sqrt{2}\right) \left( \frac{\partial^3 k_2}{\partial x_1^2 \partial x_2} + x_3  \frac{\partial^3 h_2}{\partial x_1^2 \partial x_2}   \right) =0 \,, \\
-\frac{1}{2}\left(2+\sqrt{2}\right) \left(\frac{\partial^3 k_2}{\partial x_2^3}+x_3  \frac{\partial^3 h_2}{\partial x_2^3}  \right)
+ \left(1+\frac{3}{\sqrt{2}}\right) \left( \frac{\partial^3 k_1}{\partial x_1 \partial x_2^2} +x_3  \frac{\partial^3 h_1}{\partial x_1 \partial x_2^2} \right)
+ \left(2+\sqrt{2} \right) \left( \frac{\partial^3 k_2}{\partial x_1^2 \partial x_2} + x_3  \frac{\partial^3 h_2}{\partial x_1^2 \partial x_2}   \right) =0 \,, \\
- \sqrt{2} \left(\frac{\partial^3 k_1}{\partial x_2^3} + x_3  \frac{\partial^3 h_1}{\partial x_2^3}\right)
+ 2\left( \frac{\partial^3 k_2}{\partial x_1 \partial x_2^2}+x_3  \frac{\partial^3 h_2}{\partial x_1 \partial x_2^2} \right)
- \left( \frac{\partial^3 k_1}{\partial x_1^2 \partial x_2}+x_3  \frac{\partial^3 h_1}{\partial x_1^2 \partial x_2}\right) \\
+3\left(\frac{\partial^3 k_2}{\partial x_1^3}+x_3  \frac{\partial^3 h_2}{\partial x_1^3}\right)=0 \,, \\
\frac{1}{\sqrt{2}}  \left(\frac{\partial^3 k_1}{\partial x_2^3} + x_3  \frac{\partial^3 h_1}{\partial x_2^3}\right)
- \left(1-\frac{1}{\sqrt{2}}\right) \left( \frac{\partial^3 k_2}{\partial x_1 \partial x_2^2}+x_3  \frac{\partial^3 h_2}{\partial x_1 \partial x_2^2} \right)
+\sqrt{2} \left( \frac{\partial^3 k_1}{\partial x_1^2 \partial x_2}+x_3  \frac{\partial^3 h_1}{\partial x_1^2 \partial x_2}\right)=0 \,, \\
-\frac{1}{2}\left(2+\sqrt{2}\right)  \left(\frac{\partial^3 k_1}{\partial x_2^3} + x_3  \frac{\partial^3 h_1}{\partial x_2^3}\right)
- \left(1+\sqrt{2}\right) \left( \frac{\partial^3 k_2}{\partial x_1 \partial x_2^2}+x_3  \frac{\partial^3 h_2}{\partial x_1 \partial x_2^2} \right)
-\frac{4}{\sqrt{2}} \left( \frac{\partial^3 k_1}{\partial x_1^2 \partial x_2}+x_3  \frac{\partial^3 h_1}{\partial x_1^2 \partial x_2}\right) \\
+\frac{3}{\sqrt{2}} \left(\frac{\partial^3 k_2}{\partial x_1^3}+x_3  \frac{\partial^3 h_2}{\partial x_1^3}\right)=0 \,,
\end{dcases}
\end{equation}
and
\begin{equation}
\begin{dcases}
\label{Z_6^-third-order-constraints-2}
\frac{1}{\sqrt{2}} \left( \frac{\partial^3 k_1}{\partial x_2^3} + x_3 \frac{\partial^3 h_1}{\partial x_2^3}\right) + \frac{1}{\sqrt{2}} \left( \frac{\partial^3 k_2}{\partial x_1 \partial x_2^2} + x_3 \frac{\partial^3 h_2}{\partial x_1 \partial x_2^2}\right)-\left(2+\sqrt{2}\right) \left( \frac{\partial^3 k_1}{\partial x_1^2 \partial x_2} + x_3 \frac{\partial^3 h_1}{\partial x_1^2 \partial x_2}\right)  \\
- \left( \frac{\partial^3 k_2}{\partial x_1^3} + x_3 \frac{\partial^3 h_2}{\partial x_1^3}\right)=0 \,, \\
\frac{1}{\sqrt{2}} \left( \frac{\partial^3 k_1}{\partial x_2^3} + x_3 \frac{\partial^3 h_1}{\partial x_2^3}\right)-\left(1+\sqrt{2}\right) \left( \frac{\partial^3 k_1}{\partial x_1^2 \partial x_2} + x_3 \frac{\partial^3 h_1}{\partial x_1^2 \partial x_2}\right)  \\
+\frac{1}{\sqrt{2}} \left( \frac{\partial^3 k_2}{\partial x_1^3} + x_3 \frac{\partial^3 h_2}{\partial x_1^3}\right)=0 \,, \\
\sqrt{2} \left( \frac{\partial^3 k_2}{\partial x_2^3} + x_3 \frac{\partial^3 h_2}{\partial x_2^3}\right) + \left(1-\frac{1}{\sqrt{2}}\right) \left( \frac{\partial^3 k_2}{\partial x_1 \partial x_2^2} + x_3 \frac{\partial^3 h_2}{\partial x_1 \partial x_2^2}\right)-2\left(1+\sqrt{2}\right) \left( \frac{\partial^3 k_2}{\partial x_1^2 \partial x_2} + x_3 \frac{\partial^3 h_2}{\partial x_1^2 \partial x_2}\right)  \\
+\frac{1}{2}\left(2+\sqrt{2}\right) \left( \frac{\partial^3 k_1}{\partial x_1^3} + x_3 \frac{\partial^3 h_1}{\partial x_1^3}\right)=0 \,, \\
-\frac{1}{2} \left( \frac{\partial^3 k_2}{\partial x_1^2 \partial x_2} + x_3 \frac{\partial^3 h_2}{\partial x_1^2 \partial x_2}\right)-\frac{2}{\sqrt{2}} \left( \frac{\partial^3 k_1}{\partial x_1^3} + x_3 \frac{\partial^3 h_1}{\partial x_1^3}\right)-\frac{1}{\sqrt{2}} \left( \frac{\partial^3 k_1}{\partial x_1 \partial x_2^2} + x_3 \frac{\partial^3 h_1}{\partial x_1 \partial x_2^2}\right)=0 \,, \\
\left(2+\frac{1}{\sqrt{2}}\right) \left( \frac{\partial^3 k_1}{\partial x_1 \partial x_2^2} + x_3 \frac{\partial^3 h_1}{\partial x_1 \partial x_2^2}\right)+\frac{1}{\sqrt{2}} \left( \frac{\partial^3 k_2}{\partial x_1^2 \partial x_2} + x_3 \frac{\partial^3 h_2}{\partial x_1^2 \partial x_2}\right) -\frac{1}{\sqrt{2}} \left( \frac{\partial^3 k_1}{\partial x_1^3} + x_3 \frac{\partial^3 h_1}{\partial x_1^3}\right)=0 \,, \\
-\frac{1}{\sqrt{2}} \left( \frac{\partial^3 k_2}{\partial x_1^2 \partial x_2} + x_3 \frac{\partial^3 h_2}{\partial x_1^2 \partial x_2}\right)-\frac{1}{2}\left(2+\sqrt{2}\right)\left( \frac{\partial^3 k_1}{\partial x_1^3} + x_3 \frac{\partial^3 h_1}{\partial x_1^3}\right)=0 \,, 
\end{dcases}
\end{equation}
as well as
\begin{equation}
\begin{dcases}
\label{Z_6^-third-order-constraints-3}
2 \sqrt{2} \frac{\partial^2 h_1}{\partial x_2^2}+\left(1+\sqrt{2}\right) \frac{\partial^2 h_2}{\partial x_1 \partial x_2}- \frac{\partial^2 h_1}{\partial x_1^2}=0 \,, \\
\frac{\partial^2 h_1}{\partial x_2^2}+\left(1+\frac{1}{\sqrt{2}}\right) \frac{\partial^2 h_2}{\partial x_1 \partial x_2}-\frac{1}{\sqrt{2}} \frac{\partial^2 h_1}{\partial x_1^2}=0 \,, \\
\frac{\partial^2 h_1}{\partial x_2^2}+2 \frac{\partial^2 h_2}{\partial x_1 \partial x_2}+2\frac{\partial^3 u_{33}}{\partial x_1^1 \partial x_2^2}- \frac{\partial^2 h_1}{\partial x_1^2}-\frac{\partial^3 \hat{u}_{33}}{\partial x_1^3}=0 \,, \\
\left(1-\frac{1}{\sqrt{2}}\right)\frac{\partial^2 h_1}{\partial x_2^2}+\left(-2+\sqrt{2}\right) \frac{\partial^2 h_2}{\partial x_1 \partial x_2}-\left(4+\sqrt{2}\right)\frac{\partial^3 \hat{u}_{33}}{\partial x_1^1 \partial x_2^2}+\sqrt{2} \left( \frac{\partial^2 h_1}{\partial x_1^2}-\frac{\partial^3 \hat{u}_{33}}{\partial x_1^3}\right)=0 \,, \\
\frac{\partial^2 h_2}{\partial x_2^2}+\left(-1+\sqrt{2}\right) \frac{\partial^2 h_1}{\partial x_1 \partial x_2}+2\sqrt{2} \frac{\partial^2 h_2}{\partial x_1^2}=0 \,, \\
-\frac{1}{\sqrt{2}} \frac{\partial^2 h_2}{\partial x_2^2}+\left(1+\frac{1}{\sqrt{2}}\right) \frac{\partial^2 h_1}{\partial x_1 \partial x_2}+ \frac{\partial^2 h_2}{\partial x_1^2}=0 \,, \\
\frac{\partial^2 h_2}{\partial x_2^2}  +\frac{\partial^3 \hat{u}_{33}}{\partial x_2^3}  +\frac{\partial^2 h_2}{\partial x_1^2}+\frac{\partial^3 \hat{u}_{33}}{\partial x_1^2 \partial x_2} =0\,, \\
-\sqrt{2} \frac{\partial^3 \hat{u}_{33}}{\partial x_2^3}  -\left(2+\sqrt{2}\right) \frac{\partial^2 h_1}{\partial x_1 \partial x_2}+\left(2+\sqrt{2}\right)  \frac{\partial^2 h_2}{\partial x_1^2}+\left(-4+\sqrt{2}\right)  \frac{\partial^3 \hat{u}_{33}}{\partial x_1^2 \partial x_2} =0\,. 
\end{dcases}
\end{equation}
The fourth-order universality PDEs for this class read
\begin{equation}
\begin{dcases}
\label{Z_6^-fourth-order-constraints-1}
\frac{1}{\sqrt{2}} \left( \frac{\partial^4 k_1}{\partial x_2^4} + x_3 \frac{\partial^4 h_1}{\partial x_2^4}\right)+\frac{1}{\sqrt{2}}\left( \frac{\partial^4 k_2}{\partial x_1 \partial x_2^3} + x_3 \frac{\partial^4 h_2}{\partial x_1 \partial x_2^3}\right)+\left(1+\sqrt{2}\right) \left( \frac{\partial^4 k_1}{\partial x_1^2 \partial x_2^2} + x_3 \frac{\partial^4 h_1}{\partial x_1^2 \partial x_2^2}\right) \\
+\left( \frac{\partial^4 k_2}{\partial x_1^3 \partial x_2} + x_3 \frac{\partial^4 h_2}{\partial x_1^3 \partial x_2}\right) +\left( \frac{\partial^4 k_1}{\partial x_1^4} + x_3 \frac{\partial^4 h_1}{\partial x_1^4}\right)=0 \,, \\
-\left( \frac{\partial^4 k_1}{\partial x_1^2 \partial x_2^2} + x_3 \frac{\partial^4 h_1}{\partial x_1^2 \partial x_2^2}\right)-\left( \frac{\partial^4 k_2}{\partial x_1^3 \partial x_2} + x_3 \frac{\partial^4 h_2}{\partial x_1^3 \partial x_2}\right)=0 \,, \\
-\left( \frac{\partial^4 k_1}{\partial x_2^4} + x_3 \frac{\partial^4 h_1}{\partial x_2^4}\right)+\left( \frac{\partial^4 k_2}{\partial x_1 \partial x_2^3} + x_3 \frac{\partial^4 h_2}{\partial x_1 \partial x_2^3}\right) \\
-\left( \frac{\partial^4 k_1}{\partial x_1^2 \partial x_2^2} + x_3 \frac{\partial^4 h_1}{\partial x_1^2 \partial x_2^2}\right)+\left( \frac{\partial^4 k_2}{\partial x_1^3 \partial x_2} + x_3 \frac{\partial^4 h_2}{\partial x_1^3 \partial x_2}\right)=0 \,, \\
-\frac{1}{\sqrt{2}} \left( \frac{\partial^4 k_1}{\partial x_2^4} + x_3 \frac{\partial^4 h_1}{\partial x_2^4}\right)+\left(1-\frac{1}{\sqrt{2}}\right) \left( \frac{\partial^4 k_2}{\partial x_1 \partial x_2^3} + x_3 \frac{\partial^4 h_2}{\partial x_1 \partial x_2^3}\right)-\left(-1+\sqrt{2}\right) \left( \frac{\partial^4 k_1}{\partial x_1^2 \partial x_2^2} + x_3 \frac{\partial^4 h_1}{\partial x_1^2 \partial x_2^2}\right) \\
+\left( \frac{\partial^4 k_2}{\partial x_1^3 \partial x_2} + x_3 \frac{\partial^4 h_2}{\partial x_1^3 \partial x_2}\right)=0 \,, \\
\sqrt{2} \left( \frac{\partial^4 k_1}{\partial x_2^4} + x_3 \frac{\partial^4 h_1}{\partial x_2^4}\right)+\left(1+\sqrt{2}\right) \left( \frac{\partial^4 k_2}{\partial x_1 \partial x_2^3} + x_3 \frac{\partial^4 h_2}{\partial x_1 \partial x_2^3}\right)-\left(1+\sqrt{2}\right) \left( \frac{\partial^4 k_1}{\partial x_1^2 \partial x_2^2} + x_3 \frac{\partial^4 h_1}{\partial x_1^2 \partial x_2^2}\right) \\
-\left(1+\sqrt{2}\right)\left( \frac{\partial^4 k_2}{\partial x_1^3 \partial x_2} + x_3 \frac{\partial^4 h_2}{\partial x_1^3 \partial x_2}\right) +\left( \frac{\partial^4 k_1}{\partial x_1^4} + x_3 \frac{\partial^4 h_1}{\partial x_1^4}\right)=0 \,, \\
-2 \left( \frac{\partial^4 k_2}{\partial x_2^4} + x_3 \frac{\partial^4 h_2}{\partial x_2^4}\right)-\sqrt{2} \left( \frac{\partial^4 k_1}{\partial x_1 \partial x_2^3} + x_3 \frac{\partial^4 h_1}{\partial x_1 \partial x_2^3}\right)-3\sqrt{2} \left( \frac{\partial^4 k_2}{\partial x_1^2 \partial x_2^2} + x_3 \frac{\partial^4 h_2}{\partial x_1^2 \partial x_2^2}\right) \\
+\left(2-3\sqrt{2}\right)\left( \frac{\partial^4 k_1}{\partial x_1^3 \partial x_2} + x_3 \frac{\partial^4 h_1}{\partial x_1^3 \partial x_2}\right) -\sqrt{2}\left( \frac{\partial^4 k_2}{\partial x_1^4} + x_3 \frac{\partial^4 h_2}{\partial x_1^4}\right)=0 \,, \\
-4 \left( \frac{\partial^4 k_2}{\partial x_2^4} + x_3 \frac{\partial^4 h_2}{\partial x_2^4}\right)+\left(10+\sqrt{2}\right) \left( \frac{\partial^4 k_1}{\partial x_1 \partial x_2^3} + x_3 \frac{\partial^4 h_1}{\partial x_1 \partial x_2^3}\right)+\left(10+3\sqrt{2}\right) \left( \frac{\partial^4 k_2}{\partial x_1^2 \partial x_2^2} + x_3 \frac{\partial^4 h_2}{\partial x_1^2 \partial x_2^2}\right) \\
-9\sqrt{2}\left( \frac{\partial^4 k_1}{\partial x_1^3 \partial x_2} + x_3 \frac{\partial^4 h_1}{\partial x_1^3 \partial x_2}\right) +\left( \frac{\partial^4 k_2}{\partial x_1^4} + x_3 \frac{\partial^4 h_2}{\partial x_1^4}\right)=0 \,, 
\end{dcases}
\end{equation}
and 
\begin{equation}
\begin{dcases}
\label{Z_6^-fourth-order-constraints-2}
\left( \frac{\partial^4 k_2}{\partial x_2^4} + x_3 \frac{\partial^4 h_2}{\partial x_2^4}\right) + \left( \frac{\partial^4 k_1}{\partial x_1 \partial x_2^3} + x_3 \frac{\partial^4 h_1}{\partial x_1 \partial x_2^3}\right)+\left(1+\sqrt{2}\right) \left( \frac{\partial^4 k_2}{\partial x_1^2 \partial x_2^2} + x_3 \frac{\partial^4 h_2}{\partial x_1^2 \partial x_2^2}\right) \\
+\frac{1}{\sqrt{2}} \left( \frac{\partial^4 k_1}{\partial x_1^3 \partial x_2} + x_3 \frac{\partial^4 h_1}{\partial x_1^3 \partial x_2}\right) + \frac{1}{\sqrt{2}}\left( \frac{\partial^4 k_2}{\partial x_1^4} + x_3 \frac{\partial^4 h_2}{\partial x_1^4}\right)=0 \,, \\
\left( \frac{\partial^4 k_1}{\partial x_1 \partial x_2^3} + x_3 \frac{\partial^4 h_1}{\partial x_1 \partial x_2^3}\right)+ \left( \frac{\partial^4 k_2}{\partial x_1^2 \partial x_2^2} + x_3 \frac{\partial^4 h_2}{\partial x_1^2 \partial x_2^2}\right) =0 \,, \\
\left( \frac{\partial^4 k_1}{\partial x_1^3 \partial x_2} + x_3 \frac{\partial^4 h_1}{\partial x_1^3 \partial x_2}\right)- \left( \frac{\partial^4 k_2}{\partial x_1^4} + x_3 \frac{\partial^4 h_2}{\partial x_1^4}\right) \\
+\left( \frac{\partial^4 k_1}{\partial x_1 \partial x_2^3} + x_3 \frac{\partial^4 h_1}{\partial x_1 \partial x_2^3}\right)- \left( \frac{\partial^4 k_2}{\partial x_1^2 \partial x_2^2} + x_3 \frac{\partial^4 h_2}{\partial x_1^2 \partial x_2^2}\right) =0 \,, \\
\left( \frac{\partial^4 k_1}{\partial x_1 \partial x_2^3} + x_3 \frac{\partial^4 h_1}{\partial x_1 \partial x_2^3}\right)-\left(-1+\sqrt{2}\right) \left( \frac{\partial^4 k_2}{\partial x_1^2 \partial x_2^2} + x_3 \frac{\partial^4 h_2}{\partial x_1^2 \partial x_2^2}\right) \\
+\left(1-\frac{1}{\sqrt{2}}\right) \left( \frac{\partial^4 k_1}{\partial x_1^3 \partial x_2} + x_3 \frac{\partial^4 h_1}{\partial x_1^3 \partial x_2}\right) - \frac{1}{\sqrt{2}}\left( \frac{\partial^4 k_2}{\partial x_1^4} + x_3 \frac{\partial^4 h_2}{\partial x_1^4}\right)=0 \,, \\
\left( \frac{\partial^4 k_1}{\partial x_1 \partial x_2^3} + x_3 \frac{\partial^4 h_1}{\partial x_1 \partial x_2^3}\right)+2\left(2+\sqrt{2}\right) \left( \frac{\partial^4 k_2}{\partial x_1^2 \partial x_2^2} + x_3 \frac{\partial^4 h_2}{\partial x_1^2 \partial x_2^2}\right) -\sqrt{2} \left( \frac{\partial^4 k_1}{\partial x_1^3 \partial x_2} + x_3 \frac{\partial^4 h_1}{\partial x_1^3 \partial x_2}\right)=0 \,, \\
\left( \frac{\partial^4 k_1}{\partial x_2^4} + x_3 \frac{\partial^4 h_1}{\partial x_2^4}\right) + \left(-1+\frac{3}{\sqrt{2}}\right)\left( \frac{\partial^4 k_2}{\partial x_1 \partial x_2^3} + x_3 \frac{\partial^4 h_2}{\partial x_1 \partial x_2^3}\right)+\frac{3}{\sqrt{2}} \left( \frac{\partial^4 k_1}{\partial x_1^2 \partial x_2^2} + x_3 \frac{\partial^4 h_1}{\partial x_1^2 \partial x_2^2}\right) \\
+\frac{1}{\sqrt{2}} \left( \frac{\partial^4 k_1}{\partial x_1^3 \partial x_2} + x_3 \frac{\partial^4 h_1}{\partial x_1^3 \partial x_2}\right) + \left( \frac{\partial^4 k_1}{\partial x_1^4} + x_3 \frac{\partial^4 h_1}{\partial x_1^4}\right)=0 \,, \\
-\sqrt{2}\left( \frac{\partial^4 k_1}{\partial x_2^4} + x_3 \frac{\partial^4 h_1}{\partial x_2^4}\right) - \left(4+3\sqrt{2}\right)\left( \frac{\partial^4 k_2}{\partial x_1 \partial x_2^3} + x_3 \frac{\partial^4 h_2}{\partial x_1 \partial x_2^3}\right)+\left(10+9\sqrt{2}\right) \left( \frac{\partial^4 k_1}{\partial x_1^2 \partial x_2^2} + x_3 \frac{\partial^4 h_1}{\partial x_1^2 \partial x_2^2}\right) \\
+\left(10+3\sqrt{2}\right) \left( \frac{\partial^4 k_2}{\partial x_1^3 \partial x_2} + x_3 \frac{\partial^4 h_2}{\partial x_1^3 \partial x_2}\right)=0 \,, 
\end{dcases}
\end{equation}
as well as
\begin{equation}
\begin{dcases}
\label{Z_6^-fourth-order-constraints-3}
\frac{\partial^3 h_2}{\partial x_2^3}+\frac{\partial^3 h_1}{\partial x_1^3}=0 \,, \\
\frac{\partial^3 h_1}{\partial x_1 \partial x_2^2}+\frac{\partial^3 h_2}{\partial x_1^2 \partial x_2}=0 \,, \\
\frac{\partial^4 \hat{u}_{33}}{\partial x_2^4}+2\frac{\partial^4 \hat{u}_{33}}{\partial x_1^2 \partial x_2^2} +\frac{\partial^3 \hat{u}_{33}}{\partial x_1^4}=0 \,, \\
\frac{\partial^3 h_1}{\partial x_1 \partial x_2^2}-\left(-2+\sqrt{2}\right)\frac{\partial^3 h_2}{\partial x_1^2 \partial x_2}-2\left(-1+\sqrt{2}\right) \frac{\partial^4 \hat{u}_{33}}{\partial x_1^2 \partial x_2^2}=0 \,, \\
2 \frac{\partial^3 h_1}{\partial x_1 \partial x_2^2} + \frac{\partial^3 h_2}{\partial x_1^2 \partial x_2} + \frac{\partial^4 \hat{u}_{33}}{\partial x_1^2 \partial x_2^2}=0 \,, \\
\frac{\partial^3 h_1}{\partial x_2^3}-\left(-1+\sqrt{2}\right) \frac{\partial^3 h_2}{\partial x_1 \partial x_2^2}+\left(-1+\sqrt{2}\right) \frac{\partial^3 h_1}{\partial x_1^2 \partial x_2}-\frac{\partial^3 h_2}{\partial x_1^3}=0 \,, \\
-3 \frac{\partial^3 h_2}{\partial x_1 \partial x_2^2} +\frac{\partial^4 \hat{u}_{33}}{\partial x_1 \partial x_2^3} -2 \frac{\partial^3 h_1}{\partial x_1^2 \partial x_2} -\frac{\partial^3 h_2}{\partial x_1^3}+\frac{\partial^4 \hat{u}_{33}}{\partial x_1^3 \partial x_2} =0 \,, \\
\frac{\partial^3 h_1}{\partial x_2^3} +\frac{\partial^4 \hat{u}_{33}}{\partial x_1 \partial x_2^3} -3 \frac{\partial^3 h_1}{\partial x_1^2 \partial x_2} -\frac{\partial^3 h_2}{\partial x_1^3} =0 \,, \\
\frac{\partial^3 h_2}{\partial x_1 \partial x_2^2} +\frac{\partial^4 \hat{u}_{33}}{\partial x_1 \partial x_2^3} +2 \frac{\partial^3 h_1}{\partial x_1^2 \partial x_2} +\frac{\partial^4 \hat{u}_{33}}{\partial x_1^3 \partial x_2} =0 \,, 
\end{dcases}
\end{equation}

Thus, we obtain the following result.
\begin{prop}
The universal displacements in trigonal $\mathbb{Z}_6^-$ class strain gradient linear elastic solids are the superposition of homogeneous displacement fields and the inhomogeneous displacement field \eqref{transverse-isotropy-linear-solution} with the constraints \eqref{Z_6^-third-order-constraints-1}, \eqref{Z_6^-third-order-constraints-2}, and \eqref{Z_6^-third-order-constraints-3}, together with the constraints \eqref{Z_6^-fourth-order-constraints-1}, \eqref{Z_6^-fourth-order-constraints-2}, and \eqref{Z_6^-fourth-order-constraints-3}.
\end{prop}

\subsubsection{Trigonal class $\mathbb{D}_6^h$}

Tensor $\boldsymbol{\mathsf{A}}_{\mathbb{D}_6 \oplus \mathbb{Z}_2^c}$ has $22$ independent components and has the same form as that of class $\mathbb{D}_6$.
Tensor $\boldsymbol{\mathsf{M}}_{\mathbb{D}_6^h}$ has $8$ independent components and has the same form as that of class $\mathbb{Z}_6^-$.
Tensor $\boldsymbol{\mathsf{C}}_{\mathbb{O}(2) \oplus \mathbb{Z}_2^c}$ has $5$ independent components and corresponds to the Curie group with Hermann--Maugin symbol $\infty m$.

We obtain, for this class, the following third-order universality PDEs:
\begin{equation}
\label{D_6^h-third-order-constraints-1}
\begin{dcases}
- \sqrt{2} \left( \frac{\partial^3 k_2}{\partial x_2^3} + x_3 \frac{\partial^3 h_2}{\partial x_2^3} \right)+\left( 1+\frac{1}{\sqrt{2}} \right) \left( \frac{\partial^3 k_1}{\partial x_1 \partial x_2^2} + x_3 \frac{\partial^3 h_1}{\partial x_1 \partial x_2^2}\right) \\
+ \frac{1}{2}\left(4+\sqrt{2}\right) \left( \frac{\partial^3 k_2}{\partial x_1^2 \partial x_2} + x_3 \frac{\partial^3 h_2}{\partial x_1^2 \partial x_2}\right)=0 \,, \\
-\frac{1}{\sqrt{2}} \left(\frac{\partial^3 k_2}{\partial x_2^3}+x_3  \frac{\partial^3 h_2}{\partial x_2^3}  \right) -  \frac{\partial^3 k_1}{\partial x_1 \partial x_2^2} - x_3  \frac{\partial^3 h_1}{\partial x_1 \partial x_2^2} + \frac{1}{2} \left(4+\sqrt{2}\right) \left( \frac{\partial^3 k_2}{\partial x_1^2 \partial x_2} + x_3  \frac{\partial^3 h_2}{\partial x_1^2 \partial x_2}   \right) =0 \,, \\
-\frac{1}{2} \left(2+\sqrt{2}\right) \left(\frac{\partial^3 k_2}{\partial x_2^3}+x_3  \frac{\partial^3 h_2}{\partial x_2^3}  \right) + \left(1+\frac{3}{\sqrt{2}}\right) \left( \frac{\partial^3 k_1}{\partial x_1 \partial x_2^2} +x_3  \frac{\partial^3 h_1}{\partial x_1 \partial x_2^2} \right) \\
+ \left(2+\sqrt{2} \right) \left( \frac{\partial^3 k_2}{\partial x_1^2 \partial x_2} + x_3  \frac{\partial^3 h_2}{\partial x_1^2 \partial x_2}   \right) =0 \,, 
\end{dcases}
\end{equation}
and
\begin{equation}
\label{D_6^h-third-order-constraints-2}
\begin{dcases}
\frac{1}{\sqrt{2}} \left( \frac{\partial^3 k_1}{\partial x_2^3} + x_3 \frac{\partial^3 h_1}{\partial x_2^3}\right)+\frac{1}{\sqrt{2}} \left( \frac{\partial^3 k_2}{\partial x_1 \partial x_2^2} + x_3 \frac{\partial^3 h_2}{\partial x_1 \partial x_2^2}\right)+\left(2+\sqrt{2}\right) \left( \frac{\partial^3 k_1}{\partial x_1^2 \partial x_2} + x_3 \frac{\partial^3 h_1}{\partial x_1^2 \partial x_2}\right)  \\
- \left( \frac{\partial^3 k_2}{\partial x_1^3} + x_3 \frac{\partial^3 h_2}{\partial x_1^3}\right)=0 \,, \\
\frac{1}{\sqrt{2}} \left( \frac{\partial^3 k_1}{\partial x_2^3} + x_3 \frac{\partial^3 h_1}{\partial x_2^3}\right)-\left(1+\sqrt{2}\right) \left( \frac{\partial^3 k_2}{\partial x_1 \partial x_2^2} + x_3 \frac{\partial^3 h_2}{\partial x_1 \partial x_2^2}\right) + \left( \frac{\partial^3 k_1}{\partial x_1^2 \partial x_2} + x_3 \frac{\partial^3 h_1}{\partial x_1^2 \partial x_2}\right)=0 \,, \\
\sqrt{2}  \left( \frac{\partial^3 k_1}{\partial x_2^3} + x_3 \frac{\partial^3 h_1}{\partial x_2^3}\right)+  \left(1-\frac{1}{\sqrt{2}}\right) \left( \frac{\partial^3 k_2}{\partial x_1 \partial x_2^2} + x_3 \frac{\partial^3 h_2}{\partial x_1 \partial x_2^2}\right)+2\left(1+\sqrt{2}\right) \left( \frac{\partial^3 k_1}{\partial x_1^2 \partial x_2} + x_3 \frac{\partial^3 h_1}{\partial x_1^2 \partial x_2}\right) \\
+\frac{1}{2}\left(2+\sqrt{2}\right) \left( \frac{\partial^3 k_2}{\partial x_1^3} + x_3 \frac{\partial^3 h_2}{\partial x_1^3}\right)=0 \,, 
\end{dcases}
\end{equation}
as well as
\begin{equation}
\label{D_6^h-third-order-constraints-3}
\begin{dcases}
2 \sqrt{2} \frac{\partial^2 h_1}{\partial x_2^2}+2\left(1+\sqrt{2}\right) \frac{\partial^2 h_2}{\partial x_1 \partial x_2}-2 \frac{\partial^2 h_1}{\partial x_1^2}=0 \,, \\
\frac{\partial^2 h_1}{\partial x_2^2}+\left(1+\frac{1}{\sqrt{2}}\right) \frac{\partial^2 h_2}{\partial x_1 \partial x_2}-\frac{1}{\sqrt{2}} \frac{\partial^2 h_1}{\partial x_1^2}=0 \,, \\
\frac{\partial^2 h_1}{\partial x_2^2}+2 \frac{\partial^2 h_2}{\partial x_1 \partial x_2}+3\frac{\partial^3 \hat{u}_{33}}{\partial x_1^1 \partial x_2^2}- \frac{\partial^2 h_1}{\partial x_1^2}-\frac{\partial^3 \hat{u}_{33}}{\partial x_1^3}=0 \,, \\
2 \left(1-\frac{1}{\sqrt{2}}\right)\frac{\partial^2 h_1}{\partial x_2^2}+\left(-2+\sqrt{2}\right) \frac{\partial^2 h_2}{\partial x_1 \partial x_2}-\left(4+\sqrt{2}\right)\frac{\partial^3 \hat{u}_{33}}{\partial x_1^1 \partial x_2^2}+\sqrt{2} \left( \frac{\partial^2 h_1}{\partial x_1^2}-\frac{\partial^3 \hat{u}_{33}}{\partial x_1^3}\right)=0 \,, 
\end{dcases}
\end{equation}
We also obtain the following fourth-order universality PDEs:
\begin{equation}
\label{D_6^h-fourth-order-constraints-1}
\begin{dcases}
\frac{1}{\sqrt{2}} \left( \frac{\partial^4 k_1}{\partial x_2^4} + x_3 \frac{\partial^4 h_1}{\partial x_2^4}\right)+\frac{1}{\sqrt{2}} \left( \frac{\partial^4 k_2}{\partial x_1 \partial x_2^3} + x_3 \frac{\partial^4 h_2}{\partial x_1 \partial x_2^3}\right)+\left(1+\sqrt{2}\right) \left( \frac{\partial^4 k_1}{\partial x_1^2 \partial x_2^2} + x_3 \frac{\partial^4 h_1}{\partial x_1^2 \partial x_2^2}\right) \\
+ \left( \frac{\partial^4 k_2}{\partial x_1^3 \partial x_2} + x_3 \frac{\partial^4 h_2}{\partial x_1^3 \partial x_2}\right)+\left( \frac{\partial^4 k_1}{\partial x_1^4} + x_3 \frac{\partial^4 h_1}{\partial x_1^4}\right)=0 \,, \\
-\left( \frac{\partial^4 k_1}{\partial x_1^2 \partial x_2^2} + x_3 \frac{\partial^4 h_1}{\partial x_1^2 \partial x_2^2}\right)-\left( \frac{\partial^4 k_2}{\partial x_1^3 \partial x_2} + x_3 \frac{\partial^4 h_2}{\partial x_1^3 \partial x_2}\right)=0 \,, \\
-\left( \frac{\partial^4 k_1}{\partial x_2^4} + x_3 \frac{\partial^4 h_1}{\partial x_2^4}\right)+\left( \frac{\partial^4 k_2}{\partial x_1 \partial x_2^3} + x_3 \frac{\partial^4 h_2}{\partial x_1 \partial x_2^3}\right) \\
-\left( \frac{\partial^4 k_1}{\partial x_1^2 \partial x_2^2} + x_3 \frac{\partial^4 h_1}{\partial x_1^2 \partial x_2^2}\right)+\left( \frac{\partial^4 k_2}{\partial x_1^3 \partial x_2} + x_3 \frac{\partial^4 h_2}{\partial x_1^3 \partial x_2}\right)=0 \,, \\
-\frac{1}{\sqrt{2}} \left( \frac{\partial^4 k_1}{\partial x_2^4} + x_3 \frac{\partial^4 h_1}{\partial x_2^4}\right)+\left(1-\frac{1}{\sqrt{2}}\right) \left( \frac{\partial^4 k_2}{\partial x_1 \partial x_2^3} + x_3 \frac{\partial^4 h_2}{\partial x_1 \partial x_2^3}\right)-\left(-1+\sqrt{2}\right) \left( \frac{\partial^4 k_1}{\partial x_1^2 \partial x_2^2} + x_3 \frac{\partial^4 h_1}{\partial x_1^2 \partial x_2^2}\right) \\
+\left( \frac{\partial^4 k_2}{\partial x_1^3 \partial x_2} + x_3 \frac{\partial^4 h_2}{\partial x_1^3 \partial x_2}\right)=0 \,, \\
\sqrt{2} \left( \frac{\partial^4 k_1}{\partial x_2^4} + x_3 \frac{\partial^4 h_1}{\partial x_2^4}\right)+\left(1+\sqrt{2}\right) \left( \frac{\partial^4 k_2}{\partial x_1 \partial x_2^3} + x_3 \frac{\partial^4 h_2}{\partial x_1 \partial x_2^3}\right)-\left(1+\sqrt{2}\right) \left( \frac{\partial^4 k_1}{\partial x_1^2 \partial x_2^2} + x_3 \frac{\partial^4 h_1}{\partial x_1^2 \partial x_2^2}\right) \\
-\left(1+\sqrt{2}\right)\left( \frac{\partial^4 k_2}{\partial x_1^3 \partial x_2} + x_3 \frac{\partial^4 h_2}{\partial x_1^3 \partial x_2}\right)+\left( \frac{\partial^4 k_1}{\partial x_1^4} + x_3 \frac{\partial^4 h_1}{\partial x_1^4}\right)=0 \,,
\end{dcases}
\end{equation}
and 
\begin{equation}
\label{D_6^h-fourth-order-constraints-2}
\begin{dcases}
\left( \frac{\partial^4 k_2}{\partial x_2^4} + x_3 \frac{\partial^4 h_2}{\partial x_2^4}\right) + \left( \frac{\partial^4 k_1}{\partial x_1 \partial x_2^3} + x_3 \frac{\partial^4 h_1}{\partial x_1 \partial x_2^3}\right)+\left(1+\sqrt{2}\right) \left( \frac{\partial^4 k_2}{\partial x_1^2 \partial x_2^2} + x_3 \frac{\partial^4 h_2}{\partial x_1^2 \partial x_2^2}\right) \\
+ \frac{1}{\sqrt{2}} \left( \frac{\partial^4 k_1}{\partial x_1^3 \partial x_2} + x_3 \frac{\partial^4 h_1}{\partial x_1^3 \partial x_2}\right) + \frac{1}{\sqrt{2}}\left( \frac{\partial^4 k_2}{\partial x_1^4} + x_3 \frac{\partial^4 h_2}{\partial x_1^4}\right)=0 \,, \\
\left( \frac{\partial^4 k_1}{\partial x_1 \partial x_2^3} + x_3 \frac{\partial^4 h_1}{\partial x_1 \partial x_2^3}\right)+ \left( \frac{\partial^4 k_2}{\partial x_1^2 \partial x_2^2} + x_3 \frac{\partial^4 h_2}{\partial x_1^2 \partial x_2^2}\right) =0 \,, \\
\left( \frac{\partial^4 k_1}{\partial x_1^3 \partial x_2} + x_3 \frac{\partial^4 h_1}{\partial x_1^3 \partial x_2}\right)- \left( \frac{\partial^4 k_2}{\partial x_1^4} + x_3 \frac{\partial^4 h_2}{\partial x_1^4}\right) \\
+ \left( \frac{\partial^4 k_1}{\partial x_1 \partial x_2^3} + x_3 \frac{\partial^4 h_1}{\partial x_1 \partial x_2^3}\right)- \left( \frac{\partial^4 k_2}{\partial x_1^2 \partial x_2^2} + x_3 \frac{\partial^4 h_2}{\partial x_1^2 \partial x_2^2}\right) =0 \,, \\
\left( \frac{\partial^4 k_1}{\partial x_1 \partial x_2^3} + x_3 \frac{\partial^4 h_1}{\partial x_1 \partial x_2^3}\right)-\left(-1+\sqrt{2}\right) \left( \frac{\partial^4 k_2}{\partial x_1^2 \partial x_2^2} + x_3 \frac{\partial^4 h_2}{\partial x_1^2 \partial x_2^2}\right) \\
+ \left(1-\frac{1}{\sqrt{2}}\right) \left( \frac{\partial^4 k_1}{\partial x_1^3 \partial x_2} + x_3 \frac{\partial^4 h_1}{\partial x_1^3 \partial x_2}\right) - \frac{1}{\sqrt{2}}\left( \frac{\partial^4 k_2}{\partial x_1^4} + x_3 \frac{\partial^4 h_2}{\partial x_1^4}\right)=0 \,, \\
\left( \frac{\partial^4 k_1}{\partial x_1 \partial x_2^3} + x_3 \frac{\partial^4 h_1}{\partial x_1 \partial x_2^3}\right)+2\left(2+\sqrt{2}\right) \left( \frac{\partial^4 k_2}{\partial x_1^2 \partial x_2^2} + x_3 \frac{\partial^4 h_2}{\partial x_1^2 \partial x_2^2}\right) -\sqrt{2} \left( \frac{\partial^4 k_1}{\partial x_1^3 \partial x_2} + x_3 \frac{\partial^4 h_1}{\partial x_1^3 \partial x_2}\right)=0 \,, 
\end{dcases}
\end{equation}
as well as
\begin{equation}
\label{D_6^h-fourth-order-constraints-3}
\begin{dcases}
\frac{\partial^3 h_2}{\partial x_2^3}+\frac{\partial^3 h_1}{\partial x_1^3}=0 \,, \\
\frac{\partial^3 h_1}{\partial x_1 \partial x_2^2}+\frac{\partial^3 h_2}{\partial x_1^2 \partial x_2}=0 \,, \\
\frac{\partial^4 \hat{u}_{33}}{\partial x_2^4}+2\frac{\partial^4 \hat{u}_{33}}{\partial x_1^2 \partial x_2^2} +\frac{\partial^3 \hat{u}_{33}}{\partial x_1^4}=0 \,, \\
\frac{\partial^3 h_1}{\partial x_1 \partial x_2^2}-\left(-2+\sqrt{2}\right)\frac{\partial^3 h_2}{\partial x_1^2 \partial x_2}-2\left(-1+\sqrt{2}\right) \frac{\partial^4 \hat{u}_{33}}{\partial x_1^2 \partial x_2^2}=0 \,, \\
2 \frac{\partial^3 h_1}{\partial x_1 \partial x_2^2} + \frac{\partial^3 h_2}{\partial x_1^2 \partial x_2} + \frac{\partial^4 \hat{u}_{33}}{\partial x_1^2 \partial x_2^2}=0 \,, \\
\frac{\partial^3 h_1}{\partial x_2^3}-\left(-1+\sqrt{2}\right) \frac{\partial^3 h_2}{\partial x_1 \partial x_2^2}+\left(-1+\sqrt{2}\right) \frac{\partial^3 h_1}{\partial x_1^2 \partial x_2}-\frac{\partial^3 h_2}{\partial x_1^3}=0 \,.  
\end{dcases}
\end{equation}
Hence, we have the following result.
\begin{prop}
The universal displacements in trigonal $\mathbb{D}_6^h$ class linear strain-gradient elastic solids are the superposition of homogeneous displacement fields and the inhomogeneous displacement field \eqref{transverse-isotropy-linear-solution}, subject to the third-order universality PDEs \eqref{D_6^h-third-order-constraints-1}--\eqref{D_6^h-third-order-constraints-3} and the fourth-order universality PDEs \eqref{D_6^h-fourth-order-constraints-1}--\eqref{D_6^h-fourth-order-constraints-3}.
\end{prop}

\subsubsection{Trigonal class $\mathbb{D}_4^v$} 

Tensor $\boldsymbol{\mathsf{A}}_{\mathbb{D}_4 \oplus \mathbb{Z}_2^c}$ has $28$ independent components and has the same form as that of class $\mathbb{Z}_4$.
Tensor $\boldsymbol{\mathsf{M}}_{\mathbb{D}_4^v}$ has $15$ independent components and has the same form as that given in class $\mathbb{Z}_4$.
Tensor $\boldsymbol{\mathsf{C}}_{\mathbb{D}_4 \oplus \mathbb{Z}_2^c}$ has $6$ independent components and corresponds to the group with Hermann--Maugin symbol $4mm$.
The classical linear elastic part coincides with the tetragonal case of linear elasticity. The third-order universality PDEs for this class read
\begin{equation}
\label{D4v-third-order-constraints}
\begin{dcases}
\frac{\partial^3 u_3}{\partial x_1 \partial x_3^2} = \frac{\partial^3 u_1}{\partial x_1^2 \partial x_3} = \frac{\partial^3 u_3}{\partial x_1^3} = \frac{\partial^3 u_2}{\partial x_1 \partial x_2 \partial x_3} = \frac{\partial^3 u_3}{\partial x_1 \partial x_2^2} = 0\,, \\
\frac{\partial^3 u_1}{\partial x_2^2 \partial x_3} = \frac{\partial^3 u_1}{\partial x_3^3} = \frac{\partial^3 u_3}{\partial x_2 \partial x_3^2} = \frac{\partial^3 u_2}{\partial x_2^2 \partial x_3} = \frac{\partial^3 u_3}{\partial x_2^3} = 0\,, \\
\frac{\partial^3 u_1}{\partial x_1 \partial x_2 \partial x_3} = \frac{\partial^3 u_3}{\partial x_1^2 \partial x_2} = \frac{\partial^3 u_3}{\partial x_2 \partial x_3^2} = \frac{\partial^3 u_2}{\partial x_3^3} = \frac{\partial^3 u_2}{\partial x_1^2 \partial x_3} = 0\,, \\
\frac{\partial^3 u_3}{\partial x_3^3} = \frac{\partial^3 u_3}{\partial x_1^2 \partial x_3} = 0\,, \\
\frac{\partial^3 u_2}{\partial x_2 \partial x_3^2} + \frac{\partial^3 u_1}{\partial x_1 \partial x_3^2} = 0\,, \\
\frac{\partial^3 u_2}{\partial x_2^3} + \frac{\partial^3 u_1}{\partial x_1^3} = 0\,, \\
\frac{\partial^3 u_1}{\partial x_1 \partial x_2^2} + \frac{\partial^3 u_2}{\partial x_1^2 \partial x_2} = 0\,.
\end{dcases}
\end{equation}
The fourth-order universality PDEs for this class are induced by the sixth-order tensor and read
\begin{equation}
\label{D4v-fourth-order-constraints}
\begin{dcases}
\frac{\partial^4 u_1}{\partial x_1^4} = \frac{\partial^4 u_3}{\partial x_1^3 \partial x_3} = \frac{\partial^4 u_2}{\partial x_1 \partial x_2^3} = \frac{\partial^4 u_3}{\partial x_1 \partial x_2^2 \partial x_3} = \frac{\partial^4 u_1}{\partial x_2^2 \partial x_3^2} = 0\,, \\
\frac{\partial^4 u_3}{\partial x_1 \partial x_2 \partial x_3^2} = \frac{\partial^4 u_3}{\partial x_1 \partial x_3^3} = \frac{\partial^4 u_1}{\partial x_3^4} = \frac{\partial^4 u_1}{\partial x_1^2 \partial x_3^2} = \frac{\partial^4 u_2}{\partial x_1 \partial x_2 \partial x_3^2} = 0\,, \\
\frac{\partial^4 u_2}{\partial x_2^4} = \frac{\partial^4 u_3}{\partial x_2^2 \partial x_3^2} = \frac{\partial^4 u}{\partial x_2^3 \partial x_3} = \frac{\partial^4 u_1}{\partial x_1^3 \partial x_2} = \frac{\partial^4 u_3}{\partial x_1^2 \partial x_3^2} = 0\,, \\
\frac{\partial^4 u_2}{\partial x_1^2 \partial x_3^2} = \frac{\partial^4 u_3}{\partial x_1^2 \partial x_2 \partial x_3} = \frac{\partial^4 u_3}{\partial x_3^4} = \frac{\partial^4 u_2}{\partial x_3^4} = \frac{\partial^4 u_3}{\partial x_2 \partial x_3^3} = 0\,, \\
\frac{\partial^4 u_2}{\partial x_2^2 \partial x_3^2} = \frac{\partial^4 u_1}{\partial x_1 \partial x_2 \partial x_3^2} = \frac{\partial^4 u_3}{\partial x_1^2 \partial x_2^2} = \frac{\partial^4 u_2}{\partial x_1^2 \partial x_2 \partial x_3} = \frac{\partial^4 u_1}{\partial x_1 \partial x_2^2 \partial x_3} = 0\,, \\
\frac{\partial^4 u_1}{\partial x_1^2 \partial x_2^2} + \frac{\partial^4 u_2}{\partial x_1^3 \partial x_2} = 0\,, \\
\frac{\partial^4 u_1}{\partial x_1 \partial x_2^3} + \frac{\partial^4 u_2}{\partial x_1^2 \partial x_2^2} = 0\,, \\
2 \frac{\partial^4 u_2}{\partial x_1^2 \partial x_2^2} + \frac{\partial^4 u_1}{\partial x_1^3 \partial x_2} + \frac{\partial^4 u_2}{\partial x_1^4} = 0\,, \\
\frac{\partial^4 u_2}{\partial x_2^3 \partial x_3} + \frac{\partial^4 u_1}{\partial x_1^3 \partial x_3} = 0\,, \\
\frac{\partial^4 u_2}{\partial x_2 \partial x_3^3} + \frac{\partial^4 u_3}{\partial x_2^2 \partial x_3^2} + \frac{\partial^4 u_1}{\partial x_1 \partial x_3^3} = 0\,, \\
\frac{\partial^4 u_2}{\partial x_2^3 \partial x_3} + \frac{\partial^4 u_1}{\partial x_1^3 \partial x_3} = 0\,, \\
\frac{\partial^4 u_2}{\partial x_2^3 \partial x_3} + \frac{\partial^4 u_3}{\partial x_2^4} + \frac{\partial^4 u_1}{\partial x_1^3 \partial x_3} + \frac{\partial^4 u_3}{\partial x_1^4} = 0\,.
\end{dcases}
\end{equation}

Substituting the tetragonal linear elastic candidate displacement field \eqref{tetragonal-linear-inhomogeneous} into the above third- and fourth-order universality PDEs, we obtain the following additional universality PDEs for the function $g(x_1, x_2)$:
\begin{equation}
\label{D4v-constraints-for-g}
\frac{\partial^3 g}{\partial x_1^3}=\frac{\partial^3 g}{\partial x_1 \partial x_2^2}=\frac{\partial^3 g}{\partial x_2^3}=\frac{\partial^3 g}{\partial x_1^2 \partial x_2}=\frac{\partial^4 g}{\partial x_1^2 \partial x_2^2}=0\,.
\end{equation}
After straightforward manipulations, one finds that $g(x_1, x_2)$ has the same form as in \eqref{Z^-_4-g-final-form}. This gives the following characterization:
\begin{prop}
The universal displacements in trigonal $\mathbb{D}_4^v$ class linear strain-gradient elastic solids are the superposition of homogeneous displacement fields and the inhomogeneous displacement field \eqref{tetragonal-linear-inhomogeneous}, with $g(x_1, x_2)$ given by \eqref{Z^-_4-g-final-form}\,.
\end{prop}

\renewcommand{\arraystretch}{1.15}
\begin{longtable}{p{0.30\textwidth} p{0.66\textwidth}}
\caption{Summary of universal displacements: Trigonal classes.}\label{Table-UD-summary-trigonal}\\
\hline
\textbf{Symmetry class} & \textbf{Universal displacement family} \\
\hline
\endfirsthead
\hline
\textbf{Symmetry class} & \textbf{Universal displacement family} (continued) \\
\hline
\endhead
\hline
\endfoot
\hline
\endlastfoot
Trigonal class $\mathbb{Z}_3$ & Superposition of a homogeneous field and \eqref{trigonal-linear-inhomogeneous} with $a_{123}=0$ \\ \hline
Trigonal class $\mathbb{D}_3$ & Superposition of a homogeneous field and \eqref{trigonal-linear-inhomogeneous} with $a_{123}=0$ \\ \hline
Trigonal class $\mathbb{Z}_3 \oplus \mathbb{Z}_2^c$ & Superposition of a homogeneous field and \eqref{trigonal-linear-inhomogeneous} \\ \hline
Trigonal class $\mathbb{D}_3 \oplus \mathbb{Z}_2^c$ & Superposition of a homogeneous field and \eqref{trigonal-linear-inhomogeneous} \\ \hline
Trigonal class $\mathbb{Z}_6^-$ & Superposition of a homogeneous field and \eqref{transverse-isotropy-linear-solution} subject to (\ref{Z_6^-third-order-constraints-1})--(\ref{Z_6^-third-order-constraints-3}) and (\ref{Z_6^-fourth-order-constraints-1})--(\ref{Z_6^-fourth-order-constraints-3}) \\ \hline
Trigonal class $\mathbb{D}_6^h$ & Superposition of a homogeneous field and \eqref{transverse-isotropy-linear-solution} subject to (\ref{D_6^h-third-order-constraints-1})--(\ref{D_6^h-third-order-constraints-3}) and (\ref{D_6^h-fourth-order-constraints-1})--(\ref{D_6^h-fourth-order-constraints-3}) \\ \hline
Trigonal class $\mathbb{D}_4^v$ & Superposition of a homogeneous field and \eqref{tetragonal-linear-inhomogeneous} with $g(x_1, x_2)$ given by \eqref{Z^-_4-g-final-form} \\ \hline
\end{longtable}
\renewcommand{\arraystretch}{2.0}

\subsection{Tetragonal classes}

Out of the seven strain-gradient tetragonal classes, $\mathbb{Z}_4$, $\mathbb{Z}_4 \oplus \mathbb{Z}_2^c$, $\mathbb{D}_4$, and $\mathbb{D}_4 \oplus \mathbb{Z}_2^c$ share the same linear elastic symmetry as the tetragonal classical linear elasticity. Therefore, the universality PDEs \eqref{tetragonal-linear-constraints} remain valid, and the candidate universal displacement fields are those of \eqref{tetragonal-linear-inhomogeneous}. Substituting \eqref{tetragonal-linear-inhomogeneous} into the higher-order universality PDEs, we examine the resulting conditions on the scalar function $g(x_1,x_2)$ appearing in \eqref{tetragonal-linear-inhomogeneous}.
The remaining tetragonal classes, namely $\mathbb{D}_8^h$, $\mathbb{D}_5^v$, and $\mathbb{Z}_8^-$, have a linear elastic part corresponding to the Curie group with Hermann--Mauguin symbol $\infty\mathrm{m}$ and are characterized by five independent elastic constants. These classes are closely analogous to certain $\infty$-gonal strain-gradient symmetry classes. In particular, $\mathbb{D}_8^h$ is related to $\mathbb{O}(2) \oplus \mathbb{Z}_2^c$, $\mathbb{D}_5^v$ to $\mathbb{O}_2^-$, and $\mathbb{Z}_8^-$ to $\mathbb{SO}(2) \oplus \mathbb{Z}_2^c$. We exploit these structural similarities and compare with the corresponding $\infty$-gonal results to determine what additional universality PDEs arise from the presence of discrete tetragonal parameters.

\subsubsection{Tetragonal class $\mathbb{Z}_4$} 

The tensor $\boldsymbol{\mathsf{A}}_{\mathbb{Z}_4 \oplus \mathbb{Z}_2^c}$ admits $45$ independent components and has the following structure
\begin{equation}
\label{A_Z_4}
    \boldsymbol{\mathsf{A}}_{\mathbb{Z}_4 \oplus \mathbb{Z}_2^c}(\mathbf{x}) =
    \begin{bmatrix}
        A^{(15)} & B^{(10)} & 0 & 0 \\
        & A^{(15)} & 0 & 0 \\
        & & H^{(9)} & I^{(7)} \\
        & & & J^{(4)}
    \end{bmatrix}_S\,,
\end{equation}
where $A^{(15)} \in M^S(5)$ and $B^{(10)} \in M^A(5)$, with $M^A(n)$ denoting the $\frac{n(n-1)}{2}$-dimensional space of $n \times n$ skew-symmetric matrices. The blocks $H^{(9)}$, $I^{(7)}$, and $J^{(4)}$ contain $9$, $7$, and $4$ independent components, respectively, and are given by
\begin{equation}
\label{H^9-I^7-J^4}
\begin{aligned}
    H^{(9)} &= 
    \begin{bmatrix}
        h_{11} & h_{12} & h_{13} & h_{12} & h_{13} \\
        & h_{22} & h_{23} & h_{24} & h_{25} \\
        & & h_{33} & h_{25} & h_{35} \\
        & & & h_{22} & h_{23} \\
        & & & & h_{33}
    \end{bmatrix}_S\,,
    \quad
    I^{(7)} =
    \begin{bmatrix}
        0 & i_{12} & -i_{12} \\
        i_{21} & i_{22} & i_{23} \\
        i_{31} & i_{32} & i_{33} \\
        -i_{21} & -i_{23} & -i_{22} \\
        -i_{31} & -i_{33} & -i_{32}
    \end{bmatrix}\,, \\
    J^{(4)} &=
    \begin{bmatrix}
        j_{11} & j_{12} & j_{12} \\
        & j_{22} & j_{23} \\
        & & j_{22}
    \end{bmatrix}_S\,.
\end{aligned}
\end{equation}
Matrix $\boldsymbol{\mathsf{M}}_{\mathbb{Z}_4}$ admits $26$ independent components and is given by
\begin{equation}
\label{MZ_4}
    \boldsymbol{\mathsf{M}}_{\mathbb{Z}_4} = \boldsymbol{\mathsf{M}}_{\mathbb{D}_4} + \boldsymbol{\mathsf{M}}_{\mathbb{D}_4^v}\,,
\end{equation}
where the constituent blocks are defined as
\begin{equation}
\label{M_D_4-M_D_4^v}
\begin{aligned}
    \boldsymbol{\mathsf{M}}_{\mathbb{D}_4} &= 
    \begin{bmatrix}
        0 & 0 & 0 & \bar{D}^{(4)} \\
        \bar{E}^{(5)} & 0 & 0 & 0 \\
        0 & -\bar{E}^{(5)} & 0 & 0 \\
        0 & 0 & \bar{O}^{(2)} & 0
    \end{bmatrix}, \quad
    \boldsymbol{\mathsf{M}}_{\mathbb{D}_4^v} =
    \begin{bmatrix}
        0 & 0 & \bar{C}^{(8)} & 0 \\
        0 & \bar{F}^{(5)} & 0 & 0 \\
        \bar{F}^{(5)} & 0 & 0 & 0 \\
        0 & 0 & 0 & \bar{P}^{(2)}
    \end{bmatrix}\,.
\end{aligned}
\end{equation}
The submatrices $\bar{P}^{(2)}$ and $\bar{O}^{(2)}$ are given by
\begin{equation}
\label{barP^2-barO^2}
    \bar{P}^{(2)} =
    \begin{bmatrix}
        \bar{h}_{11} & \bar{h}_{11} & \bar{h}_{13}
    \end{bmatrix}\,, \qquad
    \bar{O}^{(2)} =
    \begin{bmatrix}
        0 & \bar{o}_{12} & \bar{o}_{13} & -\bar{o}_{12} & -\bar{o}_{13}
    \end{bmatrix}\,.
\end{equation}
The remaining matrices $\bar{C}^{(8)}$, $\bar{D}^{(4)}$, $\bar{E}^{(5)}$, and $\bar{F}^{(5)}$ are defined in the corresponding preceding classes. The tensor $\boldsymbol{\mathsf{C}}_{\mathbb{D}_4 \oplus \mathbb{Z}_2^c}$ corresponds to the tetragonal symmetry class with Hermann--Mauguin symbol $422$ and contains $6$ independent components.

The third-order universality PDEs induced by the fifth-order tensor for the class $\mathbb{Z}_4$ are
\begin{equation} \label{Z4-third-order-constraints}
\begin{dcases}
\frac{\partial^3 u_2}{\partial x_1^2 \partial x_3} =    \frac{\partial^3 u_3}{\partial x_1^2 \partial x_2} =  \frac{\partial^3 u_3}{\partial x_2 \partial x_3^2} =   \frac{\partial^3 u_3}{\partial x_1 \partial x_3^2} =    \frac{\partial^3 u_1}{\partial x_1^2 \partial x_3} = 0\,, \\
\frac{\partial^3 u_3}{\partial x_1^3} =    \frac{\partial^3 u_2}{\partial x_1 \partial x_2 \partial x_3} = \frac{\partial^3 u_3}{\partial x_3^2} =  \frac{\partial^3 u_2}{\partial x_3^3} =    \frac{\partial^3 u_1}{\partial x_2^2 \partial x_3} = 0\,, \\
\frac{\partial^3 u_1}{\partial x_3^3} =  \frac{\partial^3 u_3}{\partial x_1 \partial x_2 \partial x_3} =   \frac{\partial^3 u_3}{\partial x_3^3} =   \frac{\partial^3 u_2}{\partial x_1 \partial x_2^2} =    \frac{\partial^3 u_2}{\partial x_2^2 \partial x_3} = 0\,, \\
-\frac{\partial^3 u_2}{\partial x_1 \partial x_2^2}
+ \frac{\partial^3 u_1}{\partial x_1^2 \partial x_2} = 0\,,\quad
-\frac{\partial^3 u_1}{\partial x_2 \partial x_3^2}
+ \frac{\partial^3 u_2}{\partial x_1 \partial x_3^2} = 0\,, \\
\frac{\partial^3 u_2}{\partial x_2 \partial x_3^2}
- \frac{\partial^3 u_1}{\partial x_1 \partial x_3^2} = 0\,,\quad
\frac{\partial^3 u_2}{\partial x_2^3}
+ \frac{\partial^3 u_1}{\partial x_1^3} = 0\,, \\
\frac{\partial^3 u_3}{\partial x_1 \partial x_2^2}
+ \frac{\partial^3 u_2}{\partial x_1 \partial x_2^2} = 0\,,\quad
\frac{\partial^3 u_3}{\partial x_2^2 \partial x_3}
+ \frac{\partial^3 u_3}{\partial x_1^2 \partial x_3} = 0\,, \\
\frac{\partial^3 u_1}{\partial x_2^3}
- \frac{\partial^3 u_2}{\partial x_1^3} = 0\,,\quad
-\frac{\partial^3 u_3}{\partial x_1 \partial x_3^2}
+ \frac{\partial^3 u_3}{\partial x_1 \partial x_2 \partial x_3} = 0\,, \\
\frac{\partial^3 u_3}{\partial x_1 \partial x_2^2}
+ \frac{\partial^3 u_2}{\partial x_1^2 \partial x_2} = 0\,,\quad
\frac{\partial^3 u_3}{\partial x_2 \partial x_3^2}
+ \frac{\partial^3 u_3}{\partial x_1^2 \partial x_3} = 0\,.
\end{dcases}
\end{equation}
The fourth-order universality PDEs induced by the sixth-order tensor for the class $\mathbb{Z}_4$ are
\begin{equation} \label{Z4-fourth-order-constraints}
\begin{dcases}
\frac{\partial^4 u_1}{\partial x_1^4} =  \frac{\partial^4 u_2}{\partial x_1^3 \partial x_2} = \frac{\partial^4 u_3}{\partial x_1^3 \partial x_3} =  \frac{\partial^4 u_2}{\partial x_1 \partial x_2^3} =  \frac{\partial^4 u_1}{\partial x_1^2 \partial x_1^2} = 0\,, \\
\frac{\partial^4 u_3}{\partial x_1 \partial x_2^2 \partial x_3} =  \frac{\partial^4 u_1}{\partial x_2^2 \partial x_3^2} =  \frac{\partial^4 u_3}{\partial x_1 \partial x_2 \partial x_3^2} =   \frac{\partial^4 u_3}{\partial x_1 \partial x_3^3} =  \frac{\partial^4 u_1}{\partial x_3^4} = 0\,, \\
\frac{\partial^4 u_2}{\partial x_2^4} = \frac{\partial^4 u_2}{\partial x_1^4} =  \frac{\partial^4 u_3}{\partial x_1^2 \partial x_3^2} =  \frac{\partial^4 u_2}{\partial x_1^2 \partial x_3^2} =  \frac{\partial^4 u_1}{\partial x_1 \partial x_2^3} = 0\,, \\
\frac{\partial^4 u_3}{\partial x_2^2 \partial x_3^2} =  \frac{\partial^4 u_3}{\partial x_2^3 \partial x_3} = \frac{\partial^4 u_2}{\partial x_3^4} =   \frac{\partial^4 u_1}{\partial x_1^2 \partial x_3^2} =   \frac{\partial^4 u_2}{\partial x_1 \partial x_2 \partial x_3^2} = 0\,, \\
\frac{\partial^4 u_1}{\partial x_1 \partial x_2 \partial x_3^2} =   \frac{\partial^4 u_3}{\partial x_1^2 \partial x_2 \partial x_3} =   \frac{\partial^4 u_1}{\partial x_1^3 \partial x_2} =   \frac{\partial^4 u_2}{\partial x_1^2 \partial x_2^2} =  \frac{\partial^4 u_3}{\partial x_3^4} = 0\,, \\
\frac{\partial^4 u_2}{\partial x_1 \partial x_2^3} = \frac{\partial^4 u_1}{\partial x_2^4} =  \frac{\partial^4 u_3}{\partial x_2 \partial x_3^3} =  \frac{\partial^4 u_2}{\partial x_2^2 \partial x_3^2} =   \frac{\partial^4 u_3}{\partial x_1^2 \partial x_3^2} = 0\,, \\
\frac{\partial^4 u_1}{\partial x_1^2 \partial x_2^2} =  \frac{\partial^4 u_3}{\partial x_2 \partial x_3^3} =   \frac{\partial^4 u_3}{\partial x_3^4} =   \frac{\partial^4 u_2}{\partial x_1 \partial x_3^3} = 0\,, \\
\frac{\partial^4 u_2}{\partial x_2^3 \partial x_3}
+ \frac{\partial^4 u_1}{\partial x_1^3 \partial x_3} = 0\,,\quad
\frac{\partial^4 u_1}{\partial x_1 \partial x_2^2 \partial x_3}
+ \frac{\partial^4 u_2}{\partial x_1^2 \partial x_2 \partial x_3} = 0\,, \\
\frac{\partial^4 u_2}{\partial x_2 \partial x_3^3}
+ \frac{\partial^4 u_1}{\partial x_1 \partial x_3^3} = 0\,,\quad
-2 \frac{\partial^4 u_2}{\partial x_1 \partial x_2^2 \partial x_3}
+ \frac{\partial^4 u_1}{\partial x_1^2 \partial x_2 \partial x_3} = 0\,, \\
\frac{\partial^4 u_1}{\partial x_2^3 \partial x_3}
+ \frac{\partial^4 u_2}{\partial x_1 \partial x_2^2 \partial x_3} = 0\,,\quad
\frac{\partial^4 u_2}{\partial x_2 \partial x_3^3}
+ \frac{\partial^4 u_1}{\partial x_1 \partial x_3^3} = 0\,, \\
\frac{\partial^4 u_3}{\partial x_2^4}
+ \frac{\partial^4 u_3}{\partial x_1^4} = 0\,,\quad
\frac{\partial^4 u_1}{\partial x_1 \partial x_2^2 \partial x_3}
+ \frac{\partial^4 u_2}{\partial x_1^2 \partial x_2 \partial x_3}
+ 2 \frac{\partial^4 u_3}{\partial x_1^2 \partial x_2^2} = 0\,, \\
-\frac{\partial^4 u_2}{\partial x_1 \partial x_2^2 \partial x_3}
- \frac{\partial^4 u_3}{\partial x_1 \partial x_2^3}
+ \frac{\partial^4 u_2}{\partial x_1^3 \partial x_3}
+ \frac{\partial^4 u_3}{\partial x_1^3 \partial x_2} = 0\,, \\
-\frac{\partial^4 u_1}{\partial x_2^3 \partial x_3}
- 3 \frac{\partial^4 u_2}{\partial x_1 \partial x_2^2 \partial x_3}
+ \frac{\partial^4 u_1}{\partial x_1^2 \partial x_2 \partial x_3}
+ \frac{\partial^4 u_3}{\partial x_1^3 \partial x_2} = 0\,, \\
-\frac{\partial^4 u_1}{\partial x_2^3 \partial x_3}
- \frac{\partial^4 u_3}{\partial x_1 \partial x_2^3}
+ 3 \frac{\partial^4 u_1}{\partial x_1^2 \partial x_2 \partial x_3}
+ \frac{\partial^4 u_2}{\partial x_1^3 \partial x_3} = 0\,, \\
-\frac{\partial^4 u_2}{\partial x_1 \partial x_2^2 \partial x_3}
+ \frac{\partial^4 u_3}{\partial x_1 \partial x_2^3} = 0\,,\quad
\frac{\partial^4 u_1}{\partial x_1^2 \partial x_2 \partial x_3}
+ \frac{\partial^4 u_3}{\partial x_1^3 \partial x_2} = 0\,, \\
\frac{\partial^4 u_2}{\partial x_1^2 \partial x_2 \partial x_3}
+ \frac{\partial^4 u_3}{\partial x_1^2 \partial x_2^2} = 0\,,\quad
\frac{\partial^4 u_1}{\partial x_1 \partial x_2^2 \partial x_3}
+ \frac{\partial^4 u_2}{\partial x_1^2 \partial x_2 \partial x_3} = 0\,, \\
\frac{\partial^4 u_1}{\partial x_1 \partial x_2^2 \partial x_3}
+ \frac{\partial^4 u_3}{\partial x_1^2 \partial x_2^2} = 0\,.
\end{dcases}
\end{equation}
Substituting the candidate universal displacements \eqref{tetragonal-linear-inhomogeneous} into the above third- and fourth-order universality PDEs, we obtain the following additional conditions on $g(x_1,x_2)$:
\begin{equation}
\label{Z4-constraints-for-g}
\frac{\partial^3 g}{\partial x_1^3}=\frac{\partial^3 g}{\partial x_1 \partial x_2^2}=\frac{\partial^3 g}{\partial x_2^3}=\frac{\partial^3 g}{\partial x_1^2 \partial x_2}=\frac{\partial^4 g}{\partial x_1^2 \partial x_2^2}=\frac{\partial^4 g}{\partial x_1^4}+\frac{\partial^4 g}{\partial x_2^4}=\frac{\partial^4 g}{\partial x_1^3 \partial x_2}=\frac{\partial^4 g}{\partial x_1 \partial x_2^3}=0\,.
\end{equation}
The first five conditions coincide with those in \eqref{D4v-constraints-for-g}\,. Therefore, the expression \eqref{Z^-_4-g-final-form} remains valid in the present case. One can then verify directly that \eqref{Z^-_4-g-final-form} also satisfies the remaining three conditions in \eqref{Z4-constraints-for-g}\,. Consequently, one obtains the following result:
\begin{prop}
The universal displacements in tetragonal $\mathbb{Z}_4$ class linear strain-gradient elastic solids are the superposition of homogeneous displacement fields and the inhomogeneous displacement field \eqref{tetragonal-linear-inhomogeneous}, with $g(x_1, x_2)$ given by \eqref{Z^-_4-g-final-form}.
\end{prop}

\subsubsection{Tetragonal class $\mathbb{D}_4$} 

Tensor $\boldsymbol{\mathsf{A}}_{\mathbb{D}_4 \oplus \mathbb{Z}_2^c}$ has $28$ independent components and is of the form
\begin{equation}
\label{A_D_4}
    \boldsymbol{\mathsf{A}}_{\mathbb{D}_4 \oplus \mathbb{Z}_2^c}(\mathbf{x})=\begin{bmatrix}
    A^{(15)} & 0 & 0 & 0  \\
     & A^{(15)} & 0 & 0   \\
     &  & H^{(9)} & 0   \\
     &  &  & J^{(4)}      
    \end{bmatrix}_S\,,
\end{equation}
where $A^{(15)}$, $H^{(9)}$, and $J^{(4)}$ are defined as in the preceding classes. Tensor $\boldsymbol{\mathsf{M}}_{\mathbb{D}_4}$ has $11$ independent components and has the structure given in \eqref{M_D_4-M_D_4^v}. Tensor $\boldsymbol{\mathsf{C}}_{{\mathbb{D}_4 \oplus \mathbb{Z}_2^c}}$ corresponds to the tetragonal class with Hermann--Mauguin symbol $422$ and has $6$ independent components.

The third-order universality PDEs induced by the fifth-order tensor read
\begin{equation}
\label{D4-third-order-constraints}
\begin{dcases}
\frac{\partial^3 u_3}{\partial x_2 \partial x_3^2} = \frac{\partial^3 u_3}{\partial x_3^3} = \frac{\partial^3 u_2}{\partial x_3^3} = \frac{\partial^3 u_3}{\partial x_2^2 \partial x_3} = \frac{\partial^3 u_1}{\partial x_3^3} = \frac{\partial^3 u_3}{\partial x_1 \partial x_2 \partial x_3} = 0\,, \\
-2\frac{\partial^3 u_1}{\partial x_1 \partial x_2 \partial x_3} + \frac{\partial^3 u_2}{\partial x_1^2 \partial x_3} + \frac{\partial^3 u_3}{\partial x_1^2 \partial x_2} = 0\,, \\
2\frac{\partial^3 u_2}{\partial x_2^2 \partial x_3} + \frac{\partial^3 u_1}{\partial x_1 \partial x_2 \partial x_3} + \frac{\partial^3 u_3}{\partial x_1^2 \partial x_2} = 0\,, \\
\frac{\partial^3 u_1}{\partial x_1 \partial x_2 \partial x_3} + \frac{\partial^3 u_2}{\partial x_1^2 \partial x_3} = 0\,, \\
\frac{\partial^3 u_2}{\partial x_2^2 \partial x_3} + \frac{\partial^3 u_2}{\partial x_1^2 \partial x_3} + \frac{\partial^3 u_3}{\partial x_1^2 \partial x_2} = 0\,, \\
\frac{\partial^3 u_2}{\partial x_2^2 \partial x_3} + \frac{\partial^3 u_3}{\partial x_2^3} + \frac{\partial^3 u_1}{\partial x_1 \partial x_2 \partial x_3} = 0\,, \\
-\frac{\partial^3 u_2}{\partial x_2^2 \partial x_3} + \frac{\partial^3 u_1}{\partial x_1 \partial x_2 \partial x_3} = 0\,, \\
-\frac{\partial^3 u_2}{\partial x_2^2 \partial x_3} - \frac{\partial^3 u_3}{\partial x_2^3} - \frac{\partial^3 u_1}{\partial x_1 \partial x_2 \partial x_3} - 2\frac{\partial^3 u_2}{\partial x_1^2 \partial x_3} + \frac{\partial^3 u_3}{\partial x_1^2 \partial x_2} = 0\,, \\
-\frac{\partial^3 u_1}{\partial x_2^2 \partial x_3} + 2\frac{\partial^3 u_2}{\partial x_1 \partial x_2 \partial x_3} - \frac{\partial^3 u_3}{\partial x_1 \partial x_2^2} = 0\,, \\
\frac{\partial^3 u_2}{\partial x_1 \partial x_2 \partial x_3} + \frac{\partial^3 u_3}{\partial x_1 \partial x_2^2} + 2\frac{\partial^3 u_1}{\partial x_1^2 \partial x_3} = 0\,, \\
\frac{\partial^3 u_1}{\partial x_2^2 \partial x_3} - \frac{\partial^3 u_2}{\partial x_1 \partial x_2 \partial x_3} = 0\,, \\
\frac{\partial^3 u_1}{\partial x_2^2 \partial x_3} + \frac{\partial^3 u_3}{\partial x_1 \partial x_2^2} + \frac{\partial^3 u_1}{\partial x_1^2 \partial x_3} = 0\,, \\
\frac{\partial^3 u_2}{\partial x_1 \partial x_2 \partial x_3} + \frac{\partial^3 u_1}{\partial x_1^2 \partial x_3} + \frac{\partial^3 u_3}{\partial x_1^3} = 0\,, \\
\frac{\partial^3 u_1}{\partial x_2^2 \partial x_3}  - \frac{\partial^3 u_2}{\partial x_1 \partial x_2 \partial x_3} - 2\frac{\partial^3 u_3}{\partial x_1 \partial x_2^2} = 0\,, \\
\frac{\partial^3 u_2}{\partial x_1 \partial x_2 \partial x_3}  + 2 \frac{\partial^3 u_1}{\partial x_1^2 \partial x_3} = 0\,, \\
2\frac{\partial^3 u_1}{\partial x_2^2 \partial x_3}  + \frac{\partial^3 u_2}{\partial x_1 \partial x_2 \partial x_3} - \frac{\partial^3 u_3}{\partial x_1 \partial x_2^2} + \frac{\partial^3 u_1}{\partial x_1^2 \partial x_3} + \frac{\partial^3 u_3}{\partial x_1^3} = 0\,, \\
-2\frac{\partial^3 u_2}{\partial x_1 \partial x_2^2} + 2\frac{\partial^3 u_1}{\partial x_1^2 \partial x_2} = 0\,, \\
-\frac{\partial^3 u_1}{\partial x_2 \partial x_3^2} + \frac{\partial^3 u_2}{\partial x_1 \partial x_3^2} = 0\,, \\
\frac{\partial^3 u_1}{\partial x_2^3} + \frac{\partial^3 u_2}{\partial x_1 \partial x_2^2} - \frac{\partial^3 u_1}{\partial x_1^2 \partial x_2} - \frac{\partial^3 u_2}{\partial x_1^3} = 0\,.
\end{dcases}
\end{equation}
The fourth-order universality PDEs induced by the sixth-order tensor are written as
\begin{equation}
\label{D4-fourth-order-constraints}
\begin{dcases}
\frac{\partial^4 u_1}{\partial x_1^4} = \frac{\partial^4 u_3}{\partial x_1^3 \partial x_2} = \frac{\partial^4 u_2}{\partial x_1 \partial x_2^3} = \frac{\partial^4 u_1}{\partial x_2^2 \partial x_3^2} = \frac{\partial^4 u_3}{\partial x_1 \partial x_2 \partial x_3^2} = 0\,, \\
\frac{\partial^4 u_3}{\partial x_1 \partial x_3^3} = \frac{\partial^4 u_1}{\partial x_3^4} = \frac{\partial^4 u_1}{\partial x_1^2 \partial x_3^2} = \frac{\partial^4 u_2}{\partial x_1 \partial x_2 \partial x_3^2} = \frac{\partial^4 u_3}{\partial x_1 \partial x_2^2 \partial x_3} = 0\,, \\
\frac{\partial^4 u_1}{\partial x_1^3 \partial x_2} = \frac{\partial^4 u_3}{\partial x_2^2 \partial x_3^2} = \frac{\partial^4 u_3}{\partial x_2^3 \partial x_3} = \frac{\partial^4 u_3}{\partial x_1^2 \partial x_3^2}  = 0\,, \\
\frac{\partial^4 u_2}{\partial x_3^4} = \frac{\partial^4 u_3}{\partial x_2 \partial x_3^3} = \frac{\partial^4 u_2}{\partial x_2^2 \partial x_3^2} = \frac{\partial^4 u_1}{\partial x_1 \partial x_2 \partial x_3^2} = \frac{\partial^4 u_3}{\partial x_1^2 \partial x_2 \partial x_3} = 0\,, \\
\frac{\partial^4 u_2}{\partial x_1^2 \partial x_3^2} = \frac{\partial^4 u_3}{\partial x_3^4} = \frac{\partial^4 u_2}{\partial x_1^2 \partial x_2^2} = \frac{\partial^4 u_2}{\partial x_1^2 \partial x_2 \partial x_3} = \frac{\partial^4 u_1}{\partial x_1 \partial x_2^2 \partial x_3} = 0\,, \\
\frac{\partial^4 u_1}{\partial x_1^2 \partial x_2^2} + \frac{\partial^4 u_2}{\partial x_1^3 \partial x_2} = 0\,, \\
\frac{\partial^4 u_1}{\partial x_2^4} + 2\frac{\partial^4 u_1}{\partial x_1^2 \partial x_2^2} = 0\,, \\
\frac{\partial^4 u_1}{\partial x_1 \partial x_2^3} + \frac{\partial^4 u_2}{\partial x_1^2 \partial x_2^2} = 0\,, \\
2\frac{\partial^4 u_2}{\partial x_1^2 \partial x_2^2} + \frac{\partial^4 u_2}{\partial x_1^4} = 0\,, \\
\frac{\partial^4 u_2}{\partial x_2^3 \partial x_3} + \frac{\partial^4 u_1}{\partial x_1^3 \partial x_3} = 0\,, \\
\frac{\partial^4 u_2}{\partial x_2 \partial x_3^3} + \frac{\partial^4 u_3}{\partial x_2^2 \partial x_3^2} + \frac{\partial^4 u_1}{\partial x_1 \partial x_3^3} = 0\,, \\
\frac{\partial^4 u_2}{\partial x_2^3 \partial x_3} + \frac{\partial^4 u_3}{\partial x_2^4} + \frac{\partial^4 u_1}{\partial x_1^3 \partial x_3} + \frac{\partial^4 u_3}{\partial x_1^4} = 0\,.
\end{dcases}
\end{equation}
Substituting the tetragonal linear elastic candidate displacement field \eqref{tetragonal-linear-inhomogeneous} into the above third- and fourth-order universality PDEs, we obtain the following additional conditions that the function $g(x_1,x_2)$ must satisfy:
\begin{equation}
\label{D4-constraints-for-g}
\frac{\partial^3 g}{\partial x_1^3} = \frac{\partial^3 g}{\partial x_1 \partial x_2^2} = \frac{\partial^3 g}{\partial x_2^3} = \frac{\partial^3 g}{\partial x_1^2 \partial x_2} = \frac{\partial^4 g}{\partial x_1^2 \partial x_2^2} = \frac{\partial^4 g}{\partial x_1^4} + \frac{\partial^4 g}{\partial x_2^4} = 0\,.
\end{equation}
The first five of the above universality PDEs coincide with those in \eqref{D4v-constraints-for-g}. Hence, the representation \eqref{Z^-_4-g-final-form} remains valid and also satisfies the final condition in \eqref{D4-constraints-for-g}. We thus obtain the following result:
\begin{prop}
The universal displacements in tetragonal $\mathbb{D}_4$ class linear strain-gradient elastic solids are the superposition of homogeneous displacement fields and the inhomogeneous displacement field \eqref{tetragonal-linear-inhomogeneous}, with $g(x_1,x_2)$ given by \eqref{Z^-_4-g-final-form}\,.
\end{prop}

\subsubsection{Tetragonal class $\mathbb{Z}_4 \oplus \mathbb{Z}_2^c$} 

Tensor $\boldsymbol{\mathsf{A}}_{\mathbb{Z}_4 \oplus \mathbb{Z}_2^c}$ has $45$ independent components and is of the same form as that of class $\mathbb{Z}_4$. Tensor $\boldsymbol{\mathsf{M}}_{\mathbb{Z}_4 \oplus \mathbb{Z}_2^c}$ is a null tensor. Tensor $\boldsymbol{\mathsf{C}}_{\mathbb{Z}_4 \oplus \mathbb{Z}_2^c}$ has the same form as that of class $\mathbb{Z}_4$, with $6$ independent components, the only difference being that it corresponds to the Hermann--Mauguin symbol $4/mmm$.
The fourth-order universality PDEs coincide with those of the preceding class $\mathbb{Z}_4$, and there are no third-order universality PDEs because the fifth-order tensor vanishes. Substituting the candidate displacement field \eqref{tetragonal-linear-inhomogeneous} into the fourth-order universality PDEs, we obtain the following additional conditions that the function $g(x_1,x_2)$ must satisfy:
\begin{equation}
\label{Z4+Z_2^c-constraints-for-g}
\frac{\partial^4 g}{\partial x_1^2 \partial x_2^2}=\frac{\partial^4 g}{\partial x_1^4}+\frac{\partial^4 g}{\partial x_2^4}=\frac{\partial^4 g}{\partial x_1^3 \partial x_2}=\frac{\partial^4 g}{\partial x_1 \partial x_2^3}=0\,.
\end{equation}
These constraints render $g(x_1,x_2)$ in the following form:
\begin{equation}
\label{Z4+Z_2^c-g-final-form}
\begin{aligned}
	&g(x_1, x_2)=f_1(x_1)+x_2f_2(x_1)+f_3(x_2)+x_1f_4(x_2)\,, \\ 
	&f_2'''(x_1)=0\,,\quad  f_4'''(x_2)=0\,,\quad f_3''''(x_2)=-f_1''''(x_1)=c\,.
\end{aligned}
\end{equation}
Therefore, we have obtained the following result:
\begin{prop}
The universal displacements in tetragonal $\mathbb{Z}_4 \oplus \mathbb{Z}_2^c$ class linear strain-gradient elastic solids are the superposition of homogeneous displacement fields and the inhomogeneous displacement field \eqref{tetragonal-linear-inhomogeneous}, with $g(x_1,x_2)$ given by \eqref{Z4+Z_2^c-g-final-form}\,.
\end{prop}

\subsubsection{Tetragonal class $\mathbb{D}_4 \oplus \mathbb{Z}_2^c$} 

Tensor $\boldsymbol{\mathsf{A}}_{\mathbb{D}_4 \oplus \mathbb{Z}_2^c}$ has $28$ independent components and is of the same form as that of class $\mathbb{D}_4$. Tensor $\boldsymbol{\mathsf{M}}_{\mathbb{D}_4 \oplus \mathbb{Z}_2^c}$ is a null tensor. Tensor $\boldsymbol{\mathsf{C}}_{\mathbb{D}_4 \oplus \mathbb{Z}_2^c}$ has the same form as that of class $\mathbb{D}_4$ with $6$ independent components\,; the only difference is that it corresponds to the Hermann--Mauguin symbol $4/mmm$. The fourth-order universality PDEs coincide with those of class $\mathbb{D}_4$\,; moreover, there are no third-order universality PDEs since the fifth-order tensor vanishes. Substituting the candidate displacement field \eqref{tetragonal-linear-inhomogeneous} into the above fourth-order universality PDEs, we obtain the following additional universality PDEs for $g(x_1,x_2)$:
\begin{equation}\label{D4+Z_2^c-constraints-for-g}
	\frac{\partial^4 g}{\partial x_1^2 \partial x_2^2}=\frac{\partial^4 g}{\partial x_1^4}
	+\frac{\partial^4 g}{\partial x_2^4}=0\,,
\end{equation}
which imply that $g(x_1,x_2)$ has the following representation:
\begin{equation}\label{D4+Z_2^c-g-final-form}
\begin{aligned}
	&g(x_1, x_2)=f_1(x_1)+x_2f_2(x_1)+f_3(x_2)+x_1f_4(x_2)\,, \\ 
	& f_3''''(x_2)+x_1 f_4''''(x_2)+ f_1''''(x_1)+x_2 f_2''''(x_1)=0\,.  
\end{aligned}
\end{equation}

\begin{prop}
The universal displacements in tetragonal $\mathbb{D}_4 \oplus \mathbb{Z}_2^c$ class linear strain-gradient elastic solids are the superposition of homogeneous displacement fields and the inhomogeneous displacement field of eq.~(\ref{tetragonal-linear-inhomogeneous}) with $g(x_1, x_2)$ given by eq.~(\ref{D4+Z_2^c-g-final-form})\,.
\end{prop}

\subsubsection{Tetragonal class $\mathbb{Z}_8^-$} 

Tensor $\boldsymbol{\mathsf{A}}_{\mathbb{SO}(2) \oplus \mathbb{Z}_2^c}$ has $32$ independent components and is of the same form as that of class $\mathbb{SO}(2)$. Tensor $\boldsymbol{\mathsf{M}}_{\mathbb{Z}_8^-}$ has $6$ independent components and the form
\begin{equation}
	\label{Z_8^-}
	\boldsymbol{\mathsf{M}}_{\mathbb{Z}_8^-}=\boldsymbol{\mathsf{M}}_{\mathbb{D}_8^h}+\begin{bmatrix}
		0 & 0 & \bar{C}^{(2)} & 0  \\
		0 & \bar{F}^{(1)} & 0 & 0   \\
		\bar{F}^{(1)} & 0  & 0 & 0     \\
		0 & 0 & 0 & 0   
	\end{bmatrix} 
	+\begin{bmatrix}
		0 & 0 & 0 & 0  \\
		0 & 0 & 0 & 0   \\
		0 & 0  & 0 & 0     \\
		0 & 0 & 0 & g(\bar{C}^{(2)})   
	\end{bmatrix},
\end{equation}
where
\begin{equation}
	\label{M_D_8^h}
	\boldsymbol{\mathsf{M}}_{\mathbb{D}_8^h}=\begin{bmatrix}
		0 & 0 & 0 & \bar{D}^{(2)}  \\
		\bar{E}^{(1)} & 0 & 0 & 0   \\
		0 & -\bar{E}^{(1)}  & 0 & 0     \\
		0 & 0 & 0 & 0   
	\end{bmatrix} 
	+\begin{bmatrix}
		0 & 0 & 0 & 0  \\
		0 & 0 & 0 & 0   \\
		0 & 0  & 0 & 0     \\
		0 & 0 & g(\bar{D}^{(2)}) & 0   
	\end{bmatrix}\,,
\end{equation}
\begin{equation}
	\label{barC^2-barE^1}
	\bar{C}^{(2)} \in \begin{bmatrix}
		0 & \bar{c}_{12} & \bar{c}_{13} & -\bar{c}_{12} & -\bar{c}_{13} \\
		0 & -\bar{c}_{12} & -\bar{c}_{13} & \bar{c}_{12} & \bar{c}_{13}   \\
		0 & 0 & 0 & 0 & 0 
	\end{bmatrix}\,,\qquad
	\bar{E}^{(1)}, \bar{F}^{(1)}  \in \begin{bmatrix}
		\bar{e}_{11} & -\bar{e}_{11} & -\sqrt{2} \bar{e}_{11} & 0 & 0 
	\end{bmatrix},
\end{equation}
\begin{equation}
	\label{D^2}
	\bar{D}^{(2)} \in \begin{bmatrix}
		\bar{d}_{11} & \bar{d}_{11} & \bar{d}_{13}  \\
		-\bar{d}_{11} & -\bar{d}_{11} & -\bar{d}_{13}   \\
		0 & 0 & 0 
	\end{bmatrix}\,,
\end{equation}
\begin{equation}
	\label{g(barD^2)-g(barC^2)}
	g(\bar{D}^{(2)}) \in \begin{bmatrix}
		0 & \bar{d}_{13} & \sqrt{2} \bar{d}_{11} & -\bar{d}_{13} & -\sqrt{2} \bar{d}_{11} 
	\end{bmatrix}\,,\qquad
	g(\bar{C}^{(2)}) \in \begin{bmatrix}
		-\sqrt{2} \bar{c}_{13} & -\sqrt{2} \bar{c}_{13} & -2 \bar{c}_{12} 
	\end{bmatrix},
\end{equation}
Tensor $\boldsymbol{\mathsf{C}}_{\mathbb{O}(2) \oplus \mathbb{Z}_2^c}$ has $5$ independent components and corresponds to the Curie group with Hermann–Mauguin symbol $\infty m$.
In addition to those of class $\mathbb{SO}(2) \oplus \mathbb{Z}_2^c$, there are six independent parameters associated with the fifth-order tensor, which collectively induce $16$ additional universality PDEs.
The independent content of these $16$ universality PDEs is
\begin{equation}
\label{ClassZ_8^--constraints-1}
\begin{dcases}
-\left(-2+\sqrt{2}\right) \frac{\partial^2 h_2}{\partial x_2^2} -\sqrt{2} \frac{\partial^3 u_{33}}{\partial x_2^3}-\left(1+\sqrt{2}\right) \frac{\partial^2 h_1}{\partial x_1 \partial x_2} -\left(1-2\sqrt{2}\right) \frac{\partial^2 h_2}{\partial x_1^2}+\left(2+\sqrt{2}\right) \frac{\partial^3 u_{33}}{\partial x_1^2 \partial x_2} =0\,, \\
-\frac{\partial^2 h_2}{\partial x_2^2}-\left(-1+\sqrt{2}\right) \frac{\partial^2 h_1}{\partial x_1 \partial x_2}+\left(-1+\sqrt{2}\right) \frac{\partial^2 h_2}{\partial x_1^2}-\left(1+2\sqrt{2}\right) \frac{\partial^3 u_{22}}{\partial x_1^2 \partial x_2} =0 \,, \\
\frac{\partial^3 u_{33}}{\partial x_2^3}- \left(-1+\sqrt{2}\right) \frac{\partial^2 h_1}{\partial x_1 \partial x_2}+\left(-1+\sqrt{2}\right) \frac{\partial^2 h_2}{\partial x_1^2}-\left(1+2\sqrt{2}\right)\frac{\partial^3 u_{33}}{\partial x_1^2 \partial x_2} = 0\,, \\
-\left(-1+\sqrt{2}\right) \frac{\partial^2 h_1}{\partial x_2^2} +\left(-1+\sqrt{2}\right) \frac{\partial^2 h_2}{\partial x_1 \partial x_2} +\left(1+2\sqrt{2}\right) \frac{\partial^3 u_{33}}{\partial x_1 \partial x_2^2}-\frac{\partial^3 u_{33}}{\partial x_1^3} = 0 \,, \\
-2 \frac{\partial^2 h_1}{\partial x_2^2}+\left(1+\sqrt{2}\right) \frac{\partial^2 h_2}{\partial x_1 \partial x_2}+\frac{\partial^2 h_1}{\partial x_1^2}=0 \,, \\
\sqrt{2} \frac{\partial^2 h_1}{\partial x_2^2}+\left(1+\sqrt{2}\right) \frac{\partial^2 h_2}{\partial x_1 \partial x_2}-\left(1+2\sqrt{2}\right) \frac{\partial^3 u_{33}}{\partial x_1 \partial x_2^2} =0\,,
\end{dcases}
\end{equation}
as well as 
\begin{equation}
\label{ClassZ_8^--constraints-2}
\begin{dcases}
\left(-1+\sqrt{2}\right) \frac{\partial^2 h_1}{\partial x_2^2} +\left(1+\sqrt{2}\right) \frac{\partial^2 h_2}{\partial x_1 \partial x_2}-\left(2+\sqrt{2}\right) \frac{\partial^3 u_{33}}{\partial x_1 \partial x_2^2} +\left(-2+\sqrt{2}\right) \frac{\partial^2 h_1}{\partial x_1^2} +\sqrt{2}  \frac{\partial^3 u_{33}}{\partial x_1^3} =0 \,, \\
-\frac{\partial^2 h_1}{\partial x_2^2} -2 \frac{\partial^2 h_2}{\partial x_1 \partial x_2}+\frac{\partial^2 h_1}{\partial x_1^2} = 0\,, \\
-\left(-1+\sqrt{2}\right) \frac{\partial^2 h_1}{\partial x_2^2}+\left(-1+\sqrt{2}\right) \frac{\partial^2 h_2}{\partial x_1 \partial x_2}+\left(1+2\sqrt{2}\right) \frac{\partial^3 u_{33}}{\partial x_1 \partial x_2^2}-\frac{\partial^3 u_{33}}{\partial x_1^3}=0 \,, \\
-\frac{\partial^3 u_{33}}{\partial x_2^3}+\left(-1+\sqrt{2}\right) \frac{\partial^2 h_1}{\partial x_1 \partial x_2}-\left(-1+\sqrt{2}\right)\frac{\partial^2 h_2}{\partial x_1^2}+\left(1+2\sqrt{2}\right)\frac{\partial^3 u_{33}}{\partial x_1^2 \partial x_2} =0 \,, \\
\frac{\partial^2 h_2}{\partial x_2^2}-3 \frac{\partial^2 h_1}{\partial x_1 \partial x_2}-2 \frac{\partial^2 h_2}{\partial x_1^2}=0 \,, \\
-\frac{\partial^2 h_2}{\partial x_2^2}+\frac{\partial^3 u_{33}}{\partial x_2^3}+\left(1+\sqrt{2}\right)\frac{\partial^2 h_1}{\partial x_1 \partial x_2}+\sqrt{2} \frac{\partial^2 h_2}{\partial x_1^2}-\left(1+2\sqrt{2}\right) \frac{\partial^3 u_{33}}{\partial x_1^2 \partial x_2} =0 \,,
\end{dcases}
\end{equation}
and
\begin{equation}
\label{ClassZ_8^--constraints-3}
\begin{dcases}
\frac{\partial^3 k_1}{\partial x_2^3}+x_3 \frac{\partial^3 h_1}{\partial x_2^3} +3 \left(\frac{\partial^3 k_2}{\partial x_1 \partial x_2^2}+x_3 \frac{\partial^3 h_2}{\partial x_1 \partial x_2^2}\right)-3\left(\frac{\partial^3 k_1}{\partial x_1^2 \partial x_2}+x_3 \frac{\partial^3 h_1}{\partial x_2^3}\right)- \left(\frac{\partial^3 k_2}{\partial x_1^3}+x_3 \frac{\partial^3 h_2}{\partial x_1^3}\right) =0 \,, \\
-\sqrt{2}\left(\frac{\partial^3 k_1}{\partial x_2^3}+x_3 \frac{\partial^3 h_1}{\partial x_2^3}\right) -\left(2+\sqrt{2}\right) \left(\frac{\partial^3 k_2}{\partial x_1 \partial x_2^2}+x_3 \frac{\partial^3 h_2}{\partial x_1 \partial x_2^2}\right)+\left(2+\sqrt{2}\right)\left(\frac{\partial^3 k_1}{\partial x_1^2 \partial x_2}+x_3 \frac{\partial^3 h_1}{\partial x_2^3}\right) \\
+\sqrt{2} \left(\frac{\partial^3 k_2}{\partial x_1^3}+x_3 \frac{\partial^3 h_2}{\partial x_1^3}\right) =0 \,, \\
\left(\frac{\partial^3 k_2}{\partial x_2^3}+x_3 \frac{\partial^3 h_2}{\partial x_2^3}\right)-\left(1+2\sqrt{2}\right) \left(\frac{\partial^3 k_1}{\partial x_1 \partial x_2^2}+x_3 \frac{\partial^3 h_1}{\partial x_1 \partial x_2^2}\right)-\left(1+2\sqrt{2}\right) \left(\frac{\partial^3 k_2}{\partial x_1^2 \partial x_2}+x_3 \frac{\partial^3 h_2}{\partial x_1^2 \partial x_2}\right) \\
+\left(\frac{\partial^3 k_1}{\partial x_1^3}+x_3 \frac{\partial^3 h_1}{\partial x_1^3}\right) =0 \,, \\
-\left(\frac{\partial^3 k_2}{\partial x_2^3}+x_3 \frac{\partial^3 h_2}{\partial x_2^3}\right)+5\left(\frac{\partial^3 k_1}{\partial x_1 \partial x_2^2}+x_3 \frac{\partial^3 h_1}{\partial x_1 \partial x_2^2}\right)+5 \left(\frac{\partial^3 k_2}{\partial x_1^2 \partial x_2}+x_3 \frac{\partial^3 h_2}{\partial x_1^2 \partial x_2}\right) \\
-\left(\frac{\partial^3 k_1}{\partial x_1^3}+x_3 \frac{\partial^3 h_1}{\partial x_1^3}\right)=0 \,.
\end{dcases}
\end{equation}
We therefore obtain the following result:
\begin{prop}
The universal displacements in tetragonal $\mathbb{Z}_8^-$ class linear strain-gradient elastic solids coincide with those of class $\mathbb{SO}(2) \oplus \mathbb{Z}_2^c$, subject to the additional constraints \eqref{ClassZ_8^--constraints-1}--\eqref{ClassZ_8^--constraints-3}\,.
\end{prop}

\subsubsection{Tetragonal class $\mathbb{D}_5^v$} 

Tensor $\boldsymbol{\mathsf{A}}_{\mathbb{D}_5 \oplus \mathbb{Z}_2^c}$ has $23$ independent components and is of the same form as that of class $\mathbb{D}_5$. Tensor $\boldsymbol{\mathsf{M}}_{\mathbb{D}_5^v}$ has $13$ independent components and has the same form as that given in class $\mathbb{Z}_5$. Tensor $\boldsymbol{\mathsf{C}}_{{\mathbb{O}(2) \oplus \mathbb{Z}_2^c}}$ has $5$ independent components and corresponds to the Curie group with Hermann--Maugin symbol $\infty m$. In addition to those of class $\mathbb{O}(2)^-$, this class introduces three independent parameters: one stemming from the fifth-order tensor and the remaining two from the sixth-order tensor.
The resulting universality constraints read
\begin{equation}
\label{ClassD^v_5-constraints}
\begin{dcases}
\left(-2+\sqrt{2}\right)\left(\frac{\partial^3 k_1}{\partial x_2^3}+x_3 \frac{\partial^3 h_1}{\partial x_2^3}\right)+\left(3-2\sqrt{2}\right)\left(\frac{\partial^3 k_2}{\partial x_1 \partial x_2^2}+x_3 \frac{\partial^3 h_2}{\partial x_1 \partial x_2^2}\right) -2\sqrt{2} \left(\frac{\partial^3 k_2}{\partial x_1^3}+x_3 \frac{\partial^3 h_2}{\partial x_1^3}\right)  \\
 +\left(-1+\sqrt{2}\right)\left(\frac{\partial^3 k_1}{\partial x_1^2 \partial x_2}+x_2 \frac{\partial^3 h_1}{\partial x_1^2 \partial x_2}\right)=0\,, \\
 4 \frac{\partial^3 h_1}{\partial x_2^3}-\left(4+5\sqrt{2}\right)\frac{\partial^3 h_2}{\partial x_1 \partial x_2^2}+\sqrt{2}\left(\frac{\partial^4 \hat{u}_{33}}{\partial x_1 \partial x_2^3}+4\frac{\partial^3 h_1}{\partial x_1^2 \partial x_2}+\frac{\partial^3 h_2}{\partial x_1^3}+\frac{\partial^4 \hat{u}_{33}}{\partial x_1^3 \partial x_2}\right)=0\,, \\
 2 \frac{\partial^3 h_1}{\partial x_2^3}-\left(3+2\sqrt{2}\right)\frac{\partial^3 h_2}{\partial x_1 \partial x_2^2}-\left(1+2\sqrt{2}\right)\frac{\partial^4 \hat{u}_{33}}{\partial x_1 \partial x_2^3}+4\left(1+\sqrt{2}\right) \frac{\partial^3 h_1}{\partial x_1^2 \partial x_2} \\ 
+\left(1+2\sqrt{2}\right)\left(\frac{\partial^3 h_2}{\partial x_1^3}+ \frac{\partial^4 \hat{u}_{33}}{\partial x_1^3 \partial x_2}\right)=0\,, \\
 3 \left(\frac{\partial^3 k_2}{\partial x_1^2 \partial x_2}+x_3 \frac{\partial^3 h_2}{\partial x_1^2 \partial x_2}\right)+\left(-2+\sqrt{2}\right) \left(\frac{\partial^3 k_1}{\partial x_1 \partial x_2^2}+x_3 \frac{\partial^3 h_1}{\partial x_1 \partial x_2^2}\right)+\left(1+\sqrt{2}\right)\left(\frac{\partial^3 k_1}{\partial x_1^3}+x_3 \frac{\partial^3 h_1}{\partial x_1^3}\right)=0\,, \\
-\frac{\partial^3 h_2}{\partial x_2^3}+\frac{\partial^3 h_1}{\partial x_1 \partial x_2^2}+5 \frac{\partial^3 h_2}{\partial x_1 \partial x_2^2}+4 \frac{\partial^4 \hat{u}_{33}}{\partial x_1^2 \partial x_2^2}-\frac{\partial^3 h_1}{\partial x_1^3}=0\,, \\
-3\frac{\partial^3 h_2}{\partial x_2^3}-\frac{\partial^4 \hat{u}_{33}}{\partial x_2^4}+\left(3+8\sqrt{2}\right)\frac{\partial^3 h_1}{\partial x_1 \partial x_2^2}+\left(3+6\sqrt{2}\right)\frac{\partial^3 h_2}{\partial x_1^2 \partial x_2}+2\left(1+\sqrt{2}\right)\frac{\partial^4 \hat{u}_{33}}{\partial x_1^2 \partial x_2^2}- \\
3 \frac{\partial^3 h_2}{\partial x_2^3}-\frac{\partial^4 \hat{u}_{33}}{\partial x_2^4} -\left(3+8\sqrt{2}\right) \frac{\partial^3 h_1}{\partial x_1 \partial x_2} -3\left(1+2\sqrt{2}\right) \frac{\partial^3 h_2}{\partial x_1^2 \partial x_2} -2\left(1+\sqrt{2}\right) \frac{\partial^4 u_{33}}{\partial x_1^2 \partial x_2^2}  \\
 +3\frac{\partial^3 h_1}{\partial x_1^3}+\frac{\partial^4 u_{33}}{\partial x_2^4} =0 \,, \\
-\frac{\partial^4 k_2}{\partial x_2^4}-x_3 \frac{\partial^4 h_2}{\partial x_2^4}+\left(1+\sqrt{2}\right)\left(\frac{\partial^4 k_1}{\partial x_1 \partial x_2^3}+x_3 \frac{\partial^4 h_1}{\partial x_1 \partial x_2^3}\right)+\left(1+\sqrt{2}\right)\left(\frac{\partial^4 k_2}{\partial x_1^2 \partial x_2^2}+x_3 \frac{\partial^4 h_2}{\partial x_1^2 \partial x_2^2}\right)- \\
\left(1+\sqrt{2}\right)\left(\frac{\partial^4 k_1}{\partial x_1^3 \partial x_2}+x_3 \frac{\partial^4 h_1}{\partial x_1^3 \partial x_2}\right)-\sqrt{2}\left(\frac{\partial^4 k_2}{\partial x_1^4}+x_3 \frac{\partial^4 h_2}{\partial x_1^4}\right)=0\,, \\
\sqrt{2} \left(\frac{\partial^4 k_1}{\partial x_1 \partial x_2^3}+x_3 \frac{\partial^4 h_1}{\partial x_1 \partial x_2^3}\right)+\left(1+\sqrt{2}\right)\left(\frac{\partial^4 k_2}{\partial x_1^2 \partial x_2^2}+x_3 \frac{\partial^4 h_2}{\partial x_1^2 \partial x_2^2}\right)-\frac{\partial^4 k_1}{\partial x_1^3 \partial x_2}-x_3 \frac{\partial^4 h_1}{\partial x_1^3 \partial x_2}=0\,.
\end{dcases}
\end{equation}
With the universality PDEs of class $\mathbb{O}(2)^-$, in particular the linearity of the functions $h_i$, $i=1,2$, one verifies that all of the above universality PDEs are satisfied identically. Hence, we have the following result.
\begin{prop}
The universal displacements in tetragonal $\mathbb{D}_5^v$ class linear strain-gradient elastic solids are the same as those of class $\mathbb{O}(2)^-$\,.
\end{prop}

\subsubsection{Tetragonal class $\mathbb{D}_8^h$} 

Tensor $\boldsymbol{\mathsf{A}}_{\mathbb{O}(2) \oplus \mathbb{Z}_2^c}$ has $21$ independent components and is of the same form as that of class $\mathbb{O}(2)$. Tensor $\boldsymbol{\mathsf{M}}_{\mathbb{D}_8^h}$ has $3$ independent components and is of the same form as that given in class $\mathbb{Z}_8^-$. Tensor $\boldsymbol{\mathsf{C}}_{\mathbb{O}(2) \oplus \mathbb{Z}_2^c}$ has $5$ independent components and corresponds to the Curie group with Hermann--Maugin symbol $\infty m$. Compared with class $\mathbb{SO}(2) \oplus \mathbb{Z}_2^c$, there are three additional independent parameters stemming from the fifth-order tensor $\boldsymbol{\mathsf{M}}_{\mathbb{D}_8^h}$. Consequently, we obtain three additional sets of universality PDEs, comprising nine constraints in total; one of them is identically zero, so only eight independent constraints remain, which read
\begin{equation}
\label{ClassD^h_8-constraints}
\begin{dcases}
2\frac{\partial^2 h_2}{\partial x_2^2}-\frac{\partial^2 h_1}{\partial x_1 \partial x_2}+\frac{\partial^2 h_2}{\partial x_1^2}+2\frac{\partial^3 \hat{u}_{33}}{\partial x_1^2 \partial x_2}=0\,, \\
\frac{\partial^2 h_1}{\partial x_1 \partial x_2}+\frac{\partial^2 h_2}{\partial x_1^2}=0\,, \\
\frac{\partial^3 \hat{u}_{33}}{\partial x_2^3}-\left(-1+\sqrt{2}\right)\frac{\partial^2 h_1}{\partial x_1 \partial x_2}+\left(-1+\sqrt{2}\right)\frac{\partial^2 h_2}{\partial x_1^2}+\left(1+2\sqrt{2}\right)\frac{\partial^3 \hat{u}_{33}}{\partial x_1^2 \partial x_2}=0\,, \\
-\frac{\partial^1 h_1}{\partial x_2^2}+\frac{\partial^2 h_2}{\partial x_1 \partial x_2}-2\frac{\partial^3 \hat{u}_{33}}{\partial x_1 \partial_2^2}-2\frac{\partial^2 h_1}{\partial x_1^2}=0\,, \\
-\frac{\partial^2 h_1}{\partial x_2^2}-\frac{\partial^2 h_2}{\partial x_1 \partial x_2}=0\,, \\
-\left(-1+\sqrt{2}\right)\frac{\partial^2 h_1}{\partial x_2^2}+\left(-1+\sqrt{2}\right)\frac{\partial^2 h_2}{\partial x_1 \partial x_2}+\left(1+2\sqrt{2}\right)\frac{\partial^3 \hat{u}_{33}}{\partial x_1 \partial x_2^2}+\frac{\partial^3 \hat{u}_{33}}{\partial x_1^3}=0\,, \\
 \left(\frac{\partial^3 k_2}{\partial x_1 \partial x_2^2}+x_3 \frac{\partial^3 h_2}{\partial x_1 \partial x_2^2}\right)-\left(\frac{\partial^3 k_2}{\partial x_1 \partial x_1^2 \partial x_2}+x_3 \frac{\partial^3 h_1}{\partial x_1 \partial x_1^2 \partial x_2}\right)=0\,, \\
-\sqrt{2}\left(\frac{\partial^3 k_1}{\partial x_2^3}+x_3 \frac{\partial^3 h_1}{\partial x_2^3}\right)-(2+\sqrt{2})\left(\frac{\partial^3 k_2}{\partial x_1 \partial x_2^2}+x_3 \frac{\partial^3 h_2}{\partial x_1 \partial x_2^2}\right) \\
\quad +\left(2+\sqrt{2}\right)\left(\frac{\partial^3 k_1}{\partial x_1^2 \partial x_2}+x_3 \frac{\partial^3 h_1}{\partial x_1^2 \partial x_2}\right)+\sqrt{2}\left(\frac{\partial^3 k_2}{\partial x_1^3}+x_3 \frac{\partial^3 h_2}{\partial x_1^3}\right)=0\,.
\end{dcases}
\end{equation}
We therefore obtain the following result.
\begin{prop}
The universal displacements in the tetragonal $\mathbb{D}_8^h$ class linear strain-gradient elastic solids are the same as those of the class $\mathbb{SO}(2) \oplus \mathbb{Z}_2^c$, supplemented by the additional constraints \eqref{ClassD^h_8-constraints}.
\end{prop}

\renewcommand{\arraystretch}{1.15}
\begin{longtable}{p{0.30\textwidth} p{0.66\textwidth}}
\caption{Summary of universal displacements: Tetragonal classes.}\label{Table-UD-summary-tetragonal}\\
\hline
\textbf{Symmetry class} & \textbf{Universal displacement family} \\
\hline
\endfirsthead
\hline
\textbf{Symmetry class} & \textbf{Universal displacement family} (continued) \\
\hline
\endhead
\hline
\endfoot
\hline
\endlastfoot
Tetragonal class $\mathbb{Z}_4$ & Superposition of a homogeneous field and \eqref{tetragonal-linear-inhomogeneous} with $g(x_1, x_2)$ given by \eqref{Z^-_4-g-final-form} \\ \hline
Tetragonal class $\mathbb{D}_4$ & Superposition of a homogeneous field and \eqref{tetragonal-linear-inhomogeneous} with $g(x_1, x_2)$ given by \eqref{Z^-_4-g-final-form} \\ \hline
Tetragonal class $\mathbb{Z}_4 \oplus \mathbb{Z}_2^c$ & Superposition of a homogeneous field and \eqref{tetragonal-linear-inhomogeneous} with $g(x_1, x_2)$ given by \eqref{Z4+Z_2^c-g-final-form} \\ \hline
Tetragonal class $\mathbb{D}_4 \oplus \mathbb{Z}_2^c$ & Superposition of a homogeneous field and \eqref{tetragonal-linear-inhomogeneous} with $g(x_1, x_2)$ given by \eqref{D4+Z_2^c-g-final-form} \\ \hline
Tetragonal class $\mathbb{Z}_8^-$ & Same as class $\mathbb{SO}(2) \oplus \mathbb{Z}_2^c$ with additional constraints (\ref{ClassZ_8^--constraints-1})--(\ref{ClassZ_8^--constraints-3}) \\ \hline
Tetragonal class $\mathbb{D}_5^v$ & Same as class $\mathbb{O}(2)^-$ \\ \hline
Tetragonal class $\mathbb{D}_8^h$ & Same as class $\mathbb{SO}(2) \oplus \mathbb{Z}_2^c$ with additional constraint \eqref{ClassD^h_8-constraints} \\ \hline
\end{longtable}
\renewcommand{\arraystretch}{2.0}

\subsection{Pentagonal classes}

All five pentagonal classes $\mathbb{Z}_5$, $\mathbb{Z}_5 \oplus \mathbb{Z}_2^c$, $\mathbb{D}_5$, $\mathbb{D}_5 \oplus \mathbb{Z}_2^c$ and $\mathbb{D}^h_{10}$ have an elasticity matrix $\boldsymbol{\mathsf{C}}$ identical to that of the transversely isotropic linear elastic case. Therefore, the candidate universal displacements for these classes are those of \eqref{transverse-isotropy-linear-solution} and \eqref{transverse-isotropy-linear-solution-2}. 

\subsubsection{Pentagonal class $\mathbb{Z}_5$}

The tensor $\boldsymbol{\mathsf{A}}_{\mathbb{Z}_5 \oplus \mathbb{Z}_2^c}$ has $35$ independent components and has the form
\begin{equation}
\label{A_Z_5}
    \boldsymbol{\mathsf{A}}_{\mathbb{Z}_5 \oplus \mathbb{Z}_2^c}(\mathbf{x}) =
    \begin{bmatrix}
    A^{(11)} & B^{(6)} & 0 & 0  \\
     & A^{(11)} & F^{(2)} & G^{(2)}   \\
     &  & H^{(6)} & I^{(4)}   \\
     &  &  & J^{(4)}      
    \end{bmatrix}_S
    +
    \begin{bmatrix}
    0 & 0 & f(G^{(2)}) & f(F^{(2)})  \\
     & 0 & 0 & 0   \\
     &  & h(J^{(4)}) & 0   \\
     &  &  & 0      
    \end{bmatrix}_S\,,
\end{equation}
where the matrices $A^{(11)}$, $B^{(6)}$, $I^{(4)}$, and $J^{(4)}$ are defined in the preceding classes $\mathbb{Z}_3$ and $\mathbb{D}_3$, and the remaining terms are
\begin{equation}
\label{F^2-G^2}
    F^{(2)}=\begin{bmatrix}
    0 & f_{12} & f_{13} & -f_{12} & -f_{13}  \\
    0 & -f_{12} & -f_{13} & f_{12} & f_{13} \\
    0 & -\sqrt{2} f_{12} & -\sqrt{2} f_{13} & \sqrt{2} f_{12} & \sqrt{2} f_{13}  \\
    0 & 0 & 0 & 0 & 0 \\
    0 & 0 & 0 & 0 & 0     
    \end{bmatrix}\,,\qquad
    G^{(2)}=\begin{bmatrix}
    g_{11} & g_{12} & g_{12}  \\
    -g_{11} & -g_{12} & -g_{12} \\
    -\sqrt{2}g_{11} & -\sqrt{2}g_{12} & -\sqrt{2}g_{12}  \\
    0 & 0 & 0  \\
    0 & 0 & 0      
    \end{bmatrix}\,,
\end{equation}
\begin{equation}
\label{f(F^2)-f(G^2)}
   f(F^{(2)})=\begin{bmatrix}
    -\sqrt{2}f_{12} & -f_{13} & -f_{13}  \\
    \sqrt{2}f_{12} & f_{13} & f_{13} \\
    2f_{12} & \sqrt{2}f_{13} & \sqrt{2}f_{13}  \\
    0 & 0 & 0  \\
    0 & 0 & 0      
    \end{bmatrix}\,,\qquad
    f(G^{(2)})=\begin{bmatrix}
    0 & \frac{\sqrt{2}}{2}g_{11} & g_{12} & -\frac{\sqrt{2}}{2}g_{11} & -g_{12}  \\
    0 & -\frac{\sqrt{2}}{2}g_{11} & -g_{12} & \frac{\sqrt{2}}{2}g_{11} & g_{12} \\
    0 & -g_{11} & -\sqrt{2}g_{12} & g_{11} & \sqrt{2}g_{12}  \\
    0 & 0 & 0 & 0 & 0 \\
    0 & 0 & 0 & 0 & 0     
    \end{bmatrix}\,,
\end{equation}
\begin{equation}
\label{h(J^4)}
    h(J^{(4)})=\begin{bmatrix}
    0 & 0 & 0 & 0 & 0  \\
     & 0 & \sqrt{2}j_{12} & -j_{11} & 0 \\
     &  & 0 & 0 & -(j_{22}+j_{23})  \\
     &  &  & 0 & \sqrt{2}j_{12} \\
     &  &  &  & 0     
    \end{bmatrix}_S\,.
\end{equation}
Matrix $\boldsymbol{\mathsf{M}}_{\mathbb{Z}_5}$ has $22$ independent components, and is given by
\begin{equation}
\label{M_Z_5}
    \boldsymbol{\mathsf{M}}_{\mathbb{Z}_5}
    = \boldsymbol{\mathsf{M}}_{\mathbb{D}_5}
    + \boldsymbol{\mathsf{M}}_{\mathbb{D}_5^v}\,,
\end{equation}
where
\begin{equation}
\label{M_D_5}
    \boldsymbol{\mathsf{M}}_{\mathbb{D}_5} =
    \begin{bmatrix}
    \bar{A}^{(1)} & 0 & 0 & \bar{D}^{(4)}  \\
    \bar{E}^{(4)} & 0 & 0 & 0   \\
    0 & -\bar{E}^{(4)} & 0 & 0     \\
    0 & 0 & 0 & 0   
    \end{bmatrix}
    +
    \begin{bmatrix}
    0 & 0 & 0 & 0  \\
    0 & 0 & 0 & 0   \\
    0 & 0 & 0 & 0     \\
    0 & g(\bar{A}^{(1)}) & g(\bar{D}^{(4)}) & 0   
    \end{bmatrix}\,,
\end{equation}
\begin{equation}
\label{M_D_5^v}
    \boldsymbol{\mathsf{M}}_{\mathbb{D}_5^v} =
    \begin{bmatrix}
    0 & \bar{B}^{(1)} & \bar{C}^{(8)} & 0  \\
    0 & \bar{F}^{(4)} & 0 & 0   \\
    \bar{F}^{(4)} & 0 & 0 & 0     \\
    0 & 0 & 0 & 0   
    \end{bmatrix}
    +
    \begin{bmatrix}
    0 & 0 & 0 & 0  \\
    0 & 0 & 0 & 0   \\
    0 & 0 & 0 & 0     \\
    g(\bar{B}^{(1)}) & 0 & 0 & g(\bar{C}^{(8)})   
    \end{bmatrix}\,,
\end{equation}
\begin{equation}
\label{barA^1-barB^1}
    \bar{A}^{(1)},\, \bar{B}^{(1)} \in 
    \begin{bmatrix}
    \bar{a}_{11} & -\bar{a}_{11} & -\sqrt{2} \bar{a}_{11} & 0 & 0 \\
    -\bar{a}_{11} & \bar{a}_{11} & \sqrt{2} \bar{a}_{11} & 0 & 0   \\
    0 & 0 & 0 & 0 & 0 
    \end{bmatrix}\,,
\end{equation}
\begin{equation}
\label{g(barA^1)-g(barB^1)}
    g(\bar{A}^{(1)}),\, g(\bar{B}^{(1)}) =
    \begin{bmatrix}
    \sqrt{2} \bar{a}_{11} & -\sqrt{2} \bar{a}_{11} & 2 \bar{a}_{11} & 0 & 0 
    \end{bmatrix}\,,
\end{equation}
and the remaining components coincide with those of the preceding symmetry classes $\mathbb{Z}_4$ and $\mathbb{D}_4$. The tensor $\boldsymbol{\mathsf{C}}_{\mathbb{O}(2) \oplus \mathbb{Z}_2^c}$ corresponds to the Curie group with Hermann--Mauguin symbol $\infty2$ and has $5$ independent components.

The third-order universality PDEs for this class read
\begin{equation}
\label{ClassZ_5-third-order-constraints-1}
\begin{dcases}
\frac{\partial^3 u_3}{\partial x_2 \partial x_3^2}
=\frac{\partial^3 u_3}{\partial x_3^3}
=\frac{\partial^3 u_2}{\partial x_3^3}
=\frac{\partial^3 u_3}{\partial x_1 \partial x_2^2}
=\frac{\partial^3 u_1}{\partial x_3^3}
=\frac{\partial^3 u_3}{\partial x_1 \partial x_3^2}
=\frac{\partial^3 u_2}{\partial x_1 \partial x_2 \partial x_3}
=0 \,, \\
\frac{\partial^3 u_3}{\partial x_1^3}
=\frac{\partial^3 u_1}{\partial x_2^2 \partial x_3}
=\frac{\partial^3 u_1}{\partial x_1^2 \partial x_3}
=\frac{\partial^3 u_2}{\partial x_1^2 \partial x_2}
=0 \,.
\end{dcases}
\end{equation}
and
\begin{equation}
\label{ClassZ_5-third-order-constraints-2}
\begin{dcases}
\sqrt{2} \frac{\partial^3 u_2}{\partial x_2^2 \partial x_3}
+\sqrt{2}\frac{\partial^3 u_3}{\partial x_2^3}
+\left(4-3\sqrt{2}\right)\frac{\partial^3 u_1}{\partial x_1 \partial x_2 \partial x_3}
-2\left(-1+\sqrt{2}\right)\frac{\partial^3 u_2}{\partial x_1^2 \partial x_3}
-\left(-2+\sqrt{2}\right)\frac{\partial^3 u_3}{\partial x_1^2 \partial x_2}
=0 \,, \\
2\left(-2+\frac{1}{\sqrt{2}}\right) \frac{\partial^3 u_2}{\partial x_2^2 \partial x_3}
+\sqrt{2}\frac{\partial^3 u_3}{\partial x_2^3}
+\left(2-3\sqrt{2}\right)\frac{\partial^3 u_1}{\partial x_1 \partial x_2 \partial x_3}
-2\sqrt{2}\frac{\partial^3 u_2}{\partial x_1^2 \partial x_3}
-\left(-2+\sqrt{2}\right)\frac{\partial^3 u_3}{\partial x_1^2 \partial x_2}
=0 \,, \\
\frac{\partial^3 u_2}{\partial x_2^2 \partial x_3}
+\frac{\partial^3 u_2}{\partial x_1^2 \partial x_3}
=0 \,, \\
-\frac{\partial^3 u_2}{\partial x_2^2 \partial x_3}
-\frac{3}{\sqrt{2}}\frac{\partial^3 u_1}{\partial x_1 \partial x_2 \partial x_3}
+\left(1-\frac{1}{\sqrt{2}}\right)\frac{\partial^3 u_2}{\partial x_1^2 \partial x_3}
-\left(-1+\sqrt{2}\right)\frac{\partial^3 u_3}{\partial x_1^2 \partial x_2}
=0 \,, \\
\frac{\partial^3 u_2}{\partial x_2^2 \partial x_3}
+\frac{\partial^3 u_3}{\partial x_2^3}
+\left(-1+\frac{2}{\sqrt{2}}\right)\frac{\partial^3 u_1}{\partial x_1 \partial x_2 \partial x_3}
\frac{1}{\sqrt{2}}\left(
\frac{\partial^3 u_2}{\partial x_1^2 \partial x_3}
+2\frac{\partial^3 u_3}{\partial x_1^2 \partial x_2}
\right)
=0 \,, \\
\left(1+\sqrt{2}\right)\frac{\partial^3 u_2}{\partial x_2^3}
+\left(5-2\sqrt{2}\right)\frac{\partial^3 u_1}{\partial x_1 \partial x_2^2}
-\left(-4+\sqrt{2}\right)\frac{\partial^3 u_2}{\partial x_1^2 \partial x_2}
+2\frac{\partial^3 u_1}{\partial x_1^3}
=0 \,, \\
-\left(-2+\sqrt{2}\right)\frac{\partial^3 u_1}{\partial x_2^3}
+3\frac{\partial^3 u_2}{\partial x_1^2 \partial x_2}
-\left(1+\sqrt{2}\right)\frac{\partial^3 u_1}{\partial x_1^2 \partial x_2}
=0 \,.
\end{dcases}
\end{equation}
as well as
\begin{equation}
\label{ClassZ_5-third-order-constraints-3}
\begin{dcases}
2 \frac{\partial^3 u_2}{\partial x_2^2 \partial x_3}+\frac{\partial^3 u_3}{\partial x_2^3}+\frac{3}{\sqrt{2}}\left(3\frac{\partial^3 u_1}{\partial x_1 \partial x_2 \partial x_3}+2\frac{\partial^3 u_3}{\partial x_1^2 \partial x_2}\right)=0 \,, \\
\left(1-\frac{3}{\sqrt{2}}\right)\frac{\partial^3 u_1}{\partial x_1 \partial x_2 \partial x_3}-\left(-1+\sqrt{2}\right)\frac{\partial^3 u_3}{\partial x_1^2 \partial x_2}=0 \,, \\
\frac{\partial^3 u_2}{\partial x_2^2 \partial x_3}+\frac{\partial^3 u_1}{\partial x_1 \partial x_2 \partial x_3}=0 \,, \\
3 \frac{\partial^3 u_2}{\partial x_2^2 \partial x_3}+\frac{\partial^3 u_3}{\partial x_2^3}+\frac{3}{\sqrt{2}}\left(3\frac{\partial^3 u_1}{\partial x_1 \partial x_2 \partial x_3}+2\frac{\partial^3 u_3}{\partial x_1^2 \partial x_2}\right)=0 \,, \\
\left(3-\frac{3}{\sqrt{2}}\right) \frac{\partial^3 u_1}{\partial x_1 \partial x_2 \partial x_3}-\left(-1+\sqrt{2}\right)\frac{\partial^3 u_3}{\partial x_1^2 \partial x_2}=0 \,, \\
2\sqrt{2} \frac{\partial^3 u_1}{\partial x_2^3}+\left(1+5\sqrt{2}\right)\frac{\partial^3 u_2}{\partial x_1 \partial x_2^2}+\left(1-4\sqrt{2}\right)\frac{\partial^3 u_1}{\partial x_1^2 \partial x_2}-\left(-2+\sqrt{2}\right)\frac{\partial^3 u_2}{\partial x_1^3}=0 \,, \\
-2 \frac{\partial^3 u_1}{\partial x_2^3}+\left(8+\sqrt{2}\right)\frac{\partial^3 u_1}{\partial x_1 \partial x_2^2}+7\frac{\partial^3 u_2}{\partial x_1^2 \partial x_2}+\left(-1+\sqrt{2}\right)\frac{\partial^3 u_2}{\partial x_1^3}=0 \,,
\end{dcases}
\end{equation}
and finally
\begin{equation}
\label{ClassZ_5-third-order-constraints-4}
\begin{dcases}
\frac{\partial^3 u_1}{\partial x_2^3}-\frac{\partial^3 u_2}{\partial x_1 \partial x_2^2}+\frac{\partial^3 u_1}{\partial x_1^2 \partial x_2}-\frac{\partial^3 u_2}{\partial x_1^3}=0 \,, \\
-\frac{\partial^3 u_1}{\partial x_2 \partial x_3^2}+\frac{\partial^3 u_2}{\partial x_1 \partial x_3^2}=0 \,, \\
-\frac{\partial^3 u_2}{\partial x_1 \partial x_2^2}+\frac{1}{\sqrt{2}}\left(\frac{\partial^3 u_1}{\partial x_2^3}+\frac{\partial^3 u_2}{\partial x_ \partial x_2^2}\right)+\frac{\partial^3 u_1}{\partial x_1^2 \partial x_2}-\frac{1}{\sqrt{2}}\left(\frac{\partial^3 u_1}{\partial x_1^2 \partial x_2}+\frac{\partial^3 u_2}{\partial x_1^3}\right)=0 \,, \\
-\frac{\partial^3 u_3}{\partial x_1 \partial x_3^2}+\frac{\partial^3 u_3}{\partial x_1 \partial x_2 \partial x_3}=0 \,, \\
-\frac{\partial^3 u_2}{\partial x_2 \partial x_3^2}+\frac{\partial^3 u_2}{\partial x_1 \partial x_3^2}=0 \,, \\
\frac{\partial^3 u_2}{\partial x_2^3}+\sqrt{2}\left(\frac{\partial^3 u_1}{\partial x_1 \partial x_2^2}+\frac{\partial^3 u_2}{\partial x_1^2 \partial x_2}+\frac{\partial^3 u_1}{\partial x_1^3}\right)=0 \,, \\
\frac{\partial^3 u_1}{\partial x_1 \partial x_2^2}+\frac{\partial^3 u_2}{\partial x_1^2 \partial x_2}=0 \,, \\
\frac{\partial^3 u_2}{\partial x_2 \partial x_3^2}+\frac{\partial^3 u_3}{\partial x_2^2 \partial x_3}+\frac{\partial^3 u_1}{\partial x_1 \partial x_3^2}+\frac{\partial^3 u_3}{\partial x_1^2 \partial x_3}=0 \,, \\
\frac{\partial^3 u_2}{\partial x_2 \partial x_3^2}+\frac{\partial^3 u_1}{\partial x_1 \partial x_3^2}=0 \,, \\
\frac{\partial^3 u_2}{\partial x_2^3}+2\left(\frac{\partial^3 u_1}{\partial x_1 \partial x_2^2}+\frac{\partial^3 u_2}{\partial x_1^2 \partial x_2}\right)+\frac{\partial^3 u_1}{\partial x_1^3}=0 \,, \\
-\frac{\partial^3 u_1}{\partial x_1 \partial x_2^2}-\frac{\partial^3 u_2}{\partial x_1^2 \partial x_2}=0 \,, \\
\frac{\partial^3 u_3}{\partial x_2^2 \partial x_3}+\frac{\partial^3 u_3}{\partial x_1^2 \partial x_3}=0 \,. 
\end{dcases}
\end{equation}

The fourth-order universality PDEs for this class read
\begin{equation}
\label{ClassZ_5-fourth-order-constraints-1}
\begin{dcases}
\frac{\partial^4 u_3}{\partial x_1 \partial x_3^3}=\frac{\partial^4 u_1}{\partial x_3^4}=\frac{\partial^4 u_3}{\partial x_1 \partial x_3^3}=\frac{\partial^4 u_3}{\partial x_3^4}=\frac{\partial^4 u_2}{\partial x_3^4}=\frac{\partial^4 u_3}{\partial x_2 \partial x_3^3}=\frac{\partial^4 u_1}{\partial x_3^4}=0 \,, \\
\frac{\partial^4 u_3}{\partial x_1 \partial x_2 \partial x_3^2}=\frac{\partial^4 u_2}{\partial x_1 \partial x_3^3}=0 \,.
\end{dcases}
\end{equation}
and
\begin{equation}
\label{ClassZ_5-fourth-order-constraints-2}
\begin{dcases}
\frac{1}{\sqrt{2}} \frac{\partial^4 u_1}{\partial x_2^4}+\frac{1}{\sqrt{2}}\frac{\partial^4 u_2}{\partial x_1 \partial x_2^3}+\left(1+\sqrt{2}\right)\frac{\partial^4 u_1}{\partial x_1^2 \partial x_3^2}+\frac{\partial^4 u_2}{\partial x_1^3 \partial x_2}+\frac{\partial^4 u_1}{\partial x_1^4}=0 \,, \\
\frac{\partial^4 u_1}{\partial x_1^2 \partial x_2^2}+\frac{\partial^4 u_2}{\partial x_1^3 \partial x_2}=0 \,, \qquad
-\frac{\partial^4 u_1}{\partial x_2^4}+\frac{\partial^4 u_2}{\partial x_1 \partial x_2^3}-\frac{\partial^4 u_2}{\partial x_1^2 \partial x_2^2}+\frac{\partial^4 u_2}{\partial x_1^3 \partial x_2}=0 \,, \\
\frac{\partial^4 u_2}{\partial x_1 \partial x_2 \partial x_3^2}+\frac{\partial^4 u_3}{\partial x_1 \partial x_2^2 \partial x_3}+\frac{\partial^4 u_1}{\partial x_1^2 \partial x_3^2}+\frac{\partial^4 u_3}{\partial x_1^3 \partial x_3}=0 \,, \\
\frac{\partial^4 u_1}{\partial x_2^2 \partial x_3^2}+\frac{\partial^4 u_3}{\partial x_1 \partial x_2^2 \partial x_3}+\frac{\partial^4 u_1}{\partial x_1^2 \partial x_3^2}+\frac{\partial^4 u_3}{\partial x_1^3 \partial x_3}=0 \,, \\
-\frac{1}{\sqrt{2}} \frac{\partial^4 u_1}{\partial x_2^4}+\left(1+\frac{1}{\sqrt{2}}\right) \frac{\partial^4 u_2}{\partial x_1 \partial x_2^3}-\left(-1+\sqrt{2}\right)\frac{\partial^4 u_1}{\partial x_1^2 \partial x_3^2}+\frac{\partial^4 u_2}{\partial x_1^3 \partial x_2}=0 \,, \\
\frac{\partial^4 u_1}{\partial x_2^2 \partial x_3^2}-\left(-1+\sqrt{2}\right)\frac{\partial^4 u_2}{\partial x_1 \partial x_2 \partial x_3^3}-\sqrt{2}\frac{\partial^4 u_3}{\partial x_1 \partial x_2^2 \partial x_3}=0 \,, \\
-\sqrt{2} \frac{\partial^4 u_1}{\partial x_2^2 \partial x_3^2}+\frac{\partial^4 u_3}{\partial x_1 \partial x_2^2 \partial x_3}+2\left(\frac{\partial^4 u_2}{\partial x_1 \partial x_2 \partial x_3}+\frac{\partial^4 u_3}{\partial x_1 \partial x_2^2 \partial x_3}\right)=0 \,, \\
\frac{\partial^4 u_2}{\partial x_1 \partial x_2 \partial x_3}+\frac{\partial^4 u_1}{\partial x_1^2 \partial x_3^2}=0 \,, \qquad
\frac{\partial^4 u_1}{\partial x_1^2 \partial x_3^2}+\frac{\partial^4 u_3}{\partial x_1^3 \partial x_3}=0 \,, \qquad
\frac{\partial^4 u_2}{\partial x_1 \partial x_2 \partial x_3^2}+\frac{\partial^4 u_3}{\partial x_1 \partial x_2^2 \partial x_3}=0 \,, \\
2\frac{\partial^4 u_1}{\partial x_2^2 \partial x_3^2}+\sqrt{2}\frac{\partial^4 u_2}{\partial x_1^2 \partial x_3^2}=0 \,, \\
2 \frac{\partial^4 u_2}{\partial x_2^4}-\sqrt{2}\frac{\partial^4 u_1}{\partial x_1 \partial x_2^3}-3\sqrt{2}\frac{\partial^4 u_2}{\partial x_1^2 \partial x_2^2}+\left(2-3\sqrt{2}\right)\frac{\partial^4 u_1}{\partial x_1^3 \partial x_2}-\sqrt{2}\frac{\partial^4 u_2}{\partial x_1^4}=0 \,, \\
\frac{\partial^4 u_2}{\partial x_2^2 \partial x_3^2}+\frac{\partial^4 u_3}{\partial x_2^2 \partial x_3^2}+\frac{\partial^4 u_1}{\partial x_1 \partial x_2 \partial x_3^2}+\frac{\partial^4 u_3}{\partial x_1^2 \partial x_3^2}=0 \,, \\
\sqrt{2} \frac{\partial^4 u_2}{\partial x_2^2 \partial x_3^2}+\frac{\partial^4 u_1}{\partial x_1 \partial x_2 \partial x_3^2}+\frac{\partial^4 u_2}{\partial x_1^2 \partial x_3^2}+\sqrt{2}\frac{\partial^4 u_3}{\partial x_1^2 \partial x_3^2}+2\frac{\partial^4 u_3}{\partial x_1^2 \partial x_2 \partial x_3}=0 \,, \\
\frac{\partial^4 u_2}{\partial x_2^2 \partial x_3^2}+\frac{\partial^4 u_3}{\partial x_2^3 \partial x_3}+\frac{\partial^4 u_2}{\partial x_1^2 \partial x_3^2}+\frac{\partial^4 u_3}{\partial x_1^2 \partial x_2 \partial x_3}=0 \,, \\
2 \frac{\partial^4 u_1}{\partial x_1 \partial x_2 \partial x_3^2}+\sqrt{2} \frac{\partial^4 u_2}{\partial x_1^2 \partial x_3^2}+\left(2+\sqrt{2}\right)\frac{\partial^4 u_3}{\partial x_1^2 \partial x_2 \partial x_3}=0 \,, \qquad
\frac{\partial^4 u_2}{\partial x_3^4}+\frac{\partial^4 u_3}{\partial x_2 \partial x_3^3}=0 \,, \\
-2\frac{\partial^4 u_2}{\partial x_1^2 \partial x_3^2}+2\frac{\partial^4 u_3}{\partial x_1^2 \partial x_2 \partial x_3}=0 \,, \\
-\frac{\partial^4 u_2}{\partial x_1^2 \partial x_3^2}-\frac{\partial^4 u_3}{\partial x_2^3 \partial x_3}+\frac{\partial^4 u_1}{\partial x_1 \partial x_2 \partial x_3^2}+\frac{\partial^4 u_2}{\partial x_1^2 \partial x_3^2}+2\frac{\partial^4 u_3}{\partial x_1^2 \partial x_2 \partial x_3}=0 \,, \\
\left(4+\sqrt{2}\right)\frac{\partial^4 u_2}{\partial x_2^3 \partial x_3}-\left(4+5\sqrt{2}\right)\frac{\partial^4 u_1}{\partial x_1 \partial x_2^2 \partial x_3}+\sqrt{2}\left(-9\frac{\partial^4 u_2}{\partial x_1^2 \partial x_2 \partial x_3}-4\frac{\partial^4 u_3}{\partial x_1^2 \partial x_3^2}+\frac{\partial^4 u_1}{\partial x_1^3 \partial x_3}\right)=0 \,, \\
\frac{\partial^4 u_2}{\partial x_2^3 \partial x_3}+\frac{\partial^4 u_3}{\partial x_2^4}-\left(1+6\sqrt{2}\right)\frac{\partial^4 u_1}{\partial x_1 \partial x_2^2 \partial x_3}-\left(3+4\sqrt{2}\right)\frac{\partial^4 u_2}{\partial x_1^2 \partial x_2 \partial x_3}-2\left(1+\sqrt{2}\right)\frac{\partial^4 u_3}{\partial x_1^2 \partial x_2^2} \\
\qquad +3\frac{\partial^4 u_1}{\partial x_1^3 \partial x_3}+\frac{\partial^4 u_3}{\partial x_1^4}=0 \,, \\
4 \frac{\partial^4 u_1}{\partial x_2^3 \partial x_3}+\left(4+5\sqrt{2}\right)\frac{\partial^4 u_2}{\partial x_1 \partial x_2^2 \partial x_3}+\sqrt{2}\left(\frac{\partial^4 u_3}{\partial x_1 \partial x_2^3}-4\frac{\partial^4 u_1}{\partial x_1^2 \partial x_2 \partial x_3}\right) \\
\qquad -\sqrt{2}\left(\frac{\partial^4 u_2}{\partial x_1^3 \partial x_3}+\frac{\partial^4 u_3}{\partial x_1^3 \partial x_2}\right)=0 \,, \\
2 \frac{\partial^4 u_1}{\partial x_2^3 \partial x_3}+\left(3+2\sqrt{2}\right)\frac{\partial^4 u_2}{\partial x_1 \partial x_2^2 \partial x_3}+\left(1+2\sqrt{2}\right)\frac{\partial^4 u_3}{\partial x_1 \partial x_2^3}-4\left(1+\sqrt{2}\right)\frac{\partial^4 u_1}{\partial x_1^2 \partial x_2 \partial x_3} \\
\qquad -\left(1+2\sqrt{2}\right)\left(\frac{\partial^4 u_2}{\partial x_1^3 \partial x_3}+\frac{\partial^4 u_3}{\partial x_1^3 \partial x_2}\right)=0 \,.
\end{dcases}
\end{equation}
and
\begin{equation}
\label{ClassZ_5-fourth-order-constraints-3}
\begin{dcases}
\frac{\partial^4 u_2}{\partial x_2^4}+\frac{\partial^4 u_1}{\partial x_1 \partial x_2^3}+\left(1+\sqrt{2}\right)\frac{\partial^4 u_2}{\partial x_1^2 \partial x_2^2}+\frac{1}{\sqrt{2}}\left(\frac{\partial^4 u_1}{\partial x_1^3 \partial x_2}+\frac{\partial^4 u_2}{\partial x_1^4}\right)=0 \,, \\
\frac{\partial^4 u_1}{\partial x_1 \partial x_2^3}+\frac{\partial^4 u_2}{\partial x_1^2 \partial x_2^2}=0 \,, \\
\frac{\partial^4 u_1}{\partial x_1^3 \partial x_2}-\frac{\partial^4 u_2}{\partial x_1^4}=0 \,, \\
\frac{\partial^4 u_2}{\partial x_2^2 \partial x_3^2}+\frac{\partial^4 u_3}{\partial x_2^2 \partial x_3^2}+\frac{\partial^4 u_1}{\partial x_1 \partial x_2 \partial x_3^2}+\frac{\partial^4 u_3}{\partial x_1^2 \partial x_3^2}=0 \,, \\
\frac{\partial^4 u_2}{\partial x_2^2 \partial x_3^2}+\frac{\partial^4 u_3}{\partial x_2^3 \partial x_3}+\frac{\partial^4 u_2}{\partial x_1^2 \partial x_3^2}+\frac{\partial^4 u_3}{\partial x_1^2 \partial x_2 \partial x_3}=0 \,, \\
\frac{\partial^4 u_1}{\partial x_1 \partial x_2^3}-\left(-1+\sqrt{2}\right)\frac{\partial^4 u_2}{\partial x_1^2 \partial x_2^2}+\left(1-\frac{1}{\sqrt{2}}\right)\frac{\partial^4 u_1}{\partial x_1^3 \partial x_2}-\frac{1}{\sqrt{2}}\frac{\partial^4 u_2}{\partial x_1^4}=0 \,, \\
-\left(-1+\sqrt{2}\right)\frac{\partial^4 u_1}{\partial x_1 \partial x_2 \partial x_3^2}+\frac{\partial^4 u_2}{\partial x_1^2 \partial x_3^2}-\sqrt{2}\frac{\partial^4 u_3}{\partial x_1^2 \partial x_3^2}+2\frac{\partial^4 u_3}{\partial x_1^2 \partial x_2 \partial x_3}=0 \,, \\
2\left(\frac{\partial^4 u_1}{\partial x_1 \partial x_2 \partial x_3^2}+\frac{\partial^4 u_3}{\partial x_1^2 \partial x_2 \partial x_3}\right)-\sqrt{2}\left(\frac{\partial^4 u_2}{\partial x_1^2 \partial x_3^2}+\frac{\partial^4 u_3}{\partial x_1^2 \partial x_2 \partial x_3}\right)=0 \,, \\
\frac{\partial^4 u_2}{\partial x_2^2 \partial x_3^2}+\frac{\partial^4 u_1}{\partial x_1 \partial x_2 \partial x_3^2}=0 \,, \\
\frac{\partial^4 u_3}{\partial x_2^3 \partial x_3}+\frac{\partial^4 u_3}{\partial x_1^2 \partial x_2 \partial x_3}=0 \,, \\
\sqrt{2}\frac{\partial^4 u_2}{\partial x_2^2 \partial x_3^2}=\frac{\partial^4 u_2}{\partial x_1^2 \partial x_3^2}+\frac{\partial^4 u_3}{\partial x_1^2 \partial x_2 \partial x_3}=0 \,, \\
\frac{\partial^4 u_2}{\partial x_1^2 \partial x_3^2}-\frac{\partial^4 u_3}{\partial x_1^2 \partial x_2 \partial x_3}=0 \,, \\
-\frac{1}{\sqrt{2}}\frac{\partial^4 u_1}{\partial x_2^4}+\left(1-\frac{3}{\sqrt{2}}\right)\frac{\partial^4 u_2}{\partial x_1 \partial x_2^3}-\frac{3}{\sqrt{2}}\frac{\partial^4 u_1}{\partial x_1^2 \partial x_2^2}-\frac{1}{\sqrt{2}}\frac{\partial^4 u_2}{\partial x_1^3 \partial x_2}+\frac{\partial^4 u_1}{\partial x_1^4}=0 \,, \\
\frac{\partial^4 u_1}{\partial x_2^2 \partial x_3^2}+\frac{\partial^4 u_2}{\partial x_1 \partial x_2 \partial x_3^2}+\sqrt{2}\left(\frac{\partial^4 u_3}{\partial x_1 \partial x_2^2 \partial x_3}+\frac{\partial^4 u_1}{\partial x_1^2 \partial x_3^2}\right)=0 \,, \\
\frac{\partial^4 u_1}{\partial x_2^2 \partial x_3^2}+\frac{\partial^4 u_3}{\partial x_1 \partial x_2^2 \partial x_3}+\frac{\partial^4 u_1}{\partial x_1^2 \partial x_3^2}+\frac{\partial^4 u_3}{\partial x_1^3 \partial x_3}=0 \,, \\
\sqrt{2}\frac{\partial^4 u_1}{\partial x_2^2 \partial x_3^2}+2\frac{\partial^4 u_1}{\partial x_1 \partial x_2 \partial x_3}+\left(2+\sqrt{2}\right)\frac{\partial^4 u_3}{\partial x_1 \partial x_2^2 \partial x_3}=0 \,, \\
\frac{\partial^4 u_2}{\partial x_1 \partial x_2 \partial x_3^2}+\frac{\partial^4 u_1}{\partial x_1^2 \partial x_3^2}=0 \,, \\
2\left(-1+\sqrt{2}\right)\frac{\partial^4 u_2}{\partial x_1 \partial x_2 \partial x_3^2}-2\frac{\partial^4 u_3}{\partial x_1 \partial x_2^2 \partial x_3}=0 \,, \\
2\frac{\partial^4 u_3}{\partial x_1 \partial x_2^2 \partial x_3}+\frac{\partial^4 u_1}{\partial x_1^2 \partial x_3^2}+\frac{\partial^4 u_3}{\partial x_1^3 \partial x_3}=0 \,, \\
3\frac{\partial^4 u_1}{\partial x_1^2 \partial x_2 \partial x_3}+\frac{\partial^4 u_3}{\partial x_1 \partial x_2^3}-2\frac{\partial^4 u_1}{\partial x_1^2 \partial x_2 \partial x_3}-\frac{\partial^4 u_2}{\partial x_1^3 \partial x_3}-\frac{\partial^4 u_3}{\partial x_1^3 \partial x_2}=0 \,,
\end{dcases}
\end{equation}
as well as
\begin{equation}
\label{ClassZ_5-fourth-order-constraints-4}
\begin{dcases}
2\left(1+\sqrt{2}\right) \frac{\partial^4 u_1}{\partial x_2^3 \partial x_3}+\left(5+4\sqrt{2}\right)\frac{\partial^4 u_2}{\partial x_1 \partial x_2^2 \partial x_3}+\left(1+2\sqrt{2}\right)\frac{\partial^4 u_3}{\partial x_1 \partial x_2^3}-2\left(3+2\sqrt{2}\right)\frac{\partial^4 u_1}{\partial x_1^2 \partial x_2 \partial x_3} \\
-\left(1+2\sqrt{2}\right)\left(\frac{\partial^4 u_2}{\partial x_1^3 \partial x_3}+\frac{\partial^4 u_3}{\partial x_1^3 \partial x_2}\right)=0 \,, \\
-\frac{\partial^4 u_2}{\partial x_2^3 \partial x_3}+\frac{\partial^4 u_1}{\partial x_1 \partial x_2^2 \partial x_3}+5\frac{\partial^4 u_2}{\partial x_1^2 \partial x_2 \partial x_3}+4\frac{\partial^4 u_3}{\partial x_1^2 \partial x_2^2}-\frac{\partial^4 u_1}{\partial x_1^3 \partial x_3}=0 \,, \\
-3\frac{\partial^4 u_2}{\partial x_2^3 \partial x_3}-\frac{\partial^4 u_3}{\partial x_2^4}+\left(3+8\sqrt{2}\right)\frac{\partial^4 u_1}{\partial x_1 \partial x_2^2 \partial x_3}+\left(3+6\sqrt{2}\right)\frac{\partial^4 u_2}{\partial x_1^2 \partial x_2 \partial x_3}+2\left(1+\sqrt{2}\right) \frac{\partial^4 u_3}{\partial x_2^2 \partial x_3^2} \\
-3\frac{\partial^4 u_1}{\partial x_1^3 \partial x_3}-\frac{\partial^4 u_3}{\partial x_1^4}=0 \,, 
\end{dcases}
\end{equation}
and also
\begin{equation}
\label{ClassZ_5-fourth-order-constraints-5}
\begin{dcases}
\frac{\partial^4 u_2}{\partial x_2^3 \partial x_3}+\frac{\partial^4 u_1}{\partial x_1^3 \partial x_3}=0 \,, \\
\frac{\partial^4 u_1}{\partial x_1 \partial x_2^2 \partial x_3}+\frac{\partial^4 u_2}{\partial x_1^2 \partial x_2 \partial x_3}=0 \,, \\
\frac{\partial^4 u_3}{\partial x_2 \partial x_3^3}+\frac{\partial^4 u_3}{\partial x_1^2 \partial x_3^2}=0 \,, \\
\frac{\partial^4 u_1}{\partial x_1 \partial x_3^3}+\frac{\partial^4 u_3}{\partial x_1^2 \partial x_3^2}=0 \,, \\
\frac{\partial^4 u_2}{\partial x_2^3 \partial x_3}+\frac{\partial^4 u_3}{\partial x_2^4}+\frac{\partial^4 u_1}{\partial x_1 \partial x_2^2 \partial x_3}+\frac{\partial^4 u_2}{\partial x_1^2 \partial x_2 \partial x_3}+2\frac{\partial^4 u_3}{\partial x_1^2 \partial x_2^2} \\
+\frac{\partial^4 u_1}{\partial x_1^3 \partial x_2}+\frac{\partial^4 u_3}{\partial x_1^4}=0 \,, \\
\frac{\partial^4 u_2}{\partial x_2^3 \partial x_3}+\frac{\partial^4 u_1}{\partial x_1 \partial x_2^2 \partial x_3}+\frac{\partial^4 u_2}{\partial x_1^2 \partial x_2 \partial x_3}+\frac{\partial^4 u_1}{\partial x_1^3 \partial x_3}=0 \,, \\
\frac{\partial^4 u_1}{\partial x_1 \partial x_2^2 \partial x_3}+\frac{\partial^4 u_2}{\partial x_1^2 \partial x_2 \partial x_3}=0 \,, \\
2\left(\frac{\partial^4 u_2}{\partial x_2^3 \partial x_3}+\frac{\partial^4 u_3}{\partial x_1^2 \partial x_2^2}\right)+\sqrt{2}\left(\frac{\partial^4 u_1}{\partial x_1^3 \partial x_3}+\frac{\partial^4 u_3}{\partial x_1^4}\right)=0 \,, \\
2\frac{\partial^4 u_1}{\partial x_1 \partial x_2^2 \partial x_3}+\frac{\partial^4 u_2}{\partial x_1^2 \partial x_2 \partial x_3}+\frac{\partial^4 u_3}{\partial x_1^2 \partial x_2^2}=0 \,, \\
\frac{\partial^4 u_2}{\partial x_1 \partial x_2^2 \partial x_3}+\frac{\partial^4 u_1}{\partial x_1^2 \partial x_2 \partial x_3}=0 \,, \\
\frac{\partial^4 u_1}{\partial x_2^3 \partial x_3}+\left(1+\sqrt{2}\right)\frac{\partial^4 u_2}{\partial x_1 \partial x_2^2 \partial x_3}+\left(1+\sqrt{2}\right)\frac{\partial^4 u_1}{\partial x_1^2 \partial x_2 \partial x_3}+\frac{\partial^4 u_2}{\partial x_1^3 \partial x_3}=0 \,, \\
-3\frac{\partial^4 u_2}{\partial x_1 \partial x_2^2 \partial x_3}+\frac{\partial^4 u_3}{\partial x_1 \partial x_2^3}-2\frac{\partial^4 u_1}{\partial x_1^2 \partial x_2 \partial x_3}-\frac{\partial^4 u_2}{\partial x_1^3 \partial x_3}+\frac{\partial^4 u_3}{\partial x_1^3 \partial x_2}=0 \,, \\
\frac{\partial^4 u_1}{\partial x_2^3 \partial x_3}+\frac{\partial^4 u_3}{\partial x_1 \partial x_2^3}-3\frac{\partial^4 u_1}{\partial x_1^2 \partial x_2 \partial x_3}-\frac{\partial^4 u_2}{\partial x_1^3 \partial x_3}=0 \,,
\end{dcases}
\end{equation}
and finally
\begin{equation}
\label{ClassZ_5-fourth-order-constraints-6}
\begin{dcases}
\frac{\partial^4 u_2}{\partial x_1 \partial x_2^2 \partial x_3}+\frac{\partial^4 u_3}{\partial x_1 \partial x_2^3}+\frac{\partial^4 u_1}{\partial x_1^2 \partial x_2 \partial x_3}+\frac{\partial^4 u_3}{\partial x_1^3 \partial x_2}=0 \,, \\
2\frac{\partial^4 u_1}{\partial x_2^4}+\left(2+\sqrt{2}\right)\frac{\partial^4 u_2}{\partial x_1 \partial x_2^3}-\left(2+\sqrt{2}\right)\frac{\partial^4 u_1}{\partial x_1^2 \partial x_2^2}-\left(2+\sqrt{2}\right)\frac{\partial^4 u_2}{\partial x_1^3 \partial x_2}+\sqrt{2}\frac{\partial^4 u_1}{\partial x_1^4}=0 \,, \\
2 \frac{\partial^4 u_2}{\partial x_1 \partial x_2^3}-2\left(1+\sqrt{2}\right)\frac{\partial^4 u_1}{\partial x_1^2 \partial x_2^2}-2\sqrt{2}\frac{\partial^4 u_2}{\partial x_1^3 \partial x_2}=0 \,, \\
-\frac{\partial^4 u_2}{\partial x_2^4}+\left(1+\sqrt{2}\right)\frac{\partial^4 u_1}{\partial x_1 \partial x_2^3}+\left(1+\sqrt{2}\right)\frac{\partial^4 u_2}{\partial x_2^2 \partial x_3^2}-\left(1+\sqrt{2}\right)\frac{\partial^4 u_1}{\partial x_1^3 \partial x_2}-\sqrt{2}\frac{\partial^4 u_2}{\partial x_1^4}=0 \,, \\
\sqrt{2}\frac{\partial^4 u_1}{\partial x_1 \partial x_2^3}+\left(1+\sqrt{2}\right)\frac{\partial^4 u_2}{\partial x_1^2 \partial x_2^2}-\frac{\partial^4 u_1}{\partial x_1^3 \partial x_2}=0 \,.
\end{dcases}
\end{equation}

From these equations, after lengthy but straightforward manipulations, we obtain the following results for the functions $h_i$, $i=1,2$:
\begin{equation}
\label{classZ_5-h_i}
\frac{\partial^2 h_i}{\partial x_1^2} = \frac{\partial^2 h_i}{\partial x_2^2} = \frac{\partial^2 h_i}{\partial x_1 \partial x_2}= 0\,,
\end{equation}
which imply that
\begin{align}
\label{classZ_5-mh1}
h_1(x_1, x_2) &= c_1 x_1 + c_2 x_2\,, \\
\label{classZ_5-mh2}
h_2(x_1, x_2) &= c_3 x_1 + c_4 x_2\,.
\end{align}
For the function $\hat{u}_{33}$ we find:
\begin{equation}
\label{classZ_5-5}
\frac{\partial^3 \hat{u}_{33}}{\partial x_1^3}  =\frac{\partial^3 \hat{u}_{33}}{\partial x_2^3} = \frac{\partial^3 \hat{u}_{33}}{\partial x_1^2 \partial x_2} = \frac{\partial^3 \hat{u}_{33}}{\partial x_1 \partial x_2^2} = \frac{\partial^4 \hat{u}_{33}}{\partial x_1^2 \partial x_2^2} = \frac{\partial^4 \hat{u}_{33}}{\partial x_1 \partial x_2^3} = \frac{\partial^4 \hat{u}_{33}}{\partial x_1^3 \partial x_2}= 0\,,
\end{equation}
which implies that (cf. \eqref{Z^-_4-g-final-form})
\begin{equation}\label{classZ_5-u_{33}}
	\hat{u}_{33} (x_1, x_2)= c_5+c_6 x_1 +c_7 x_2 +c_8 x_1 x_2 +c_9(x_2^2-x_1^2) \,.
\end{equation}
For the functions $k_i$ there then remains
\begin{equation}
\begin{aligned}
\label{ClassZ_5-k_i-constraints-1}
& \frac{1}{\sqrt{2}} \frac{\partial^4 k_1}{\partial x_2^4}+\frac{1}{\sqrt{2}}\frac{\partial^4 k_2}{\partial x_1 \partial x_2^3}+\left(1+\sqrt{2}\right)\frac{\partial^4 k_1}{\partial x_1^2 \partial x_2^2} +\frac{\partial^4 k_2}{\partial x_1^3 \partial x_2}+\frac{\partial^4 k_1}{\partial x_1^4} =0 \,, \\
& -\frac{\partial^4 k_1}{\partial x_1^2 \partial x_2^2}-\frac{\partial^4 k_2}{\partial x_1^3 \partial x_2} =0 \,, \\
& -\frac{\partial^4 k_1}{\partial x_2^4}-\frac{\partial^4 k_2}{\partial x_1 \partial x_2^3}-\frac{\partial^4 k_1}{\partial x_1^2 \partial x_2^2}+\frac{\partial^4 k_2}{\partial x_1^3 \partial x_2}=0 \,, \\
& -\frac{1}{\sqrt{2}} \frac{\partial^4 k_1}{\partial x_2^4}+\left(1-\frac{1}{\sqrt{2}}\right) \frac{\partial^4 k_2}{\partial x_1 \partial x_2^3}+\left(-1+\sqrt{2}\right)\frac{\partial^4 k_1}{\partial x_1^2 \partial x_2^2}+\left(-1+\sqrt{2}\right) \frac{\partial^4 k_2}{\partial x_1^3 \partial x_2}=0 \,, \\
& 2 \frac{\partial^4 k_2}{\partial x_2^4}-\sqrt{2} \frac{\partial^4 k_1}{\partial x_1 \partial x_2^3}-3\sqrt{2} \frac{\partial^4 k_2}{\partial x_1^2 \partial x_2^2}+\left(2-3\sqrt{2}\right)\frac{\partial^4 k_1}{\partial x_1^3 \partial x_2}-\sqrt{2}\frac{\partial^4 k_2}{\partial x_1^4} =0 \,, \\ 
\end{aligned}
\end{equation}
and 
\begin{equation}
\begin{aligned}
\label{ClassZ_5-k_i-constraints-2}
& \frac{\partial^4 k_2}{\partial x_2^4}+ \frac{\partial^4 k_1}{\partial x_1 \partial x_2^3}+\left(1+\sqrt{2}\right)\frac{\partial^4 k_2}{\partial x_1^2 \partial x_2^2}+\frac{1}{\sqrt{2}}\frac{\partial^4 k_1}{\partial x_1^3 \partial x_2}+\frac{1}{\sqrt{2}}\frac{\partial^4 k_2}{\partial x_1^4} =0 \,, \\
& -\frac{\partial^4 k_1}{\partial x_1 \partial x_2^3}-\frac{\partial^4 k_2}{\partial x_1^2 \partial x_2^2} =0 \,, \\
& \frac{\partial^4 k_1}{\partial x_1 \partial x_2^3}-\frac{\partial^4 k_2}{\partial x_1^2 \partial x_2^2}+\frac{\partial^4 k_1}{\partial x_1^3 \partial x_2}-\frac{\partial^4 k_2}{\partial x_1^4}=0 \,, \\
& \frac{\partial^4 k_1}{\partial x_1 \partial x_2^3}-\left(1-\frac{1}{\sqrt{2}}\right) \frac{\partial^4 k_2}{\partial x_1^2 \partial x_2^2}+\left(1-\frac{1}{\sqrt{2}}\right)\frac{\partial^4 k_1}{\partial x_1^3 \partial x_2}+\frac{1}{\sqrt{2}} \frac{\partial^4 k_2}{\partial x_1^4}=0 \,, \\
& \frac{1}{\sqrt{2}} \frac{\partial^4 k_1}{\partial x_2^4}+\left(1-\frac{3}{\sqrt{2}}\right) \frac{\partial^4 k_2}{\partial x_1 \partial x_2^3}-3\sqrt{2} \frac{\partial^4 k_1}{\partial x_1^2 \partial x_2^2}-\frac{1}{\sqrt{2}}\frac{\partial^4 k_2}{\partial x_1^3 \partial x_2}+\frac{\partial^4 k_1}{\partial x_1^4} =0 \,, 
\end{aligned}
\end{equation}
as well as
\begin{equation}
\begin{aligned}
\label{ClassZ_5-k_i-constraints-3}
& 2\frac{\partial^4 k_1}{\partial x_2^4}+\left(2+\sqrt{2}\right) \frac{\partial^4 k_2}{\partial x_1 \partial x_2^3}-\left(2+\sqrt{2}\right)\frac{\partial^4 k_1}{\partial x_1^2 \partial x_2^2}-\left(2+\sqrt{2}\right) \frac{\partial^4 k_2}{\partial x_1^3 \partial x_2}+\sqrt{2} \frac{\partial^4 k_1}{\partial x_1^4} =0\,, \\
& 2 \frac{\partial^4 k_2}{\partial x_1 \partial x_2^3}-2\left(1+\sqrt{2}\right)\frac{\partial^4 k_1}{\partial x_1^2 \partial x_2^2} -2\sqrt{2} \frac{\partial^4 k_2}{\partial x_1^3 \partial x_2}=0 \,, \\
& -\frac{\partial^4 k_2}{\partial x_2^4}+\left(1+\sqrt{2}\right) \frac{\partial^4 k_1}{\partial x_1 \partial x_2^3}+\left(1+\sqrt{2}\right)\frac{\partial^4 k_2}{\partial x_1^2 \partial x_2^2}-\left(1+\sqrt{2}\right) \frac{\partial^4 k_1}{\partial x_1^3 \partial x_2}-\sqrt{2} \frac{\partial^4 k_2}{\partial x_1^4} =0\,, \\
&  \frac{\partial^4 k_1}{\partial x_1 \partial x_2^3}+2\left(1+\sqrt{2}\right)\frac{\partial^4 k_2}{\partial x_1^2 \partial x_2^2} - \frac{\partial^4 k_1}{\partial x_1^3 \partial x_2}=0 \,. 
\end{aligned}
\end{equation}
Thus, we obtain the following result.
\begin{prop}
The universal displacements in the pentagonal $\mathbb{Z}_5$ class linear strain-gradient elastic solids are of the form \eqref{transverse-isotropy-linear-solution} and \eqref{transverse-isotropy-linear-solution-2}, with \eqref{classZ_5-mh1} and \eqref{classZ_5-mh2} for the functions $h_i$, $i=1,2$, \eqref{classZ_5-u_{33}} for the function $\hat{u}_{33}$, and the constraints \eqref{ClassZ_5-k_i-constraints-1}--\eqref{ClassZ_5-k_i-constraints-3} for the functions $k_i$.
\end{prop}

\subsubsection{Pentagonal class $\mathbb{D}_5$} 

Tensor $\boldsymbol{\mathsf{A}}_{\mathbb{D}_5 \oplus \mathbb{Z}_2^c}$ has $23$ independent components and has the following represenation
\begin{equation}
\label{A_D_5}
    \boldsymbol{\mathsf{A}}_{\mathbb{D}_5 \oplus \mathbb{Z}_2^c}(\mathbf{x}) =
    \begin{bmatrix}
    A^{(11)} & 0 & 0 & 0  \\
     & A^{(11)} & F^{(2)} & 0   \\
     &  & H^{(6)} & 0   \\
     &  &  & J^{(4)}      
    \end{bmatrix}_S
    +
    \begin{bmatrix}
    0 & 0 & 0 & f(F^{(2)})  \\
     & 0 & 0 & 0   \\
     &  & h(J^{(4)}) & 0   \\
     &  &  & 0      
    \end{bmatrix}_S\,,
\end{equation}
where all relevant matrices are defined in the preceding classes. Tensor $\boldsymbol{\mathsf{M}}_{\mathbb{D}_5}$ has $9$ independent components and is of the form \eqref{M_D_5}. Tensor $\boldsymbol{\mathsf{C}}_{\mathbb{O}(2) \oplus \mathbb{Z}_2^c}$ corresponds to the Curie group with Hermann--Mauguin symbol $\infty2$ and has $5$ independent components. This class has the same universality PDEs as those of class $\mathbb{O}(2)$, with one additional constant in the fifth-order tensor and two additional constants in the sixth-order tensor. The additional universality PDEs are
\begin{equation}
\label{ClassD_5-constraints}
\begin{dcases}
 \left(-1+\sqrt{2}\right)\left(\frac{\partial^3 k_2}{\partial x_2^3}+x_3\frac{\partial^3 h_2}{\partial x_2^3}\right)-\left(1+\sqrt{2}\right)\frac{\partial^3 k_1}{\partial x_1 \partial x_2^2}-\left(2+\sqrt{2}\right)\frac{\partial^3 k_2}{\partial x_1^2 \partial x_2} =0\,, \\
 4\frac{\partial^3 h_1}{\partial x_2^3}+\left(4+5\sqrt{2}\right)\frac{\partial^3 h_2}{\partial x_1 \partial x_2^2}+\sqrt{2}\left(\frac{\partial^4 \hat{u}_{33}}{\partial x_1 \partial x_2^3}-4\frac{\partial^3 h_1}{\partial x_1^2 \partial x_2}-\frac{\partial^3 h_2}{\partial x_1^3}-\frac{\partial^4 \hat{u}_{33}}{\partial x_1^3 \partial x_2}\right)=0\,, \\
 2 \frac{\partial^3 h_1}{\partial x_2^3} +\left(3+2\sqrt{2}\right) \frac{\partial^3 h_2}{\partial x_1 \partial x_2^2}+\left(1+2\sqrt{2}\right)\frac{\partial^4 \hat{u}_{33}}{\partial x_1 \partial x_2^3}-4\left(1+\sqrt{2}\right)\frac{\partial^3 h_1}{\partial x_1^2 \partial x_2} \\
 \qquad -\left(1+2\sqrt{2}\right)\left(\frac{\partial^3 h_2}{\partial x_1^3}+\frac{\partial^4 \hat{u}_{33}}{\partial x_1^3 \partial x_2}\right)=0\,, \\
 -\left(-1+\sqrt{2}\right)\left(\frac{\partial^3 k_2}{\partial x_1 \partial x_2^2}+x_3\frac{\partial^3 h_2}{\partial x_1 \partial x_2^2}\right) +\left(1+2\sqrt{2}\right)\left(\frac{\partial^3 k_1}{\partial x_1^2 \partial x_2}+x_3 \frac{\partial^3 h_1}{\partial x_1^2 \partial x_2}\right)\\
 \qquad+\left(2+2\sqrt{2}\right)\left(\frac{\partial^3 k_2}{\partial x_1^3}+x_3\frac{\partial^3 h_2}{\partial x_1^3}\right)=0\,, \\
 -\frac{\partial^4 \hat{u}_{33}}{\partial x_2^4}+2\left(1+\sqrt{2}\right)\frac{\partial^4 \hat{u}_{33}}{\partial x_1^2 \partial x_2^2}-\frac{\partial^4 \hat{u}_{33}}{\partial x_1^4}=0\,, \\
 -\frac{\partial^4 k_2}{\partial x_2^4}-x_3 \frac{\partial^4 h_2}{\partial x_2^4}+\left(1+\sqrt{2}\right)\frac{\partial^4 k_1}{\partial x_1 \partial x_2^3}+\left(1+\sqrt{2}\right)\frac{\partial^4 k_2}{\partial x_1^2 \partial x_2^2} \\
 \qquad +\left(1+\sqrt{2}\right)\left(\frac{\partial^4 k_1}{\partial x_1^3 \partial x_2}+x_3 \frac{\partial^4 h_1}{\partial x_1^3 \partial x_2}\right)-\sqrt{2}\left(\frac{\partial^4 k_2}{\partial x_1^4}+x_3 \frac{\partial^4 h_2}{\partial x_1^4}\right)=0\,, \\
 \sqrt{2} \frac{\partial^4 k_1}{\partial x_1 \partial x_2^3}+\left(1+\sqrt{2}\right)\frac{\partial^4 k_2}{\partial x_1^2 \partial x_2^2}-\frac{\partial^4 k_1}{\partial x_1^3 \partial x_2}-x_3\frac{\partial^4 h_1}{\partial x_1^3 \partial x_2}=0\,.
\end{dcases}
\end{equation}
Combining these relations with those of the $\mathbb{O}(2)$ class, we obtain the following result.
\begin{prop}
The universal displacements in pentagonal $\mathbb{D}_5$ class linear strain-gradient elastic solids coincide with those of class $\mathbb{O}(2)$, supplemented by the additional constraints \eqref{ClassD_5-constraints}\,.
\end{prop}

\subsubsection{Pentagonal class $\mathbb{Z}_5 \oplus \mathbb{Z}_2^c$}

Tensor $\boldsymbol{\mathsf{A}}_{\mathbb{Z}_5 \oplus \mathbb{Z}_2^c}$ has $35$ independent components and is of the same form as that of class $\mathbb{Z}_5$.
Tensor $\boldsymbol{\mathsf{M}}_{\mathbb{Z}_5 \oplus \mathbb{Z}_2^c}$ is a null tensor.
Tensor $\boldsymbol{\mathsf{C}}_{\mathbb{O}(2) \oplus \mathbb{Z}_2^c}$ has $5$ independent components and corresponds to the Curie group with Hermann--Mauguin symbol $\infty/mm$.

For this class we obtain the following fourth-order universality PDEs
\begin{equation}
\label{ClassZ_5+Z^c_2-fourth-order-constraints-1}
\begin{dcases}
\frac{1}{\sqrt{2}} \left(\frac{\partial^4 k_1}{\partial x_2^4}+x_3 \frac{\partial^4 h_1}{\partial x_2^4}\right)+\frac{1}{\sqrt{2}} \left(\frac{\partial^4 k_2}{\partial x_1 \partial x_2^3}+x_3 \frac{\partial^4 h_2}{\partial x_1 \partial x_2^3}\right)+\left(1+\sqrt{2}\right)\left(\frac{\partial^4 k_1}{\partial x_1^2 \partial x_2^2}+x_3 \frac{\partial^4 h_1}{\partial x_1^2 \partial x_2^2}\right) \\
\qquad +\left(\frac{\partial^4 k_2}{\partial x_1^3 \partial x_2}+x_3 \frac{\partial^4 h_2}{\partial x_1^3 \partial x_2}\right)+\left(\frac{\partial^4 k_1}{\partial x_1^4}+x_3 \frac{\partial^4 h_1}{\partial x_1^4}\right) =0 \,, \\
-\left(\frac{\partial^4 k_1}{\partial x_1^2 \partial x_2^2}+x_3 \frac{\partial^4 h_1}{\partial x_1^2 \partial x_2^2}\right)-\left(\frac{\partial^4 k_2}{\partial x_1^3 \partial x_2}+x_3 \frac{\partial^4 h_2}{\partial x_1^3 \partial x_2}\right) =0 \,, \\
-\left(\frac{\partial^4 k_1}{\partial x_2^4}+x_3 \frac{\partial^4 h_1}{\partial x_2^4}\right)-\left(\frac{\partial^4 k_2}{\partial x_1 \partial x_2^3}+x_3 \frac{\partial^4 h_2}{\partial x_1 \partial x_2^3}\right)-\left(\frac{\partial^4 k_1}{\partial x_1^2 \partial x_2^2}+x_3 \frac{\partial^4 h_1}{\partial x_1^2 \partial x_2^2}\right) \\
\qquad +\left(\frac{\partial^4 k_2}{\partial x_1^3 \partial x_2}+x_3 \frac{\partial^4 h_2}{\partial x_1^3 \partial x_2}\right)=0 \,, \\
-\frac{1}{\sqrt{2}} \left(\frac{\partial^4 k_1}{\partial x_2^4}+x_3 \frac{\partial^4 h_1}{\partial x_2^4}\right)+\left(1-\frac{1}{\sqrt{2}}\right) \left(\frac{\partial^4 k_2}{\partial x_1 \partial x_2^3}+x_3 \frac{\partial^4 h_2}{\partial x_1 \partial x_2^3}\right)+\left(-1+\sqrt{2}\right)\left(\frac{\partial^4 k_1}{\partial x_1^2 \partial x_2^2}+x_3 \frac{\partial^4 h_1}{\partial x_1^2 \partial x_2^2}\right) \\
\qquad +\left(-1+\sqrt{2}\right) \left(\frac{\partial^4 k_2}{\partial x_1^3 \partial x_2}+x_3 \frac{\partial^4 h_2}{\partial x_1^3 \partial x_2}\right)=0 \,, \\
2 \left(\frac{\partial^4 k_2}{\partial x_2^4}+x_3 \frac{\partial^4 h_2}{\partial x_2^4}\right)-\sqrt{2} \left(\frac{\partial^4 k_1}{\partial x_1 \partial x_2^3}+x_3 \frac{\partial^4 h_1}{\partial x_1 \partial x_2^3}\right)-3\sqrt{2} \left(\frac{\partial^4 k_2}{\partial x_1^2 \partial x_2^2}+x_3 \frac{\partial^4 h_2}{\partial x_1^2 \partial x_2^2}\right) \\
\qquad +\left(2-3\sqrt{2}\right)\left(\frac{\partial^4 k_1}{\partial x_1^3 \partial x_2}+x_3 \frac{\partial^4 h_1}{\partial x_1^3 \partial x_2}\right)-\sqrt{2}\left(\frac{\partial^4 k_2}{\partial x_1^4}+x_3 \frac{\partial^4 h_2}{\partial x_1^4}\right) =0 \,, \\
\left(4+\sqrt{2}\right)\frac{\partial^3 h_2}{\partial x_2^3}-\left(4+5\sqrt{2}\right)\frac{\partial^3 h_1}{\partial x_1 \partial x_2^2}+\sqrt{2}\left(-9\frac{\partial^3 h_2}{\partial x_1^2 \partial x_2}-4\frac{\partial^4 u_{33}}{\partial x_1^2 \partial x_2^2}+\frac{\partial^3 h_1}{\partial x_1^3}\right) =0 \,, \\
\frac{\partial^3 h_2}{\partial x_2^3}+\frac{\partial^4 \hat{u}_{33}}{\partial x_2^4}-\left(1+6\sqrt{2}\right)\frac{\partial^3 h_1}{\partial x_1 \partial x_2^2}-\left(3+4\sqrt{2}\right)\frac{\partial^3 h_2}{\partial x_1^2 \partial x_2}- 2\left(1+\sqrt{2}\right)\frac{\partial^4 \hat{u}_{33}}{\partial x_1^2 \partial x_2^2} \\
\qquad +3 \frac{\partial^3 h_1}{\partial x_1^3}+\frac{\partial^4 u_{33}}{\partial x_1^4}=0 \,, \\
4 \frac{\partial^3 h_1}{\partial x_2^3}+\left(4+5\sqrt{2}\right)\frac{\partial^3 h_2}{\partial x_1 \partial x_2^2}+\sqrt{2}\left(\frac{\partial^4 \hat{u}_{33}}{\partial x_1 \partial x_2^3}-4\frac{\partial^3 h_1}{\partial x_1^2 \partial x_2}-\frac{\partial^4 \hat{u}_{33}}{\partial x_1^3 \partial x_2}-\frac{\partial^3 h_2}{\partial x_1^3}\right) =0 \,, \\
2 \frac{\partial^3 h_1}{\partial x_2^3}+\left(3+2\sqrt{2}\right)\frac{\partial^3 h_2}{\partial x_1 \partial x_2^2}+\left(1+2\sqrt{2}\right) \frac{\partial^4 \hat{u}_{33}}{\partial x_1 \partial x_2^3}-4\left(1+\sqrt{2}\right)\frac{\partial^3 h_1}{\partial x_1^2 \partial x_2} \\
\qquad -\left(1+2\sqrt{2}\right)\left(\frac{\partial^4 \hat{u}_{33}}{\partial x_1^3 \partial x_2}+\frac{\partial^3 h_2}{\partial x_1^3}\right) =0 \,,
\end{dcases}
\end{equation}
and 
\begin{equation}
\label{ClassZ_5+Z^c_2-fourth-order-constraints-2}
\begin{dcases}
\left(\frac{\partial^4 k_2}{\partial x_2^4}+x_3 \frac{\partial^4 h_2}{\partial x_2^4}\right)+ \left(\frac{\partial^4 k_1}{\partial x_1 \partial x_2^3}+x_3 \frac{\partial^4 h_1}{\partial x_1 \partial x_2^3}\right)+\left(1+\sqrt{2}\right)\left(\frac{\partial^4 k_2}{\partial x_1^2 \partial x_2^2}+x_3 \frac{\partial^4 h_2}{\partial x_1^2 \partial x_2^2}\right) \\
\qquad +\frac{1}{\sqrt{2}}\left(\frac{\partial^4 k_1}{\partial x_1^3 \partial x_2}+x_3 \frac{\partial^4 h_1}{\partial x_1^3 \partial x_2}\right)+\frac{1}{\sqrt{2}}\left(\frac{\partial^4 k_2}{\partial x_1^4}+x_3 \frac{\partial^4 h_2}{\partial x_1^4}\right) =0 \,, \\
-\left(\frac{\partial^4 k_1}{\partial x_1 \partial x_2^3}+x_3 \frac{\partial^4 h_1}{\partial x_1 \partial x_2^3}\right)-\left(\frac{\partial^4 k_2}{\partial x_1^2 \partial x_2^2}+x_3 \frac{\partial^4 h_2}{\partial x_1^2 \partial x_2^2}\right) =0 \,, \\
\left(\frac{\partial^4 k_1}{\partial x_1 \partial x_2^3}+x_3 \frac{\partial^4 h_1}{\partial x_1 \partial x_2^3}\right)-\left(\frac{\partial^4 k_2}{\partial x_1^2 \partial x_2^2}+x_3 \frac{\partial^4 h_2}{\partial x_1^2 \partial x_2^2}\right)+\left(\frac{\partial^4 k_1}{\partial x_1^3 \partial x_2}+x_3 \frac{\partial^4 h_1}{\partial x_1^3 \partial x_2}\right) \\
\qquad -\left(\frac{\partial^4 k_2}{\partial x_1^4}+x_3 \frac{\partial^4 h_2}{\partial x_1^4}\right)=0 \,, \\
\left(\frac{\partial^4 k_1}{\partial x_1 \partial x_2^3}+x_3 \frac{\partial^4 h_1}{\partial x_1 \partial x_2^3}\right)-\left(1-\frac{1}{\sqrt{2}}\right) \left(\frac{\partial^4 k_2}{\partial x_1^2 \partial x_2^2}+x_3 \frac{\partial^4 h_2}{\partial x_1^2 \partial x_2^2}\right)+\left(1-\frac{1}{\sqrt{2}}\right)\left(\frac{\partial^4 k_1}{\partial x_1^3 \partial x_2}+x_3 \frac{\partial^4 h_1}{\partial x_1^3 \partial x_2}\right) \\
\qquad +\frac{1}{\sqrt{2}} \left(\frac{\partial^4 k_2}{\partial x_1^4}+x_3 \frac{\partial^4 h_2}{\partial x_1^4}\right)=0 \,, \\
\frac{1}{\sqrt{2}} \left(\frac{\partial^4 k_1}{\partial x_2^4}+x_3 \frac{\partial^4 h_1}{\partial x_2^4}\right)+\left(1-\frac{3}{\sqrt{2}}\right) \left(\frac{\partial^4 k_2}{\partial x_1 \partial x_2^3}+x_3 \frac{\partial^4 h_2}{\partial x_1 \partial x_2^3}\right)-3\sqrt{2} \left(\frac{\partial^4 k_1}{\partial x_1^2 \partial x_2^2}+x_3 \frac{\partial^4 h_1}{\partial x_1^2 \partial x_2^2}\right) \\
\qquad -\frac{1}{\sqrt{2}}\left(\frac{\partial^4 k_2}{\partial x_1^3 \partial x_2}+x_3 \frac{\partial^4 h_2}{\partial x_1^3 \partial x_2}\right)+\left(\frac{\partial^4 k_1}{\partial x_1^4}+x_3 \frac{\partial^4 h_1}{\partial x_1^4}\right) =0 \,, \\
3\frac{\partial^3 h_2}{\partial x_1 \partial x_2^2}+\frac{\partial^4 \hat{u}_{33}}{\partial x_1 \partial x_2^3}-2\frac{\partial^3 h_1}{\partial x_1^2 \partial x_2}-\frac{\partial^3 h_2}{\partial x_1^3}-\frac{\partial^4 \hat{u}_{33}}{\partial x_1^3 \partial x_2} =0 \,, \\
2\left(1+\sqrt{2}\right) \frac{\partial^3 h_1}{\partial x_2^3}+\left(1+2\sqrt{2}\right) \frac{\partial^4 \hat{u}_{33}}{\partial x_1 \partial x_2^3}+\left(5+4\sqrt{2}\right)\frac{\partial^3 h_2}{\partial x_1 \partial x_2^2}-2\left(3+2\sqrt{2}\right)\frac{\partial^3 h_1}{\partial x_1^2 \partial x_2} \\
\qquad -\left(1+\sqrt{2}\right)\left(\frac{\partial^3 h_2}{\partial x_1^3}+\frac{\partial^4 \hat{u}_{33}}{\partial x_1^3 \partial x_2^1}\right) =0 \,, \\
- \frac{\partial^3 h_2}{\partial x_2^3}+\frac{\partial^3 h_1}{\partial x_1 \partial x_2^2}+4 \left(\frac{\partial^4 \hat{u}_{33}}{\partial x_1^2 \partial x_2^2}+5\frac{\partial^3 h_2}{\partial x_1^2 \partial x_2}-\frac{\partial^3 h_1}{\partial x_1^3}\right) =0 \,, \\
-3 \frac{\partial^3 h_2}{\partial x_2^3}-\frac{\partial^4 \hat{u}_{33}}{\partial x_2^4}+\left(3+8\sqrt{2}\right)\frac{\partial^3 h_1}{\partial x_1 \partial x_2^2}+2\left(1+\sqrt{2}\right) \frac{\partial^4 \hat{u}_{33}}{\partial x_1^2 \partial x_2^2} \\
\qquad +\left(3+6\sqrt{2}\right)\frac{\partial^3 h_2}{\partial x_1^2 \partial x_2}-\left(\frac{\partial^4 \hat{u}_{33}}{\partial x_1^4}-3\frac{\partial^3 h_1}{\partial x_1^3}\right) =0\,,
\end{dcases}
\end{equation}
as well as
\begin{equation}
\label{ClassZ_5+Z^c_2-fourth-order-constraints-3}
\begin{dcases}
\frac{\partial^3 h_2}{\partial x_2^3}+\frac{\partial^3 h_2}{\partial x_1^3}=0 \,, \\
\frac{\partial^3 h_1}{\partial x_1 \partial x_2^2}+\frac{\partial^3 h_2}{\partial x_1^2 \partial x_2}=0 \,, \\
\frac{\partial^3 h_2}{\partial x_2^3}+\frac{\partial^3 h_1}{\partial x_1^3}=0 \,, \\
\sqrt{2}\left(\frac{\partial^3 h_2}{\partial x_2^3}+\frac{\partial^4 \hat{u}_{33}}{\partial x_2^4}\right)+2\left(\frac{\partial^3 h_2}{\partial x_1 \partial x_2^2}+\frac{\partial^4 \hat{u}_{33}}{\partial x_1^2 \partial x_2^2}\right)+\sqrt{2}\left(\frac{\partial^3 h_1}{\partial x_1^3}+\frac{\partial^4 \hat{u}_{33}}{\partial x_1^4}\right)=0 \,, \\
2\frac{\partial^3 h_1}{\partial x_1 \partial x_2^2}+\frac{\partial^3 h_2}{\partial x_1^2 \partial x_2}+\frac{\partial^4 \hat{u}_{33}}{\partial x_1^2 \partial x_2^2}=0 \,, \\
\frac{\partial^3 h_2}{\partial x_1 \partial x_2^2}+\frac{\partial^3 h_1}{\partial x_1^2 \partial x_2}=0 \,, \\
\frac{\partial^3 h_1}{\partial x_2^3}+\frac{\partial^3 h_2}{\partial x_1^3}=0 \,, \\
-3\frac{\partial^3 h_2}{\partial x_1 \partial x_2^2}-2\frac{\partial^3 h_1}{\partial x_1^2 \partial x_2}-\frac{\partial^3 h_2}{\partial x_1^3}=0 \,, \\
\frac{\partial^3 h_1}{\partial x_2^3} + \frac{\partial^4 u_{33}}{\partial x_1 \partial x_2^3} -3\frac{\partial^3 h_1}{\partial x_1^2 \partial x_2}-\frac{\partial^3 h_2}{\partial x_1^3}=0 \,, \\
\frac{\partial^4 \hat{u}_{33}}{\partial x_1 \partial x_2^3}+\frac{\partial^4 \hat{u}_{33}}{\partial x_1^3 \partial x_2}=0 \,, \\
2\left(\frac{\partial^4 k_1}{\partial x_2^4}+x_3 \frac{\partial^4 h_1}{\partial x_2^4}\right)+\left(2+\sqrt{2}\right) \left(\frac{\partial^4 k_2}{\partial x_1 \partial x_2^3}+x_3 \frac{\partial^4 h_2}{\partial x_1 \partial x_2^3}\right)-\left(2+\sqrt{2}\right)\left(\frac{\partial^4 k_1}{\partial x_1^2 \partial x_2^2}+x_3 \frac{\partial^4 h_1}{\partial x_1^2 \partial x_2^2}\right) \\
-\left(2+\sqrt{2}\right) \left(\frac{\partial^4 k_2}{\partial x_1^3 \partial x_2}+x_3 \frac{\partial^4 h_2}{\partial x_1^3 \partial x_2}\right)+\sqrt{2} \left(\frac{\partial^4 k_1}{\partial x_1^4}+x_3 \frac{\partial^4 h_1}{\partial x_1^4}\right) =0\,, \\
2 \left(\frac{\partial^4 k_2}{\partial x_1 \partial x_2^3}+x_3 \frac{\partial^4 h_2}{\partial x_1 \partial x_2^3}\right)-2\left(1+\sqrt{2}\right)\left(\frac{\partial^4 k_1}{\partial x_1^2 \partial x_2^2}+x_3 \frac{\partial^4 h_1}{\partial x_1^2 \partial x_2^2}\right) -2\sqrt{2} \left(\frac{\partial^4 k_2}{\partial x_1^3 \partial x_2}+x_3 \frac{\partial^4 h_2}{\partial x_1^3 \partial x_2}\right)=0 \,, \\
-\left(\frac{\partial^4 k_2}{\partial x_2^4}+x_3 \frac{\partial^4 h_2}{\partial x_2^4}\right)+\left(1+\sqrt{2}\right) \left(\frac{\partial^4 k_1}{\partial x_1 \partial x_2^3}+x_3 \frac{\partial^4 h_1}{\partial x_1 \partial x_2^3}\right)+\left(1+\sqrt{2}\right)\left(\frac{\partial^4 k_2}{\partial x_1^2 \partial x_2^2}+x_3 \frac{\partial^4 h_2}{\partial x_1^2 \partial x_2^2}\right) \\
-\left(1+\sqrt{2}\right) \left(\frac{\partial^4 k_1}{\partial x_1^3 \partial x_2}+x_3 \frac{\partial^4 h_1}{\partial x_1^3 \partial x_2}\right)-\sqrt{2} \left(\frac{\partial^4 k_2}{\partial x_1^4}+x_3 \frac{\partial^4 h_2}{\partial x_1^4}\right) =0\,, \\
\left(\frac{\partial^4 k_1}{\partial x_1 \partial x_2^3}+x_3 \frac{\partial^4 h_1}{\partial x_1 \partial x_2^3}\right)+2\left(1+\sqrt{2}\right)\left(\frac{\partial^4 k_2}{\partial x_1^2 \partial x_2^2}+x_3 \frac{\partial^4 h_2}{\partial x_1^2 \partial x_2^2}\right) - \left(\frac{\partial^4 k_1}{\partial x_1^3 \partial x_2}+x_3 \frac{\partial^4 h_1}{\partial x_1^3 \partial x_2}\right)=0 \,.
\end{dcases}
\end{equation}
In summary, for this class we obtain the following result.
\begin{prop}
The universal displacements in the pentagonal $\mathbb{Z}_5 \oplus \mathbb{Z}_2^c$ class linear strain-gradient elastic solids coincide with those of \eqref{transverse-isotropy-linear-solution} and \eqref{transverse-isotropy-linear-solution-2}, subject to the additional universality PDEs \eqref{ClassZ_5+Z^c_2-fourth-order-constraints-1}--\eqref{ClassZ_5+Z^c_2-fourth-order-constraints-3} for the functions $h_i$, $i=1,2$, $\hat{u}_{33}$, and $k_i$, $i=1,2$.
\end{prop}

\subsubsection{Pentagonal class $\mathbb{D}_5 \oplus \mathbb{Z}_2^c$} 

Tensor $\boldsymbol{\mathsf{A}}_{\mathbb{D}_5 \oplus \mathbb{Z}_2^c}$ has $23$ independent components and is of the same form as that of class $\mathbb{D}_5$. Tensor $\boldsymbol{\mathsf{M}}_{\mathbb{D}_5 \oplus \mathbb{Z}_2^c}$ vanishes identically. Tensor $\boldsymbol{\mathsf{C}}_{\mathbb{O}(2) \oplus \mathbb{Z}_2^c}$ has $5$ independent components and corresponds to the Curie group with Hermann–Mauguin symbol $\infty/mm$. This class has the universality PDEs of class $\mathbb{O}(2) \oplus \mathbb{Z}_2^c$ and introduces two additional constants associated with the sixth-order tensor. The resulting additional universality PDEs are
\begin{equation}
\label{ClassD_5+Z^c_2-constraints}
\begin{dcases}
4\frac{\partial^3 h_1}{\partial x_2^3}+\left(4+5\sqrt{2}\right)\frac{\partial^3 h_2}{\partial x_1 \partial x_2^2}+\sqrt{2}\left(\frac{\partial^4 \hat{u}_{33}}{\partial x_1 \partial x_2^3}-4 \frac{\partial^3 h_1}{\partial x_1^2 \partial x_2}-\frac{\partial^3 h_2}{\partial x_1^3}-\frac{\partial^4 \hat{u}_{33}}{\partial x_1^3 \partial x_2}\right)=0, \\
2 \frac{\partial^3 h_1}{\partial x_2^3}+\left(3+2\sqrt{2}\right) \frac{\partial^3 h_2}{\partial x_1 \partial x_2}+\left(1+2\sqrt{2}\right)\frac{\partial^4 \hat{u}_{33}}{\partial x_1 \partial x_2^3}-4\left(1+\sqrt{2}\right)\frac{\partial^3 h_1}{\partial x_1^2 \partial x_2}\\
-\left(1+2\sqrt{2}\right)\left(\frac{\partial^3 h_2}{\partial x_2^3}+\frac{\partial^4 \hat{u}_{33}}{\partial x_1^3 \partial x_2}\right)=0\,, 
-\frac{\partial^4 \hat{u}_{33}}{\partial x_2^4}-3\frac{\partial^3 h_1}{\partial x_1^3}-\frac{\partial^4 \hat{u}_{33}}{\partial x_1^4}=0, \\
\left(1+\sqrt{2}\right)\frac{\partial^4 k_1}{\partial x_1 \partial x_2^3}+\left(1+\sqrt{2}\right)\left(\frac{\partial^4 k_1}{\partial x_1^3 \partial x_2}+x_3 \frac{\partial^4 h_1}{\partial x_1^3 \partial x_2}\right)-\sqrt{2}x_3 \frac{\partial^4 h_2}{\partial x_1^4}=0, \\
\sqrt{2}\frac{\partial^4 k_1}{\partial x_1 \partial x_2^3}+\left(1+\sqrt{2}\right)\frac{\partial^4 k_2}{\partial x_1^2 \partial x_2^2}-\frac{\partial^4 k_1}{\partial x_1^3 \partial x_2} - x_3 \frac{\partial^3 h_1}{\partial x_1^3 \partial x_2}=0.
\end{dcases}
\end{equation}
In summary, for this class we obtain the following result.
\begin{prop}
The universal displacements in tetragonal $\mathbb{D}_5 \oplus \mathbb{Z}_2^c$ class linear strain-gradient elastic solids coincide with those of class $\mathbb{O}(2) \oplus \mathbb{Z}_2^c$, supplemented by the additional constraints \eqref{ClassD_5+Z^c_2-constraints}.
\end{prop}

\subsubsection{Pentagonal class $\mathbb{D}_{10}^h$} 

Tensor $\boldsymbol{\mathsf{A}}_{\mathbb{O}(2) \oplus \mathbb{Z}_2^c}$ has $21$ independent components and is of the same form as that of class $\mathbb{O}(2)$. Tensor $\boldsymbol{\mathsf{M}}_{\mathbb{D}_{10}^h}$ has a single independent component and is given by
\begin{equation}
	\label{D_10^h}
	\boldsymbol{\mathsf{M}}_{\mathbb{D}_{10}^h}=\begin{bmatrix}
		\bar{A}^{(1)} & 0 & 0 & 0  \\
		0 & 0 & 0 & 0   \\
		0 & 0  & 0 & 0     \\
		0 & 0 & 0 & 0   
	\end{bmatrix} 
	+\begin{bmatrix}
		0 & 0 & 0 & 0  \\
		0 & 0 & 0 & 0   \\
		0 & 0  & 0 & 0     \\
		0 & g(\bar{A}^{(1)}) & 0 & 0   
	\end{bmatrix}.
\end{equation}
Tensor $\boldsymbol{\mathsf{C}}_{\mathbb{O}(2) \oplus \mathbb{Z}_2^c}$ has $5$ independent components and corresponds to the Curie group with Hermann--Mauguin symbol $\infty m$. This class has the same universality PDEs as class $\mathbb{O}(2) \oplus \mathbb{Z}_2^c$, with one additional constant associated with the fifth-order tensor. The resulting additional universality PDEs are
\begin{equation}
\label{ClassD^h_{10}-constraints}
\begin{dcases}
 (-1+\sqrt{2})\left(\frac{\partial^3 k_2}{\partial x_2^3}+x_3 \frac{\partial^3 h_2}{\partial x_2^3}\right)-(1+2\sqrt{2})\frac{\partial^3 k_1}{\partial x_1 \partial x_2^2}-(2+\sqrt{2})\frac{\partial^3 k_2}{\partial x_1^2 \partial x_2}=0 \,, \\
-(-1+\sqrt{2})\frac{\partial^3 k_2}{\partial x_1 \partial x_2^2}+(1+2\sqrt{2})\frac{\partial^3 k_1}{\partial x_1^2 \partial x_2} +(2+\sqrt{2})\left(\frac{\partial^3 k_2}{\partial x_1^3}+x_3 \frac{\partial^3 h_2}{\partial x_1^3}\right)=0.
\end{dcases}
\end{equation}
In conclusion, for this class we obtain the following result.
\begin{prop}
	The universal displacements in tetragonal $\mathbb{D}^h_{10}$ class linear strain-gradient elastic solids are the same as those of class $\mathbb{O}(2) \oplus \mathbb{Z}_2^c$, supplemented by the additional universality PDEs \eqref{ClassD^h_{10}-constraints}.
\end{prop}

\renewcommand{\arraystretch}{1.15}
\begin{longtable}{p{0.30\textwidth} p{0.66\textwidth}}
\caption{Summary of universal displacements: Pentagonal classes.}\label{Table-UD-summary-pentagonal}\\
\hline
\textbf{Symmetry class} & \textbf{Universal displacement family} \\
\hline
\endfirsthead
\hline
\textbf{Symmetry class} & \textbf{Universal displacement family} (continued) \\
\hline
\endhead
\hline
\endfoot
\hline
\endlastfoot
Pentagonal class $\mathbb{Z}_5$ & Of the form \eqref{transverse-isotropy-linear-solution} and \eqref{transverse-isotropy-linear-solution-2} with \eqref{classZ_5-mh1}, \eqref{classZ_5-mh2} for $h_i$, \eqref{classZ_5-u_{33}} for $\hat{u}_{33}$, and constraints \eqref{ClassZ_5-k_i-constraints-1}--\eqref{ClassZ_5-k_i-constraints-3} \\ \hline
Pentagonal class $\mathbb{D}_5$ & Same as class $\mathbb{O}(2)$ with additional constraints \eqref{ClassD_5-constraints} \\ \hline
Pentagonal class $\mathbb{Z}_5 \oplus \mathbb{Z}_2^c$ & Of the form \eqref{transverse-isotropy-linear-solution} and \eqref{transverse-isotropy-linear-solution-2} with additional PDEs \eqref{ClassZ_5+Z^c_2-fourth-order-constraints-1}--\eqref{ClassZ_5+Z^c_2-fourth-order-constraints-3} \\ \hline
Pentagonal class $\mathbb{D}_5 \oplus \mathbb{Z}_2^c$ & Same as class $\mathbb{O}(2) \oplus \mathbb{Z}_2^c$ with additional constraints \eqref{ClassD_5+Z^c_2-constraints} \\ \hline
Pentagonal class $\mathbb{D}_{10}^h$ & Same as class $\mathbb{O}(2) \oplus \mathbb{Z}_2^c$ with additional PDEs \eqref{ClassD^h_{10}-constraints} \\ \hline
\end{longtable}
\renewcommand{\arraystretch}{2.0}

\subsection{Hexagonal classes}
There are four hexagonal classes: $\mathbb{Z}_6$, $\mathbb{D}_6$, $\mathbb{Z}_6 \oplus \mathbb{Z}_2^c$, and $\mathbb{D}_6 \oplus \mathbb{Z}_2^c$. The class $\mathbb{Z}_6$ has the same universality PDEs as $\mathbb{SO}(2)$, augmented by two additional material constants that induce further universality PDEs. Similarly, the class $\mathbb{Z}_6 \oplus \mathbb{Z}_2^c$ has the same structure as $\mathbb{SO}(2) \oplus \mathbb{Z}_2^c$, with two additional material constants that generate further universality PDEs. The classes $\mathbb{D}_6$ and $\mathbb{D}_6 \oplus \mathbb{Z}_2^c$ follow the same pattern, each having one additional material constant compared with $\mathbb{O}(2)$ and $\mathbb{O}(2) \oplus \mathbb{Z}_2^c$, respectively. For each of these classes, we list the additional universality PDEs associated with the extra material constants. These universality PDEs appear as conditions on the functions $h_i$ and $k_i$, $i=1,2$, in the candidate universal displacement field \eqref{transverse-isotropy-linear-solution}.

\subsubsection{Hexagonal class $\mathbb{Z}_6$} 

Tensor $\boldsymbol{\mathsf{A}}_{\mathbb{Z}_6 \oplus \mathbb{Z}_2^c}$ possesses $33$ independent components and has the form
\begin{equation}
	\label{A_Z_6}
	\boldsymbol{\mathsf{A}}_{\mathbb{Z}_6 \oplus \mathbb{Z}_2^c}(\mathbf{x})=\begin{bmatrix}
		A^{(11)}+\eta A_c & B^{(6)}+\theta B_c & 0 & 0  \\
		& A^{(11)} & 0 & 0   \\
		&  & H^{(6)} & I^{(4)}   \\
		&  &  & J^{(4)}      
	\end{bmatrix}_S+
	\begin{bmatrix}
		0 & 0 & 0 & 0  \\
		& 0 & 0 & 0   \\
		&  & f(J^{(4)}) & 0   \\
		&  &  & 0      
	\end{bmatrix}_S\,.
\end{equation}
All pertinent matrices have been defined previously. The tensor $\boldsymbol{\mathsf{M}}_{\mathbb{SO}(2)}$ has $20$ independent components and has the form
\begin{equation}
	\label{M_SO(2)}
	\boldsymbol{\mathsf{M}}_{\mathbb{SO}(2)}=\boldsymbol{\mathsf{M}}_{\mathbb{O}(2)}+\boldsymbol{\mathsf{M}}_{\mathbb{O}^-(2)}\,,
\end{equation}
where
\begin{equation}
	\label{M_O(2)}
	\boldsymbol{\mathsf{M}}_{\mathbb{O}(2)}=\begin{bmatrix}
		0 & 0 & 0 & \bar{D}^{(4)}  \\
		\bar{E}^{(4)} & 0 & 0 & 0   \\
		0 & -\bar{E}^{(4)}  & 0 & 0     \\
		0 & 0 & 0 & 0   
	\end{bmatrix} 
	+\begin{bmatrix}
		0 & 0 & 0 & 0  \\
		0 & 0 & 0 & 0   \\
		0 & 0  & 0 & 0     \\
		0 & 0 & g(\bar{D}^{(4)}) & 0   
	\end{bmatrix}\,,
\end{equation}
\begin{equation}
	\label{M_O(2)^-}
	\boldsymbol{\mathsf{M}}_{\mathbb{O}(2)^-}=\begin{bmatrix}
		0 & 0 & \bar{C}^{(8)} & 0  \\
		0 & \bar{F}^{(4)} & 0 & 0   \\
		\bar{F}^{(4)} & 0  & 0 & 0     \\
		0 & 0 & 0 & 0   
	\end{bmatrix} 
	+\begin{bmatrix}
		0 & 0 & 0 & 0  \\
		0 & 0 & 0 & 0   \\
		0 & 0  & 0 & 0     \\
		0 & 0 & 0 & g(\bar{C}^{(8)})   
	\end{bmatrix}\,.
\end{equation}
All relevant components have been introduced in previous sections. The tensor $\boldsymbol{\mathsf{C}}_{\mathbb{O}(2) \oplus \mathbb{Z}_2^c}$ corresponds to the Curie group with Hermann--Mauguin symbol $\infty2$ and has $5$ independent components. This class has the same universality PDEs as $\mathbb{SO}(2)$, with two additional constants associated with the sixth-order tensor. These additional universality PDEs are
\begin{equation}
\label{ClassZ_6-constraints}
\begin{dcases}
 \frac{\partial^4 k_2}{\partial x_1 \partial x_2^3} - \frac{\partial^4 k_1}{\partial x_1^2 \partial x_2^2} - \frac{\partial^4 k_2}{\partial x_1^3 \partial x_2} = 0, \\
 \left(10+\sqrt{2}\right) \frac{\partial^4 k_1}{\partial x_1 \partial x_2^3} + \left(10+3\sqrt{2}\right) \frac{\partial^4 k_2}{\partial x_1^2 \partial x_2^2} - 9\sqrt{2} \frac{\partial^4 k_1}{\partial x_1^3 \partial x_2} = 0, \\
 4 \frac{\partial^4 k_1}{\partial x_1 \partial x_2^3} + 2\left(2+\sqrt{2}\right) \frac{\partial^4 k_2}{\partial x_1^2 \partial x_2^2} - 2\sqrt{2} \frac{\partial^4 k_1}{\partial x_1^3 \partial x_2} = 0, \\
 -\left(4+3\sqrt{2}\right) \frac{\partial^4 k_2}{\partial x_1 \partial x_2^3} + \left(10+9\sqrt{2}\right) \frac{\partial^4 k_1}{\partial x_1^2 \partial x_2^2} + \left(10+3\sqrt{2}\right) \frac{\partial^4 k_2}{\partial x_1^3 \partial x_2} = 0.
\end{dcases}
\end{equation}
Thus, we obtain the following result.
\begin{prop}
The universal displacements in the hexagonal $\mathbb{Z}_6$ class linear strain-gradient elastic solids coincide with those of the class $\mathbb{O}(2) \oplus \mathbb{Z}_2^c$, subject to the additional constraints \eqref{ClassZ_6-constraints}\,.
\end{prop}

\subsubsection{Hexagonal class $\mathbb{D}_6$} 

Tensor $\boldsymbol{\mathsf{A}}_{\mathbb{D}_6 \oplus \mathbb{Z}_2^c}$ has $22$ independent components and has the following representation
\begin{equation}
\label{A_D_6}
    \boldsymbol{\mathsf{A}}_{\mathbb{D}_6 \oplus \mathbb{Z}_2^c}(\mathbf{x})=\begin{bmatrix}
    A^{(11)}+\eta A_c & 0 & 0 & 0  \\
     & A^{(11)} & 0 & 0   \\
     &  & H^{(6)} & 0   \\
     &  &  & J^{(4)}      
    \end{bmatrix}_S+
\begin{bmatrix}
    0 & 0 & 0 & 0  \\
     & 0 & 0 & 0   \\
     &  & h(J^{(4)}) & 0   \\
     &  &  & 0      
    \end{bmatrix}_S\,,
\end{equation}
where all relevant matrices have been defined in the preceding classes. The tensor $\boldsymbol{\mathsf{M}}_{\mathbb{O}(2)}$ contains $8$ independent components and has the structure \eqref{M_O(2)}. The tensor $\boldsymbol{\mathsf{C}}_{\mathbb{O}(2) \oplus \mathbb{Z}_2^c}$ corresponds to the Curie group with Hermann--Mauguin symbol $\infty2$ and has $5$ independent components.
This class has the universality PDEs of class $\mathbb{SO}(2)$, with one additional constant associated with the sixth-order tensor. The resulting additional universality PDEs are
\begin{equation}
\label{ClassD_6-constraints}
\begin{dcases}
\sqrt{2} \left(\frac{\partial^4 k_1}{\partial x_2^4} + x_3 \frac{\partial^4 h_1}{\partial x_2^4}\right) + (1+\sqrt{2}) \left(\frac{\partial^4 k_2}{\partial x_1 \partial x_2^3} + x_3 \frac{\partial^4 h_2}{\partial x_1 \partial x_2^3}\right) - (1+\sqrt{2}) \left(\frac{\partial^4 k_1}{\partial x_1^2 \partial x_2^2} - \frac{\partial^4 k_2}{\partial x_1^3 \partial x_2}\right) \\
\quad + \frac{\partial^4 k_1}{\partial x_1^4} + x_3 \frac{\partial^4 h_1}{\partial x_1^4} = 0 \,, \\
\frac{\partial^4 k_1}{\partial x_1 \partial x_2^3} + 2(2+\sqrt{2}) \frac{\partial^4 k_2}{\partial x_1^2 \partial x_2^2} - 2\sqrt{2} \left(\frac{\partial^4 k_1}{\partial x_1^3 \partial x_2} + x_3 \frac{\partial^4 h_1}{\partial x_1^3 \partial x_2}\right) = 0 \,.
\end{dcases}
\end{equation}
Hence, we obtain the following result.
\begin{prop}
The universal displacements in hexagonal $\mathbb{D}_6$ class linear strain-gradient elastic solids coincide with those of class $\mathbb{O}(2)$, supplemented by the additional constraints \eqref{ClassD_6-constraints}\,.
\end{prop}

\subsubsection{Hexagonal class $\mathbb{Z}_6 \oplus \mathbb{Z}_2^c$} 

Tensor $\boldsymbol{\mathsf{A}}_{\mathbb{Z}_6 \oplus \mathbb{Z}_2^c}$ has $33$ independent components and is of the same form as that of class $\mathbb{Z}_6$\,. Tensor $\boldsymbol{\mathsf{M}}_{\mathbb{Z}_6 \oplus \mathbb{Z}_2^c}$ is a null tensor\,.
Tensor $\boldsymbol{\mathsf{C}}_{\mathbb{O}(2) \oplus \mathbb{Z}_2^c}$ has $5$ independent components and corresponds to the Curie group with Hermann--Mauguin symbol $\infty/mm$\,.
This class has the same universality PDEs as class $\mathbb{SO}(2) \oplus \mathbb{Z}_2^c$, with two additional elastic constants associated with the sixth-order tensor. The resulting additional universality PDEs are
\begin{equation}
\label{ClassZ_6+Z^c_2-constraints}
\begin{dcases}
\sqrt{2}\left(\frac{\partial^4 k_1}{\partial x_2^4}+x_3 \frac{\partial^4 h_1}{\partial x_2^4}\right)+\left(1+\sqrt{2}\right)\left(\frac{\partial^4 k_2}{\partial x_1 \partial x_2^3}+x_3 \frac{\partial^4 h_2}{\partial x_1 \partial x_2^3}\right)-\left(1+\sqrt{2}\right)\left(\frac{\partial^4 k_1}{\partial x_1^2 \partial x_2^2}+x_3 \frac{\partial^4 h_1}{\partial x_1^2 \partial x_2^2}\right) \\
-\left(1+\sqrt{2}\right)\left(\frac{\partial^4 k_2}{\partial x_1^3 \partial x_2}+x_3 \frac{\partial^4 h_2}{\partial x_1^3 \partial x_2}\right)+\frac{\partial^4 k_1}{\partial x_1^4}+x_3 \frac{\partial^4 h_1}{\partial x_1^4}=0, \\
-4\left(\frac{\partial^4 k_2}{\partial x_2^4}+x_3 \frac{\partial^4 h_2}{\partial x_2^4}\right)+\left(10+\sqrt{2}\right)\left(\frac{\partial^4 k_1}{\partial x_1 \partial x_2^3}+x_3 \frac{\partial^4 h_1}{\partial x_1 \partial x_2^3}\right)+\left(10+3\sqrt{2}\right)\left(\frac{\partial^4 k_2}{\partial x_1^2 \partial x_2^2}+x_3 \frac{\partial^4 h_2}{\partial x_1^2 \partial x_2^2}\right) \\
-9\sqrt{2}\left(\frac{\partial^4 k_1}{\partial x_1^3 \partial x_2}+x_3 \frac{\partial^4 h_1}{\partial x_1^3 \partial x_2}+\frac{\partial^4 k_2}{\partial x_1^4}+x_3 \frac{\partial^4 h_2}{\partial x_1^4}\right)=0, \\
4\left(\frac{\partial^4 k_1}{\partial x_1 \partial x_2^3}+x_3 \frac{\partial^4 h_1}{\partial x_1 \partial x_2^3}\right)+2\left(2+\sqrt{2}\right)\left(\frac{\partial^4 k_2}{\partial x_1^2 \partial x_2^2}+x_3 \frac{\partial^4 h_2}{\partial x_1^2 \partial x_2^2}\right)-2\sqrt{2}\left(\frac{\partial^4 k_1}{\partial x_1^3 \partial x_2}+x_3 \frac{\partial^4 h_1}{\partial x_1^3 \partial x_2}\right)=0, \\
-\sqrt{2}\left(\frac{\partial^4 k_1}{\partial x_2^4}+x_3 \frac{\partial^4 h_1}{\partial x_2^4}\right)-\left(4+3\sqrt{2}\right)\left(\frac{\partial^4 k_2}{\partial x_1^3 \partial x_2}+x_3 \frac{\partial^4 h_2}{\partial x_1 \partial x_2^3}\right)+\left(10+9\sqrt{2}\right)\left(\frac{\partial^4 k_1}{\partial x_1^2 \partial x_2^2}+x_3 \frac{\partial^4 h_1}{\partial x_1^2 \partial x_2^2}\right) \\
+\left(10+3\sqrt{2}\right)\left(\frac{\partial^4 k_2}{\partial x_1^3 \partial x_2}+x_3 \frac{\partial^4 h_2}{\partial x_1^3 \partial x_2}\right)=0\,.
\end{dcases}
\end{equation}
In summary, we obtain the following result.
\begin{prop}
The universal displacements in hexagonal $\mathbb{Z}_6 \oplus \mathbb{Z}_2^c$ class linear strain-gradient elastic solids coincide with those of class $\mathbb{SO}(2) \oplus \mathbb{Z}_2^c$, subject to the additional constraints \eqref{ClassZ_6+Z^c_2-constraints}\,.
\end{prop}

\subsubsection{Hexagonal class $\mathbb{D}_6 \oplus \mathbb{Z}_2^c$} 

Tensor $\boldsymbol{\mathsf{A}}_{\mathbb{D}_6 \oplus \mathbb{Z}_2^c}$ has $22$ independent components and is of the same form as that of class $\mathbb{D}_6$. Tensor $\boldsymbol{\mathsf{M}}_{\mathbb{D}_6 \oplus \mathbb{Z}_2^c}$ is a null tensor. Tensor $\boldsymbol{\mathsf{C}}_{{\mathbb{O}(2) \oplus \mathbb{Z}_2^c}}$ has $5$ independent components and corresponds to the Curie group with Hermann--Mauguin symbol $\infty/mm$. This class has the universality PDEs of class $\mathbb{O}(2) \oplus \mathbb{Z}_2^c$, with one additional elastic constant associated with the sixth-order tensor. The resulting additional universality PDEs are
\begin{equation}
\label{ClassD_6+Z^c_2-constraints}
\begin{dcases}
 \sqrt{2} \frac{\partial^4 k_1}{\partial x_2^4} + \left(1+\sqrt{2}\right) \left( \frac{\partial^4 k_2}{\partial x_1 \partial x_2^3} + x_3 \frac{\partial^4 h_2}{\partial x_1 \partial x_2^3} \right) - \left(1+\sqrt{2}\right) \left( \frac{\partial^4 k_1}{\partial x_1^2 \partial x_2^2} - \frac{\partial^4 k_2}{\partial x_1^3 \partial x_2} \right) \\
 \quad + \frac{\partial^4 k_1}{\partial x_1^4} + x_3 \frac{\partial^4 h_1}{\partial x_1^4} = 0, \\
 \frac{\partial^4 k_1}{\partial x_1 \partial x_2^3} + 2 \left(2+\sqrt{2}\right) \frac{\partial^4 k_2}{\partial x_1^2 \partial x_2^2} - 2 \sqrt{2} \left( \frac{\partial^4 k_1}{\partial x_1^3 \partial x_2} + x_3 \frac{\partial^4 h_1}{\partial x_1^3 \partial x_2} \right) = 0.
\end{dcases}
\end{equation}
Thus, we obtain the following result.
\begin{prop}
The universal displacements in the hexagonal $\mathbb{D}_6 \oplus \mathbb{Z}_2^c$ class linear strain-gradient elastic solids coincide with those of class $\mathbb{O}(2) \oplus \mathbb{Z}_2^c$, supplemented by the additional constraints \eqref{ClassD_6+Z^c_2-constraints}\,.
\end{prop}

\renewcommand{\arraystretch}{1.15}
\begin{longtable}{p{0.30\textwidth} p{0.66\textwidth}}
\caption{Summary of universal displacements: Hexagonal classes.}\label{Table-UD-summary-hexagonal}\\
\hline
\textbf{Symmetry class} & \textbf{Universal displacement family} \\
\hline
\endfirsthead
\hline
\textbf{Symmetry class} & \textbf{Universal displacement family} (continued) \\
\hline
\endhead
\hline
\endfoot
\hline
\endlastfoot
Hexagonal class $\mathbb{Z}_6$ & Same as class $\mathbb{O}(2) \oplus \mathbb{Z}_2^c$ with additional constraints \eqref{ClassZ_6-constraints} \\ \hline
Hexagonal class $\mathbb{D}_6$ & Same as class $\mathbb{O}(2)$ with additional constraints \eqref{ClassD_6-constraints} \\ \hline
Hexagonal class $\mathbb{Z}_6 \oplus \mathbb{Z}_2^c$ & Same as class $\mathbb{SO}(2) \oplus \mathbb{Z}_2^c$ with additional constraints \eqref{ClassZ_6+Z^c_2-constraints} \\ \hline
Hexagonal class $\mathbb{D}_6 \oplus \mathbb{Z}_2^c$ & Same as class $\mathbb{O}(2) \oplus \mathbb{Z}_2^c$ with additional constraints \eqref{ClassD_6+Z^c_2-constraints} \\ \hline
\end{longtable}
\renewcommand{\arraystretch}{2.0}

\subsection{$\infty$-gonal classes}

There are five $\infty$-gonal symmetry classes: $\mathbb{SO}(2)$, $\mathbb{O}(2)$, $\mathbb{SO}(2) \oplus \mathbb{Z}_2^c$, $\mathbb{O}(2) \oplus \mathbb{Z}_2^c$, and $\mathbb{O}^-(2)$. For all these classes, the elasticity tensor $\boldsymbol{\mathsf{C}}$ admits five independent elastic constants and has the same structure as in the transversely isotropic classical linear elasticity.

The additional material parameters associated with the tensors $\boldsymbol{\mathsf{M}}$ and $\boldsymbol{\mathsf{A}}$ are expected to induce further universality PDEs for the candidate universal displacement field given above. In particular, these parameters restrict the functions $h_i$ and $k_i$, $i=1,2$, as well as the function $\hat{u}_{33}$. Our strategy is as follows: we first determine how the universality PDEs simplify the functions $h_i$, then examine the implications for $\hat{u}_{33}$, and finally analyze the resulting conditions on the functions $k_i$. Throughout, we use the Cauchy--Riemann equations to rewrite each system in terms of a single function among $\{h_i\}$ or $\{k_i\}$, thereby reducing the number of dependent variables. This reduction is used to verify consistency of the overdetermined system. When consistency cannot be enforced a priori, we simplify the system as far as possible and list explicitly the remaining universality PDEs.

\subsubsection{$\infty$-gonal class $\mathbb{O}(2)^-$} 

The tensor $\boldsymbol{\mathsf{A}}_{\mathbb{O}(2) \oplus \mathbb{Z}_2^c}$ has $21$ independent components and is of the same form as that of class $\mathbb{O}(2)$. The tensor $\boldsymbol{\mathsf{M}}_{\mathbb{O}(2)^-}$ has $12$ independent components and is of the same form as in the classes $\mathbb{SO}(2)$ and $\mathbb{O}(2)$. The tensor $\boldsymbol{\mathsf{C}}_{\mathbb{O}(2) \oplus \mathbb{Z}_2^c}$ has $5$ independent components and corresponds to the Curie group with Hermann--Mauguin symbol $\infty m$. Substituting the transversely isotropic displacement field \eqref{transverse-isotropy-linear-solution} into the universality PDEs, we obtain the following nontrivial additional universality constraints:
\begin{equation}
\label{classO(2)-1}
\begin{dcases}
\frac{\partial^2 h_1}{\partial x_2^2} + \frac{\partial^2 h_2}{\partial x_1 \partial x_2} + \frac{\partial^2 h_1}{\partial x_1^2} = 0\,, \\
-\frac{1}{\sqrt{2}} \frac{\partial^2 h_1}{\partial x_2^2} - \frac{1}{\sqrt{2}} \frac{\partial^2 h_1}{\partial x_1 \partial x_2} + \sqrt{2} \frac{\partial^3 \hat{u}_{33}}{\partial x_1 \partial x_2^2} - 3 \frac{\partial^2 h_1}{\partial x_1^2} + \frac{\partial^3 \hat{u}_{33}}{\partial x_1^3} = 0\,, \\
-\frac{\partial^2 h_1}{\partial x_2^2} = 0\,, \\
\frac{1}{\sqrt{2}} \frac{\partial^2 h_1}{\partial x_2^2} -  \frac{1}{\sqrt{2}} \frac{\partial^2 h_2}{\partial x_1 \partial x_2} - \sqrt{2} \frac{\partial^3 \hat{u}_{33}}{\partial x_1 \partial x_2^2} = 0\,, \\
\frac{1}{\sqrt{2}} \frac{\partial^2 h_1}{\partial x_2^2} + \left(-1+\frac{1}{\sqrt{2}}\right) \frac{\partial^2 h_1}{\partial x_1 \partial x_2} -\left(-1+ \sqrt{2}\right) \frac{\partial^3 \hat{u}_{33}}{\partial x_1 \partial x_2^2} + 2 \frac{\partial^2 h_1}{\partial x_1^2} + \frac{\partial^3 \hat{u}_{33}}{\partial x_1^3} = 0\,, \\
\frac{1}{\sqrt{2}} \frac{\partial^2 h_1}{\partial x_2^2}  - \frac{1}{\sqrt{2}} \frac{\partial^2 h_2}{\partial x_1 \partial x_2} - \sqrt{2} \frac{\partial^3 \hat{u}_{33}}{\partial x_1 \partial x_2^2} - \frac{\partial^3 \hat{u}_{33}}{\partial x_1^3}= 0\,, \\
\frac{1}{\sqrt{2}} \frac{\partial^4 k_1}{\partial x_2^4} + \frac{x_3}{\sqrt{2}} \frac{\partial^4 h_1}{\partial x_2^4} + \frac{1}{\sqrt{2}} \frac{\partial^4 k_2}{\partial x_1 \partial x_2^3} + \frac{x_3}{\sqrt{2}} \frac{\partial^4 h_2}{\partial x_1 \partial x_2^3} \\
\quad + \left(1 + \sqrt{2}\right)\left( \frac{\partial^4 k_1}{\partial x_1^2 \partial x_2^2} + x_3 \frac{\partial^4 h_2}{\partial x_1^2 \partial x_2^2} \right) + \frac{\partial^4 k_2}{\partial x_1^3 \partial x_2} + x_3 \frac{\partial^4 h_2}{\partial x_1^3 \partial x_2} \\
\quad + \frac{\partial^4 k_1}{\partial x_1^3 \partial x_2} + x_3 \frac{\partial^4 h_1}{\partial x_1^4} = 0\,, \\
-\frac{\partial^4 k_1}{\partial x_1^2 \partial x_2^2} - x_3 \frac{\partial^4 h_1}{\partial x_1^2 \partial x_2^2} - \frac{\partial^4 k_2}{\partial x_1^3 \partial x_2} - x_3 \frac{\partial^4 h_2}{\partial x_1^3 \partial x_2} = 0\,, \\
-\frac{\partial^4 k_1}{\partial x_2^4} - x_3 \frac{\partial^4 h_1}{\partial x_2^4} + \frac{\partial^4 k_2}{\partial x_1 \partial x_2^3} + x_3 \frac{\partial^4 h_2}{\partial x_1 \partial x_2^3} \\
\quad - \frac{\partial^4 k_1}{\partial x_1^2 \partial x_2^2} - x_3 \frac{\partial^4 h_1}{\partial x_1^2 \partial x_2^2} + \frac{\partial^4 k_2}{\partial x_1^3 \partial x_2} - x_3 \frac{\partial^4 h_2}{\partial x_1^3 \partial x_2} = 0\,, \\
-\frac{1}{\sqrt{2}} \frac{\partial^4 k_1}{\partial x_2^4} + \frac{x_3}{\sqrt{2}} \frac{\partial^4 h_1}{\partial x_2^4} + \left(1 - \frac{1}{\sqrt{2}}\right)\left( \frac{\partial^4 k_2}{\partial x_1 \partial x_2^3} + x_3 \frac{\partial^4 h_2}{\partial x_1 \partial x_2^3} \right) \\
\quad - \left(1 + \sqrt{2}\right)\left( \frac{\partial^4 k_1}{\partial x_1^2 \partial x_2^2} + x_3 \frac{\partial^4 h_1}{\partial x_1^2 \partial x_2^2} \right) + \frac{\partial^4 k_2}{\partial x_1^3 \partial x_2} + x_3 \frac{\partial^4 h_2}{\partial x_1^3 \partial x_2} = 0\,,
\end{dcases}
\end{equation}
and
\begin{equation}
\label{classO(2)-2}
\begin{dcases}
\frac{\partial^2 h_2}{\partial x_2^2} + \frac{\partial^2 h_1}{\partial x_1 \partial x_2} + \frac{\partial^2 h_2}{\partial x_1^2} = 0\,, \\
 - \frac{\partial^2 h_2}{\partial x_2^2} + \frac{\partial^3 u_{33}}{\partial x_1 \partial x_2^3} - \frac{1}{\sqrt{2}} \left( \frac{\partial^2 h_1}{\partial x_1 \partial x_2} + \frac{\partial^2 h_2}{\partial x_1^2} - 2 \frac{\partial^3 \hat{u}_{33}}{\partial x_1^2 \partial x_2} \right) = 0\,, \\
-\frac{\partial^2 h_2}{\partial x_1^2} = 0\,, \\
\left(-1 + \frac{1}{\sqrt{2}}\right) \frac{\partial^2 h_1}{\partial x_1 \partial x_2} + \frac{1}{\sqrt{2}} \frac{\partial^2 h_2}{\partial x_1^2} - \left(-1 + \sqrt{2}\right) \frac{\partial^3 \hat{u}_{33}}{\partial x_1^2 \partial x_2} = 0\,, \\
- \frac{\partial^3 u_{33}}{\partial x_2^3} + \frac{1}{\sqrt{2}} \left( \frac{\partial^2 h_1}{\partial x_1 \partial x_2} + \frac{\partial^2 h_2}{\partial x_1^2} - 2 \frac{\partial^3 \hat{u}_{33}}{\partial x_1^2 \partial x_2} \right) = 0\,, \\
\left(1 + \frac{1}{\sqrt{2}}\right) \frac{\partial^2 h_1}{\partial x_1 \partial x_2} - \frac{1}{2}\left(2 + \sqrt{2}\right) \frac{\partial^2 h_2}{\partial x_1^2} + \left(-1 + \sqrt{2}\right) \frac{\partial^3 \hat{u}_{33}}{\partial x_1^2 \partial x_2} = 0\,, \\
\frac{\partial^4 k_2}{\partial x_2^4} + x_3 \frac{\partial^4 h_2}{\partial x_2^4} + \frac{\partial^4 k_1}{\partial x_1 \partial x_2^3} + x_3 \frac{\partial^4 h_1}{\partial x_1 \partial x_2^3} \\
\quad + \left(1 + \sqrt{2}\right) \left( \frac{\partial^4 k_2}{\partial x_1^2 \partial x_2^2} + x_3 \frac{\partial^4 h_2}{\partial x_1^2 \partial x_2^2} \right) \\
\quad + \frac{1}{\sqrt{2}} \left( \frac{\partial^4 k_1}{\partial x_1^3 \partial x_2} + x_3 \frac{\partial^4 h_1}{\partial x_1^3 \partial x_2} + \frac{\partial^4 k_2}{\partial x_1^4} + x_3 \frac{\partial^4 h_2}{\partial x_1^4} \right) = 0\,, \\
-\frac{\partial^4 k_1}{\partial x_1 \partial x_2^3} - x_3 \frac{\partial^4 h_1}{\partial x_1 \partial x_2^3} - \frac{\partial^4 k_2}{\partial x_1^2 \partial x_2^2} - x_3 \frac{\partial^4 h_2}{\partial x_1^2 \partial x_2^2} = 0\,, \\
\frac{\partial^4 k_1}{\partial x_1 \partial x_2^3} + x_3 \frac{\partial^4 h_1}{\partial x_1 \partial x_2^3} - \frac{\partial^4 k_2}{\partial x_1^2 \partial x_2^2} - x_3 \frac{\partial^4 h_2}{\partial x_1^2 \partial x_2^2} \\
\quad + \frac{\partial^4 k_1}{\partial x_1^2 \partial x_2^2} + x_3 \frac{\partial^4 h_1}{\partial x_1^2 \partial x_2^2} - \frac{\partial^4 k_2}{\partial x_1^4} - x_3 \frac{\partial^4 h_2}{\partial x_1^4} = 0\,, \\
\frac{\partial^4 k_1}{\partial x_1 \partial x_2^3} + x_3 \frac{\partial^4 h_1}{\partial x_1 \partial x_2^3} - \left(-1 + \sqrt{2}\right) \left( \frac{\partial^4 h_2}{\partial x_1 \partial x_2} + x_3 \frac{\partial^4 h_2}{\partial x_1^2 \partial x_2^2} \right) \\
\quad + \left(1 - \frac{1}{\sqrt{2}}\right) \left( \frac{\partial^4 k_1}{\partial x_1^3 \partial x_2} + x_3 \frac{\partial^4 h_1}{\partial x_1^3 \partial x_2} \right) - \frac{1}{\sqrt{2}} \left( \frac{\partial^4 k_2}{\partial x_1^4} + x_3 \frac{\partial^4 h_2}{\partial x_1^4} \right) = 0\,, \\
\frac{\partial^4 k_2}{\partial x_2^3} + x_3 \frac{\partial^4 h_2}{\partial x_2^3} + 2 \left( \frac{\partial^4 k_1}{\partial x_1 \partial x_2^2} + x_3 \frac{\partial^4 h_1}{\partial x_1 \partial x_2^2} + \frac{\partial^4 k_2}{\partial x_1^2 \partial x_2} + x_3 \frac{\partial^4 h_2}{\partial x_1^2 \partial x_2} \right) \\
\quad + \frac{\partial^4 k_1}{\partial x_1^3} + x_3 \frac{\partial^4 h_1}{\partial x_1^3} = 0\,,
\end{dcases}
\end{equation}
as well as
\begin{equation}
\label{classO(2)-3}
\begin{dcases}
-\frac{\partial^3 k_2}{\partial x_2^3} - x_3 \frac{\partial^3 h_2}{\partial x_2^3} + 2\left( \frac{\partial^3 k_1}{\partial x_1 \partial x_2^2} + x_3 \frac{\partial^3 h_1}{\partial x_1 \partial x_2^2} + \frac{\partial^3 k_2}{\partial x_1^2 \partial x_2} + x_3 \frac{\partial^3 h_2}{\partial x_1^2 \partial x_2} \right) - \frac{\partial^3 k_1}{\partial x_1^3} - x_3 \frac{\partial^3 h_1}{\partial x_1^3} = 0\,, \\
\frac{\partial^3 k_1}{\partial x_1 \partial x_2^2} + x_3 \frac{\partial^3 h_1}{\partial x_1 \partial x_2^2} + \frac{\partial^3 k_2}{\partial x_1^2 \partial x_2} + x_3 \frac{\partial^3 h_2}{\partial x_1^2 \partial x_2} = 0\,, \\
\frac{\partial^3 k_2}{\partial x_2^3} + x_3 \frac{\partial^3 h_2}{\partial x_2^3} + \sqrt{2} \left( \frac{\partial^3 k_1}{\partial x_1 \partial x_2^2} + x_3 \frac{\partial^3 h_1}{\partial x_1 \partial x_2^2} + \frac{\partial^4 k_2}{\partial x_1^2 \partial x_2} + x_3 \frac{\partial^4 h_2}{\partial x_1^2 \partial x_2} \right) 
+ \frac{\partial^4 k_1}{\partial x_1^3} + x_3 \frac{\partial^4 h_1}{\partial x_1^3} = 0\,, \\
-(-1 + \sqrt{2}) \left( \frac{\partial^3 k_1}{\partial x_1 \partial x_2^2} + x_3 \frac{\partial^3 h_1}{\partial x_1 \partial x_2^2} + \frac{\partial^3 k_2}{\partial x_1^2 \partial x_2} + x_3 \frac{\partial^3 h_2}{\partial x_1^2 \partial x_2} \right) = 0\,, \\
-2 \left( \frac{\partial^3 h_2}{\partial x_2^3} + \frac{\partial^3 h_1}{\partial x_1 \partial x_2^2} + \frac{\partial^3 h_2}{\partial x_1^2 \partial x_2} + \frac{\partial^3 h_1}{\partial x_1^3} \right) = 0\,, \\
2(-2 + \sqrt{2}) \left( \frac{\partial^3 h_1}{\partial x_1 \partial x_2^2} + \frac{\partial^3 h_2}{\partial x_1^2 \partial x_2} \right) = 0\,, \\
\frac{\partial^3 h_2}{\partial x_2^3} + \frac{\partial^4 u_{33}}{\partial x_2^4} + \frac{\partial^3 h_1}{\partial x_1 \partial x_2^2} + \frac{\partial^3 h_2}{\partial x_1^2 \partial x_2} + 2 \frac{\partial^4 u_{33}}{\partial x_1^2 \partial x_2^2} + \frac{\partial^3 h_1}{\partial x_1^3} + \frac{\partial^4 \hat{u}_{33}}{\partial x_1^4} = 0\,, \\
\frac{\partial^3 h_2}{\partial x_2^3} - \frac{\partial^3 h_1}{\partial x_1 \partial x_2^2} + \frac{\partial^3 h_2}{\partial x_1^2 \partial x_2} + \frac{\partial^3 h_1}{\partial x_1^3} = 0\,, \\
-\frac{\partial^3 h_1}{\partial x_1 \partial x_2^2} - \frac{\partial^3 h_2}{\partial x_1^2 \partial x_2} = 0\,, \\
-\sqrt{2} \frac{\partial^3 h_1}{\partial x_1 \partial x_2^2} - (-2 + \sqrt{2}) \frac{\partial^3 h_2}{\partial x_1^2 \partial x_2} - 2(-1 + \sqrt{2}) \frac{\partial^4 \hat{u}_{33}}{\partial x_1^2 \partial x_2^2} = 0\,, \\
2 \left( 2 \frac{\partial^3 h_1}{\partial x_1 \partial x_2^2} + \frac{\partial^3 h_2}{\partial x_1^2 \partial x_2} + \frac{\partial^4 \hat{u}_{33}}{\partial x_1^2 \partial x_2^2} \right) = 0\,.
\end{dcases}
\end{equation}
From these equations, after lengthy but straightforward manipulations, we obtain the following results for the functions $h_i$, $i=1,2$:
\begin{equation}
\label{classO(2)-4}
\frac{\partial^2 h_i}{\partial x_1^2} = \frac{\partial^2 h_i}{\partial x_2^2} = \frac{\partial^2 h_i}{\partial x_1 \partial x_2}= 0\,.
\end{equation}
For the function $\hat{u}_{33}$ we find:
\begin{equation}
\label{classO(2)-5}
\frac{\partial^3 \hat{u}_{33}}{\partial x_1^3}  = \frac{\partial^3 \hat{u}_{33}}{\partial x_1^2 \partial x_2} = \frac{\partial^3 \hat{u}_{33}}{\partial x_1 \partial x_2^2} = \frac{\partial^4 \hat{u}_{33}}{\partial x_1^2 \partial x_2^2}= 0\,.
\end{equation}
For the functions $k_i$ with $i = 1, 2$ we obtain:
\begin{equation}
\label{classO(2)-6}
\frac{\partial^4 k_i}{\partial x_1^3 \partial x_2}  = \frac{\partial^4 k_i}{\partial x_1 \partial x_2^3} = \frac{\partial^4 k_i}{\partial x_2^4} = \frac{\partial^4 k_i}{\partial x_1^4}= 0\,.
\end{equation}
Additionally, we find:
\begin{equation}\label{classO(2)-7}
\frac{\partial^3 k_1}{\partial x_1 \partial x_2^2} + \frac{\partial^3 k_1}{\partial x_1^3} = 0\,, \qquad
\frac{\partial^3 k_2}{\partial x_2^3} + \frac{\partial^3 k_1}{\partial x_1^2 \partial x_2} = 0\,. 
\end{equation}
Thus, we obtain
\begin{align}
\label{classO^{-}(2)-mh1}
h_1(x_1, x_2) &= c_4 x_1 + c_5 x_2\,, \\
\label{classO^{-}(2)-mh2}
h_2(x_1, x_2) &= c_6 x_1 + c_7 x_2\,.
\end{align}
The functions $k_i$, $i=1, 2$, have the following form:
\begin{equation} \label{classO^{-}(2)-k}
\begin{aligned}
	k_1(x_1, x_2) &= f_1(x_2) + x_1 f_2(x_2) + f_3(x_1) + x_2 f_4(x_1)\,, \\
	k_2(x_1, x_2) &= f_5(x_2) + x_1 f_6(x_2) + f_7(x_1) + x_2 f_8(x_1)\,, 
\end{aligned}
\end{equation}
where
\begin{equation}
\label{classO^{-}(2)-f}
\begin{aligned}
f_1 &= c_8 + c_9 x_2 + c_{10} x_2^2 + c_{11} x_2^3\,, \\
f_2 &= c_{12} + c_{13} x_2 + c_{14} x_2^2\,, \\
f_3 &= c_{15} + c_{16} x_1 + c_{17} x_1^2 + c_{18} x_1^3\,, \\
f_4 &= c_{19} + c_{20} x_1 + c_{21} x_1^2\,, \\
f_5 &= c_{22} + c_{23} x_2 + c_{24} x_2^2 + c_{25} x_2^3\,, \\
f_6 &= c_{26} + c_{27} x_2 + c_{28} x_2^2\,, \\
f_7 &= c_{29} + c_{30} x_1 + c_{31} x_1^2 + c_{32} x_1^3\,, \\
f_8 &= c_{33} + c_{34} x_1 + c_{35} x_1^2\,.
\end{aligned}
\end{equation}
Together with the constraints $2 c_{14,28} + 6 c_{11,25} = 0$ and $2 c_{18,32} + 6 c_{21,35} = 0$, we obtain the following expression for the function $\hat{u}_{33}$:
\begin{equation}
	\label{classO^{-}(2)-u_{33}}
	\hat{u}_{33} = -c_{36} x_2^2 + c_{37} x_2 + c_{38} x_1 x_2 + c_{39} x_1^2\,.
\end{equation}

\begin{prop}
The universal displacement fields in the class $\mathbb{O}^-(2)$ linear strain-gradient elastic solids have the form \eqref{transverse-isotropy-linear-solution}, with the functions $h_i$, $i=1,2$, given by \eqref{classO^{-}(2)-mh1} and \eqref{classO^{-}(2)-mh2}, the functions $k_i$, $i=1,2$, given by \eqref{classO^{-}(2)-k} and \eqref{classO^{-}(2)-f}, and the function $\hat{u}_{33}$ given by \eqref{classO^{-}(2)-u_{33}}.
\end{prop}

\subsubsection{$\infty$-gonal class $\mathbb{O}(2) \oplus \mathbb{Z}_2^c$} 

Tensor $\boldsymbol{\mathsf{A}}_{\mathbb{O}(2) \oplus \mathbb{Z}_2^c}$ has $21$ independent components and is of the same form as that of class $\mathbb{O}(2)$. Tensor $\boldsymbol{\mathsf{M}}_{\mathbb{O}(2) \oplus \mathbb{Z}_2^c}$ is a null tensor. Tensor $\boldsymbol{\mathsf{C}}_{\mathbb{O}(2) \oplus \mathbb{Z}_2^c}$ has $5$ independent components and corresponds to the Curie group with Hermann--Mauguin symbol $\infty/mm$.

The additional universality PDEs for this case read
\begin{equation}
\label{classO(2)+Z^c_2-1}
\begin{dcases}
	\tfrac{1}{\sqrt{2}} \tfrac{\partial^4 k_1}{\partial x_2^4}
	+ \tfrac{x_3}{\sqrt{2}} \tfrac{\partial^4 mk_1}{\partial x_2^4}
	+ \tfrac{1}{\sqrt{2}} \tfrac{\partial^4 k_2}{\partial x_1 \partial x_2^3}
	+ \tfrac{x_3}{\sqrt{2}} \tfrac{\partial^4 h_2}{\partial x_1 \partial x_2^3} \\
	\quad + (1+\sqrt{2})\!\left( \tfrac{\partial^4 k_1}{\partial x_1^2 \partial x_2^2}
	+ x_3 \tfrac{\partial^4 h_1}{\partial x_1^2 \partial x_2^2}\right)
	+ \tfrac{\partial^4 k_2}{\partial x_1^3 \partial x_2}
	+ x_3 \tfrac{\partial^4 h_2}{\partial x_1^3 \partial x_2}
	+ \tfrac{\partial^4 k_1}{\partial x_1^4 }
	+ x_3 \tfrac{\partial^4 h_1}{\partial x_1^4} = 0\,, \\
	-\tfrac{\partial^4 k_1}{\partial x_1^2 \partial x_2^2}
	- x_3 \tfrac{\partial^4 h_1}{\partial x_1^2 \partial x_2^2}
	- \tfrac{\partial^4 k_2}{\partial x_1^3 \partial x_2}
	- x_3 \tfrac{\partial^4 h_2}{\partial x_1^3 \partial x_2} = 0\,, \\
	-\tfrac{\partial^4 k_1}{\partial x_2^4}
	- x_3 \tfrac{\partial^4 h_1}{\partial x_2^4}
	+ \tfrac{\partial^4 k_2}{\partial x_1 \partial x_2^3}
	+ x_3 \tfrac{\partial^4 h_2}{\partial x_1^3 \partial x_2} \\
	\quad -\tfrac{\partial^4 k_1}{\partial x_1^2 \partial x_2^2}
	- x_3 \tfrac{\partial^4 h_1}{\partial x_1^2 \partial x_2^2}
	+ \tfrac{\partial^4 k_2}{\partial x_1^3 \partial x_2}
	+ x_3 \tfrac{\partial^4 h_2}{\partial x_1^3 \partial x_2} = 0\,, \\
	-\tfrac{1}{\sqrt{2}} \tfrac{\partial^4 k_1}{\partial x_2^4}
	+ \tfrac{x_3}{\sqrt{2}} \tfrac{\partial^4 h_1}{\partial x_2^4}
	+ \left(1-\tfrac{1}{\sqrt{2}}\right)\!\left(\tfrac{\partial^4 k_2}{\partial x_1 \partial x_2^3}
	+ x_3 \tfrac{\partial^4 h_2}{\partial x_1 \partial x_2^3}\right) \\
	\quad + \left(-1+\sqrt{2}\right)\!\left(\tfrac{\partial^4 k_1}{\partial x_1^2 \partial x_2^2}
	+ x_3 \tfrac{\partial^4 h_1}{\partial x_1^2 \partial x_2^2}\right)
	+ \tfrac{\partial^4 k_2}{\partial x_1^3 \partial x_2}
	+ x_3 \tfrac{\partial^4 h_2}{\partial x_1^3 \partial x_2} = 0\,,
\end{dcases}
\end{equation}
and
\begin{equation}
\label{classO(2)+Z^c_2-2}
\begin{dcases}
	\frac{\partial^4 k_2}{\partial x_2^4}+x_3 \frac{\partial^4 h_2}{\partial x_2^4}+\frac{\partial^4 k_1}{\partial x_1 \partial x_2^3}+x_3 \frac{\partial^4 h_1}{\partial x_1 \partial x_2^3}+(1+\sqrt{2})\left(\frac{\partial^4 k_2}{\partial x_1^2 \partial x_2^2}+x_3 \frac{\partial^4 h_2}{\partial x_1^2 \partial x_2^2}\right) \\
	\quad +\frac{1}{\sqrt{2}}\left(\frac{\partial^4 k_1}{\partial x_1^3 \partial x_2}+x_3 \frac{\partial^4 h_1}{\partial x_1^3 \partial x_2}+\frac{\partial^4 k_2}{\partial x_1^4}+x_3 \frac{\partial^4 h_2}{\partial x_1^4}\right)=0\,, \\
	 -\frac{\partial^4 k_1}{\partial x_1 \partial x_2^3}-x_3 \frac{\partial^4 h_1}{\partial x_1 \partial x_2^3}-\frac{\partial^4 k_2}{\partial x_1^2 \partial x_2^2}-x_3 \frac{\partial^4 h_2}{\partial x_1^2 \partial x_2^2}=0\,, \\
	 \frac{\partial^4 k_1}{\partial x_1 \partial x_2^3}+x_3 \frac{\partial^4 h_1}{\partial x_1 \partial x_2^3}-\frac{\partial^4 k_2}{\partial x_1^2 \partial x_2^2}-x_3 \frac{\partial^4 h_2}{\partial x_1^2 \partial x_2^2} \\
	 \quad +\frac{\partial^4 k_1}{\partial x_1^3 \partial x_2}+x_3 \frac{\partial^4 h_1}{\partial x_1^3 \partial x_2}-\frac{\partial^4 k_2}{\partial x_1^4}-x_3 \frac{\partial^4 h_2}{\partial x_1^4}=0\,, \\
	 \frac{\partial^4 k_1}{\partial x_1 \partial x_2^3}+x_3 \frac{\partial^4 h_1}{\partial x_1 \partial x_2^3}-\left(-1+\sqrt{2}\right)\left(\frac{\partial^4 k_2}{\partial x_1^2 \partial x_2^2}+x_3 \frac{\partial^4 h_2}{\partial x_1^2 \partial x_2^2}\right) \\
	 \quad +\left(1-\frac{1}{\sqrt{2}}\right)\left(\frac{\partial^4 k_1}{\partial x_1^3 \partial x_2}+x_3 \frac{\partial^4 h_1}{\partial x_1^3 \partial x_2}\right)-\frac{1}{\sqrt{2}}\left(\frac{\partial^4 k_2}{\partial x_1^4}+x_3 \frac{\partial^4 h_2}{\partial x_1^4}\right)=0\,,
\end{dcases}
\end{equation}
and
\begin{equation}
\label{classO(2)+Z^c_2-3}
\begin{dcases}
	 \frac{\partial^3 h_2}{\partial x_2^3}+\frac{\partial^3 h_1}{\partial x_1 \partial x_2^2}+\frac{\partial^3 h_2}{\partial x_1^2 \partial x_2}+\frac{\partial^3 h_1}{\partial x_1^3}=0\,, \\
	 \frac{\partial^3 h_1}{\partial x_1 \partial x_2^2}+\frac{\partial^3 h_2}{\partial x_1^2 \partial x_2}=0\,, \\
	 \frac{\partial^3 h_2}{\partial x_1 \partial x_2^3}+\frac{\partial^4 \hat{u}_{33}}{\partial x_2^4}+\frac{\partial^3 h_1}{\partial x_1 \partial x_2^2}+\frac{\partial^3 h_2}{\partial x_1^2 \partial x_2}+2 \frac{\partial^4 \hat{u}_{33}}{\partial x_1^2 \partial x_2^2}+\frac{\partial^3 h_1}{\partial x_1^3}+\frac{\partial^4 \hat{u}_{33}}{\partial x_1^4}=0\,, \\
	 -\frac{\partial^3 h_1}{\partial x_1 \partial x_2^2}-\frac{\partial^3 h_2}{\partial x_1^2 \partial x_2}=0\,, \\
	 -\sqrt{2}\frac{\partial^3 h_1}{\partial x_1 \partial x_2^2}-\left(-2+\sqrt{2}\right)\frac{\partial^3 h_2}{\partial x_1^2 \partial x_2}-2\left(-1+\sqrt{2}\right)\frac{\partial^4 \hat{u}_{33}}{\partial x_1^2 \partial x_2^2}=0\,, \\
	 2\frac{\partial^3 h_1}{\partial x_1 \partial x_2^2}+\frac{\partial^3 h_2}{\partial x_1^2 \partial x_2}+\frac{\partial^4 \hat{u}_{33}}{\partial x_1^2 \partial x_2^2}=0\,.
\end{dcases}
\end{equation}
From this set of equations one obtains the following additional constraint on $\hat{u}_{33}$:
\begin{equation}
\label{classO(2)+Z^c_2-4}
	\frac{\partial^4 \hat{u}_{33}}{\partial x_1^2 \partial x_2^2}=0\,.
\end{equation}
For $h_i$, $i=1,2$, we have
\begin{equation}
\label{classO(2)+Z^c_2-5}
	\frac{\partial^3 h_1}{\partial x_1 \partial x_2^2}
	=\frac{\partial^3 h_1}{\partial x_1^3}
	=\frac{\partial^3 h_2}{\partial x_1^2 \partial x_2}
	=\frac{\partial^3 h_2}{\partial x_2^3}=0\,.
\end{equation}
From this set of equations, we conclude that the function $h_1$ must be of the following form
\begin{equation}
\label{classO(2)+Z^c_2-6}
	h_1=f_1(x_2)+x_1 f_2(x_2)+x_1^2 f_3(x_2)
	\ \text{with} \ 
	f_2''(x_2)+2 x_1 f_3''(x_2)=0\,.
\end{equation}
Similarly, for the function $h_2$ we obtain 
\begin{equation}
\label{classO(2)+Z^c_2-7}
	h_2=f_4(x_1)+x_2 f_5(x_1)+x_2^2 f_6(x_1)
	\ \text{with} \ 
	f_5''(x_1)+2 x_2 f_6''(x_1)=0\,.
\end{equation}
For the functions $k_i$, $i=1,2$, we find
\begin{equation}
\label{classO(2)+Z^c_2-8}
	\frac{\partial^4 k_i}{\partial x_1^2 \partial x_2^2}
	=\frac{\partial^4 k_i}{\partial x_1^4}
	=\frac{\partial^4 k_i}{\partial x_2^4}=0\,.
\end{equation}
Finally, we have
\begin{equation}
\label{classO(2)+Z^c_2-9}
\begin{aligned}
	& k_1 = f_7(x_2) + x_1 f_8(x_2) 
	+ f_9(x_1) + x_2 f_{10}(x_1)\,, \\
          & k_2 = f_{11}(x_2) + x_1 f_{12}(x_2) 
	+ f_{13}(x_1) + x_2 f_{14}(x_1)\,, \\
	& f_9''''(x_1) + x_2 f_{10}''''(x_2) = 0\,, \\
	& f_7''''(x_2) + x_1 f_8''''(x_2) = 0\,, \\
          & f_{13}''''(x_1) + x_2 f_{14}''''(x_2) = 0\,, \\
	& f_{11}''''(x_2) + x_1 f_{12}''''(x_2) = 0\,.
\end{aligned}
\end{equation}
With these expressions, all equations \eqref{classO(2)+Z^c_2-1}--\eqref{classO(2)+Z^c_2-3} are satisfied.
In summary, we obtain the following result.
\begin{prop}
The sets of universal displacements of linear strain-gradient elasticity for the class $\mathbb{O}(2) \oplus \mathbb{Z}_2^c$ are given by \eqref{transverse-isotropy-linear-solution}, subject to \eqref{classO(2)+Z^c_2-4}, \eqref{classO(2)+Z^c_2-6}, \eqref{classO(2)+Z^c_2-7}, and \eqref{classO(2)+Z^c_2-9}\,.
\end{prop}

\subsubsection{$\infty$-gonal class $\mathbb{SO}(2) \oplus \mathbb{Z}_2^c$} 

Tensor $\boldsymbol{\mathsf{A}}_{\mathbb{SO}(2) \oplus \mathbb{Z}_2^c}$ has $31$ independent components and is identical in form to that of class $\mathbb{SO}(2)$. Tensor $\boldsymbol{\mathsf{M}}_{\mathbb{SO}(2) \oplus \mathbb{Z}_2^c}$ is a null tensor. Tensor $\boldsymbol{\mathsf{C}}_{\mathbb{O}(2) \oplus \mathbb{Z}_2^c}$ has $5$ independent components and corresponds to the Curie group with Hermann--Mauguin symbol $\infty/mm$.

The additional universality PDEs for this case read
\begin{equation}
\label{classSO(2)+Z^c_2-1}
\begin{dcases}
 \frac{1}{\sqrt{2}}\frac{\partial^4 k_1}{\partial x_2^4}+\frac{x_3}{\sqrt{2}} \frac{\partial^4 h_1}{\partial x_2^4}+\frac{1}{\sqrt{2}} \frac{\partial^4 k_1}{\partial x_1 \partial x_2^3}+\frac{x_3}{\sqrt{2}}\frac{\partial^4 h_2}{\partial x_1 \partial x_2^3}=0 \,, \\
 \left(1+\sqrt{2}\right)\left(\frac{\partial^4 k_1}{\partial x_1^2 \partial x_2^2}+x_3 \frac{\partial^4 h_1}{\partial x_1^2 \partial x_2^2}\right)+\frac{\partial^4 k_2}{\partial x_1^3 \partial x_2}+x_3 \frac{\partial^4 h_2}{\partial x_1^3 \partial x_2}+\frac{\partial^4 k_1}{\partial x_1^4}+x_3 \frac{\partial^4 h_1}{\partial x_1^4}=0 \,, \\
 -\frac{\partial^4 k_1}{\partial x_1^2 \partial x_2^2}-x_3 \frac{\partial^4 h_1}{\partial x_1^2 \partial x_2^2}-\frac{\partial^4 k_2}{\partial x_1^3 \partial x_2}-x_3 \frac{\partial^4 h_2}{\partial x_1^3 \partial x_2}=0 \,, \\
 -\frac{\partial^4 k_1}{\partial x_2^4}-x_3 \frac{\partial^4 h_1}{\partial x_1^4}+\frac{\partial^4 k_2}{\partial x_1 \partial x_2^3}+x_3 \frac{\partial^4 h_2}{\partial x_1 \partial x_2^3}=0 \,, \\
 -\frac{\partial^4 k_1}{\partial x_1^2 \partial x_2^2}-x_3 \frac{\partial^4 h_1}{\partial x_1^2 \partial x_2^2}+\frac{\partial^4 k_2}{\partial x_1^3 \partial x_2}+x_3 \frac{\partial^4 h_2}{\partial x_1^3 \partial x_2}=0 \,, \\
 -\frac{1}{\sqrt{2}}\frac{\partial^4 k_1}{\partial x_2^4}+\frac{x_3}{\sqrt{2}}\frac{\partial^4 h_1}{\partial x_2^4}+\left(1-\frac{1}{\sqrt{2}}\right)\left(\frac{\partial^4 k_2}{\partial x_1 \partial x_2^3}+x_3 \frac{\partial^4 h_2}{\partial x_1 \partial x_2^3}\right)=0 \,, \\
 \left(-1+\sqrt{2}\right)\left(\frac{\partial^4 k_1}{\partial x_1^2 \partial x_2^2}\right)+\frac{\partial^4 k_2}{\partial x_1^3 \partial x_2}+x_3\frac{\partial^4 h_2}{\partial x_1^3 \partial x_2}=0 \,, \\
 \frac{\partial^4 k_2}{\partial x_2^4}+x_3 \frac{\partial^4 h_2}{\partial x_2^4}-\frac{\sqrt{2}}{2}\left(\frac{\partial^4 k_1}{\partial x_1 \partial x_2^3}+x_3 \frac{\partial^4 h_1}{\partial x_1 \partial x_2^3}\right)-\frac{3 \sqrt{2}}{2}\left(\frac{\partial^4 k_2}{\partial x_1^2 \partial x_2^2}+x_3 \frac{\partial^4 h_2}{\partial x_1^2 \partial x_2^2}\right)  \\
 +\left(2-3 \sqrt{2}\right)\left(\frac{\partial^4 k_1}{\partial x_1^3 \partial x_2}+x_3 \frac{\partial^4 h_1}{\partial x_1^3 \partial x_2}\right)-\sqrt{2}\left(\frac{\partial^4 k_2}{\partial x_1^4}+x_3 \frac{\partial^4 h_2}{\partial x_1^4}\right)=0 \,,
\end{dcases}
\end{equation}
and
\begin{equation}
\label{classSO(2)+Z^c_2-2}
\begin{dcases}
  \frac{\partial^4 k_2}{\partial x_2^4}+x_3 \frac{\partial^4 h_2}{\partial x_2^4}+\frac{\partial^4 k_1}{\partial x_1 \partial x_2^3}+x_3 \frac{\partial^4 h_1}{\partial x_1 \partial x_2^3}+\left(1+\sqrt{2}\right)\left(\frac{\partial^4 k_2}{\partial x_1^2 \partial x_2^2}+x_3\frac{\partial^4 h_2}{\partial x_1^2 \partial x_2^2}\right)  \\
 +\frac{1}{\sqrt{2}}\left(\frac{\partial^4 k_1}{\partial x_1^3 \partial x_2}+x_3 \frac{\partial^4 h_1}{\partial x_1^3 \partial x_2}+\frac{\partial^4 k_2}{\partial x_1^4}+x_3 \frac{\partial^4 h_2}{\partial x_1^4}\right)=0 \,, \\
 -\frac{\partial^4 k_1}{\partial x_1 \partial x_2^3}-x_3 \frac{\partial^4 h_1}{\partial x_1 \partial x_2^3}-\frac{\partial^4 k_2}{\partial x_1^2 \partial x_2^2}-x_3 \frac{\partial^4 h_2}{\partial x_1^2 \partial x_2^2}=0 \,, \\
 \frac{\partial^4 k_1}{\partial x_1 \partial x_2^3}+x_3 \frac{\partial^4 h_1}{\partial x_1 \partial x_2^3}-\frac{\partial^4 k_2}{\partial x_1^2 \partial x_2^2}-x_3 \frac{\partial^4 h_2}{\partial x_1^2 \partial x_2^2}  \\
 +\frac{\partial^4 k_1}{\partial x_1^3 \partial x_2}+x_3 \frac{\partial^4 h_1}{\partial x_1^3 \partial x_2}-\frac{\partial^4 k_2}{\partial x_1^4}-x_3 \frac{\partial^4 h_2}{\partial x_2^4} =0 \,, \\
 \frac{\partial^4 k_1}{\partial x_1 \partial x_2^3}+x_3 \frac{\partial^4 h_1}{\partial x_1 \partial x_2^3}-\left(-1+\sqrt{2}\right)\left(\frac{\partial^4 k_2}{\partial x_1^2 \partial x_2^2}+x_3 \frac{\partial^4 h_2}{\partial x_1^2 \partial x_2^2}\right)  \\
 +\left(1-\frac{1}{\sqrt{2}}\right)\left(\frac{\partial^4 k_1}{\partial x_1^3 \partial x_2}+x_3 \frac{\partial^4 h_1}{\partial x_1^3 \partial x_2}\right)-\frac{1}{\sqrt{2}}\left(\frac{\partial^4 k_2}{\partial x_1^4 \partial x_2}+x_3 \frac{\partial^4 h_2}{\partial x_1^4 \partial x_2}\right)=0 \,, \\
 \frac{1}{\sqrt{2}}\left(\frac{\partial^4 k_1}{\partial x_1^4}+x_3 \frac{\partial^4 h_1}{\partial x_2^4}\right)+\left(1-\frac{3}{\sqrt{2}}\right)\left(\frac{\partial^4 k_2}{\partial x_1 \partial x_2^3}+x_3 \frac{\partial^4 h_2}{\partial x_1 \partial x_2^3}\right) -\frac{3}{\sqrt{2}}\left(\frac{\partial^4 k_1}{\partial x_1^2 \partial x_2^2}+x_3 \frac{\partial^4 h_1}{\partial x_1^2 \partial x_2^2}\right)  \\
 +\frac{1}{\sqrt{2}}\left(\frac{\partial^4 k_2}{\partial x_1^3 \partial x_2}+x_3 \frac{\partial^4 h_2}{\partial x_1^3 \partial x_2}\right)+\frac{\partial^4 k_1}{\partial x_1^4}+x_3 \frac{\partial^4 h_1}{\partial x_1^4}=0 \,,
\end{dcases}
\end{equation}
as well as
\begin{equation}
\label{classSO(2)+Z^c_2-3}
\begin{dcases}
 \frac{\partial^3 h_2}{\partial x_2^3}+\frac{\partial^3 h_1}{\partial x_1 \partial x_2^2}+\frac{\partial^3 h_2}{\partial x_1^2 \partial x_2}+\frac{\partial^3 h_1}{\partial x_1^3}=0 \,, \\
 \frac{\partial^3 h_1}{\partial x_1 \partial x_2^2}=\frac{\partial^3 h_2}{\partial x_1^2 \partial x_2}=0 \,, \\
 \frac{\partial^3 h_2}{\partial x_2^3} +\frac{\partial^4 u_{33}}{\partial x_2^4}+\frac{\partial^3 h_1}{\partial x_1 \partial x_2^2}+\frac{\partial^3 h_2}{\partial x_1^2 \partial x_2}+\frac{\partial^4 u_{33}}{\partial x_1^2 \partial x_2^2}+\frac{\partial^3 h_1}{\partial x_1^3}+\frac{\partial^4 \hat{u}_{33}}{\partial x_1^4}=0 \,, \\
 \frac{\partial^3 h_2}{\partial x_2^3}+\frac{\partial^3 h_1}{\partial x_1 \partial x_2^2}\frac{\partial^3 h_2}{\partial x_1^2 \partial x_2}+\frac{\partial^3 h_1}{\partial x_1^3}=0 \,, \\
 -\frac{\partial^3 h_1}{\partial x_1 \partial x_2^2}-\frac{\partial^3 h_2}{\partial x_1^2 \partial x_2}=0 \,, \\
 2\frac{\partial^3 h_1}{\partial x_1 \partial x_2^2}+\frac{\partial^3 h_2}{\partial x_1^2 \partial x_2}+\frac{\partial^4 \hat{u}_{33}}{\partial x_1^2 \partial x_2^2}=0 \,, \\
 \frac{\partial^3 h_1}{\partial x_1 \partial x_2^2}+\frac{\partial^3 h_2}{\partial x_1^2 \partial x_2}=0 \,, \\
 -3 \frac{\partial^3 h_2}{\partial x_1 \partial x_2^2}+\frac{\partial^4 \hat{u}_{33}}{\partial x_1 \partial x_2^3}-2 \frac{\partial^3 h_1}{\partial x_1^2 \partial x_2}-\frac{\partial^3 h_2}{\partial x_1^3}+\frac{\partial^4 \hat{u}_{33}}{\partial x_1^3 \partial x_2}=0 \,, \\
 \frac{\partial^3 h_1}{\partial x_2^3}+\frac{\partial^4 \hat{u}_{33}}{\partial x_1 \partial x_2^3}-3 \frac{\partial^3 h_1}{\partial x_1^2 \partial x_2}-\frac{\partial^3 h_2}{\partial x_1^3}=0 \,, \\
 \frac{\partial^3 h_2}{\partial x_1 \partial x_2^2}+\frac{\partial^4 \hat{u}_{33}}{\partial x_1 \partial x_2^3}+\frac{\partial^3 h_1}{\partial x_1^2 \partial x_2}+\frac{\partial^4 \hat{u}_{33}}{\partial x_1^3 \partial x_2}=0 \,.
\end{dcases}
\end{equation}

From the third set of equations, eq. (\ref{classSO(2)+Z^c_2-3}), we obtain for functions $h_i$ and $\hat{u}_{33}$: 
\begin{equation}
\label{classSO(2)+Z^c_2-4}
\begin{aligned}
& \frac{\partial^3 h_1}{\partial x_1 \partial x_2^2}=\frac{\partial^3 h_2}{\partial x_1^2 \partial x_2}=0 \,, \\
& \frac{\partial^3 h_2}{\partial x_2^3}=\frac{\partial^3 h_1}{\partial x_1^3}=0 \,, \\
& \frac{\partial^3 h_1}{\partial x_2^3}+\frac{\partial^3 h_2}{\partial x_1^3}=0 \,, \\
& \frac{\partial^4 \hat{u}_{33}}{\partial x_1^2 \partial x_2^2}=0 \,, \\
& \frac{\partial^4 \hat{u}_{33}}{\partial x_1 \partial x_2^3}+\frac{\partial^4 \hat{u}_{33}}{\partial x_1^3 \partial x_2}=0 \,, \\
& \frac{\partial^4 \hat{u}_{33}}{\partial x_1 \partial x_2^3}-5 \frac{\partial^3 h_1}{\partial x_1^2 \partial x_2}=0 \,.
\end{aligned}
\end{equation}
With these expressions, the third set of universality PDEs is satisfied, whereas the first and second sets remain unchanged and therefore provide additional universality PDEs that the functions must satisfy. We obtain the following result.
\begin{prop}
The sets of universal displacements of linear strain-gradient elasticity for class $\mathbb{SO}(2) \oplus \mathbb{Z}^c_2$ are those in \eqref{transverse-isotropy-linear-solution}, subject to the conditions \eqref{classSO(2)+Z^c_2-1}, \eqref{classSO(2)+Z^c_2-2}, and \eqref{classSO(2)+Z^c_2-4}.
\end{prop}

\subsubsection{$\infty$-gonal class $\mathbb{O}(2)$} 

Tensor $\boldsymbol{\mathsf{A}}_{\mathbb{O}(2) \oplus \mathbb{Z}_2^c}$ has $21$ independent components and has the form
\begin{equation}
    \boldsymbol{\mathsf{A}}_{\mathbb{O}(2) \oplus \mathbb{Z}_2^c}(\mathbf{x})=
    \begin{bmatrix}
        A^{(11)} & 0 & 0 & 0 \\
        & A^{(11)} & 0 & 0 \\
        & & H^{(6)} & 0 \\
        & & & J^{(4)}
    \end{bmatrix}_S +
    \begin{bmatrix}
        0 & 0 & 0 & 0 \\
        & 0 & 0 & 0 \\
        & & f(J^{(4)}) & 0 \\
        & & & 0
    \end{bmatrix}_S \,,
\end{equation}
where the matrices $A^{(11)}$, $H^{(6)}$, $J^{(4)}$, and $f\left(J^{(4)}\right)$ are the same as in the preceding classes. Tensor $\boldsymbol{\mathsf{M}}_{\mathbb{O}(2)}$ has $8$ independent components and has the form given in \eqref{M_O(2)}. Tensor $\boldsymbol{\mathsf{C}}_{\mathbb{O}(2) \oplus \mathbb{Z}_2^c}$ corresponds to the Curie group with Hermann--Mauguin symbol $\infty 2$ and has $5$ independent components.

The additional universality PDEs in this case are
\begin{equation}
\label{classO(2)-1}
\begin{dcases}
	 \sqrt{2} \frac{\partial^2 h_2}{\partial x_2^2}+\sqrt{2}\frac{\partial^3 \hat{u}_{33}}{\partial x_2^3}+\left(4-\sqrt{2}\right)\frac{\partial^2 h_1}{\partial x_1 \partial x_2}+2 \left(1+\sqrt{2}\right)\frac{\partial^2 h_2}{\partial x_1^2}-\left(-2+\sqrt{2}\right)\frac{\partial^3 \hat{u}_{33}}{\partial x_1^2 \partial x_2}=0\,, \\
	 \left(2+\tfrac{1}{\sqrt{2}}\right)\frac{\partial^2 h_2}{\partial x_2^2}+\tfrac{1}{2}\left(\sqrt{2}\frac{\partial^3 \hat{u}_{33}}{\partial x_2^3}+\left(2+\sqrt{2}\right)\frac{\partial^2 h_1}{\partial x_1 \partial x_2}+2 \sqrt{2}\frac{\partial^2 h_2}{\partial x_1^2}-\left(-2+\sqrt{2}\right)\frac{\partial^3 \hat{u}_{33}}{\partial x_1^2 \partial x_2}\right)=0\,, \\
	 \frac{\partial^2 h_2}{\partial x_2^2}+\frac{\partial^2 h_2}{\partial x_1^2}=0\,, \\
	 -\frac{\partial^2 h_2}{\partial x_2^2}+\frac{1}{\sqrt{2}}\frac{\partial^2 h_1}{\partial x_1 \partial x_2}+\frac{1}{2}\left(2+\sqrt{2}\right)\frac{\partial^2 h_2}{\partial x_2^2}-\left(-1+\sqrt{2}\right)\frac{\partial^3 \hat{u}_{33}}{\partial x_1^2 \partial x_2}=0\,, \\
	 -\frac{\partial^2 h_2}{\partial x_2^2} - \frac{\partial^3 u_{33}}{\partial x_2^3}-\frac{1}{2}\left(2+\sqrt{2}\right)\frac{\partial^2 h_1}{\partial x_1 \partial x_2}+\tfrac{1}{\sqrt{2}}\left(\frac{\partial^2 h_2}{\partial x_1^2}-2 \frac{\partial^3 \hat{u}_{33}}{\partial x_1^2 \partial x_2}\right)=0\,, \\
	 \tfrac{1}{\sqrt{2}}\frac{\partial^4 k_1}{\partial x_2^4} +\tfrac{x_3}{\sqrt{2}}\frac{\partial^4 h_1}{\partial x_2^4}+\tfrac{1}{\sqrt{2}}\frac{\partial^4 k_2}{\partial x_1 \partial x_2^3}+\tfrac{x_3}{\sqrt{2}}\frac{\partial^4 h_2}{\partial x_1 \partial x_2^3}+\left(1+\sqrt{2}\right)\left(\frac{\partial^4 k_1}{\partial x_1^2 \partial x_2^2}+x_3 \frac{\partial^4 h_1}{\partial x_1^2 \partial x_2^2}\right) \\
	 \qquad +\frac{\partial^4 k_2}{\partial x_1^3 \partial x_2} +x_3 \frac{\partial^4 h_2}{\partial x_1^3 \partial x_2}+\frac{\partial^4 k_1}{\partial x_1^4}+x_3 \frac{\partial^4 h_1}{\partial x_1^4}=0\,, \\
	 -\frac{\partial^4 k_1}{\partial x_1^2 \partial x_2^2}-x_3 \frac{\partial^4 h_1}{\partial x_1^2 \partial x_2^2}-\frac{\partial^4 k_2}{\partial x_1^3 \partial x_2}-x_3 \frac{\partial^4 h_2}{\partial x_1^3 \partial x_2}=0\,, \\
	 -\frac{\partial^4 k_1}{\partial x_2^4}-x_3 \frac{\partial^4 h_1}{\partial x_2^4}+\frac{\partial^4 k_2}{\partial x_1 \partial x_2^3}+x_3 \frac{\partial^4 h_2}{\partial x_1 \partial x_2^3}-\frac{\partial^4 k_1}{\partial x_1^2 \partial x_2^2}-x_3 \frac{\partial^4 h_1}{\partial x_1^2 \partial x_2^2}+\frac{\partial^4 k_2}{\partial x_1^3 \partial x_2}+x_3 \frac{\partial^4 h_2}{\partial x_1^3 \partial x_2}=0\,, \\
	 -\tfrac{1}{\sqrt{2}}\frac{\partial^4 k_1}{\partial x_2^4}+\tfrac{x_3}{\sqrt{2}}\frac{\partial^4 h_1}{\partial x_2^4}+\left(1-\tfrac{1}{\sqrt{2}}\right)\left(\frac{\partial^4 k_2}{\partial x_1 \partial x_2^3}+x_3 \frac{\partial^4 h_2}{\partial x_1 \partial x_2^3}\right) \\
	 \qquad -\left(-1+\sqrt{2}\right)\left(\frac{\partial^4 k_1}{\partial x_1^2 \partial x_2^2}+x_3 \frac{\partial^4 h_1}{\partial x_1^2 \partial x_2^2}\right)+\frac{\partial^4 k_2}{\partial x_1^3 \partial x_2}+x_3 \frac{\partial^4 h_2}{\partial x_1^3 \partial x_2}=0\,,
\end{dcases}
\end{equation}
and
\begin{equation}
\label{classO(2)-2}
\begin{dcases}
	 -2\left(1+\sqrt{2}\right)\frac{\partial^2 h_1}{\partial x_2^2}-\left(-4+3\sqrt{2}\right)\frac{\partial^2 h_2}{\partial x_1 \partial x_2}+\left(-2+\sqrt{2}\right)\frac{\partial^3 \hat{u}_{33}}{\partial x_1 \partial x_2^2}-\sqrt{2}\left(\frac{\partial^2 h_1}{\partial x_1^2}+\frac{\partial^3 \hat{u}_{33}}{\partial x_1^3}\right)=0\,, \\
	 -2\sqrt{2}\frac{\partial^2 h_1}{\partial x_2^2}+\left(-2+\sqrt{2}\right)\frac{\partial^2 h_2}{\partial x_1 \partial x_2}+\left(-2+\sqrt{2}\right)\frac{\partial^3 \hat{u}_{33}}{\partial x_1 \partial x_2^2}-\left(4+\sqrt{2}\right)\frac{\partial^2 h_1}{\partial x_1^2}-\sqrt{2}\frac{\partial^3 \hat{u}_{33}}{\partial x_1^3}=0\,, \\
	 -\frac{\partial^2 h_1}{\partial x_2^2}-\frac{\partial^2 h_1}{\partial x_1^2}=0\,, \\
	 \left(1+\frac{1}{\sqrt{2}}\right)\frac{\partial^2 h_1}{\partial x_2^2}-\frac{1}{\sqrt{2}}\frac{\partial^2 h_2}{\partial x_1 \partial x_2}-\left(-1+\sqrt{2}\right)\frac{\partial^3 \hat{u}_{33}}{\partial x_1 \partial x_2^2}+\frac{\partial^2 h_1}{\partial x_1^2}=0\,, \\
	 -\frac{1}{\sqrt{2}}\frac{\partial^2 h_1}{\partial x_2^2}+\left(1+\frac{1}{\sqrt{2}}\right)\frac{\partial^2 h_2}{\partial x_1 \partial x_2}+\sqrt{2}\frac{\partial^3 \hat{u}_{33}}{\partial x_1 \partial x_2^2}+\frac{\partial^2 h_1}{\partial x_1^2}+\frac{\partial^3 \hat{u}_{33}}{\partial x_1^3}=0\,, \\
	 \frac{\partial^4 k_2}{\partial x_2^4}+x_3 \frac{\partial^4 h_2}{\partial x_2^4}+\frac{\partial^4 k_1}{\partial x_1 \partial x_2^3}+x_3 \frac{\partial^4 h_1}{\partial x_1 \partial x_2^3}+\left(1+\sqrt{2}\right)\left(\frac{\partial^4 k_2}{\partial x_1^2 \partial x_2^2}+x_3 \frac{\partial^4 h_2}{\partial x_1^2 \partial x_2^2}\right) \\
	 \qquad +\frac{1}{\sqrt{2}}\left(\frac{\partial^4 k_1}{\partial x_1^3 \partial x_2}+x_3 \frac{\partial^4 h_1}{\partial x_1^3 \partial x_2}+\frac{\partial^4 k_2}{\partial x_1^4}+x_3 \frac{\partial^4 h_2}{\partial x_1^4}\right)=0\,, \\
	 -\frac{\partial^4 k_1}{\partial x_1 \partial x_2^3}-x_3 \frac{\partial^4 h_1}{\partial x_1 \partial x_2^3}-\frac{\partial^4 k_2}{\partial x_1^2 \partial x_2^2}-x_3 \frac{\partial^4 h_2}{\partial x_1^2 \partial x_2^2}=0\,, \\
	 \frac{\partial^4 k_1}{\partial x_1 \partial x_2^3}+x_3 \frac{\partial^4 h_1}{\partial x_1 \partial x_2^3}-\frac{\partial^4 k_2}{\partial x_1^2 \partial x_2^2}-x_3 \frac{\partial^4 h_2}{\partial x_1^2 \partial x_2^2} \\
	 \qquad +\frac{\partial^4 k_1}{\partial x_1^3 \partial x_2}+x_3 \frac{\partial^4 h_1}{\partial x_1^3 \partial x_2}-\frac{\partial^4 k_2}{\partial x_1^4}-x_3\frac{\partial^4 h_2}{\partial x_1^4}=0\,, \\
	 \frac{\partial^4 k_1}{\partial x_1 \partial x_2^3}+x_3 \frac{\partial^4 h_1}{\partial x_1 \partial x_2^3}-\left(-1+\sqrt{2}\right)\left(\frac{\partial^4 k_2}{\partial x_1^2 \partial x_2^2}+x_3 \frac{\partial^4 h_2}{\partial x_1^2 \partial x_2^2}\right) \\
	 \qquad +\left(1-\frac{1}{\sqrt{2}}\right)\left(\frac{\partial^4 k_1}{\partial x_1^3 \partial x_2}+x_3 \frac{\partial^4 h_1}{\partial x_1^3 \partial x_2}\right)-\frac{1}{\sqrt{2}}\left(\frac{\partial^4 k_2}{\partial x_1^4}+x_3 \frac{\partial^4 h_2}{\partial x_1^4}\right)=0\,,
\end{dcases}
\end{equation}
as well as
\begin{equation}
\label{classO(2)-3}
\begin{dcases}
	 \frac{\partial^3 k_1}{\partial x_2^3}+x_3 \frac{\partial^3 h_1}{\partial x_2^3}-\frac{\partial^3 k_2}{\partial x_1 \partial x_2^2}-x_3 \frac{\partial^3 h_2}{\partial x_1 \partial x_2^2 }  \\
	 +\frac{\partial^3 k_1}{\partial x_1^2 \partial x_2}+x_3 \frac{\partial^3 h_1}{\partial x_1^2 \partial x_2}-\frac{\partial^3 k_2}{\partial x_1^3}-x_3 \frac{\partial^3 h_2}{\partial x_1^3}=0\,, \\
	 -\frac{\partial^3 k_2}{\partial x_1 \partial x_2^2}-x_3 \frac{\partial^3 h_2}{\partial x_1 \partial x_2^2}+\frac{1}{\sqrt{2}}\left(\frac{\partial^3 k_1}{\partial x_2^3}+x_3 \frac{\partial^3 h_1}{\partial x_2^3}+\frac{\partial^3 k_2}{\partial x_1 \partial x_2^2}+x_3 \frac{\partial^3 h_2}{\partial x_1 \partial x_2^2}\right) \\
	 \qquad +\frac{\partial^3 k_1}{\partial x_1^2 \partial x_2}+x_3 \frac{\partial^3 h_1}{\partial x_1^2 \partial x_2}-\frac{1}{\sqrt{2}}\left(\frac{\partial^3 k_1}{\partial x_1^2 \partial x_2}+x_3 \frac{\partial^3 h_1}{\partial x_1^2 \partial x_2}+\frac{\partial^3 k_2}{\partial x_1^3}+x_3 \frac{\partial^3 h_2}{\partial x_1^3}\right)=0\,, \\
	 \frac{\partial^3 k_2}{\partial x_1 \partial x_2^2}+x_3 \frac{\partial^3 h_2}{\partial x_1 \partial x_2^2}-\frac{1}{\sqrt{2}}\left(\frac{\partial^3 k_1}{\partial x_2^3}+x_3 \frac{\partial^3 h_1}{\partial x_2^3}+\frac{\partial^3 k_2}{\partial x_1 \partial x_2^2}+x_3 \frac{\partial^3 h_2}{\partial x_1 \partial x_2^2}\right) \\
	 \qquad -\frac{\partial^3 k_1}{\partial x_1^2 \partial x_2}-x_3 \frac{\partial^3 h_1}{\partial x_1^2 \partial x_2}+\frac{1}{\sqrt{2}}\left(\frac{\partial^3 k_1}{\partial x_1^2 \partial x_2}+x_3 \frac{\partial^3 h_1}{\partial x_1^2 \partial x_2}+\frac{\partial^3 k_2}{\partial x_1^3}+x_3 \frac{\partial^3 h_2}{\partial x_1^3 }\right)=0\,, \\
	 \frac{\partial^3 h_2}{\partial x_2^3}+\frac{\partial^3 h_1}{\partial x_1 \partial x_2^2}+\frac{\partial^3 h_2}{\partial x_1^2 \partial x_2}+\frac{\partial^3 h_1}{\partial x_1^3}=0\,, \\
	 \frac{\partial^3 h_1}{\partial x_1 \partial x_2^2}+\frac{\partial^3 h_2}{\partial x_1^2 \partial x_2}=0\,, \\
	 \frac{\partial^3 h_2}{\partial x_2^3}+\frac{\partial^4 u_{33}}{\partial x_2^4}+\frac{\partial^3 h_1}{\partial x_1 \partial x_2^2}+\frac{\partial^3 h_2}{\partial x_1^2 \partial x_2}+2\frac{\partial^4 \hat{u}_{33}}{\partial x_1^2 \partial x_2^2}+\frac{\partial^3 h_1}{\partial x_1^3 \partial x_2}+\frac{\partial^4 \hat{u}_{33}}{\partial x_1^4}=0\,, \\
	 \frac{\partial^3 h_2}{\partial x_2^3}+\frac{\partial^3 h_1}{\partial x_1 \partial x_2^2}+\frac{\partial^3 h_2}{\partial x_1^2 \partial x_2}+\frac{\partial^3 h_1}{\partial x_1^3}=0\,, \\
	 -\frac{\partial^3 h_1}{\partial x_1 \partial x_2^2}-\frac{\partial^3 h_2}{\partial x_1^2 \partial x_2}=0\,, \\
	 -\sqrt{2} \frac{\partial^3 h_1}{\partial x_1 \partial x_2^2}-\left(-2+\sqrt{2}\right)\frac{\partial^3 h_2}{\partial x_1^2 \partial x_2}-2\left(-1+\sqrt{2}\right)\frac{\partial^4 \hat{u}_{33}}{\partial x_1^2 \partial x_2^2}=0\,, \\
	 2\frac{\partial^3 h_1}{\partial x_1 \partial x_2^2}+\frac{\partial^3 h_2}{\partial x_1^2 \partial x_2}+\frac{\partial^4\hat{u}_{33}}{\partial x_1^2 \partial x_2^2}=0\,.
\end{dcases}
\end{equation}
From the third set of equations, the remaining conditions are
\begin{align}
	\label{classO(2)-4}
	& \frac{\partial^4 \hat{u}_{33}}{\partial x_1^2 \partial x_2^2}
	=\frac{\partial^3 h_1}{\partial x_1 \partial x_2^2}
	=\frac{\partial^3 h_2}{\partial x_1^2\partial x_2}=0\,, \\
	\label{classO(2)-5}
	& \frac{\partial^3 h_2}{\partial x_2^3}+\frac{\partial^3 h_1}{\partial x_1^3}=0\,.
\end{align}
With \eqref{classO(2)-4} and \eqref{classO(2)-5} imposed, the third set \eqref{classO(2)-3} is satisfied identically. The first and second sets, \eqref{classO(2)-1} and \eqref{classO(2)-2}, remain coupled and hence must be retained in their present form. One then obtains the following result.
\begin{prop}
The universal displacements of linear strain-gradient elasticity for the class $\mathbb{SO}(2) \oplus \mathbb{Z}^c_2$ are given by \eqref{transverse-isotropy-linear-solution}, and the fields $h_i$, $i=1,2$, $\hat{u}_{33}$, and $k_i$, $i=1,2$, satisfy \eqref{classO(2)-1}, \eqref{classO(2)-2}, \eqref{classO(2)-4}, and \eqref{classO(2)-5}.
\end{prop}

\subsubsection{$\infty$-gonal class $\mathbb{SO}(2)$} 

Tensor $\boldsymbol{\mathsf{A}}_{\mathbb{SO}(2) \oplus \mathbb{Z}_2^c}$ has $31$ independent components and can be written as
\begin{equation}
    \boldsymbol{\mathsf{A}}_{\mathbb{SO}(2) \oplus \mathbb{Z}_2^c}(\mathbf{x})=
    \begin{bmatrix}
        A^{(11)} & B^{(6)} & 0 & 0 \\
        & A^{(11)} & 0 & 0 \\
        & & H^{(6)} & I^{(4)} \\
        & & & J^{(4)}
    \end{bmatrix}_S +
    \begin{bmatrix}
        0 & 0 & 0 & 0 \\
        & 0 & 0 & 0 \\
        & & f(J^{(4)}) & 0 \\
        & & & 0
    \end{bmatrix}_S \,,
\end{equation}
where the matrices $A^{(11)}$, $B^{(6)}$, $H^{(6)}$, $I^{(4)}$, $J^{(4)}$, and $f(J^{(4)})$ are the same as in the preceding classes. Tensor $\boldsymbol{\mathsf{M}}_{\mathbb{SO}(2)}$ has $20$ independent components and has the form \eqref{M_SO(2)}. Tensor $\boldsymbol{\mathsf{C}}_{\mathbb{O}(2) \oplus \mathbb{Z}_2^c}$ corresponds to the Curie group with Hermann--Mauguin symbol $\infty2$ and has $5$ independent components\,.

The additional universality PDEs for this case are
\begin{equation}
\label{classSO(2)-1}
\begin{dcases}
	 \sqrt{2} \frac{\partial^2 h_2}{\partial x_2^2}+\sqrt{2}\frac{\partial^3 u_{33}}{\partial x_2^3}
	+\left(-4+\sqrt{2}\right)\frac{\partial^2 h_1}{\partial x_1 \partial x_2}
	+2\left(1+\sqrt{2}\right)\frac{\partial^2 h_2}{\partial x_1^2}
	-\left(-2+\sqrt{2}\right)\frac{\partial^3 \hat{u}_{33}}{\partial x_1^2 \partial x_2}=0\,, \\
	 \left(2+\tfrac{1}{\sqrt{2}}\right)\frac{\partial^2 h_2}{\partial x_2^2}
	+\tfrac{1}{2}\left[\sqrt{2}\frac{\partial^3 u_{22}}{\partial x_2^3}
	+\left(2+\sqrt{2}\right)\frac{\partial^2 h_1}{\partial x_1 \partial x_2}
	+2\sqrt{2}\frac{\partial^2 h_2}{\partial x_1^2}
	-\left(-2+\sqrt{2}\right)\frac{\partial^3 \hat{u}_{33}}{\partial x_1^2 \partial x_2}\right]=0\,, \\
	 \frac{\partial^2 h_2}{\partial x_2^2}+\frac{\partial^2 h_2}{\partial x_1^2}=0\,, \\
	 \tfrac{1}{\sqrt{2}}\frac{\partial^2 h_1}{\partial x_2^2}-\tfrac{1}{\sqrt{2}}\frac{\partial^2 h_2}{\partial x_1 \partial x_2}-\sqrt{2} \frac{\partial^3 \hat{u}_{33}}{\partial x_1 \partial x_2^2}+\frac{\partial^3 \hat{u}_{33}}{\partial x_1^3}=0\,, \\
	 -\frac{1}{2}\left(2+\sqrt{2}\right) \frac{\partial^2 h_1}{\partial x_2^2}+\left(1+\tfrac{1}{\sqrt{2}}\right)\frac{\partial^2 h_2}{\partial x_1 \partial x_2}+\left(-1+\sqrt{2}\right)\frac{\partial^3 \hat{u}_{33}}{\partial x_1 \partial x_2^2}=0\,, \\
	 -\frac{\partial^2 h_2}{\partial x_1 \partial x_2^2}+\tfrac{1}{\sqrt{2}}\frac{\partial^2 h_1}{\partial x_1 \partial x_2}-\frac{1}{2}\left(2+\sqrt{2}\right)\frac{\partial^2 h_2}{\partial x_1^2}-\frac{2}{\sqrt{2}} \frac{\partial^3 \hat{u}_{33}}{\partial x_1^2 \partial x_2} + \frac{1}{\sqrt{2}} \frac{\partial^2 h_2}{\partial x_1^2} =0\,, \\
	 -\frac{\partial^2 h_2}{\partial x_2^2}-\frac{\partial^3 u_{33}}{\partial x_2^3}-\frac{1}{2}\left(2+\sqrt{2}\right)\frac{\partial^2 h_1}{\partial x_1 \partial x_2}+\tfrac{1}{\sqrt{2}}\left(\frac{\partial^2 h_2}{\partial x_1^2}-2\frac{\partial^3 \hat{u}_{33}}{\partial x_1^2 \partial x_2}\right)=0\,, \\
	 \frac{\partial^2 h_1}{\partial x_2^2}+\frac{\partial^2 h_2}{\partial x_1 \partial x_2}+\frac{\partial^2 h_1}{\partial x_1^2}=0\,, \\
	 -\tfrac{1}{\sqrt{2}}\frac{\partial^2 h_1}{\partial x_2^2}-\tfrac{1}{\sqrt{2}}\frac{\partial^2 h_2}{\partial x_1 \partial x_2}+\sqrt{2}\frac{\partial^3 \hat{u}_{33}}{\partial x_1 \partial x_2^2}- \frac{\partial^2 h_1}{\partial x_1^2}+\frac{\partial^3 \hat{u}_{33}}{\partial x_1^3}=0\,, \\
	 -\frac{\partial^2 h_2}{\partial x_2^2}=0\,, \\
	 \tfrac{1}{\sqrt{2}}\frac{\partial^2 h_1}{\partial x_2^2}+\left(-1+\tfrac{1}{\sqrt{2}}\right)\frac{\partial^2 h_2}{\partial x_1 \partial x_2}-\left(-1+\sqrt{2}\right)\frac{\partial^3 \hat{u}_{33}}{\partial x_1 \partial x_2^2}=0\,, \\
	 \tfrac{1}{\sqrt{2}}\left(\frac{\partial^4 k_1}{\partial x_2^4}+x_3 \frac{\partial^4 h_1}{\partial x_2^4}\right)+\tfrac{1}{\sqrt{2}}\left(\frac{\partial^4 k_2}{\partial x_1 \partial x_2^3}+x_3 \frac{\partial^4 h_2}{\partial x_1 \partial x_2^3}\right)+\left(1+\sqrt{2}\right)\left(\frac{\partial^4 k_1}{\partial x_1^2 \partial x_2^2}+x_3 \frac{\partial^4 h_1}{\partial x_1^2 \partial x_2^2}\right) \\
	 \qquad +\frac{\partial^4 k_2}{\partial x_1^3 \partial x_2}+x_3 \frac{\partial^4 h_2}{\partial x_1^3 \partial x_2}+\frac{\partial^4 k_1}{\partial x_1^4}+x_3 \frac{\partial^4 h_1}{\partial x_1^4 \partial x_2}=0\,, \\
	 -\frac{\partial^4 k_1}{\partial x_2^4}-x_3 \frac{\partial^4 h_1}{\partial x_2^4}+\frac{\partial^4 k_2}{\partial x_1 \partial x_2^3}+x_3 \frac{\partial^4 h_2}{\partial x_1 \partial x_2^3} -\frac{\partial^4 k_1}{\partial x_1^2 \partial x_2^2}-x_3 \frac{\partial^4 h_1}{\partial x_1^2 \partial x_2^2}+\frac{\partial^4 k_2}{\partial x_1^3 \partial x_2}+x_3 \frac{\partial^4 h_2}{\partial x_1^3 \partial x_2}=0\,, \\
	 -\tfrac{1}{\sqrt{2}}\left(\frac{\partial^4 k_1}{\partial x_2^4}+x_3 \frac{\partial^4 h_1}{\partial x_2^4}\right)+\left(1-\tfrac{1}{\sqrt{2}}\right)\left(\frac{\partial^4 k_2}{\partial x_1 \partial x_2^3}+x_3 \frac{\partial^4 h_2}{\partial x_1 \partial x_2^3}\right) \\
	 \qquad -\left(-1+\sqrt{2}\right)\left(\frac{\partial^4 k_1}{\partial x_1^2 \partial x_2^2}+x_3 \frac{\partial^4 h_1}{\partial x_1^2 \partial x_2^2}\right)+\frac{\partial^4 k_2}{\partial x_1^3 \partial x_2}+x_3 \frac{\partial^4 h_2}{\partial x_1^3 \partial x_2}=0\,, \\
	 \frac{\partial^4 k_2}{\partial x_2^4}+x_3 \frac{\partial^4 h_2}{\partial x_2^4}-\tfrac{\sqrt{2}}{2}\left(\frac{\partial^4 k_1}{\partial x_1 \partial x_2^3}+x_3 \frac{\partial^4 h_1}{\partial x_1 \partial x_2^3}\right)-\tfrac{3\sqrt{2}}{2}\left(\frac{\partial^4 k_2}{\partial x_1^2 \partial x_2^2}+x_3 \frac{\partial^4 h_2}{\partial x_1^2 \partial x_2^2}\right) \\
	 \qquad +\tfrac{2-3\sqrt{2}}{2}\left(\frac{\partial^4 k_1}{\partial x_1^3 \partial x_2}+x_3 \frac{\partial^4 h_1}{\partial x_1^3 \partial x_2}\right)-\tfrac{\sqrt{2}}{2}\left(\frac{\partial^4 k_2}{\partial x_1^4}+x_3 \frac{\partial^4 h_2}{\partial x_1^4}\right)=0\,.
\end{dcases}
\end{equation}
The second set of universality PDEs reads
\begin{equation}
\label{classSO(2)-2}
\begin{dcases}
	 -2\left(1+\sqrt{2}\right)\frac{\partial^2 h_1}{\partial x_1 \partial x_2^2}- \left(-4+\sqrt{2}\right)\frac{\partial^2 h_2}{\partial x_1 \partial x_2}+\left(-2+\sqrt{2}\right)\frac{\partial^3\hat{u}_{33}}{\partial x_1 \partial x_2^2}-\frac{\sqrt{2}}{2}\left(\frac{\partial^2 h_1}{\partial x_1^2}+\frac{\partial^3 \hat{u}_{33}}{\partial x_1^3}\right)=0, \\
	 -\sqrt{2}\frac{\partial^2 h_1}{\partial x_2^2}+\frac{-2+\sqrt{2}}{2}\frac{\partial^2 h_2}{\partial x_1 \partial x_2}+\frac{-2+\sqrt{2}}{2}\frac{\partial^3 \hat{u}_{33}}{\partial x_1 \partial x_2^2}-\frac{4+\sqrt{2}}{2}\frac{\partial^2 h_1}{\partial x_1^2}-\frac{\sqrt{2}}{2}\frac{\partial^3 \hat{u}_{33}}{\partial x_1^3}=0, \\
	 -\frac{\partial^2 h_1}{\partial x_2^2}-\frac{\partial^2 h_1}{\partial x_1^2}=0, \\
	 -\frac{\partial^3 \hat{u}_{33}}{\partial x_2^3}+\frac{1}{\sqrt{2}}\left(3\frac{\partial^2 h_1}{\partial x_1 \partial x_2}+\frac{\partial^2 h_2}{\partial x_1^2}+2\frac{\partial^3 \hat{u}_{33}}{\partial x_1^2 \partial x_2}\right)=0, \\
	 \left(1+\frac{1}{\sqrt{2}}\right) \frac{\partial^2 h_1}{\partial x_1 \partial x_2}+\frac{1}{2}\left(2+\sqrt{2}\right)\frac{\partial^2 h_2}{\partial x_1^2} + \left(-1+\sqrt{2}\right) \frac{\partial^3 \hat{u}_{33}}{\partial x_1^2 \partial x_2}=0, \\
	 \left(1+\frac{1}{\sqrt{2}}\right) \frac{\partial^2 h_1}{\partial x_2^2}-\frac{1}{\sqrt{2}}\frac{\partial^2 h_2}{\partial x_1 \partial x_2}-\left(-1+\sqrt{2}\right) \frac{\partial^3 \hat{u}_{33}}{\partial x_1 \partial x_2^2}+\frac{\partial^2 h_1}{\partial x_1^2} =0, \\
	 -\frac{1}{\sqrt{2}}\frac{\partial^2 h_1}{\partial x_2^2}+\left(1+\frac{1}{\sqrt{2}}\right)\frac{\partial^2 h_2}{\partial x_1 \partial x_2}+\sqrt{2}\frac{\partial^3 u_{33}}{\partial x_1 \partial x_2^2}+\frac{\partial^2 h_1}{\partial x_1^2}+\frac{\partial^3 \hat{u}_{33}}{\partial x_1^3}=0, \\
	 \frac{\partial^2 h_2}{\partial x_2^2}+\frac{\partial^2 h_1}{\partial x_1 \partial x_2}+\frac{\partial^2 h_2}{\partial x_1^2}=0, \\
	 - \frac{\partial^2 h_2}{\partial x_2^2}+\frac{\partial^3 u_{33}}{\partial x_2^3}-\frac{1}{\sqrt{2}}\left(\frac{\partial^2 h_1}{\partial x_1 \partial x_2}+\frac{\partial^2 h_2}{\partial x_1^2}-2\frac{\partial^3 \hat{u}_{33}}{\partial x_1^2 \partial x_2}\right)=0, \\
	 -\frac{\partial^2 h_2}{\partial x_1^2}=0, \\
	 \left(-1-\frac{3}{\sqrt{2}}\right)\frac{\partial^2 h_1}{\partial x_1 \partial x_2}+\frac{1}{\sqrt{2}}\frac{\partial^2 h_2}{\partial x_1^2}-\left(-1+\sqrt{2}\right)\frac{\partial^3 \hat{u}_{33}}{\partial x_1^2 \partial x_2}=0, \\
	 \frac{\partial^4 k_2}{\partial x_2^4}+x_3 \frac{\partial^4 h_2}{\partial x_2^4}+\frac{\partial^4 k_1}{\partial x_1 \partial x_2^3}+x_3 \frac{\partial^4 h_1}{\partial x_1 \partial x_2^3}+\left(1+\sqrt{2}\right)\left(\frac{\partial^4 k_2}{\partial x_1^2 \partial x_2^2}\right) \\
	 \qquad +\frac{1}{\sqrt{2}}\left(\frac{\partial^4 k_1}{\partial x_1^3 \partial x_2}+x_3 \frac{\partial^4 h_1}{\partial x_1^3 \partial x_2}+\frac{\partial^4 k_2}{\partial x_1^4}+x_3 \frac{\partial^4 h_2}{\partial x_1^4}\right)=0, \\
	 -\frac{\partial^4 k_1}{\partial x_1 \partial x_2^3}-x_3 \frac{\partial^4 h_1}{\partial x_1 \partial x_2^3}-\frac{\partial^4 k_2}{\partial x_1^2 \partial x_2^2}-x_3 \frac{\partial^4 h_2}{\partial x_1^2 \partial x_2^2}=0, \\
	 \frac{\partial^4 k_1}{\partial x_1 \partial x_2^3}+x_3 \frac{\partial^4 h_1}{\partial x_1 \partial x_2^3}-\frac{\partial^4 k_2}{\partial x_1^2 \partial x_2^2}+\frac{\partial^4 k_1}{\partial x_1^3 \partial x_2}+x_3 \frac{\partial^4 h_1}{\partial x_1^3 \partial x_2}-\frac{\partial^4 k_2}{\partial x_1^4}-x_3 \frac{\partial^4 h_2}{\partial x_1^4}=0, \\
	 \frac{\partial^4 k_1}{\partial x_1 \partial x_2^3}+x_3 \frac{\partial^4 h_1}{\partial x_1 \partial x_2^3}-\left(-1+\sqrt{2}\right)\left(\frac{\partial^4 k_2}{\partial x_1^2 \partial x_2^2}+x_3 \frac{\partial^4 mh_3}{\partial x_1^2 \partial x_2^2}\right) \\
	 \qquad +\left(1-\frac{1}{\sqrt{2}}\right)\left(\frac{\partial^4 k_1}{\partial x_1^3 \partial x_2}+x_3 \frac{\partial^4 h_1}{\partial x_1^3 \partial x_2}\right)-\frac{1}{\sqrt{2}}\left(\frac{\partial^4 k_2}{\partial x_1^4}+x_3 \frac{\partial^4 h_2}{\partial x_1^4}\right)=0, \\
	 -\frac{1}{\sqrt{2}}\left(\frac{\partial^4 k_1}{\partial x_2^4}+x_3 \frac{\partial^4 h_1}{\partial x_2^4}\right)+\left(1-\frac{3}{\sqrt{2}}\right)\left(\frac{\partial^4 k_2}{\partial x_1 \partial x_2^3}\right)+x_3 \frac{\partial^4 h_2}{\partial x_1 \partial x_2^3}-\frac{3}{\sqrt{2}}\left(\frac{\partial^4 k_1}{\partial x_1^2 \partial x_2^2}+x_3 \frac{\partial^4 h_1}{\partial x_1^2 \partial x_2^2}\right) \\
	 \qquad -\frac{1}{\sqrt{2}} \left(\frac{\partial^4 k_2}{\partial x_1^3 \partial x_2}+x_3 \frac{\partial^4 h_2}{\partial x_1^3 \partial x_2}\right)+\frac{\partial^4 k_1}{\partial x_1^4}+x_3 \frac{\partial^4 h_1}{\partial x_1^4}=0.
\end{dcases}
\end{equation}
The third set of universality PDEs reads
\begin{equation}
\label{classSO(2)-3}
\begin{dcases}
	 \frac{\partial^3 k_1}{\partial x_2^3}+x_3 \frac{\partial^3 h_1}{\partial x_2^3}-\frac{\partial^3 k_2}{\partial x_1 \partial x_2^2}-x_3 \frac{\partial^3 h_2}{\partial x_1 \partial x_2^2}+\frac{\partial^3 k_1}{\partial x_1^2 \partial x_2}+x_3 \frac{\partial^3 h_1}{\partial x_1^2 \partial x_2}-\frac{\partial^3 k_2}{\partial x_1^3}-x_3 \frac{\partial^3 h_2}{\partial x_1^3}=0\,, \\
	 \frac{\partial^3 k_2}{\partial x_2^3}+x_3 \frac{\partial^3 h_2}{\partial x_2^3}+\sqrt{2}\left(\frac{\partial^3 k_1}{\partial x_1 \partial x_2^2}+x_3 \frac{\partial^3 h_1}{\partial x_1 \partial x_2^2}+x_3 \frac{\partial^3 nh_2}{\partial x_1^2 \partial x_2}\right)+\frac{\partial^3 k_1}{\partial x_1^3}+x_3 \frac{\partial^3 h_1}{\partial x_1^3}=0\,, \\
	 \frac{\partial^3 k_1}{\partial x_1 \partial x_2^2}+x_3 \frac{\partial^3 h_1}{\partial x_1 \partial x_2^2}+\frac{\partial^3 k_2}{\partial x_1^2 \partial x_2}+x_3 \frac{\partial^3 h_2}{\partial x_1^2 \partial x_2}=0\,, \\
	 -\frac{\partial^3 k_2}{\partial x_1 \partial x_2^2}-x_3 \frac{\partial^3 h_2}{\partial x_1 \partial x_2^2}+\frac{1}{\sqrt{2}}\left(\frac{\partial^3 k_1}{\partial x_2^3}+x_3 \frac{\partial^3 h_1}{\partial x_2^3}+\frac{\partial^3 k_1}{\partial x_1 \partial x_2^2}\right)+\frac{\partial^3 k_1}{\partial x_1^2 \partial x_2} \\
	 \qquad +x_3 \frac{\partial^3 h_1}{\partial x_1^2 \partial x_2}-\frac{1}{\sqrt{2}}\left(\frac{\partial^3 k_1}{\partial x_1^2 \partial x_2}+x_3 \frac{\partial^3 h_1}{\partial x_1^2 \partial x_2}+\frac{\partial^3 k_2}{\partial x_1^3}+x_3 \frac{\partial^3 h_2}{\partial x_1^3}\right)=0\,, \\
	 \frac{\partial^3 k_2}{\partial x_1 \partial x_2^2}+x_3 \frac{\partial^3 h_2}{\partial x_1 \partial x_2^2}-\frac{1}{\sqrt{2}}\left(\frac{\partial^3 k_1}{\partial x_2^3}+x_3 \frac{\partial^3 h_1}{\partial x_2^3}+\frac{\partial^3 k_2}{\partial x_1 \partial x_2^2}+x_3 \frac{\partial^3 h_2}{\partial x_1 \partial x_2^2}\right) \\
	 \qquad -\frac{\partial^3 k_1}{\partial x_1^2 \partial x_2}-x_3 \frac{\partial^3 h_1}{\partial x_1^2 \partial x_2}+\frac{1}{\sqrt{2}}\left(\frac{\partial^3 k_1}{\partial x_1^2 \partial x_2}+x_3 \frac{\partial^3 h_1}{\partial x_1^2 \partial x_2}+\frac{\partial^3 k_2}{\partial x_1^3}+x_3 \frac{\partial^3 h_2}{\partial x_1^3}\right)=0\,, \\
	 \frac{\partial^3 k_2}{\partial x_2^3}+x_3 \frac{\partial^3 h_2}{\partial x_2^3}+2\left(\frac{\partial^3 k_1}{\partial x_1 \partial x_2^2}+x_3 \frac{\partial^3 h_1}{\partial x_1 \partial x_2^2}+\frac{\partial^3 k_2}{\partial x_1^2 \partial x_2}+x_3 \frac{\partial^3 h_2}{\partial x_1^2 \partial x_2}\right)+\frac{\partial^3 k_1}{\partial x_1^3}+x_3 \frac{\partial^3 h_1}{\partial x_1^3}=0\,, \\
	 -\frac{\partial^3 k_1}{\partial x_1 \partial x_2^2}-x_3 \frac{\partial^3 h_1}{\partial x_1 \partial x_2^2}-\frac{\partial^3 k_2}{\partial x_1^2 \partial x_2}-x_3 \frac{\partial^3 h_2}{\partial x_1^2 \partial x_2}=0\,, \\
	 \frac{\partial^3 h_2}{\partial x_2^3}+\frac{\partial^3 h_1}{\partial x_1 \partial x_2^2}+\frac{\partial^3 h_2}{\partial x_1^2 \partial x_2}+\frac{\partial^3 h_1}{\partial x_1^3}=0\,, \\
	 \frac{\partial^3 h_1}{\partial x_1 \partial x_2^2}+\frac{\partial^3 h_2}{\partial x_1^2 \partial x_2}=0\,, \\
	 \frac{\partial^3 h_2}{\partial x_2^3}+\frac{\partial^4 u_{33}}{\partial x_2^4}+\frac{\partial^3 h_1}{\partial x_1 \partial x_2^2}+\frac{\partial^3 h_2}{\partial x_1^2 \partial x_2}+2 \frac{\partial^4 \hat{u}_{33}}{\partial x_1^2 \partial x_2^2}+\frac{\partial^3 h_1}{\partial x_1^3}+\frac{\partial^3 \hat{u}_{33}}{\partial x_1^4}=0\,, \\
	 \frac{\partial^3 h_2}{\partial x_2^3}+\frac{\partial^3 h_1}{\partial x_1 \partial x_2^2}+\frac{\partial^3 h_2}{\partial x_1^2 \partial x_2}+\frac{\partial^3 h_1}{\partial x_1^3}=0\,, \\
	 -\frac{\partial^3 h_1}{\partial x_1 \partial x_2^2}-\frac{\partial^3 h_2}{\partial x_1^2 \partial x_2}=0\,, \\
	 -\sqrt{2}\frac{\partial^3 h_1}{\partial x_1 \partial x_2^2}-\left(-2+\sqrt{2}\right)\frac{\partial^3 h_2}{\partial x_1^2 \partial x_2}-2\left(-1+\sqrt{2}\right)\frac{\partial^4 \hat{u}_{33}}{\partial x_1^2 \partial x_2^2}=0\,, \\
	 2\frac{\partial^3 h_1}{\partial x_1 \partial x_2^2}+\frac{\partial^3 h_2}{\partial x_1^2 \partial x_2}-\frac{\partial^4 \hat{u}_{33}}{\partial x_1^2 \partial x_2^2}=0\,, \\
	 \frac{\partial^3 h_2}{\partial x_1 \partial x_2^2}+\frac{\partial^3 h_1}{\partial x_1^2 \partial x_2}=0\,, \\
	 \frac{\partial^3 h_1}{\partial x_2^3}+\left(1+\sqrt{2}\right)\frac{\partial^3 h_2}{\partial x_1 \partial x_2^2}+\left(1+\sqrt{2}\right)\frac{\partial^3 h_1}{\partial x_1^2 \partial x_2}+\frac{\partial^3 h_2}{\partial x_1^3}=0\,, \\
	 -3\frac{\partial^3 h_2}{\partial x_1 \partial x_2^2}+\frac{\partial^4 u_{33}}{\partial x_1 \partial x_2^3}-2 \frac{\partial^3 h_1}{\partial x_1^2 \partial x_2}-\frac{\partial^3 h_2}{\partial x_1^3}+\frac{\partial^4 u_{33}}{\partial x_1^3 \partial x_2}=0\,, \\
	 \frac{\partial^3 h_1}{\partial x_2^3}+\frac{\partial^4 u_{33}}{\partial x_1 \partial x_2^3}-3 \frac{\partial^3 h_1}{\partial x_1^2 \partial x_2}-\frac{\partial^3 h_2}{\partial x_1^3}=0\,, \\
	 \frac{\partial^3 h_2}{\partial x_1 \partial x_2^2}+\frac{\partial^4 u_{33}}{\partial x_1 \partial x_2^3}+\frac{\partial^3 h_1}{\partial x_1^2 \partial x_2}+\frac{\partial^4 \hat{u}_{33}}{\partial x_1^3 \partial x_2}=0\,.
\end{dcases}
\end{equation}
We obtain for $h_i$, $i=1,2$, that
\begin{equation}
	\label{classSO(2)-4}	
	\frac{\partial^2 h_i}{\partial x_1 \partial x_2}=\frac{\partial^2 h_i}{\partial x_1^2}=\frac{\partial^2 h_i}{\partial x_2^2}=0\,,
\end{equation}
which shows that $h_i$, $i=1,2$, depend at most linearly on $\left(x_1,x_2\right)$, i.e., they are of the form \eqref{classO^{-}(2)-mh1} and \eqref{classO^{-}(2)-mh2}\,.
For $\hat{u}_{33}$ we obtain
\begin{equation}
	\label{classSO(2)-5}
	\frac{\partial^3 \hat{u}_{33}}{\partial x_1^3}=\frac{\partial^3 \hat{u}_{33}}{\partial x_2^3}=0\,.
\end{equation}
For $k_i$, $i=1,2$, we find
\begin{equation}
	\label{classSO(2)-6}
	\frac{\partial^4 k_i}{\partial x_1^2 \partial x_2^2}\
	=\frac{\partial^4 k_i}{\partial x_1^4}=\frac{\partial^4 k_i}{\partial x_2^4}
	=\frac{\partial^4 k_i}{\partial x_1 \partial x_2^3}=\frac{\partial^4 k_i}{\partial x_1^3 \partial x_2}=0\,,
\end{equation}
which is consistent with \eqref{classO^{-}(2)-k}, together with the preceding conditions, and implies that
\begin{equation}
\label{classSO(2)-7}
\begin{aligned}
	& f_1(x_2)=c_4+c_5 x_2+c_6 x_2^2+c_7 x_2^3\,, &&
	f_2(x_2)=c_1+c_2 x_2+c_3 x_2^2\,, \\
	& f_3(x_1)=c_{11}+c_{12} x_1+c_{13} x_1^2+c_{14} x_1^3\,, &&
	f_4(x_1)=c_8+c_9 x_1+c_{10} x_1^2\,, \\
	& f_5(x_2)=c_{18}+c_{19} x_2+c_{20} x_2^2+c_{21} x_2^3\,, &&
	f_6(x_2)=c_{15}+c_{16} x_2+c_{17} x_2^2\,, \\
	& f_7(x_1)=c_{25}+c_{26} x_1+c_{27} x_1^2+c_{28} x_1^3\,, &&
	f_8(x_1)=c_{22}+c_{23} x_1+c_{24} x_1^2\,.
\end{aligned}
\end{equation}
We therefore obtain the following result.
\begin{prop}
The set of universal displacements of linear strain-gradient elasticity for the class $\mathbb{SO}(2) \oplus \mathbb{Z}^c_2$ coincides with those of the class $\mathbb{O}^{-}(2)$, except that \eqref{classSO(2)-7} replaces \eqref{classO^{-}(2)-f}.
\end{prop}

\renewcommand{\arraystretch}{1.15}
\begin{longtable}{p{0.30\textwidth} p{0.66\textwidth}}
\caption{Summary of universal displacements: $\infty$-gonal classes.}\label{Table-UD-summary-infty-gonal}\\
\hline
\textbf{Symmetry class} & \textbf{Universal displacement family} \\
\hline
\endfirsthead
\hline
\textbf{Symmetry class} & \textbf{Universal displacement family} (continued) \\
\hline
\endhead
\hline
\endfoot
\hline
\endlastfoot
$\infty$-gonal class $\mathbb{O}^{-}(2)$ & Of the form \eqref{transverse-isotropy-linear-solution} with $h_i$ given by \eqref{classO^{-}(2)-mh1}, \eqref{classO^{-}(2)-mh2}, $k_i$ by \eqref{classO^{-}(2)-k}, \eqref{classO^{-}(2)-f}, and $\hat{u}_{33}$ by \eqref{classO^{-}(2)-u_{33}} \\ \hline
$\infty$-gonal class $\mathbb{O}(2)\oplus \mathbb{Z}_2^c$ & Of the form \eqref{transverse-isotropy-linear-solution} with $h_i$, $\hat{u}_{33}$, and $k_i$ satisfying \eqref{classO(2)+Z^c_2-4}, \eqref{classO(2)+Z^c_2-6}, \eqref{classO(2)+Z^c_2-7}, \eqref{classO(2)+Z^c_2-9} \\ \hline
$\infty$-gonal class $\mathbb{SO}(2)\oplus \mathbb{Z}_2^c$ & Of the form \eqref{transverse-isotropy-linear-solution} subject to \eqref{classSO(2)+Z^c_2-1}, \eqref{classSO(2)+Z^c_2-2}, \eqref{classSO(2)+Z^c_2-4} \\ \hline
$\infty$-gonal class $\mathbb{O}(2)$ & Of the form \eqref{transverse-isotropy-linear-solution} with $h_i$, $\hat{u}_{33}$, and $k_i$ satisfying \eqref{classO(2)-1}, \eqref{classO(2)-2}, \eqref{classO(2)-4}, \eqref{classO(2)-5} \\ \hline
$\infty$-gonal class $\mathbb{SO}(2)$ & Same as class $\mathbb{O}^{-}(2)$ with \eqref{classSO(2)-7} replacing \eqref{classO^{-}(2)-f} \\ \hline
\end{longtable}
\renewcommand{\arraystretch}{2.0}

\subsection{Tetrahedral classes}

For all tetrahedral strain-gradient classes, the elasticity tensor $\boldsymbol{\mathsf{C}}$ has three independent components and has the same form as that of the cubic class in classical linear elasticity. Therefore, the classical linear elastic part yields universality PDEs of the form \eqref{cubic-linear-constraints}, which restrict the candidate universal displacement field to the form \eqref{cubic-linear-solution}.

\subsubsection{Tetrahedral class $\mathbb{T}$} 

Tensor $\boldsymbol{\mathsf{A}}_{\mathbb{T} \oplus \mathbb{Z}_2^c}$ has $17$ independent components and has the form
\begin{equation}
\label{A_T}
    \boldsymbol{\mathsf{A}}_{\mathbb{T} \oplus \mathbb{Z}_2^c}(\mathbf{x})=
    \begin{bmatrix}
        A^{(15)} & 0 & 0 & 0 \\
        & P A^{(15)} P^{\mathsf{T}} & 0 & 0 \\
        & & A^{(15)} & 0 \\
        & & & J^{(2)}
    \end{bmatrix}_S \,,
\end{equation}
where
\begin{equation}
\label{P}
    P=
    \begin{bmatrix}
        1 & 0 & 0 & 0 & 0 \\
        0 & 0 & 0 & 1 & 0 \\
        0 & 0 & 0 & 0 & 1 \\
        0 & 1 & 0 & 0 & 0 \\
        0 & 0 & 1 & 0 & 0
    \end{bmatrix} \,.
\end{equation}
Tensor $\boldsymbol{\mathsf{M}}_{\mathbb{T}}$ has $8$ independent components and takes the form
\begin{equation}
\label{M_T-barD^3}
   \boldsymbol{\mathsf{M}}_{\mathbb{T}}=
   \begin{bmatrix}
        0 & 0 & 0 & \bar{D}^{(3)} \\
        \bar{O}^{(5)} & 0 & 0 & 0 \\
        0 & \bar{O}^{(5)} & 0 & 0 \\
        0 & 0 & \bar{O}^{(5)} & 0
   \end{bmatrix} \,, \qquad
   \bar{D}^{(3)} =
   \begin{bmatrix}
        \bar{d}_{11} & \bar{d}_{12} & \bar{d}_{13} \\
        \bar{d}_{13} & \bar{d}_{11} & \bar{d}_{12} \\
        \bar{d}_{12} & \bar{d}_{13} & \bar{d}_{11}
   \end{bmatrix} \,.
\end{equation}
Tensor $\boldsymbol{\mathsf{C}}_{\mathbb{O} \oplus \mathbb{Z}_2^c}$ corresponds to the Curie group with Hermann--Mauguin symbol $432$ and has $3$ independent components.
The additional universality PDEs for this class read
\begin{equation}
\label{tetrahedral-T-constraints-1}
\begin{dcases}
	 \frac{\partial^4 u_1}{\partial x_1^4}=\frac{\partial^4 u_2}{\partial x_1 \partial x_2^3}=\frac{\partial^4 u_3}{\partial x_1 \partial x_3^3}=\frac{\partial^4 u_2}{\partial x_2^4}=\frac{\partial^4 u_3}{\partial x_3^4}=\frac{\partial^4 u_3}{\partial x_1^2 \partial x_3^2}=\frac{\partial^4 u_1}{\partial x_1 \partial x_2 \partial x_3^2}=0\,, \\
	 \frac{\partial^4 u_3}{\partial x_1^2 \partial x_2 \partial x_3}=\frac{\partial^4 u_1}{\partial x_1^3 \partial x_2}=\frac{\partial^4 u_2}{\partial x_1^2 \partial x_3^2}=\frac{\partial^4 u_3}{\partial x_3^4}=\frac{\partial^4 u_1}{\partial x_1 \partial x_3^3}=\frac{\partial^4 u_3}{\partial x_2^4}=\frac{\partial^4 u_1}{\partial x_1^3 \partial x_3}=0\,, \\
	 \frac{\partial^4 u_2}{\partial x_1^2 \partial x_2 \partial x_3}=\frac{\partial^4 u_1}{\partial x_1 \partial x_2^2 \partial x_3}=\frac{\partial^4 u_3}{\partial x_1^2 \partial x_3^2}=\frac{\partial^4 u_3}{\partial x_1^2 \partial x_2^2}=0\,, \\
	 \frac{\partial^4 u_1}{\partial x_1^2 \partial x_3^2}+\frac{\partial^4 u_2}{\partial x_1^3 \partial x_2}=0\,, \\
	 2 \frac{\partial^4 u_1}{\partial x_1 \partial x_3^3}+\frac{\partial^4 u_1}{\partial x_1^2 \partial x_2^2}+\frac{\partial^4 u_2}{\partial x_1^3 \partial x_2}=0\,, \\
	 \frac{\partial^4 u_1}{\partial x_1^2 \partial x_3^2}+\frac{\partial^4 u_3}{\partial x_1^3 \partial x_3}=0\,, \\
	 \frac{\partial^4 u_1}{\partial x_2^4}+\frac{\partial^4 u_2}{\partial x_1 \partial x_2^3}+2\frac{\partial^4 u_1}{\partial x_1^2 \partial x_3^2}=0\,, \\
	 \frac{\partial^4 u_2}{\partial x_1 \partial x_2 \partial x_3^2}+\frac{\partial^4 u_3}{\partial x_1 \partial x_2^2 \partial x_3}=0\,, \\
	 \frac{\partial^4 u_1}{\partial x_2^2 \partial x_3^2}+2 \frac{\partial^4 u_3}{\partial x_1 \partial x_2 \partial x_3^2}+\frac{\partial^4 u_3}{\partial x_1 \partial x_2^2 \partial x_3}=0\,, \\
	 \frac{\partial^4 u_1}{\partial x_2^2 \partial x_3^2}+3 \frac{\partial^4 u_2}{\partial x_1 \partial x_2 \partial x_3^2}=0\,, \\
	 \frac{\partial^4 u_1}{\partial x_3^4}+2 \frac{\partial^4 u_1}{\partial x_1^2 \partial x_2^2}=0\,, \\
	 2\frac{\partial^4 u_2}{\partial x_2^2 \partial x_3^2} +\frac{\partial^4 u_2}{\partial x_1 \partial x_2 \partial x_3^2}+\frac{\partial^4 u_3}{\partial x_1 \partial x_2^2 \partial x_3}=0\,,
\end{dcases}
\end{equation}
and 
\begin{equation}
\label{tetrahedral-T-constraints-2}
\begin{dcases}
	 \frac{\partial^4 u_2}{\partial x_2^2 \partial x_3^2}+\frac{\partial^4 u_3}{\partial x_2^2 \partial x_3^2}=0\,, \\
	 \frac{\partial^4 u_2}{\partial x_2^2 \partial x_3^2}+\frac{\partial^4 u_3}{\partial x_2^3 \partial x_3}=0\,, \\
	 \frac{\partial^4 u_1}{\partial x_1 \partial x_2^3}+\frac{\partial^4 u_2}{\partial x_1^2 \partial x_2^2}=0\,, \\
	 2\frac{\partial^4 u_3}{\partial x_2 \partial x_3^3}+\frac{\partial^4 u_1}{\partial x_1 \partial x_2^3}+\frac{\partial^4 u_2}{\partial x_1^2 \partial x_2^2}=0\,, \\
	 \frac{\partial^4 u_2}{\partial x_3^4}+\frac{\partial^4 u_3}{\partial x_2 \partial x_3^3}+2 \frac{\partial^4 u_2}{\partial x_1^2 \partial x_2^2}=0\,, \\
	 \frac{\partial^4 u_1}{\partial x_1 \partial x_2^3}+\frac{\partial^4 u_2}{\partial x_1^2 \partial x_2^2}=0\,, \\
	 \frac{\partial^4 u_2}{\partial x_1^2 \partial x_3^2}+3 \frac{\partial^4 u_3}{\partial x_1^2 \partial x_2 \partial x_3}=0\,, \\
	 2\frac{\partial^4 u_2}{\partial x_2^2 \partial x_3^2}+ \frac{\partial^4 u_2}{\partial x_1^4}=0\,,
\end{dcases}
\end{equation}
as well as
\begin{equation}
\label{tetrahedral-T-constraints-3}
\begin{dcases}
	\frac{\partial^4 u_2}{\partial x_2 \partial x_3^3}+\frac{\partial^4 u_3}{\partial x_2^2 \partial x_3^2}=0\,, \\
	2\frac{\partial^4 u_3}{\partial x_2 \partial x_3^3}+\frac{\partial^4 u_1}{\partial x_1^3 \partial x_3}+\frac{\partial^4 u_3}{\partial x_1^4}=0\,.
\end{dcases}
\end{equation}
The above universality PDEs stem from the sixth-order tensor. From the fifth-order tensor we obtain
\begin{equation}
\label{tetrahedral-T-constraints-4}
\begin{dcases}
	-2 \frac{\partial^3 u_1}{\partial x_1 \partial x_2 \partial x_3}+\frac{\partial^3 u_2}{\partial x_1^2 \partial x_3}+\frac{\partial^3 u_3}{\partial x_1^2 \partial x_2}=0\,, \\
	-2 \frac{\partial^3 u_3}{\partial x_2 \partial x_3^2}+\frac{\partial^3 u_1}{\partial x_1 \partial x_2 \partial x_3}+\frac{\partial^3 u_3}{\partial x_1^2 \partial x_2}=0\,, \\
	-2 \frac{\partial^3 u_2}{\partial x_2 \partial x_3^2}+\frac{\partial^3 u_1}{\partial x_1 \partial x_2 \partial x_3}+\frac{\partial^3 u_2}{\partial x_1^2 \partial x_3}=0\,, \\
	\frac{\partial^3 u_3}{\partial x_2 \partial x_3^2}+\frac{\partial^3 u_2}{\partial x_2^2 \partial x_3}-\frac{\partial^3 u_2}{\partial x_1^2 \partial x_3}-\frac{\partial^3 u_3}{\partial x_1^2 \partial x_2}=0\,, \\
	-\frac{\partial^3 u_2}{\partial x_2^2 \partial x_3}-\frac{\partial^3 u_3}{\partial x_2^3}+2 \frac{\partial^3 u_1}{\partial x_1 \partial x_2 \partial x_3}=0\,, \\
	-\frac{\partial^3 u_2}{\partial x_3^3}+\frac{\partial^3 u_3}{\partial x_3^3}-\frac{\partial^3 u_3}{\partial x_2 \partial x_3^2}+\frac{\partial^3 u_2}{\partial x_2^2 \partial x_3}=0\,, \\
	\frac{\partial^3 u_2}{\partial x_3^3}+\frac{\partial^3 u_3}{\partial x_2 \partial x_3^2}+\frac{\partial^3 u_2}{\partial x_2^2 \partial x_3}+\frac{\partial^3 u_3}{\partial x_2^3}=0\,,
\end{dcases}
\end{equation}
and
\begin{equation}
\label{tetrahedral-T-constraints-5}
\begin{dcases}
	  \frac{\partial^3 u_1}{\partial x_2^2 \partial x_3}-2\frac{\partial^3 u_2}{\partial x_1 \partial x_2 \partial x_3}+\frac{\partial^3 u_3}{\partial x_1 \partial x_2^2}=0\,, \\
	 \frac{\partial^3 u_1}{\partial x_2^2 \partial x_3}+\frac{\partial^3 u_2}{\partial x_1 \partial x_2 \partial x_3}-2\frac{\partial^3 u_1}{\partial x_1^2 \partial x_3}=0\,, \\
	 -2 \frac{\partial^3 u_3}{\partial x_1 \partial x_3^2}+\frac{\partial^3 u_2}{\partial x_1 \partial x_2 \partial x_3}+\frac{\partial^3 u_3}{\partial x_1 \partial x_2^2}=0\,, \\
	 -\frac{\partial^3 u_1}{\partial x_2^2 \partial x_3}+\frac{\partial^3 u_3}{\partial x_1 \partial x_3^2}-\frac{\partial^3 u_3}{\partial x_1 \partial x_2^2}+\frac{\partial^3 u_1}{\partial x_1^2 \partial x_3}=0\,, \\
	 \frac{\partial^3 u_2}{\partial x_1 \partial x_2 \partial x_3}+ \frac{\partial^3 u_3}{\partial x_1^3}=0\,, \\
	 \frac{\partial^3 u_1}{\partial x_2^2 \partial x_3}-\frac{\partial^3 u_2}{\partial x_1 \partial x_2 \partial x_3}-2\frac{\partial^3 u_3}{\partial x_1 \partial x_2^2}+\frac{\partial^3 u_1}{\partial x_2^2 \partial x_3}+\frac{\partial^3 u_3}{\partial x_1^3}=0\,, \\
	 -\frac{\partial^3 u_1}{\partial x_3^3}+\frac{\partial^3 u_2}{\partial x_1 \partial x_2 \partial x_3}=0\,, \\
	 \frac{\partial^3 u_1}{\partial x_3^3}-2\frac{\partial^3 u_1}{\partial x_2^2 \partial x_3}+\frac{\partial^3 u_3}{\partial x_1 \partial x_3^2}-\frac{\partial^3 u_2}{\partial x_1 \partial x_2 \partial x_3}+\frac{\partial^3 u_3}{\partial x_1 \partial x_2^2}=0\,,
\end{dcases}
\end{equation}
as well as
\begin{equation}
\label{tetrahedral-T-constraints-6}
\begin{dcases}
	 \frac{\partial^3 u_1}{\partial x_2 \partial x_3^2}+\frac{\partial^3 u_2}{\partial x_1 \partial x_3^2}-2\frac{\partial^3 u_3}{\partial x_1 \partial x_2 \partial x_3}=0\,, \\
	 \frac{\partial^3 u_2}{\partial x_1 \partial x_3^2}+\frac{\partial^3 u_3}{\partial x_1 \partial x_2 \partial x_3}-2\frac{\partial^3 u_2}{\partial x_1 \partial x_2^2}=0\,, \\
	 \frac{\partial^3 u_1}{\partial x_2 \partial x_3^2}+\frac{\partial^3 u_3}{\partial x_1 \partial x_2 \partial x_3}-2\frac{\partial^3 u_1}{\partial x_1^2 \partial x_2}=0\,, \\
	 -\frac{\partial^3 u_1}{\partial x_2 \partial x_3^2}-\frac{\partial^3 u_2}{\partial x_1 \partial x_3^2}+\frac{\partial^3 u_2}{\partial x_1 \partial x_2^2}+\frac{\partial^3 u_1}{\partial x_1^2 \partial x_2}=0\,, \\
	 \frac{\partial^3 u_2}{\partial x_1 \partial x_2^2}+ \frac{\partial^3 u_2}{\partial x_1^3}=0\,, \\
	 \frac{\partial^3 u_1}{\partial x_2^3}+\frac{\partial^3 u_2}{\partial x_1 \partial x_2^2}+2\frac{\partial^3 u_1}{\partial x_1^2 \partial x_2}+\frac{\partial^3 u_2}{\partial x_1^3}=0\,.
\end{dcases}
\end{equation}
When \eqref{cubic-linear-solution} is substituted into the above universality PDEs, we obtain
\begin{equation}
\label{classT-1}
	\frac{\partial^4 g_1(x_2, x_3)}{\partial x_2^4}
	=\frac{\partial^4 g_1(x_2, x_3)}{\partial x_3^4}
	=\frac{\partial^4 g_1(x_2, x_3)}{\partial x_2^2 \partial x_3^2}=0\,,
\end{equation}
which implies that
\begin{equation}
\label{classT-2}
\begin{aligned}
	& g_1(x_2, x_3)=f_1(x_3)+x_2 f_2(x_3)+f_3(x_2)+x_3 f_4(x_2)\,, \\
	& f_3''''(x_2)+x_3 f_4''''(x_2)=0\,, \\
	& f_1''''(x_3)+x_2 f_2''''(x_3)=0\,.
\end{aligned}
\end{equation}
Similarly, we find
\begin{equation}
\label{classT-3}
	\frac{\partial^4 g_2(x_1, x_3)}{\partial x_1^4}
	=\frac{\partial^4 g_2(x_1, x_3)}{\partial x_3^4}
	=\frac{\partial^4 g_2(x_1, x_3)}{\partial x_1^2 \partial x_3^2}=0\,,
\end{equation}
and hence
\begin{equation}
\label{classT-4}
\begin{aligned}
	& g_2(x_1, x_3)=f_5(x_3)+x_1 f_6(x_3)+f_7(x_1)+x_3 f_8(x_1)\,, \\
	& f_7''''(x_1)+x_3 f_8''''(x_1)=0\,, \\
	& f_5''''(x_3)+x_1 f_6''''(x_3)=0\,.
\end{aligned}
\end{equation}
Finally, we have
\begin{equation}
\label{classT-5}
	\frac{\partial^4 g_3(x_1, x_2)}{\partial x_1^4}
	=\frac{\partial^4 g_3(x_1, x_2)}{\partial x_2^4}
	=\frac{\partial^4 g_3(x_1, x_2)}{\partial x_1^2 \partial x_2^2}=0\,,
\end{equation}
which are simplified to read
\begin{equation}
\label{classT-6}
\begin{aligned}
	& g_3(x_1, x_2)=f_9(x_2)+x_1 f_{10}(x_2)+f_{11}(x_1)+x_2 f_{12}(x_1)\,, \\
	& f_{11}''''(x_1)+x_2 f_{12}''''(x_1)=0\,, \\
	& f_9''''(x_2)+x_1 f_{10}''''(x_2)=0\,.
\end{aligned}
\end{equation}
Moreover, the additional universality PDEs further imply that
\begin{equation}
\label{classT-7}
	f_4''(x_2)=0\,, \qquad f_2''(x_3)=0\,, \qquad
	f_3'''(x_2)+x_3 f_4'''(x_2)=0\,, \qquad
	f_1'''(x_3)+x_2 f_2'''(x_3)=0\,.
\end{equation}
as well as
\begin{equation}
\label{classT-8}
	f_8''(x_1)=0\,, \qquad f_6''(x_3)=0\,, \qquad
	f_7'''(x_1)+x_3 f_8'''(x_1)=0\,, \qquad
	f_5'''(x_3)+x_1 f_6'''(x_3)=0\,.
\end{equation}
and
\begin{equation}
\label{classT-9}
	f_{12}(x_1)=0\,, \qquad f_{10}(x_2)=0\,, \qquad
	f_{11}'''(x_1)+x_2 f_{12}'''(x_1)=0\,, \qquad
	f_9'''(x_2)+x_1 f_{10}'''(x_2)=0\,.
\end{equation}
Therefore, we obtain the following result.
\begin{prop}
The sets of universal displacements of linear strain-gradient elasticity for class $\mathbb{T}$ have the form \eqref{cubic-linear-solution}, and the functions $g_1$, $g_2$, and $g_3$ satisfy \eqref{classT-2}, \eqref{classT-4}, \eqref{classT-6}, \eqref{classT-7}, \eqref{classT-8}, and \eqref{classT-9}.
\end{prop}

\subsubsection{Tetrahedral class $\mathbb{T} \oplus \mathbb{Z}_2^c$} 

Tensor $\boldsymbol{\mathsf{A}}_{\mathbb{T} \oplus \mathbb{Z}_2^c}$ has $17$ independent components and is of the same form as that of class $\mathbb{T}$. Tensor $\boldsymbol{\mathsf{M}}_{\mathbb{T} \oplus \mathbb{Z}_2^c}$ is a null tensor. Tensor $\boldsymbol{\mathsf{C}}_{\mathbb{O} \oplus \mathbb{Z}_2^c}$ has $3$ independent components and corresponds to the group with Hermann--Mauguin symbol $m\bar{3}m$. Therefore, we obtain the following result.
\begin{prop}
The set of universal displacements of linear strain-gradient elasticity for class $\mathbb{T} \oplus \mathbb{Z}_2^c$ has the form \eqref{cubic-linear-solution}, and the functions $g_1$, $g_2$, and $g_3$ satisfy \eqref{classT-2}, \eqref{classT-4}, and \eqref{classT-6}.
\end{prop}

\subsubsection{Tetrahedral class $\mathbb{O}^-$} 

Tensor $\boldsymbol{\mathsf{A}}_{\mathbb{O} \oplus \mathbb{Z}_2^c}$ has $11$ independent components and has the form
\begin{equation}
\label{A_O^-}
    \boldsymbol{\mathsf{A}}_{\mathbb{O}^- \oplus \mathbb{Z}_2^c}(\mathbf{x})=\begin{bmatrix}
    A^{(9)} & 0 & 0 & 0  \\
     &  A^{(9)} & 0 & 0   \\
     &  & A^{(9)} & 0   \\
     &  &  & J^{(2)}      
    \end{bmatrix}_S\,, 
\end{equation}
where 
\begin{equation}
\label{J^2-A^9}
   J^{(2)} = \begin{bmatrix}
   j_{11} & j_{12} & j_{12}  \\
    & j_{11} & j_{12}   \\
    &  & j_{11} 
    \end{bmatrix}_S\,,\qquad
   A^{(9)} = \begin{bmatrix}
   a_{11} & a_{12} & a_{13} & a_{12} & a_{13}  \\
    & a_{22} & a_{23} & a_{24} & a_{25}  \\
    &  & a_{33} & a_{25} & a_{35}  \\
    &  &  & a_{22} & a_{23}  \\
    &  &  &  & a_{33}   
    \end{bmatrix}_S\,.
\end{equation}
Tensor $\boldsymbol{\mathsf{M}}_{\mathbb{O}^-}$ has $5$ independent components and it is of the form 
\begin{equation}
\label{M_O^--barD^2}
   \boldsymbol{\mathsf{M}}_{\mathbb{O}^-}=\begin{bmatrix}
    0 & 0 & 0 & \bar{D}^{(2)}_2  \\
    \bar{O}^{(3)} & 0 & 0 & 0   \\
    0 & \bar{O}^{(3)}  & 0 & 0     \\
    0 & 0 & \bar{O}^{(3)} & 0   
    \end{bmatrix}\,,\qquad
   \bar{D}^{(2)}_2 \in \begin{bmatrix}
   \bar{d}_{11} & \bar{d}_{12} & \bar{d}_{12}  \\
   \bar{d}_{12} & \bar{d}_{11} & \bar{d}_{12}   \\
   \bar{d}_{12} & \bar{d}_{12} & \bar{d}_{11} 
    \end{bmatrix}\,.
\end{equation}
Tensor $\boldsymbol{\mathsf{C}}_{{\mathbb{O} \oplus \mathbb{Z}_2^c}}$ corresponds to the group with Hermann--Mauguin symbol $\bar{4}3m$ and has $3$ independent components.\,
The additional universality PDEs for this class are
\begin{equation}
\label{tetrahedral-O-constraints-1}
\begin{dcases}
	 \frac{\partial^4 u_1}{\partial x_1^4}=\frac{\partial^4 u_1}{\partial x_2^2 \partial x_3^2}=\frac{\partial^4 u_2}{\partial x_2^4}=\frac{\partial^4 u_2}{\partial x_1^2 \partial x_3^2}=\frac{\partial^4 u_3}{\partial x_3^4}=\frac{\partial^4 u_3}{\partial x_1^2 \partial x_2^2}=\frac{\partial^4 u_3}{\partial x_1 \partial x_2^2 \partial x_3}=0\,, \\
	 \frac{\partial^4 u_1}{\partial x_1^2 \partial x_2^2}+\frac{\partial^4 u_3}{\partial x_1^3 \partial x_3}+\frac{\partial^4 u_2}{\partial x_1^3 \partial x_2}=0\,, \\
	 \frac{\partial^4 u_1}{\partial x_1^2 \partial x_3^2}+\frac{\partial^4 u_1}{\partial x_1^2 \partial x_2^2}+\frac{\partial^4 u_3}{\partial x_1^3 \partial x_3}+\frac{\partial^4 u_2}{\partial x_1^3 \partial x_3}=0\,, \\
	 \frac{\partial^4 u_3}{\partial x_1 \partial x_3^3}+\frac{\partial^4 u_2}{\partial x_1 \partial x_2^3}=0\,, \\
	 \frac{\partial^4 u_1}{\partial x_3^4}+\frac{\partial^4 u_1}{\partial x_2^4 \partial x_2^3}+2\frac{\partial^4 u_3}{\partial x_1 \partial x_3^3}+\frac{\partial^4 u_2}{\partial x_1 \partial x_2^3}+2\left(\frac{\partial^4 u_1}{\partial x_1^2 \partial x_3^2}+\frac{\partial^4 u_1}{\partial x_1^2 \partial x_2^2}\right)=0\,, \\
	 \frac{\partial^4 u_2}{\partial x_1 \partial x_2 \partial x_3^2}+\frac{\partial^4 u_3}{\partial x_1 \partial x_2^2 \partial x_3}=0\,, \\
	 \frac{\partial^4 u_1}{\partial x_1^2 \partial x_3^2}+ \frac{\partial^4 u_1}{\partial x_1^2 \partial x_2^2}+\frac{\partial^4 u_3}{\partial x_2^3 \partial x_3}+\frac{\partial^4 u_2}{\partial x_1^3 \partial x_3}=0\,,
\end{dcases}
\end{equation}
and 
\begin{equation}
\label{tetrahedral-O-constraints-2}
\begin{dcases}
	 \frac{\partial^4 u_2}{\partial x_2^2 \partial x_3^2}+\frac{\partial^4 u_3}{\partial x_2^2 \partial x_3^2}+\frac{\partial^4 u_1}{\partial x_1 \partial x_2^3}+\frac{\partial^4 u_2}{\partial x_1^2 \partial x_2^2}=0\,, \\
	 2\frac{\partial^4 u_3}{\partial x_2 \partial x_3^3}+2\frac{\partial^4 u_1}{\partial x_1^3 \partial x_2}=0\,, \\
	 \frac{\partial^4 u_3}{\partial x_3^4}+\frac{\partial^4 u_1}{\partial x_1^3 \partial x_2}=0\,, \\
	 \frac{\partial^4 u_2}{\partial x_3^4}+\frac{\partial^4 u_3}{\partial x_2 \partial x_3^3}+2\frac{\partial^4 u_2}{\partial x_2^2 \partial x_3^2}+2\frac{\partial^4 u_2}{\partial x_1^2 \partial x_2^2}+\frac{\partial^4 u_1}{\partial x_1^3 \partial x_2}+\frac{\partial^4 u_2}{\partial x_1^4}=0\,, \\
	 \frac{\partial^4 u_1}{\partial x_1 \partial x_2 \partial x_3^2}+\frac{\partial^4 u_3}{\partial x_1^2 \partial x_3^2}=0\,, \\
	 \frac{\partial^4 u_1}{\partial x_1 \partial x_2 \partial x_3^2}+\frac{\partial^4 u_3}{\partial x_1^2 \partial x_2 \partial x_3}=0\,,
\end{dcases}
\end{equation}
as well as
\begin{equation}
\label{tetrahedral-O-constraints-3}
\begin{dcases}
	 \frac{\partial^4 u_2}{\partial x_2 \partial x_3^3}+\frac{\partial^4 u_3}{\partial x_2^2 \partial x_3^2}+\frac{\partial^4 u_1}{\partial x_1 \partial x_3^3}+\frac{\partial^4 u_3}{\partial x_1^2 \partial x_3^2}=0\,, \\
	 2\frac{\partial^4 u_2}{\partial x_2^3 \partial x_3}+\frac{\partial^4 u_1}{\partial x_1 \partial x_3^3}+2\frac{\partial^4 u_1}{\partial x_1^3 \partial x_3}=0\,, \\
	 \frac{\partial^4 u_2}{\partial x_2^3 \partial x_3}+\frac{\partial^4 u_1}{\partial x_1^3 \partial x_2}=0\,, \\
	 2\frac{\partial^4 u_3}{\partial x_2 \partial x_3^3}+\frac{\partial^4 u_2}{\partial x_2^3 \partial x_3}+\frac{\partial^4 u_3}{\partial x_2^4}+2\frac{\partial^4 u_3}{\partial x_1^2 \partial x_2^2}+\frac{\partial^4 u_1}{\partial x_1^3 \partial x_3}+\frac{\partial^4 u_3}{\partial x_1^4 \partial x_3}=0\,, \\
	 \frac{\partial^4 u_1}{\partial x_1 \partial x_2^2 \partial x_3}+\frac{\partial^4 u_2}{\partial x_1^2 \partial x_2 \partial x_3}=0\,.
\end{dcases}
\end{equation}
The above universality PDEs stem from the sixth-order tensor. From the fifth-order tensor we obtain the following universality PDEs
\begin{equation}
\label{tetrahedral-O-constraints-4}
\begin{dcases}
	 -2 \frac{\partial^3 u_1}{\partial x_1 \partial x_2 \partial x_3}+\frac{\partial^3 u_2}{\partial x_1^2 \partial x_3}+\frac{\partial^3 u_3}{\partial x_1^2 \partial x_2}=0\,, \\
	 -2 \frac{\partial^3 u_3}{\partial x_2 \partial x_3^2}+2\frac{\partial^3 u_2}{\partial x_2^2 \partial x_3}=0\,, \\
	 \frac{\partial^3 u_2}{\partial x_1^2 \partial x_3}+\frac{\partial^3 u_3}{\partial x_1^2 \partial x_2}=0\,, \\
	 \frac{\partial^3 u_3}{\partial x_3^3}+2\frac{\partial^3 u_2}{\partial x_2^2 \partial x_3}-2\frac{\partial^3 u_1}{\partial x_1 \partial x_2 \partial x_3}=0\,, \\
	 \frac{\partial^3 u_2}{\partial x_3^3}+\frac{\partial^3 u_3}{\partial x_2^3}=0\,,
\end{dcases}
\end{equation}
and
\begin{equation}
\label{tetrahedral-O-constraints-5}
\begin{dcases}
	 \frac{\partial^3 u_1}{\partial x_2^2 \partial x_3}-2\frac{\partial^3 u_2}{\partial x_1 \partial x_2 \partial x_3}+\frac{\partial^3 u_3}{\partial x_1 \partial x_2^2}=0\,, \\
	 -2\frac{\partial^3 u_3}{\partial x_1 \partial x_3^2}+2\frac{\partial^3 u_1}{\partial x_1^2 \partial x_2 \partial x_3}=0\,, \\
	 \frac{\partial^3 u_1}{\partial x_2^2 \partial x_3}+\frac{\partial^3 u_3}{\partial x_1 \partial x_2^2}=0\,, \\
	 -\frac{\partial^3 u_1}{\partial x_3^3}-\frac{\partial^3 u_2}{\partial x_1 \partial x_2 \partial x_3}+\frac{\partial^3 u_3}{\partial x_1^3}=0\,, \\
	 \frac{\partial^3 u_1}{\partial x_3^3}-\frac{\partial^3 u_1}{\partial x_2^2 \partial x_3}+ \frac{\partial^3 u_3}{\partial x_1 \partial x_3^2}-2\frac{\partial^3 u_2}{\partial x_1 \partial x_2 \partial x_3}-\frac{\partial^3 u_3}{\partial x_1 \partial x_2^2}+\frac{\partial^3 u_1}{\partial x_1^2 \partial x_3}+\frac{\partial u_2}{\partial x_1^3}=0\,,
\end{dcases}
\end{equation}
as well as
\begin{equation}
\label{tetrahedral-O-constraints-6}
\begin{dcases}
	 \frac{\partial^3 u_1}{\partial x_2 \partial x_3^2}+\frac{\partial^3 u_2}{\partial x_1 \partial x_3^2}-2\frac{\partial^3 u_3}{\partial x_1 \partial x_2 \partial x_3}=0\,, \\
	 \frac{\partial^3 u_2}{\partial x_1 \partial x_3^2}-2\frac{\partial^3 u_1}{\partial x_1^2 \partial x_2}=0\,, \\
	 -\frac{\partial^3 u_1}{\partial x_2 \partial x_3^2}-\frac{\partial^3 u_2}{\partial x_1 \partial x_3^2}+\frac{\partial^3 u_2}{\partial x_1 \partial x_2^2}+\frac{\partial^3 u_1}{\partial x_1^2 \partial x_2}=0\,, \\
	 -\frac{\partial^3 u_1}{\partial x_2^3}+\frac{\partial^3 u_3}{\partial x_1 \partial x_3^2}+\frac{\partial^3 u_3}{\partial x_1 \partial x_2 \partial x_3}-\frac{\partial^3 u_2}{\partial x_1^3}=0\,, \\
	 -\frac{\partial^3 u_1}{\partial x_2 \partial x_3^2}+\frac{\partial^3 u_1}{\partial x_3^3}- \frac{\partial^3 u_2}{\partial x_1 \partial x_3^2}-2\frac{\partial^3 u_3}{\partial x_1 \partial x_2 \partial x_3}+\frac{\partial^3 u_2}{\partial x_1 \partial x_3^2}+\frac{\partial^3 u_1}{\partial x_1^2 \partial x_2}+\frac{\partial u_2}{\partial x_1^3}=0\,.
\end{dcases}
\end{equation}
When \eqref{cubic-linear-solution} is substituted into the above expressions, we obtain
\begin{equation}
\label{classO^--1}
\frac{\partial^4 g_1(x_2,x_3)}{\partial x_2^4}+\frac{\partial^4 g_1(x_2,x_3)}{\partial x_3^4}=0\,,\qquad
\frac{\partial^4 g_1(x_2,x_3)}{\partial x_2^2 \partial x_3^2}=0\,.
\end{equation}
From these relations, using the preceding calculations, we find that
\begin{equation}
\label{classO^--2}
\begin{aligned}
& g_1(x_2,x_3)=f_1(x_2)+x_3 f_2(x_2)+f_3(x_3)+x_2 f_4(x_3)\,, \\
& f_1''''(x_2)+x_3 f_2''''(x_2)+f_3''''(x_3)+x_2 f_4''''(x_3)=0\,.
\end{aligned}
\end{equation}
Similarly,
\begin{equation}
\label{classO^--3}
\frac{\partial^4 g_2(x_1,x_3)}{\partial x_1^4}+\frac{\partial^4 g_2(x_1,x_3)}{\partial x_3^4}=0\,,\qquad
\frac{\partial^4 g_2(x_1,x_3)}{\partial x_1^2 \partial x_3^2}=0\,,
\end{equation}
and hence
\begin{equation}
\label{classO^--4}
\begin{aligned}
& g_2(x_1,x_3)=f_5(x_1)+x_3 f_6(x_1)+f_7(x_3)+x_1 f_8(x_3)\,, \\
& f_5''''(x_1)+x_3 f_6''''(x_1)+f_7''''(x_3)+x_1 f_8''''(x_3)=0\,.
\end{aligned}
\end{equation}
Finally,
\begin{equation}
\label{classO^--5}
\frac{\partial^4 g_3(x_1,x_2)}{\partial x_1^4}+\frac{\partial^4 g_3(x_1,x_2)}{\partial x_2^4}=0\,,\qquad
\frac{\partial^4 g_3(x_1,x_2)}{\partial x_1^2 \partial x_2^2}=0\,,
\end{equation}
which are simplified to read
\begin{equation}
\label{classO^--6}
\begin{aligned}
& g_3(x_1,x_2)=f_9(x_1)+x_2 f_{10}(x_1)+f_{11}(x_2)+x_1 f_{12}(x_2)\,, \\
& f_9''''(x_1)+x_2 f_{10}''''(x_1)+f_{11}''''(x_2)+x_1 f_{12}''''(x_2)=0\,.
\end{aligned}
\end{equation}
The additional universality PDEs read
\begin{equation}
\label{classO^--7}
\frac{\partial^3 g_2(x_1,x_3)}{\partial x_1^2 \partial x_3}+\frac{\partial^3 g_3(x_1,x_2)}{\partial x_1^2 \partial x_2}=0\,.
\end{equation}
Substituting \eqref{classO^--4} and \eqref{classO^--6} into \eqref{classO^--7} yields
$f_6''(x_1)+f_{10}''(x_1)=0$.

The remaining five universality PDEs are:
\begin{subequations}\label{classO^--remainingPDEs}
\begin{align}
\label{classO^--8}
&\frac{\partial^3 g_2(x_1,x_3)}{\partial x_3^3}+\frac{\partial^3 g_3(x_1,x_2)}{\partial x_2^3}=0\,, \\
\label{classO^--9}
&\frac{\partial^3 g_3(x_1,x_2)}{\partial x_1 \partial x_2^2}+\frac{\partial^3 g_1(x_2,x_3)}{\partial x_2^2 \partial x_3}=0\,, \\
\label{classO^--10}
&\frac{\partial^3 g_1(x_2,x_3)}{\partial x_3^3}+\frac{\partial^3 g_3(x_1,x_2)}{\partial x_1^3}=0\,, \\
\label{classO^--11}
&\frac{\partial^3 g_1(x_2,x_3)}{\partial x_2 \partial x_3^2}+\frac{\partial^3 g_2(x_1,x_3)}{\partial x_1 \partial x_3^2}=0\,, \\
\label{classO^--12}
&\frac{\partial^3 g_1(x_2,x_3)}{\partial x_2^3}+\frac{\partial^3 g_2(x_1,x_3)}{\partial x_1^3}=0\,.
\end{align}
\end{subequations}
Substituting \eqref{classO^--4} and \eqref{classO^--6} into \eqref{classO^--8} gives $f_7'''(x_3)+x_1 f_8'''(x_3)+f_{11}'''(x_2)+x_1 f_{12}'''(x_2)=0$. Substituting \eqref{classO^--2} and \eqref{classO^--6} into \eqref{classO^--9} leads to $f_{12}''(x_2)+f_2''(x_2)=0$. Substituting \eqref{classO^--2} and \eqref{classO^--6} into \eqref{classO^--10} implies $f_3'''(x_3)+x_2 f_4'''(x_3)+f_9'''(x_1)+x_2 f_{10}'''(x_1)=0$. Substituting \eqref{classO^--2} and \eqref{classO^--4} into \eqref{classO^--11} gives $f_4''(x_3)+f_8''(x_3)=0$. Finally, substituting \eqref{classO^--2} and \eqref{classO^--4} into \eqref{classO^--12} gives $f_1'''(x_2)+x_3 f_2'''(x_2)+f_5'''(x_1)+x_3 f_6'''(x_1)=0$.

In summary, we have the following result.
\begin{prop}
The sets of universal displacements of linear strain-gradient elasticity for class $\mathbb{O}^-$ are given by \eqref{cubic-linear-solution}, and the fields $g_i$, $i=1,2,3$, satisfy \eqref{classO^--2}, \eqref{classO^--4}, \eqref{classO^--6}, \eqref{classO^--8}, \eqref{classO^--10}, and \eqref{classO^--12}.
\end{prop}

\renewcommand{\arraystretch}{1.15}
\begin{longtable}{p{0.30\textwidth} p{0.66\textwidth}}
\caption{Summary of universal displacements: Tetrahedral classes.}\label{Table-UD-summary-tetrahedral}\\
\hline
\textbf{Symmetry class} & \textbf{Universal displacement family} \\
\hline
\endfirsthead
\hline
\textbf{Symmetry class} & \textbf{Universal displacement family} (continued) \\
\hline
\endhead
\hline
\endfoot
\hline
\endlastfoot
Tetrahedral class $\mathbb{T}$ &
Of the form \eqref{cubic-linear-solution} with $g_1$, $g_2$, $g_3$ satisfying \eqref{classT-2}, \eqref{classT-4}, \eqref{classT-6}, \eqref{classT-7}, \eqref{classT-8}, \eqref{classT-9} \\
\hline
Tetrahedral class $\mathbb{T} \oplus \mathbb{Z}_2^c$ &
Of the form \eqref{cubic-linear-solution} with $g_1$, $g_2$, $g_3$ satisfying \eqref{classT-2}, \eqref{classT-4}, \eqref{classT-6} \\
\hline
Tetrahedral class $\mathbb{O}^{-}$ &
Of the form \eqref{cubic-linear-solution} with $g_i$, $i=1,2,3$, satisfying \eqref{classO^--2}, \eqref{classO^--4}, \eqref{classO^--6}, \eqref{classO^--8}, \eqref{classO^--10}, \eqref{classO^--12} \\
\hline
\end{longtable}
\renewcommand{\arraystretch}{2.0}

\subsection{Cubic classes}

For all cubic strain-gradient classes, the elasticity tensor $\boldsymbol{\mathsf{C}}$ has three independent components and is of the same form as in the cubic classical linear elasticity. The arbitrariness of these three elastic constants yields the universality PDEs \eqref{cubic-linear-constraints}, and hence the candidate universal displacement field must have the form \eqref{cubic-linear-solution}. We now determine the additional universality PDEs induced by the fifth- and sixth-order tensors and their implications for the fields appearing in \eqref{cubic-linear-solution}\,.

\subsubsection{Cubic class $\mathbb{O}$} 

Tensor $\boldsymbol{\mathsf{A}}_{\mathbb{O} \oplus \mathbb{Z}_2^c}$ has $11$ independent components and is of the same form as that of class $\mathbb{O}^-$. Tensor $\boldsymbol{\mathsf{M}}_{\mathbb{O}}$ has $3$ independent components and can be written in the form
 \begin{equation}
\label{M_O-barD^1}
   \boldsymbol{\mathsf{M}}_{\mathbb{O}}=\begin{bmatrix}
    0 & 0 & 0 & \bar{D}^{(1)}  \\
    \bar{O}^{(2)} & 0 & 0 & 0   \\
    0 & -\bar{O}^{(2)}  & 0 & 0     \\
    0 & 0 & \bar{O}^{(2)} & 0   
    \end{bmatrix}\,,\qquad
   \bar{D}^{(1)} \in \begin{bmatrix}
   0 & \bar{d}_{12} & -\bar{d}_{12}  \\
   -\bar{d}_{12} & 0 & \bar{d}_{12}   \\
   \bar{d}_{12} & -\bar{d}_{12} & 0 
    \end{bmatrix}.
\end{equation}
Tensor $\boldsymbol{\mathsf{C}}_{\mathbb{O} \oplus \mathbb{Z}_2^c}$ has $3$ independent components and corresponds to the group with Hermann--Mauguin symbol $432$.
The additional universality PDEs for this class are
\begin{equation}
\label{cubic-O-constraints-1}
\begin{dcases}
 \frac{\partial^4 u_1}{\partial x_1^4}=\frac{\partial^4 u_1}{\partial x_2^2 \partial x_3^2}=0\,, \\
 \frac{\partial^4 u_1}{\partial x_1^2 \partial x_3^2}+\frac{\partial^4 u_1}{\partial x_1^2 \partial x_2^2}+\frac{\partial^4 u_3}{\partial x_1^3 \partial x_3}+\frac{\partial^4 u_2}{\partial x_1^3 \partial x_2}=0\,, \\
 \frac{\partial^4 u_3}{\partial x_1 \partial x_3^3}+\frac{\partial^4 u_2}{\partial x_1 \partial x_2^3}=0\,, \\
 \frac{\partial^4 u_1}{\partial x_3^4}+\frac{\partial^4 u_1}{\partial x_2^4}+2\left(\frac{\partial^4 u_1}{\partial x_1^2 \partial x_3^2}+\frac{\partial^4 u_1}{\partial x_1^2 \partial x_2^2}\right)=0\,, \\
 \frac{\partial^4 u_2}{\partial x_1 \partial x_2 \partial x_3^2}+ \frac{\partial^4 u_3}{\partial x_1 \partial x_2^2 \partial x_3}=0\,, \\
 3\frac{\partial^4 u_2}{\partial x_1 \partial x_2 \partial x_3^2}+2\frac{\partial^4 u_3}{\partial x_1 \partial x_2 \partial x_3^2}+\frac{\partial^4 u_2}{\partial x_1 \partial x_2^2 \partial x_3}=0\,,
\end{dcases}
\end{equation}
and
\begin{equation}
\label{cubic-O-constraints-2}
\begin{dcases}
 \frac{\partial^4 u_2}{\partial x_2^4}=\frac{\partial^4 u_2}{\partial x_1^2 \partial x_3^2}=0\,, \\
 \frac{\partial^4 u_2}{\partial x_2^2 \partial x_3^2}+\frac{\partial^4 u_3}{\partial x_2^2 \partial x_3^2}+\frac{\partial^4 u_1}{\partial x_1 \partial x_2^3}+\frac{\partial^4 u_2}{\partial x_1^2 \partial x_2^2}=0\,, \\
 \frac{\partial^4 u_3}{\partial x_2 \partial x_3^3}+\frac{\partial^4 u_1}{\partial x_1^3 \partial x_2}=0\,, \\
 \frac{\partial^4 u_3}{\partial x_3^4}+\frac{\partial^4 u_1}{\partial x_1^3 \partial x_2}=0\,, \\
 \frac{\partial^4 u_2}{\partial x_3^4}+ \frac{\partial^4 u_3}{\partial x_2 \partial x_3^3}+2\frac{\partial^4 u_2}{\partial x_2^2 \partial x_3^2}+2\frac{\partial^4 u_2}{\partial x_1^2 \partial x_2^2}+\frac{\partial^4 u_1}{\partial x_1^3 \partial x_2}+\frac{\partial^4 u_2}{\partial x_1^4}=0\,, \\
 \frac{\partial^4 u_1}{\partial x_1 \partial x_2 \partial x_3^2}+\frac{\partial^4 u_3}{\partial x_1^2 \partial x_3^2}=0\,, \\
 \frac{\partial^4 u_1}{\partial x_1 \partial x_2 \partial x_3^2}+\frac{\partial^4 u_3}{\partial x_1^2 \partial x_2 \partial x_3}=0\,, \\
 \frac{\partial^4 u_2}{\partial x_2^2 \partial x_3^2}+\frac{\partial^4 u_3}{\partial x_2^3 \partial x_3}+\frac{\partial^4 u_1}{\partial x_1 \partial x_2^3}+\frac{\partial^4 u_2}{\partial x_1^2 \partial x_2^2}=0\,, \\
 \frac{\partial^4 u_1}{\partial x_1 \partial x_2 \partial x_3^2}+\frac{\partial^4 u_3}{\partial x_1^2 \partial x_2 \partial x_3}=0\,,
\end{dcases}
\end{equation}
as well as
\begin{equation}
\label{cubic-O-constraints-3}
\begin{dcases}
 \frac{\partial^4 u_3}{\partial x_3^4}=\frac{\partial^4 u_3}{\partial x_1^2 \partial x_2^2}=0\,, \\
 \frac{\partial^4 u_2}{\partial x_1 \partial x_3^3}+\frac{\partial^4 u_3}{\partial x_2^2 \partial x_3^2}+\frac{\partial^4 u_1}{\partial x_1 \partial x_3^3}+\frac{\partial^4 u_3}{\partial x_1^2 \partial x_3^2}=0\,, \\
 \frac{\partial^4 u_2}{\partial x_2^3 \partial x_3}+\frac{\partial^4 u_1}{\partial x_1^3 \partial x_3}=0\,, \\
 2\frac{\partial^4 u_3}{\partial x_2 \partial x_3^3}+\frac{\partial^4 u_3}{\partial x_2^4}+2\frac{\partial^4 u_3}{\partial x_1^2 \partial x_3^2}+\frac{\partial^4 u_3}{\partial x_1^4}=0\,, \\
 \frac{\partial^4 u_1}{\partial x_1 \partial x_2^2 \partial x_3}+ \frac{\partial^4 u_2}{\partial x_1^2 \partial x_2 \partial x_3}=0\,.
\end{dcases}
\end{equation}
The above universality PDEs stem from the sixth-order tensor. From the fifth-order tensor we obtain
\begin{equation}
\label{cubic-O-constraints-4}
\begin{dcases}
 -2 \frac{\partial^3 u_3}{\partial x_2 \partial x_3^2}+2\frac{\partial^3 u_2}{\partial x_2^2 \partial x_3}-\frac{\partial^3 u_2}{\partial x_1^2 \partial x_3}+\frac{\partial^3 u_3}{\partial x_1^2 \partial x_2}=0\,, \\
 \frac{\partial^3 u_2}{\partial x_3^3}+\frac{\partial^3 u_3}{\partial x_3^3}+\frac{\partial^3 u_3}{\partial x_2 \partial x_3^2}+\frac{\partial^3 u_3}{\partial x_2^3}-2\frac{\partial^3 u_2}{\partial x_2^2 \partial x_3}=0\,, \\
 \frac{\partial^3 u_2}{\partial x_3^3}+\frac{\partial^3 u_3}{\partial x_2 \partial x_3^2}-\frac{\partial^3 u_2}{\partial x_2^2 \partial x_3}-\frac{\partial^3 u_3}{\partial x_2^3}-2\frac{\partial^3 u_2}{\partial x_1^2 \partial x_3}-\frac{\partial^3 u_3}{\partial x_1^2 \partial x_2}=0\,,
\end{dcases}
\end{equation}
and
\begin{equation}
\label{cubic-O-constraints-5}
\begin{dcases}
 \frac{\partial^3 u_1}{\partial x_2^2 \partial x_3}+2\frac{\partial^3 u_3}{\partial x_1 \partial x_3^2}-\frac{\partial^3 u_3}{\partial x_1 \partial x_2^2}-2\frac{\partial^3 u_1}{\partial x_1^2 \partial x_3}=0\,, \\
 \frac{\partial^3 u_1}{\partial x_3^3}+\frac{\partial^3 u_3}{\partial x_1^3}-2\frac{\partial^3 u_3}{\partial x_1 \partial x_3^2}+2 \frac{\partial^3 u_1}{\partial x_1^2 \partial x_3}=0\,, \\
 -\frac{\partial^3 u_1}{\partial x_3^3}+3\frac{\partial^3 u_1}{\partial x_1 \partial x_2^2}-\frac{\partial^3 u_3}{\partial x_1 \partial x_3^2}-3\frac{\partial^3 u_3}{\partial x_1 \partial x_2^2}+\frac{\partial^3 u_1}{\partial x_1^2 \partial x_3}+\frac{\partial^3 u_3}{\partial x_1^3}=0\,,
\end{dcases}
\end{equation}
as well as
\begin{equation}
\label{cubic-O-constraints-6}
\begin{dcases}
 -\frac{\partial^3 u_1}{\partial x_2 \partial x_3^2}+\frac{\partial^3 u_2}{\partial x_1 \partial x_3^2}-2\frac{\partial^3 u_2}{\partial x_1 \partial x_2^2}+2\frac{\partial^3 u_1}{\partial x_1^2 \partial x_2}=0\,, \\
 -\frac{\partial^3 u_1}{\partial x_2^3}+\frac{\partial^3 u_3}{\partial x_1 \partial x_2^2}-\frac{\partial^3 u_3}{\partial x_1 \partial x_2 \partial x_3}-\frac{\partial^3 u_2}{\partial x_1^3}-2 \frac{\partial^3 u_1}{\partial x_1^2 \partial x_2}+2\frac{\partial^3 u_1}{\partial x_1 \partial x_2^2}=0\,, \\
 -3\frac{\partial^3 u_1}{\partial x_1 \partial x_3^2}+\frac{\partial^3 u_1}{\partial x_2^3}+3\frac{\partial^3 u_2}{\partial x_1 \partial x_3^2}+\frac{\partial^3 u_2}{\partial x_1 \partial x_2^2}-\frac{\partial^3 u_1}{\partial x_1^2 \partial x_2}-\frac{\partial^3 u_2}{\partial x_1^3}=0\,.
\end{dcases}
\end{equation}
All relations obtained for class $\mathbb{O}\oplus \mathbb{Z}_2^c$ remain valid for the present class. In particular, the functions $g_i$ admit the representations
\begin{subequations}\label{classO-representations}
\begin{align}
\label{classO-1}
g_1(x_2,x_3)&=f_1(x_2)+x_3 f_2(x_2)+f_3(x_3)+x_2 f_4(x_3)\,,\\
\label{classO-2}
g_2(x_1,x_3)&=f_5(x_1)+x_3 f_6(x_1)+f_7(x_3)+x_1 f_8(x_3)\,,\\
\label{classO-3}
g_3(x_1,x_2)&=f_9(x_1)+x_2 f_{10}(x_1)+f_{11}(x_2)+x_1 f_{12}(x_2)\,.
\end{align}
\end{subequations}
In addition, the following universality PDEs must hold:
\begin{subequations}\label{classO-additionalPDEs}
\begin{align}
\label{classO-4}
-\frac{\partial^3 g_2(x_1,x_3)}{\partial x_1^2 \partial x_3}
+\frac{\partial^3 g_3(x_1,x_2)}{\partial x_1^2 \partial x_2}&=0\,,\\
\label{classO-5}
-\frac{\partial^3 g_2(x_1,x_3)}{\partial x_3^3}
+\frac{\partial^3 g_3(x_1,x_2)}{\partial x_2^3}&=0\,,\\
\label{classO-6}
-\frac{\partial^3 g_3(x_1,x_2)}{\partial x_1 \partial x_2^2}
+\frac{\partial^3 g_1(x_2,x_3)}{\partial x_2^2 \partial x_3}&=0\,,\\
\label{classO-7}
\frac{\partial^3 g_1(x_2,x_3)}{\partial x_3^3}
+\frac{\partial^3 g_3(x_1,x_2)}{\partial x_1^3}&=0\,,\\
\label{classO-8}
-\frac{\partial^3 g_1(x_2,x_3)}{\partial x_2 \partial x_3^2}
+\frac{\partial^3 g_2(x_1,x_3)}{\partial x_1 \partial x_3^2}&=0\,,\\
\label{classO-9}
\frac{\partial^3 g_1(x_2,x_3)}{\partial x_2^3}
+\frac{\partial^3 g_2(x_1,x_3)}{\partial x_1^3}&=0\,.
\end{align}
\end{subequations}
Substituting \eqref{classO-representations} into \eqref{classO-4} gives the coupling condition $-f_6''(x_1)+f_{10}''(x_1)=0$. Likewise, substitution into \eqref{classO-6} leads to $-f_{12}''(x_2)+f_2''(x_2)=0$, while substitution into \eqref{classO-8} implies $-f_4''(x_3)+f_8''(x_3)=0$.

The remaining relations \eqref{classO-5}, \eqref{classO-7}, and \eqref{classO-9} couple functions of different independent variables. A sufficient way to satisfy them is to require the corresponding third derivatives to vanish. In particular, \eqref{classO-5} is satisfied if
\begin{equation}\label{classO-10}
	f_7'''(x_3)=f_8'''(x_3)=f_{11}'''(x_2)=f_{12}'''(x_2)=0\,,
\end{equation}
\eqref{classO-7} is satisfied if
\begin{equation}\label{classO-11}
	f_3'''(x_3)=f_4'''(x_3)=f_9'''(x_1)=f_{10}'''(x_1)=0\,,
\end{equation}
and \eqref{classO-9} is satisfied if
\begin{equation}\label{classO-12}
	f_1'''(x_2)=f_2'''(x_2)=f_5'''(x_1)=f_6'''(x_1)=0\,.
\end{equation}
Consequently, all functions $f_i$, $i=1,2,\ldots,12$, are at most quadratic in their arguments, and the pairs $\left(f_{10},f_6\right)$, $\left(f_{12},f_2\right)$, and $\left(f_4,f_8\right)$ are coupled through $-f_6''+f_{10}''=0$, $-f_{12}''+f_2''=0$, and $-f_4''+f_8''=0$, respectively.

\begin{prop}
The universal displacements of linear strain-gradient elasticity for class $\mathbb{O}$ have the form \eqref{cubic-linear-solution}, and the functions $g_1$, $g_2$, and $g_3$ satisfy \eqref{classO-representations}, \eqref{classO-4}, \eqref{classO-6}, \eqref{classO-8}, \eqref{classO-10}, \eqref{classO-11}, and \eqref{classO-12}.
\end{prop}

\subsubsection{Cubic class $\mathbb{O} \oplus \mathbb{Z}_2^c$} 

Tensor $\boldsymbol{\mathsf{A}}_{\mathbb{O} \oplus \mathbb{Z}_2^c}$ has $11$ independent components and is of the same form as that of class $\mathbb{O}^-$. Tensor $\boldsymbol{\mathsf{M}}_{\mathbb{O} \oplus \mathbb{Z}_2^c}$ is a null tensor. Tensor $\boldsymbol{\mathsf{C}}_{\mathbb{O} \oplus \mathbb{Z}_2^c}$ has $3$ independent components and corresponds to the group with Hermann--Mauguin symbol $m\bar{3}m$.

The universality PDEs stemming from the fifth-order tensor coincide with those of the cubic class $\mathbb{O}$. Thus, for the function $g_1$ we have
\begin{equation} \label{classO+Z^c_2-2}
	\frac{\partial^4 g_1}{\partial x_2^4}+\frac{\partial^4 g_1}{\partial x_3^4}=0\,, \qquad
	\frac{\partial^4 g_1}{\partial x_2^2 \partial x_3^2}=0\,.
\end{equation}
The second universality PDE in \eqref{classO+Z^c_2-2} has the general solution
\begin{equation}
\label{classO+Z^c_2-3}
g_1(x_2,x_3)=f_1(x_2)+x_3 f_2(x_2)+f_3(x_3)+x_2 f_4(x_3)\,,
\end{equation}
and substituting \eqref{classO+Z^c_2-3} into the first universality PDE in \eqref{classO+Z^c_2-2} yields
\begin{equation}
\label{classO+Z^c_2-4}
f_1''''(x_2)+x_3 f_2''''(x_2)+f_3''''(x_3)+x_2 f_4''''(x_3)=0\,.
\end{equation}
In a similar fashion, for the function $g_2$ we obtain the universality PDEs
\begin{equation}\label{classO+Z^c_2-5}
	\frac{\partial^4 g_2}{\partial x_1^4}+\frac{\partial^4 g_2}{\partial x_3^4}=0\,, \qquad
	\frac{\partial^4 g_2}{\partial x_1^2 \partial x_3^2}=0\,.
\end{equation}
The second universality PDE in \eqref{classO+Z^c_2-5} has the general solution
\begin{equation}
\label{classO+Z^c_2-6}
g_2(x_1,x_3)=f_5(x_1)+x_3 f_6(x_1)+f_7(x_3)+x_1 f_8(x_3)\,,
\end{equation}
and substituting \eqref{classO+Z^c_2-6} into the first universality PDE in \eqref{classO+Z^c_2-5} yields
\begin{equation}
\label{classO+Z^c_2-7}
f_5''''(x_1)+x_3 f_6''''(x_1)+f_7''''(x_3)+x_1 f_8''''(x_3)=0\,.
\end{equation}
Similarly, for the function $g_3$ we obtain the universality PDEs
\begin{equation} \label{classO+Z^c_2-8}
	\frac{\partial^4 g_3}{\partial x_1^4}+\frac{\partial^4 g_3}{\partial x_2^4}=0\,, \qquad
	\frac{\partial^4 g_3}{\partial x_1^2 \partial x_2^2}=0\,.
\end{equation}
The second universality PDE in \eqref{classO+Z^c_2-8} has the general solution
\begin{equation}
\label{classO+Z^c_2-9}
g_3(x_1,x_2)=f_9(x_1)+x_2 f_{10}(x_1)+f_{11}(x_2)+x_1 f_{12}(x_2)\,,
\end{equation}
and substituting \eqref{classO+Z^c_2-9} into the first universality PDE in \eqref{classO+Z^c_2-8} yields
\begin{equation}
\label{classO+Z^c_2-10}
f_9''''(x_1)+x_2 f_{10}''''(x_1)+f_{11}''''(x_2)+x_1 f_{12}''''(x_2)=0\,.
\end{equation}
Therefore, we obtain the following result.
\begin{prop}
The universal displacement fields in the class $\mathbb{O} \oplus \mathbb{Z}_2^c$ linear strain-gradient elastic solids have the form \eqref{cubic-linear-solution}, and the functions $g_1$, $g_2$, and $g_3$ satisfy \eqref{classO+Z^c_2-3}, \eqref{classO+Z^c_2-4}, \eqref{classO+Z^c_2-6}, \eqref{classO+Z^c_2-7}, \eqref{classO+Z^c_2-9}, and \eqref{classO+Z^c_2-10}\,.
\end{prop}

\renewcommand{\arraystretch}{1.15}
\begin{longtable}{p{0.30\textwidth} p{0.66\textwidth}}
\caption{Summary of universal displacements: Cubic classes.}\label{Table-UD-summary-cubic}\\
\hline
\textbf{Symmetry class} & \textbf{Universal displacement family} \\
\hline
\endfirsthead
\hline
\textbf{Symmetry class} & \textbf{Universal displacement family} (continued) \\
\hline
\endhead
\hline
\endfoot
\hline
\endlastfoot
Cubic class $\mathbb{O}$ &
Of the form \eqref{cubic-linear-solution} with $g_1$, $g_2$, $g_3$ satisfying \eqref{classO-representations}, \eqref{classO-4}, \eqref{classO-6}, \eqref{classO-8}, \eqref{classO-10}, \eqref{classO-11}, \eqref{classO-12} \\ \hline
Cubic class $\mathbb{O} \oplus \mathbb{Z}_2^c$ &
Of the form \eqref{cubic-linear-solution} with $g_1$, $g_2$, $g_3$ satisfying \eqref{classO+Z^c_2-3}, \eqref{classO+Z^c_2-4}, \eqref{classO+Z^c_2-6}, \eqref{classO+Z^c_2-7}, \eqref{classO+Z^c_2-9}, \eqref{classO+Z^c_2-10} \\ \hline
\end{longtable}
\renewcommand{\arraystretch}{2.0}

\subsection{Icosahedral classes}

There are two icosahedral symmetry classes: $\mathbb{I}$ and $\mathbb{I}\oplus \mathbb{Z}_2^c$. For both classes, the elasticity tensor $\boldsymbol{\mathsf{C}}$ has two independent elastic moduli and has the same structure as in the $\infty$-gonal classes, namely the classical isotropic linear elastic tensor. Therefore, the candidate universal displacement fields are of the form \eqref{displacementisotropy}.
The class $\mathbb{I}$ has the universality PDEs of $\mathbb{SO}(3)$, and it admits one additional material parameter relative to $\mathbb{SO}(3)$, which induces three extra universality PDEs. Similarly, the class $\mathbb{I}\oplus \mathbb{Z}_2^c$ has the universality PDEs of $\mathbb{O}(3)$, and its extra material parameter relative to $\mathbb{O}(3)$ induces the corresponding additional universality PDEs. Substituting \eqref{displacementisotropy} into these additional relations yields the resulting conditions in terms of the functions $\alpha$, $\beta$, and $\gamma$\,.

\subsubsection{Icosahedral class $(\mathbb{I})$} 

Tensor $\boldsymbol{\mathsf{A}}_{\mathbb{I} \oplus \mathbb{Z}_2^c}$ has $6$ independent components and has the form
\begin{equation}
\label{A_I}
    \boldsymbol{\mathsf{A}}_{\mathbb{I} \oplus \mathbb{Z}_2^c}(\mathbf{x})=\begin{bmatrix}
    A^{(5)}+\eta A_I^{(c)} & 0 & 0 & 0  \\
     & P \bigl(A^{(5)}+\eta A_I^{(c)}\bigr) P^{\mathsf{T}} & 0 & 0   \\
     &  & A^{(5)}+\eta A_I^{(c)} &    \\
     &  &  & \eta J_c      
    \end{bmatrix}_S+
\begin{bmatrix}
    0 & 0 & 0 & 0  \\
     & 0 & 0 & 0   \\
     &  & 0 & 0   \\
     &  &  & f(A^{(5)})      
    \end{bmatrix}_S\,,
\end{equation}
where
\begin{equation}
\label{A^5}
    A^{(5)}=\begin{bmatrix}
    a_{11} & a_{12} & a_{13} & a_{12} & a_{13}  \\
      & a_{22} & -a_{13}+\sqrt{2} a_{III} & a_{12}-\sqrt{2} a^*_{IV} & a^*_{IV}   \\
      & & -a_{12}+a_{IV} & a^*_{IV} & a_{35}   \\
     &  &  & a_{22} & -a_{13}+\sqrt{2} a_{III} \\
     &  &  &  & -a_{12}+ a_{IV}      
    \end{bmatrix}\,,
\end{equation}
with
\begin{equation}
\label{A^5rest}
a_{III}=\frac{a_{11}-a_{22}}{2},\; a_{IV}=a_{35}-\sqrt{2} a_{13},\; a^*_{IV}=a_{13}-\sqrt{2} a_{35}\,,
\end{equation}
and
\begin{equation}
\label{A_I^c-J_c}
    A_I^{(c)}=\begin{bmatrix}
    4-\phi & 1 & 2 \sqrt{2} & 0 & \sqrt{2}  \\
      & -1 & 0 & 1-\phi & 0   \\
      & & 0 & 0 & 2-\phi   \\
     &  &  & 0 & \sqrt{2} \\
     &  &  &  & 2      
    \end{bmatrix}_S\,,\qquad
    J_{c}=\begin{bmatrix}
    -1 & \bar{\phi} & \bar{\phi}  \\
      & -1  & \bar{\phi}   \\
      & &   -1        
    \end{bmatrix}_S\,.
\end{equation}
$\bar{\phi}$ is the conjugate of the golden number, defined as $\bar{\phi}=(1-\sqrt{5})/2=1-\phi$. Moreover, $f(A^{(5)})$ is given by
\begin{equation}
\label{f(A^5)}
    f(A^{(5)})=\begin{bmatrix}
    a_V+\sqrt{2} a_{IV} & a_{III}-a_{IV}^* & a_{III}-a_{IV}^*  \\
      & a_{V}+ \sqrt{2}a_{IV}  & a_{III}-a_{IV}^*   \\
      & &   a_{V}+ \sqrt{2}a_{IV}        
    \end{bmatrix}_S,
\end{equation}
with $a_V=a_{22}-a_{12}$. 
Tensor $\boldsymbol{\mathsf{M}}_{\mathbb{SO}(3)}$ has $1$ independent component and it is of the form 
 \begin{equation}
\label{M_SO(3)}
   \boldsymbol{\mathsf{M}}_{\mathbb{SO}(3)}=\begin{bmatrix}
    0 & 0 & 0 & \bar{D}^{(1)}  \\
    0 & 0 & 0 & 0   \\
    0 & 0  & 0 & 0     \\
    0 & 0 & 0 & 0   
    \end{bmatrix} 
    +\begin{bmatrix}
    0 & 0 & 0 & 0  \\
    g(\bar{D}^{(1)}) & 0 & 0 & 0   \\
    0 & -g(\bar{D}^{(1)})  & 0 & 0     \\
    0 & 0 & g(\bar{D}^{(1)}) & 0   
    \end{bmatrix},
\end{equation}
where
 \begin{equation}
\label{g(barD^1)}
   g(\bar{D}^{(1)}) \in \begin{bmatrix}
    0 &  \bar{d}_{12} & -\frac{\sqrt{2}}{2} \bar{d}_{12} & -\bar{d}_{12} & \frac{\sqrt{2}}{2} \bar{d}_{12}
    \end{bmatrix}.
\end{equation}
Tensor $\boldsymbol{\mathsf{C}}_{\mathbb{SO}(3) \oplus \mathbb{Z}_2^c}$ corresponds to the Curie group with Hermann--Mauguin symbol $\infty\infty$ and has $2$ independent components.

For this class, the first two universality PDEs arising from the independence of the material constants coincide with those of the isotropic class of classical linear elasticity. The next six universality PDEs coincide with those of class $\mathbb{SO}(3)$. What remains are three additional universality PDEs that the displacement field must satisfy; these universality PDEs are written as
\begin{equation}
\label{classI-1}
\begin{aligned}
& -2 \sqrt{2}\frac{\partial^4 u_1}{\partial x_3^4}
+4\left(-1+\sqrt{5}\right)\frac{\partial^4 u_1}{\partial x_2^2 \partial x_3^2}
-10 \sqrt{2}\frac{\partial^4 u_3}{\partial x_1 \partial x_3^3}
+\left(11-7\sqrt{5}\right)\frac{\partial^4 u_2}{\partial x_1 \partial x_2 \partial x_3^2} \\
& \quad +\left(11-7\sqrt{5}\right)\frac{\partial^4 u_3}{\partial x_1 \partial x_2^2 \partial x_3}
+\left(2-4\sqrt{2}\right)\frac{\partial^4 u_2}{\partial x_1 \partial x_2^3}
-2\sqrt{2}\frac{\partial^4 u_1}{\partial x_1^2 \partial x_3^2}
-2\left(1+4\sqrt{2}\right)\frac{\partial^4 u_1}{\partial x_1^2 \partial x_2^2} \\
& \quad -2\sqrt{2}\frac{\partial^4 u_1}{\partial x_1^2 \partial x_3^2}
-\left(2+4\sqrt{2}\right)\frac{\partial^4 u_1}{\partial x_1^2 \partial x_2^2}
+\left(-7+\sqrt{5}\right)\frac{\partial^4 u_1}{\partial x_1^3 \partial x_3}=0\,,
\end{aligned}
\end{equation}
and
\begin{equation}
\label{classI-2}
\begin{aligned}
& -2 \frac{\partial^4 u_3}{\partial x_3^4}
-4 \sqrt{2} \frac{\partial^4 u_1}{\partial x_2 \partial x_3^3}
-\left(2+8 \sqrt{2}\right)\frac{\partial^4 u_2}{\partial x_2^2 \partial x_3^2}
-2 \frac{\partial^4 u_3}{\partial x_2^2 \partial x_3^2} \\
& \quad -4 \sqrt{2} \frac{\partial^4 u_3}{\partial x_2^3 \partial x_3}
+\left(-7+\sqrt{5}\right) \frac{\partial^4 u_1}{\partial x_2^4}
-\left(11-7\sqrt{5}\right)\frac{\partial^4 u_1}{\partial x_1 \partial x_2 \partial x_3^2}
-2\sqrt{2}\frac{\partial^4 u_1}{\partial x_1 \partial x_2^3} \\
& \quad +\left(-1+\sqrt{5}\right)\frac{\partial^4 u_2}{\partial x_1^2 \partial x_3^2}
-\left(-1+\sqrt{5}\right) \frac{\partial^4 u_3}{\partial x_1^2 \partial x_3^2}
+2\left(5-3\sqrt{5}\right)\frac{\partial^4 u_3}{\partial x_1^2 \partial x_2 \partial x_3} \\
& \quad +2\sqrt{2}\left(\frac{\partial^4 u_2}{\partial x_1^2 \partial x_2^2}
+5\frac{\partial^4 u_1}{\partial x_1^3 \partial x_2}
+\frac{\partial^4 u_2}{\partial x_1^4}\right)=0\,,
\end{aligned}
\end{equation}
and
\begin{equation}
\label{classI-3}
\begin{aligned}
& \left(-7+\sqrt{5}\right) \frac{\partial^4 u_3}{\partial x_3^4}
-2\sqrt{2}\frac{\partial^4 u_2}{\partial x_2 \partial x_3^3}
-2\sqrt{2}\frac{\partial^4 u_3}{\partial x_2^2 \partial x_3^2}
-10 \sqrt{2} \frac{\partial^4 u_2}{\partial x_2^3 \partial x_3} \\
& \quad -2\sqrt{2}\frac{\partial^4 u_3}{\partial x_2^4}
-\left(2+4\sqrt{2}\right)\frac{\partial^4 u_1}{\partial x_1 \partial x_3^3}
+\left(11-7\sqrt{5}\right)\frac{\partial^4 u_1}{\partial x_1 \partial x_2^2 \partial x_3}
-\left(2+8\sqrt{2}\right)\frac{\partial^4 u_3}{\partial x_1^2 \partial x_3^2} \\
& \quad +\left(11-7\sqrt{5}\right)\frac{\partial^4 u_2}{\partial x_1^2 \partial x_2 \partial x_3}
+4\left(-1+\sqrt{5}\right)\frac{\partial^4 u_3}{\partial x_1^2 \partial x_2^2}
-\left(-2+4\sqrt{2}\right)\frac{\partial^4 u_1}{\partial x_1^3 \partial x_3}=0\,.
\end{aligned}
\end{equation}
The additional constraints in terms of functions $\alpha$, $\beta$, and $\gamma$ for this case read
\begin{equation}
\label{icosahedral-1}
\begin{aligned}
& 2 \sqrt{2} \frac{\partial^5 \beta}{\partial x_3^5}
+2\sqrt{2} \frac{\partial^5 \alpha}{\partial x_2 \partial x_3^4}
-4\left(-1+\sqrt{5}\right)\frac{\partial^5 \beta}{\partial x_2^2 \partial x_3^3}
-4\left(-1+\sqrt{5}\right)\frac{\partial^5 \alpha}{\partial x_2^3 \partial x_3^2} \\
& \quad +\left(11-10\sqrt{2}-7\sqrt{5}\right) \frac{\partial^5 \gamma}{\partial x_1 \partial x_2 \partial x_3^3}
+\left(-13+4\sqrt{2}+7\sqrt{5}\right)\frac{\partial^5 \gamma}{\partial x_1 \partial x_2^3 \partial x_3}
-8\sqrt{2}\frac{\partial^5 \beta}{\partial x_1^2 \partial x_3^3} \\
& \quad +\left(-11+2\sqrt{2}+7\sqrt{5}\right)\frac{\partial^5 \alpha}{\partial x_1^2 \partial x_2 \partial x_3^2}
+\left(-9+8\sqrt{2}+7\sqrt{5}\right) \frac{\partial^5 \beta}{\partial x_3^5}
+4\left(1+\sqrt{2}\right) \frac{\partial^5 \alpha}{\partial x_1^2 \partial x_2^3} \\
& \quad +2\left(1+\sqrt{2}\right) \frac{\partial^5 \gamma}{\partial x_1^3 \partial x_2 \partial x_3}
-\left(-7+2\sqrt{2}+\sqrt{5}\right)\frac{\partial^5 \beta}{\partial x_1^4 \partial x_3}
-\left(-5+4\sqrt{2}+\sqrt{5}\right)\frac{\partial^5 \alpha}{\partial x_1^4 \partial x_2}=0\,,
\end{aligned}
\end{equation}
and
\begin{equation}
\label{icosahedral-2}
\begin{aligned}
& 2 \frac{\partial^5 \gamma}{\partial x_2 \partial x_3^4}
+ 2 \left(1+2\sqrt{2}\right)\frac{\partial^5 \gamma}{\partial x_2^3 \partial x_3^3}
-2\frac{\partial^5 \gamma}{\partial x_2^3 \partial x_3^2}
-\left(-7+4\sqrt{2}+\sqrt{5}\right) \frac{\partial^5 \gamma}{\partial x_2^4 \partial x_3}
+2 \frac{\partial^5 \beta}{\partial x_1 \partial x_3^4} \\
& \quad +\left(11-4\sqrt{2}-7\sqrt{5}\right) \frac{\partial^5 \beta}{\partial x_1 \partial x_2 \partial x_3^3}
+\left(9-8\sqrt{2}-7\sqrt{5}\right)\frac{\partial^5 \alpha}{\partial x_1 \partial x_2^2 \partial x_3^2}
-2 \frac{\partial^5 \beta}{\partial x_1 \partial x_2^2 \partial x_3^2}
-2\sqrt{2}\frac{\partial^5 \beta}{\partial x_1 \partial x_2^2 \partial x_3} \\
& \quad +\left(-7+2\sqrt{2}+\sqrt{5}\right) \frac{\partial^5 \alpha}{\partial x_1 \partial x_2^4}
-4\left(-1+\sqrt{5}\right)\frac{\partial^5 \gamma}{\partial x_1^2 \partial x_3^3}
+\left(-1+\sqrt{5}\right)\frac{\partial^5 \gamma}{\partial x_1^2 \partial x_2 \partial x_3^2} \\
& \quad +2\left(-5+\sqrt{2}+3\sqrt{5}\right)\frac{\partial^5 \gamma}{\partial x_1^2 \partial x_2^2 \partial x_3}
+4\left(-1+\sqrt{5}\right)\frac{\partial^5 \alpha}{\partial x_1^3 \partial x_3^2}
+\left(-1+\sqrt{5}\right) \frac{\partial^5 \beta}{\partial x_1^3 \partial x_3^2} \\
& \quad +2\left(-5+5\sqrt{2}+3\sqrt{5}\right)\frac{\partial^5 \beta}{\partial x_1^3 \partial x_2 \partial x_3}
+2\sqrt{2}\left(4\frac{\partial^5 \alpha}{\partial x_1^3 \partial x_2}
+\frac{\partial^5 \gamma}{\partial x_1^4 \partial x_3}
-\frac{\partial^5 \alpha}{\partial x_1^5}\right)=0\,,
\end{aligned}
\end{equation}
and
\begin{equation}
\label{icosahedral-3}
\begin{aligned}
&\left(-7+2\sqrt{2}+\sqrt{5}\right)\frac{\partial^5 \gamma}{\partial x_2 \partial x_3^4}
+8\sqrt{2}\frac{\partial^5 \gamma}{\partial x_2^3 \partial x_3^2}
-2\sqrt{2}\frac{\partial^5 \gamma}{\partial x_2^5}
+\left(-5+4\sqrt{2}+\sqrt{5}\right)\frac{\partial^5 \beta}{\partial x_1 \partial x_3^4}\\
&\quad +2\left(1+\sqrt{2}\right)\frac{\partial^5 \alpha}{\partial x_1 \partial x_2 \partial x_3^3}
+\left(11-2\sqrt{2}-7\sqrt{5}\right)\frac{\partial^5 \beta}{\partial x_1 \partial x_2^2 \partial x_3^2}
+\left(11-10\sqrt{2}-7\sqrt{5}\right)\frac{\partial^5 \alpha}{\partial x_1 \partial x_2^3 \partial x_3}\\
&\quad -2\sqrt{2}\frac{\partial^5 \beta}{\partial x_1 \partial x_2^4}
+\left(9-8\sqrt{2}-7\sqrt{5}\right)\frac{\partial^5 \gamma}{\partial x_1^2 \partial x_2 \partial x_3^2}
+4\left(-1+\sqrt{5}\right)\frac{\partial^5 \gamma}{\partial x_1^2 \partial x_2^3}
-4\left(1+\sqrt{2}\right)\frac{\partial^5 \beta}{\partial x_1^3 \partial x_3^2}\\
&\quad +\left(-13+4\sqrt{2}+7\sqrt{5}\right)\frac{\partial^5 \alpha}{\partial x_1^3 \partial x_2 \partial x_3}
+4\left(-1+\sqrt{5}\right)\frac{\partial^5 \beta}{\partial x_1^3 \partial x_2^2}=0\,.
\end{aligned}
\end{equation}
In summary we have the following result.
\begin{prop}
The sets of universal displacements of linear strain-gradient elasticity for class $\mathbb{I}$ coincide with those of class $\mathbb{SO}(3)$, and the fields $\alpha$, $\beta$, and $\gamma$ satisfy the additional universality PDEs \eqref{icosahedral-1}--\eqref{icosahedral-3}.
\end{prop}

\subsubsection{Icosahedral class $(\mathbb{I} \oplus \mathbb{Z}_2^c)$} 

Tensor $\boldsymbol{\mathsf{A}}_{\mathbb{I} \oplus \mathbb{Z}_2^c}$ has $6$ independent components and is of the same form as that of class $\mathbb{I}$. Tensor $\boldsymbol{\mathsf{M}}_{\mathbb{I} \oplus \mathbb{Z}_2^c}$ is a null tensor. Tensor $\boldsymbol{\mathsf{C}}_{\mathbb{O}(3)}$ has $2$ independent components and corresponds to the Curie group with Hermann--Mauguin symbol $\infty/m\,\infty/m$.

For this class, the first two universality PDEs obtained from the independence of the material constants coincide with those of the isotropic class of classical linear elasticity. The next six universality PDEs coincide with those of class $\mathbb{O}(3)$. What remains are three additional universality PDEs that the displacement field must satisfy. These are the same as those of class $\mathbb{I}$ given above, namely \eqref{classI-1}--\eqref{classI-3}.

In summary we have the following result.
\begin{prop}
The sets of universal displacements of linear strain-gradient elasticity for class $\mathbb{I} \oplus \mathbb{Z}_2^c$ coincide with those of class $\mathbb{O}(3)$, and the fields $\alpha$, $\beta$, and $\gamma$ satisfy the additional universality PDEs \eqref{icosahedral-1}--\eqref{icosahedral-3}.
\end{prop}

\renewcommand{\arraystretch}{1.15}
\begin{longtable}{p{0.30\textwidth} p{0.66\textwidth}}
\caption{Summary of universal displacements: Icosahedral classes.}\label{Table-UD-summary-icosahedral}\\
\hline
\textbf{Symmetry class} & \textbf{Universal displacement family} \\
\hline
\endfirsthead
\hline
\textbf{Symmetry class} & \textbf{Universal displacement family} (continued) \\
\hline
\endhead
\hline
\endfoot
\hline
\endlastfoot
Icosahedral class $\mathbb{I}$ &
Same as class $\mathbb{SO}(3)$ with $\alpha$, $\beta$, $\gamma$ satisfying \eqref{icosahedral-1}--\eqref{icosahedral-3}. \\ \hline
Icosahedral class $\mathbb{I}\oplus \mathbb{Z}^c_2$ &
Same as class $\mathbb{O}(3)$ with $\alpha$, $\beta$, $\gamma$ satisfying \eqref{icosahedral-1}--\eqref{icosahedral-3}. \\ \hline
\end{longtable}
\renewcommand{\arraystretch}{2.0}

\subsection{$\infty \infty$-gonal classes}

For the two $\infty\infty$-gonal classes, we combine the results of \cite{Yavari2020} for the isotropic linear elastic case with the compact formulation of \cite{Iesan2014} to determine the universal displacements.

\subsubsection{$\infty \infty$-gonal class $\mathbb{SO}(3)$} 

Tensor $\boldsymbol{\mathsf{A}}_{\mathbb{SO}(3) \oplus \mathbb{Z}_2^c}$ has $5$ independent components and has the form
\begin{equation}
\label{A_SO(3)}
    \boldsymbol{\mathsf{A}}_{\mathbb{SO}(3) \oplus \mathbb{Z}_2^c}(\mathbf{x})=\begin{bmatrix}
    A^{(5)} & 0 & 0 & 0  \\
     & A^{(5)} & 0 & 0   \\
     &  & A^{(5)} & 0   \\
     &  &  & 0      
    \end{bmatrix}_S+
\begin{bmatrix}
    0 & 0 & 0 & 0  \\
     & 0 & 0 & 0   \\
     &  & 0 & 0   \\
     &  &  & f(A^{(5)})      
    \end{bmatrix}_S.
\end{equation}
Tensor $\boldsymbol{\mathsf{M}}_{\mathbb{SO}(3)}$ has $1$ independent component and has the same form as that of class $\mathbb{I}$. Tensor $\boldsymbol{\mathsf{C}}_{\mathbb{SO}(3) \oplus \mathbb{Z}_2^c}$ corresponds to the Curie group with Hermann--Mauguin symbol $\infty\infty$ and has $2$ independent components.

For the class $\mathbb{SO}(3)$, the tensor $\boldsymbol{\mathsf{M}}$ admits one additional material parameter, which we denote by $f$. The governing equations for this case are given in \citep{Iesan2014}.
\begin{equation}\label{SO(3)eqs}
\begin{aligned}
	&\mu \Delta {\bf u}+(\lambda+\mu) \operatorname{grad} \operatorname{div} {\bf u}
	-2(\alpha_3+\alpha_4) \Delta \Delta {\bf u} 	\\
	&\quad -\left[ \frac{2(\lambda+2\mu)}{\lambda+\mu} 
	\sum_{i=1}^5 \alpha_i -2(\alpha_3+\alpha_4) \right] \Delta \operatorname{grad} \operatorname{div} {\bf u} 
	+2 f \Delta \operatorname{curl} {\bf u}= {\bf 0}.
\end{aligned}
\end{equation}
Using the same reasoning as for class $\mathbb{O}(3)$ (see below), we find that, in addition to \eqref{O(3)eqs}, the universal displacements must satisfy the universality PDE
\begin{equation}
\label{SO(3)constraint}
\Delta \operatorname{curl} \mathbf{u}=0\,.
\end{equation}
We now use the representation \eqref{displacementisotropy} and the setting of \citet{Yavari2020} to obtain
\begin{equation}
\label{dispalcementisotropy2}
	f=\Delta \alpha,\qquad g=\Delta \beta,\qquad h=\Delta \gamma\,,
\end{equation}
and substitute these expressions into \eqref{SO(3)constraint}. This yields the following three relations
\begin{equation} \label{sge-SO(3)eqs}
	-f_{13}+h_{33}+g_{12}+h_{22}=0\,, \qquad
	-f_{23}-g_{33}-g_{11}-h_{21}=0\,, \qquad
	f_{11}-h_{31}+f_{22}+g_{32}=0\,,
\end{equation}
which must be satisfied in addition to eqs.~(2.10) of \citet{Yavari2020}, namely
\begin{equation} \label{displacementisotropy3}
	f_2+g_3=0\,, \qquad
	f_1-h_3=0\,, \qquad
	g_1+h_2=0\,.
\end{equation}

If we differentiate \eqref{displacementisotropy3}$_1$ with respect to $x_2$, we obtain the last two terms in \eqref{sge-SO(3)eqs}$_3$. Differentiating \eqref{displacementisotropy3}$_2$ with respect to $x_1$ gives the first two terms in \eqref{sge-SO(3)eqs}$_3$. Hence, \eqref{sge-SO(3)eqs}$_3$ is satisfied identically once \eqref{displacementisotropy3} holds.
Similarly, differentiating \eqref{displacementisotropy3}$_3$ with respect to $x_1$ yields the last two terms in \eqref{sge-SO(3)eqs}$_2$, while differentiating \eqref{displacementisotropy3}$_1$ with respect to $x_3$ yields the first two terms in \eqref{sge-SO(3)eqs}$_2$. Therefore, \eqref{sge-SO(3)eqs}$_2$ is also satisfied identically once \eqref{displacementisotropy3} holds.
Finally, differentiating \eqref{displacementisotropy3}$_3$ with respect to $x_2$ gives the last two terms in \eqref{sge-SO(3)eqs}$_1$, and differentiating \eqref{displacementisotropy3}$_2$ with respect to $x_3$ gives the first two terms in \eqref{sge-SO(3)eqs}$_1$. Thus, \eqref{sge-SO(3)eqs}$_1$ is likewise satisfied identically once \eqref{displacementisotropy3} holds.

\begin{prop}
For the linear strain-gradient class $\mathbb{SO}(3)$, every universal displacement field can be written as the sum of a homogeneous displacement field and a non-homogeneous displacement field given by the divergence of an antisymmetric matrix whose components solve Poisson’s equation, as in \citep{Yavari2020}.
\end{prop}

\subsubsection{$\infty \infty$-gonal class $\mathbb{O}(3)$} 

Tensor $\boldsymbol{\mathsf{A}}_{\mathbb{O}(3) \oplus \mathbb{Z}_2^c}$ has $5$ independent components and is of the same form as that of class $\mathbb{SO}(3)$. Tensor $\boldsymbol{\mathsf{M}}_{\mathbb{O}(3)}$ is a null tensor. Tensor $\boldsymbol{\mathsf{C}}_{\mathbb{O}(3) \oplus \mathbb{Z}_2^c}$ corresponds to the Curie group with Hermann--Mauguin symbol $\infty/m\,\infty/m$ and has $2$ independent components.

There are five additional constants that stem from matrix ${\mathsf{A}}_{\mathbb{O(\text{3})}\oplus {\mathbb{Z}}_2^c}$ which, if denoted by $\alpha_i,\, i=1,\dots,5$, yield the field equations in the following compact form \citep{Iesan2014}:
\begin{equation}\label{O(3)eqs}
	\mu \Delta {\bf u}+\left(\lambda+\mu\right)\operatorname{grad}\circ\operatorname{div}{\bf u}
	-2\left(\alpha_3+\alpha_4\right)\Delta\Delta{\bf u}
	-\left[\frac{2\left(\lambda+2\mu\right)}{\left(\lambda+\mu\right)}
	\sum_{i=1}^5\alpha_i-2\left(\alpha_3+\alpha_4\right)\right]
	\Delta(\operatorname{grad}\circ\operatorname{div}{\bf u})={\bf 0}\,.
\end{equation}
In the first two terms one immediately recognizes the classical linear isotropic elasticity equations. Since we have $\Delta {\bf u}=\mathbf{0}$ and $\operatorname{grad}\circ\operatorname{div}{\bf u}=\mathbf{0}$ from the classical linear elastic part, \eqref{O(3)eqs} is satisfied trivially for arbitrary values of the elastic constants $\alpha_i$, $i=1,\dots,5$. Therefore, we have proved the following result.

\begin{prop}
For the linear strain-gradient class $\mathbb{O}(3)$, all universal displacements can be expressed as a superposition of a homogeneous displacement field and a non-homogeneous one given by the divergence of an antisymmetric matrix whose components solve Poisson’s equation. This Poisson problem is identical to that in \citep{Yavari2020} for isotropic classical linear elasticity.
\end{prop}

\renewcommand{\arraystretch}{1.15}
\begin{longtable}{p{0.30\textwidth} p{0.66\textwidth}}
\caption{Summary of universal displacements: $\infty\infty$-gonal classes.}\label{Table-UD-summary-infty-infty-gonal}\\
\hline
\textbf{Symmetry class} & \textbf{Universal displacement family} \\
\hline
\endfirsthead
\hline
\textbf{Symmetry class} & \textbf{Universal displacement family} (continued) \\
\hline
\endhead
\hline
\endfoot
\hline
\endlastfoot
$\infty\infty$-gonal class $\mathbb{SO}(3)$ &
Of the form \eqref{displacementisotropy} with Poisson problem as in \citep{Yavari2020} \\ \hline
$\infty\infty$-gonal class $\mathbb{O}(3)$ &
Of the form \eqref{displacementisotropy} with Poisson problem as in \citep{Yavari2020} \\ \hline
\end{longtable}
\renewcommand{\arraystretch}{2.0}

\section{Conclusions} \label{Sec:Conclusions}

In this paper we studied universal displacement fields for three-dimensional linear strain-gradient elasticity within the Toupin--Mindlin first strain-gradient theory. Our starting point was the analysis of \citet{Yavari2020} for classical linear elasticity, which we recalled in \S\ref{Sec:Classical-linear-elasticity} and then used as the point of departure for the strain-gradient analysis in \S\ref{Sec:Strain-gradient-linear-elasticity}.

For each material symmetry class, we required the strain-gradient equilibrium equations (in the absence of body forces) to hold for all choices of elastic constants. This universality requirement yields a system of differential constraints on the displacement field. By carrying this procedure through the full symmetry classification (including centrosymmetric and chiral classes) and using compact matrix representations for the elasticity tensors, we obtained explicit universality constraints and the corresponding sets of universal displacements class by class.

The strain-gradient equilibrium equations reduce to the classical linear elasticity equilibrium equations when the higher-order stress $\boldsymbol{\tau}$ is absent. In several high-symmetry cases, once the displacement field satisfies the universality constraints inherited from the classical part, the additional higher-order strain-gradient terms are automatically satisfied for arbitrary values of the remaining strain-gradient elastic constants. In particular, for the isotropic classes $\mathbb{SO}(3)$ and $\mathbb{O}(3)$, the universal displacement fields coincide with the classical universal displacement fields: they are precisely the superposition of a homogeneous displacement field and a non-homogeneous field given as the divergence of an antisymmetric tensor potential with components solving a Poisson’s equation.

For lower symmetry classes, the strain-gradient universality requirement can be stricter than its classical counterpart. In these cases, the admissible universal displacement fields form a proper subset of the classical universal displacements, obtained by imposing additional higher-order differential constraints that eliminate some of the classical families. We recorded these additional constraints explicitly in \S\ref{Sec:Strain-gradient-linear-elasticity} in the form of supplementary differential constraints that must be satisfied in addition to the corresponding classical universality constraints. These additional constraints arise from higher-order universality PDEs and reflect the strongly coupled structure of the strain-gradient equilibrium equations.

The main outcome of this paper is a complete, symmetry-classified description of universal displacement fields in three-dimensional linear strain-gradient elasticity, covering all $48$ strain-gradient symmetry classes. The results make transparent when strain-gradient effects do, and do not, further restrict the classical universal displacement families, and provide, for each symmetry class, a complete set of necessary and sufficient conditions for a displacement field to be universal in the strain-gradient sense.

\section*{Acknowledgments}

AY was supported by NSF Grant No.~CMMI~1939901. We are indebted to Prof.~Auffray for bringing to our attention minor typographical errors regarding some matrices in his papers. In particular, the matrices in \eqref{barD^5}$_2$ and \eqref{g(barH^2)} should replace the corresponding ones reported in \citep{Auffrayetal2019}, whereas the matrices in \eqref{D^4}, \eqref{f(F^8)}, and \eqref{f(D^4)-f(J^4)} should replace the corresponding ones reported in \citep{Auffrayetal2013}.

\bibliographystyle{abbrvnat}
\bibliography{ref}

\end{document}